\documentclass[twoside,openright,frontopenright]{ip3thesis}
\usepackage[utf8]{inputenc} % utf8 input good for non-ascii characters
\usepackage[T1]{fontenc} % Good for hyphenating non-ascii characters
\usepackage{lmodern}
\usepackage[british]{babel}
\usepackage[nottoc]{tocbibind} % Includes bibliography in table of contents, but not the table of contents itself
\usepackage[figuresright]{rotating} % Allows for sideways figures and tables
\usepackage[hidelinks,pdfusetitle]{hyperref} % hidelinks removes the boxes around links when viewed on screen, pdfusetitle option includes the thesis title and author in the PDF metadata
\usepackage[printonlyused,withpage]{acronym}
\usepackage{cite} % Combine multiple citations in range
\usepackage{multirow} 
\usepackage{graphicx}
\usepackage[shrink=10,stretch=10]{microtype}
\makeatletter
\@ifpackageloaded{microtype}{%
	\providecommand{\disableprotrusion}{\microtypesetup{protrusion=false}}
	\providecommand{\enableprotrusion}{\microtypesetup{protrusion=true}}
}{%
	\providecommand{\disableprotrusion}{}
	\providecommand{\enableprotrusion}{}
}
\makeatother

\usepackage{booktabs} % Better tables

\usepackage[margin=15mm,hang]{caption}
\usepackage{subcaption}
\usepackage{perpage} % Allows to reset counters on each page
\MakePerPage{footnote} % Resets footnote counters on each page

\graphicspath{{img/}}

\newenvironment{itemize*}%
{\begin{itemize}%
	\setlength{\itemsep}{0pt}%
	\setlength{\parskip}{0pt}}%
{\end{itemize}}
\newenvironment{enumerate*}%
{\begin{enumerate}%
	\setlength{\itemsep}{0pt}%
	\setlength{\parskip}{0pt}}%
{\end{enumerate}}

\usepackage{amsmath,amssymb}
\numberwithin{equation}{section}
\allowdisplaybreaks % Allow display equations to break over different pages
\usepackage{amssymb}
\usepackage{bm} % More bold maths symbols
\usepackage{bbm} % More blackboard bold characters, via \mathbbm. Mostly for blackboard bold 1
\usepackage{xfrac} % Nice small fractions
\usepackage{array} % Allows us to define custom column types for tables
\newcolumntype{L}{>{$}l<{$}} % a left aligned maths column type

\usepackage{cleveref} % Add automatic reference type text via \cref command

\usepackage[italic]{hepnames} % Add particle name macros (e.g. \PBs)
\usepackage{braket} % Adds Dirac bra-ket notation
\usepackage{slashed} % Adds Dirac slash notation
\usepackage{siunitx} % Add support for units
\DeclareSIUnit\fb{\femto\barn}
\newcommand{\azul}[1]{{\color{blue}#1}}

\newcommand{\nn}{\nonumber} % Define \nn to be equivalent to \nonumber
\usepackage{float}

\usepackage{longtable} % Required for longtable environment
\usepackage{xcolor}
\usepackage[frozencache,cachedir=_minted-main]{minted}
\newcounter{code}
\newcommand{\codecaption}[1]{%
    \refstepcounter{code}%
    Code Block~\thecode: #1%
}

\definecolor{bgcolor}{rgb}{0.95,0.95,0.95}

\usepackage{subfiles}

\begin{document}
\title{Phenomenology of scalar particles assisted by machine learning}
%\subtitle{An optional subtitle}
\author{Edwin Ali Herrera-Chacon}
\researchgroup{Institute for Particle Physics Phenomenology}
\maketitlepage*

	\begin{abstract}
In this thesis, we explore the phenomenology of scalar particles within Beyond Standard Model (BSM) frameworks, using Machine Learning techniques to enhance sensitivity and discovery potential at current and future collider experiments, the Large Hadron Collider (LHC) and the High-Luminosity LHC (HL-LHC). Specifically, we study scalar extensions of the Standard Model (SM) such as the Two Higgs Doublet Model Type-III (2HDM-III) and the Froggatt-Nielsen Flavon model.

We perform a detailed collider analysis focusing on charged Higgs boson pair production within the 2HDM-III, examining final states involving muons, neutrinos and quark jets. Our studies identify parameter regions consistent with recent experimental anomalies reported by the A Toroidal LHC Apparatus (ATLAS) collaboration, particularly in charged Higgs decays involving charm-bottom quark transitions, and suggest concrete scenarios for achieving statistically significant signals of 5$\sigma$ at future luminosities.

In the context of the Flavon model, we analyse potential signatures of a new scalar called Flavon decaying into a Higgs boson and a pair of bottom quarks, followed by the channels where the Higgs decays into a pair of bottom quarks or a pair of photons. Additionally, we analyse Lepton-Flavour-Violating (LFV) processes, both of them achieving discovery level significances of up to $5\sigma$ at the HL-LHC.

Using multivariate analysis techniques, specifically Boosted Decision Trees (BDTs), we demonstrate a significant improvement in signal discrimination. Throughout this thesis, Machine Learning methodologies have been integral, notably enhancing the signal from background separation and significantly improving the robustness of phenomenological predictions. The methods and analyses presented here contribute to clarifying the flavour structure mysteries of the SM and offer actionable targets for future experimental searches.
\end{abstract}

\disableprotrusion
\tableofcontents*
\listoffigures
\listoftables
\enableprotrusion
\chapter*{List of Abbreviations}
\addcontentsline{toc}{chapter}{List of Abbreviations}

\begin{acronym}[DGLAP]
\acro{2HDM-III}{Two-Higgs Doublet Model of type III}
\acro{ATLAS}{A Toroidal LHC Apparatus}
\acro{AUC}{Area Under a ROC Curve}
\acro{BDT}{Boosted Decision Tree}
\acroplural{BDT}[BDTs]{Boosted Decision Trees}
\acro{BR}{Branching Ratio}
\acro{BSM}{Beyond the Standard Model}
\acro{C}{Charge Conjugation}
\acro{CKM}{Cabibbo-Kobayashi-Maskawa}
\acro{CP}{Charge Conjugation Parity}
\acro{EW}{Electroweak}
\acro{EWSB}{Electroweak Symmetry Breaking}
\acro{FCNCs}{Flavour-changing Neutral Currents}
\acro{FNSM}{Froggatt-Nielsen Singlet Model}
\acro{FN}{Froggatt-Nielsen}
\acro{HEP}{High Energy Physics}
\acro{HL-LHC}{High-Luminosity LHC}
\acro{KS}{Kolmogorov-Smirnov}
\acro{LHC}{Large Hadron Collider}
\acro{LHCO}{LHC Olympics}
\acro{LFV}{Lepton-Flavour-Violating}
\acro{LFVp}{Lepton Flavour Violating Processes}
\acro{LO}{Leading Order}
\acro{MET}{Missing Transverse Energy}
\acro{PMNS}{Ponte-corvo-Maki-Nakagawa-Sakata}
\acro{QCD}{Quantum Chromodynamics}
\acro{ROC}{Receiver Operating Characteristic}
\acro{SM}{Standard Model}
\acro{SSB}{Spontaneous Symmetry Breaking}
\acro{VEV}{Vacuum Expectation Value}
\acro{XGBoost}{Extreme Gradient Boosting}

\end{acronym}

\begin{declaration*}
	The work in this thesis is based on research carried out in the Department of
	Physics at Durham University. No part of this thesis has been
	submitted elsewhere for any degree or qualification. Unless otherwise stated, all original work is my own in collaboration with my co-supervisor Marco Antonio Arroyo-Ure\~na.

Parts of this thesis are based on joint work, presented in the publications:

\begin{itemize}
  \item M. A. Arroyo‑Ureña, E. A. Herrera‑Chacón, S. Rosado‑Navarro and H. Salazar, “Hunting for a charged Higgs boson pair in proton–proton collisions,” \textit{Phys. Rev. D} \textbf{111} (2025) 015023, doi:\href{https://doi.org/10.1103/PhysRevD.111.015023}{10.1103/PhysRevD.111.015023}. Available at: \href{https://journals.aps.org/prd/abstract/10.1103/PhysRevD.111.015023}{APS Journal}.
  
    \item M. A. Arroyo‑Ureña, J. Lorenzo Díaz‑Cruz, E. A. Herrera‑Chacón, T. A. Valencia‑Pérez and J. Mejia Guisao, “Flavoring the production of Higgs boson pairs,” \textit{Phys. Rev. D} \textbf{111} (2025) 055028, doi:\href{https://doi.org/10.1103/PhysRevD.111.055028}{10.1103/PhysRevD.111.055028}. Available at: \href{https://journals.aps.org/prd/abstract/10.1103/PhysRevD.111.055028}{APS Journal}.

\end{itemize}
\end{declaration*}

\begin{acknowledgements*}
This thesis is a witness to the long journey I have taken to complete this stage of my life, which essentially began in 2020 when I applied to join Durham University. Along the way, I have been blessed to be associated with many people who, in different ways, have supported me throughout these years. I would like to say thank you to most of them.

To my supervisor Frank Krauss, for giving me the freedom to work on this project.

To my co-supervisor and friend Marco Arroyo, who has supported me for years with deep discussions about physics and who always inspires me with his passion for his work. Many thanks for believing in me!

To my colleagues Tomas Valencia and Sebastian Rosado, with whom I closely collaborated on the projects presented in this thesis, thank you for your excellent teamwork and enriching discussions.

To Christoph Englert and Michael Spannowsky, for their valuable comments and corrections.

To my mum Elizabeth, who is always there for me at any moment, even from a distance, and who taught me that hard work and integrity will always help us achieve our goals. To my dad Jose, for his constant support and for showing me that it is we who shape our destiny, not our circumstances. I love you both so much! 

To my \textit{bosses} from Bhaktivedanta Manor, Parama and Acarya, who always made sure everything was going well with my studies. To the devotees at Bhaktivedanta Institute and from the Pranama course for their association. To Bhakti Vijnana Goswami for his inspiration and instructions, and to A.C. Bhaktivedanta Swami Srila Prabhupada for all the facilities he provided to allow us to learn about all types of matter and their interactions.

To Adriana, who has supported me from the very beginning of my degree, shared her experiences inspiring me to study in the UK, and proofread this thesis at the end; to Mia, who has always been very supportive in both my personal and academic life, even sacrificing a lovely Friday going out to proofread this thesis; and to James, who probably can't imagine how useful it was for me when he asked about my work.

To Ryan, Magan, and Reza; their hospitality in their office, kitchen, and flat respectively, made my working hours productive and joyful. Thank you for sharing this time with me.

To David (Damodar Das), Nita, Ram, Samuel, Alexandre, Jorge, and Liam, who have enriched this journey with their words of support.

To Diana, Niels, Emma, and Anthony, with whom I had the privilege to work online during the pandemic at the beginning of my PhD.

To all the community at University College, especially the Cricket Club 2023-2025 and Rugby Football Club 2024-2025, from whom I learnt so much (I swear more than how to chop a drink!). To all the KCSoc members, in particular Riya, Ankit, Shatakshi, Alec, and Shamayita, who have always supported me in my efforts to share Krishna consciousness. All of you were definitely my spiritual support in this process. Thank you, Kirtida and Bhaktirasa, for your service and for supporting me and the KCSoc! Thank you also to Wendy, Ellen, Sukanya, David, Nicola, Janet, Stephanie, Jamie, and all the porters. You have all contributed to making me feel part of the Castle community.

To all my housemates, especially Yue, Joe, Lisa, Pier, Adam, Mathew, Jake, and Joseph, sharing this experience with you has been truly special.

To all my colleagues at IPPP: Sofie, Lois, Yannick, Peter, Tommy, Guillaume, Ansh, Jack, Yunji, Ery, Elliot, Thomas, Malina, Hitham, and Livia; and special thanks to Despoina for her kindness in proofreading my thesis. To Helen, Trudy and Joanne, for their constant help and support in administrative matters.

To Across the Pond, Monica in particular, for her valuable guidance during my application process to Durham University and her support with my visa paperwork.

Finally, I gratefully acknowledge financial support from SECIHTI (previously known as CONACYT) as well as from the Student Support Fund at Durham University. 

\raggedleft Hare Krishna!
\end{acknowledgements*}

\begin{epigraph*}

	The plurality that we perceive is only an appearance; it is not real. Vedantic philosophy, in which this is a fundamental dogma, has sought to clarify it by a number of analogies, one of the most attractive being the many-faceted crystal which, while showing hundreds of little pictures of what is in reality a single existent object, does not really multiply that object.
	\source{My View of the World}{Erwin Schrödinger}
\end{epigraph*}

%\dedicationtext{This thesis is dedicated to}
\begin{dedication*}
Sri Sri Radha Gokulananda\\ 
from \\
Bhaktivedanta Manor
%	\also[and]{my parents Elizabeth Chacon and Jose Herrera}
%
\end{dedication*}

\cleardoublepage

\chapter{Introduction}
\label{chap:SMIntroduction}
\noindent\rule{\linewidth}{0.4pt}
%==============================================================================
\section{The Standard Model}
%==============================================================================
The \ac{SM} unifies three of the four fundamental forces of nature: electromagnetic, weak, and strong, through a renormalisable quantum field theory \cite{Peskin}. This model successfully explains the behaviour of all known elementary particles, from quarks and leptons to gauge bosons like gluons ($g$), \(W^\pm\), \(Z\), and the photon (\(\gamma\)). At the core of this framework lies the Higgs mechanism, which introduces a single scalar Higgs doublet. Through \ac{SSB}, the Higgs field acquires a \ac{VEV}, providing the weak bosons and charged fermions with mass.

Despite the successes in explaining the most fundamental interactions of nature, the SM has certain limitations. It leaves several questions unanswered, hinting at the existence of physics beyond its scope. These unresolved issues include:

\begin{itemize}
\item \textbf{Neutrino Masses and Oscillations:} The SM predicts that neutrinos are massless. However, experimental evidence from neutrino oscillation experiments shows that they possess non-zero masses. This discrepancy requires extensions to the SM, such as the introduction of right-handed neutrinos or higher dimensional operators, \cite{PhysRevD.111.015005}.

\item \textbf{Matter-Antimatter Asymmetry:} The observed dominance of matter over antimatter in the universe cannot be fully explained with the SM. According to the Sakharov criteria, baryon asymmetry in the early universe can only emerge if three conditions are satisfied: (i) baryon number is not an exact symmetry; (ii) \ac{C} and \ac{CP} invariance are violated, which manifests as unequal partial probabilities for charge conjugate reactions; and (iii) the universe experiences a strong departure from thermal equilibrium during the superdense stage of its non-stationary expansion. \cite{Sakharov:1967dj}. These are conditions that the SM does not fulfil simultaneously. The amount of \ac{CP} violation in the SM through the \ac{CKM} matrix is therefore insufficient to account for the observed baryon asymmetry, and new sources of CP violation or alternative mechanisms are then required. \cite{10.1088/978-0-7503-3571-3ch2}

\item \textbf{Flavour Structure:} The SM does not explain the hierarchical pattern of fermion masses or the specific values of the mixing angles in the CKM and \ac{PMNS} matrices. The origin of the extensive hierarchy in Yukawa couplings remains as one of the major open questions in the SM. 
\item \textbf{Dark Matter and Unification:} The SM lacks a candidate for dark matter. This mysterious `substance' constitutes about 27\% of the universe's energy density \cite{Finkenberger_2022}. Additionally, the gauge couplings of the SM do not unify at high energies  \cite{Schwichtenberg_2019}.
\end{itemize}

These limitations indicate that the SM is likely an approximation of a more fundamental theory. Therefore, extending the SM with additional Higgs doublets or scalar fields like Flavons offers promising options for solving some of these issues. The presence of extra scalars can provide new sources of CP violation for the matter-antimatter asymmetry, and a Flavon framework naturally explains the significant hierarchies in fermion masses.

\section{Research Aims and Key Contributions}

In this work, we explore the phenomenology of scalar extensions of the SM, focusing on potential signatures of hypothetical scalar particles at the \ac{LHC} and its upgrade, the \ac{HL-LHC}. We analyse two \ac{BSM} frameworks: the \ac{FN} mechanism~\cite{Froggatt:1978nt} which introduces a new scalar particle known as the Flavon $(H_F)$, and the \ac{2HDM-III}, which extends the scalar sector by adding a heavy CP-even scalar $(H)$, a CP-odd scalar $(A)$, and a charged scalar pair $(H^\pm)$.
We investigate the production and possible detection of prospects of charged scalar pairs within the 2HDM-III framework. We identify viable regions in the model parameter space that accommodate current experimental observations, particularly the reported excess in the process $\mathcal{BR}(t\to H^{\pm}b)\times \mathcal{BR}(H^{\pm}\to cb)$ at $M_{H^{\pm}}=130$ GeV, as observed by the \ac{ATLAS} collaboration. 

Our analysis incorporates both theoretical and experimental constraints, proposing scenarios where charged scalar bosons with masses in the $100-350$ GeV range can be probed with significances up to $5\sigma$ at the HL-LHC.
Similarly, for the Flavon framework, we study the production and decay of scalar resonances, including processes where the SM-like Higgs boson $(h)$ decays into pairs of photons or $b$ quarks, as well as \ac{LFV} decays such as $(h \to e\mu)$. Using multivariate analysis techniques and Machine Learning tools, we demonstrate that, under specific parameter scenarios, these processes can also reach discovery level sensitivity at the HL-LHC, with predicted significances up to $5\sigma$.
Overall, the key contributions of this thesis include a comprehensive collider analysis of these scalar extensions, demonstrating their potential observability at current and future collider experiments. Throughout this work, we also highlight the relevance of Machine Learning methodologies in improving the signal from background separation to enhance the discovery prospects of BSM scalar particles.

\section{Structure of This Thesis}

In Chapter~\ref{chap:GaugeSym}, we review the theoretical framework of the SM, discussing its gauge structure, analysing the field content, the mechanisms behind \ac{EWSB}, and going deeper on the motivations to extend its scalar sector. Chapter~\ref{chap:Beyond Standard Models} introduces the scalar extensions of the SM considered in this thesis. We give a detailed description of the 2HDM-III and the FN Flavon model, focusing on their scalar sectors, Yukawa structures, and relevant experimental constraints. Following this, in Chapter~\ref{chap:methodology}, we present the multivariate analysis techniques that we used. We describe the setup of the event simulation, the \ac{LHCO} file format, and the implementation of \acp{BDT} to improve signal background discrimination. In Chapter~\ref{chap:collyder_analysis}, we perform detailed collider analyses for both models. For the 2HDM-III, we study the production of charged Higgs boson pairs and their decays into $\mu\nu_\mu\,cb$ final states, identifying parameter regions compatible with current ATLAS observations. For the Flavon model, we analyse the production of heavy CP‐even Flavon resonances that decay into pairs of SM-like Higgs bosons, which subsequently decay into $b\bar{b}$ or $\gamma\gamma$ final states; in addition, we study the LFV process, $h\to e\mu$. Finally, we conclude in Chapter~\ref{chap:conclusions_future_work}, summarising the key findings of this thesis and discussing possible future directions.

\chapter{Theoretical Framework}
\label{chap:GaugeSym}

\noindent\rule{\linewidth}{0.4pt}
%==============================================================================
This chapter provides a comprehensive review of the SM structure, from its gauge symmetries and field content to the mechanism of EWSB, setting the stage for exploring these extensions in greater detail further on.

\section{Gauge Symmetries and the Field Content}
%==============================================================================

\subsection{Local Gauge Group}

The SM is built upon the gauge group
\[
   \mathrm{SU}(3)_c \,\times\, \mathrm{SU}(2)_L \,\times\, \mathrm{U}(1)_Y,
\]
which governs the interactions of elementary particles. This product of Lie groups follows the Yang–Mills principle of local gauge invariance, originally formulated for \(\mathrm{SU}(2)\)~\cite{Yang:1954ek}. The associated Lie algebras fix the number and quantum numbers of the gauge bosons that mediate each interaction in the SM. Each component of this group plays a role as follows:
\begin{itemize}
\item \(\mathrm{SU}(3)_c\) describes the strong interactions, \ac{QCD}, mediated by eight massless gluons that couple to quarks, which carry colour charge.
\item \(\mathrm{SU}(2)_L\) governs the weak interactions, acting on left-handed fermion doublets through three gauge bosons: \(W^1\), \(W^2\), and \(W^3\).
\item \(\mathrm{U}(1)_Y\) corresponds to weak hypercharge Y, mediated by the gauge boson \(B_\mu\). After EWSB, \(B_\mu\) is mixed with \(W^3_\mu\) to form the \(\gamma\) and the \(Z\)-boson.\cite{Peskin}
\end{itemize} 

The fermions of the SM are organised into three generations, each consisting of:
\[
(u,\, d)_L,\quad (c,\,s)_L,\quad (t,\,b)_L \quad\text{for quarks,}\]
\[
(e,\,\nu_e)_L,\quad (\mu,\,\nu_\mu)_L,\quad (\tau,\,\nu_\tau)_L \quad\text{for leptons.}
\]
right-handed fermions, such as \(u_R\), \(d_R\), and \(e_R\), are singlets under \(\mathrm{SU}(2)_L\), they do not transform under that symmetry. The gauge principle holds that the physical laws must remain invariant under certain local transformations, gauge symmetries. Alternatively, in non-Abelian gauge groups like \(\mathrm{SU}(3)_c\) and \(\mathrm{SU}(2)_L\), the gauge bosons themselves carry an associated charge. Consequently, these bosons can self interact, except in an Abelian theory such as electromagnetism, where the \(Z\) gauge boson and the \(\gamma\) do not carry electric charge.

\subsection{Standard Model Lagrangian}
The full \ac{EW} Lagrangian can be decomposed into four distinct sectors \cite{jaime,papaqui,hunters}:
\begin{equation}
\label{eq:EWdecompose}
\mathcal{L}_{SU (2 )_L \times U (1 )_Y}
~=~
\mathcal{L}_{\mathrm{Gauge}}
~+~
\mathcal{L}_{\Phi}
~+~
\mathcal{L}_{F}
~+~
\mathcal{L}_{Yukawa}.
\end{equation}

\subsubsection{Gauge Sector \(\boldsymbol{\mathcal{L}_{\mathrm{Gauge}}}\)}
The gauge sector describes the gauge fields of \(\mathrm{SU}(2)_L\) and \(\mathrm{U}(1)_Y\) and their interactions, determined by field strength tensors that describe how these fields evolve across space and time and, for the non-Abelian cases, incorporate their self-interactions.

This sector is described by the Lagrangian:
\[
\mathcal{L}_{\mathrm{Gauge}}
~=~
-\,\frac{1}{4}\,F^\alpha_{\mu \nu}\,F^{\alpha\,\mu \nu}
~-~
\frac{1}{4}\,B_{\mu \nu}\,B^{\mu \nu},
\]
where the field strength tensors are defined as:
\[
B_{\mu \nu} 
~=~ 
\partial_\mu B_\nu 
~-~
\partial_\nu B_\mu,
\quad\]
\[F^\alpha_{\mu \nu}
~=~
\partial_\mu W^\alpha_\nu 
~-~
\partial_\nu W^\alpha_\mu
~-~
g_2\,\epsilon^{\alpha\beta\gamma}\,W^\beta_\mu\,W^\gamma_\nu.
\]
Here, \(g_2\) is the \(\mathrm{SU}(2)_L\) coupling constant and \(\epsilon^{\alpha\beta\gamma}\) is the Levi-Civita symbol in the adjoint indices.

\subsubsection{Fermion Sector \(\boldsymbol{\mathcal{L}_F}\)}
In the fermionic sector, the kinetic terms for quarks and leptons are written as left-handed doublets and right-handed singlets, respectively, along with their interactions through covariant derivatives that introduce the gauge fields in accordance with the gauge symmetries of the theory.

For each family \(m\), the left-handed quarks and lepton doublets are defined as:
\[
Q'_{Lm}
=
\begin{pmatrix}
u'_m \\ d'_m
\end{pmatrix}_L,\quad
L'_{Lm}
=
\begin{pmatrix}
\nu'_m\\ l'_m
\end{pmatrix}_L.
\]
The right-handed fields \(u'_R, d'_R, l'_R\) are singlets under \(\mathrm{SU}(2)_L\). 
Then the fermionic Lagrangian is given by:
\begin{align}
\mathcal{L}_{F} 
&= 
\sum_{m=1}^{3}
\Bigl(
\overline{Q'}_{Lm}\,i\gamma^\mu D_\mu\,Q'_{Lm}
~+~
\overline{L'}_{Lm}\,i\gamma^\mu D_\mu\,L'_{Lm}
\nonumber\\
&\qquad +~
\overline{u'}_{Lm}\,i\gamma^\mu D_\mu\,u'_{Lm}
~+~
\overline{d'}_{Lm}\,i\gamma^\mu D_\mu\,d'_{Lm}
~+~
\overline{l'}_{Lm}\,i\gamma^\mu D_\mu\,l'_{Lm}
\Bigr),
\end{align}
where \(D_\mu\) is the covariant derivative, which includes the gauge fields. We define the SU(2)$_L$ generators in the fundamental representation 
as $\tau^a \equiv \tfrac{1}{2}\sigma^a$. Hence, the covariant derivative acting on 
the left-handed quark doublet is \cite{Peskin}:
\[
D_\mu Q'_{Lm}
=
\Bigl(
\partial_\mu
~+~
i\,g_2\,\tau^a\,W^a_\mu
~+~
i\,g_1 \,Y \,B_\mu
\Bigr)\,
Q'_{Lm},
\]
In this expression, \( Y \) denotes the weak hypercharge of the corresponding fermion multiplet. For instance, for the left-handed quark doublet, we have \( Y = \frac{1}{6} \); for the right-handed up- and down-type singlets, \( Y = \frac{2}{3} \) and \( Y = -\frac{1}{3} \), respectively. Similarly, for the left-handed lepton doublet, \( Y = -\frac{1}{2} \), and for the right-handed charged lepton singlet, \( Y = -1 \) \cite{pich2012standardmodelelectroweakinteractions}.

\subsection*{Scalar Sector \texorpdfstring{$\boldsymbol{\mathcal{L}_{\Phi}}$}{Lphi}}
The scalar sector includes the Higgs doublet \(\Phi\) and its potential. This sector is the responsible for EWSB and the generation of particle masses.

This sector is described by the Lagrangian:
\[
\mathcal{L}_{\Phi}
~=~
(D_{\mu}\Phi)^\dagger(D^{\mu}\Phi)
~-~
V(\Phi),
\]
where \(\Phi\) is an \(\mathrm{SU}(2)_L\) doublet with hypercharge \(+\tfrac12\):
\[
\Phi
~=~
\begin{pmatrix}
\Phi^+\\
\Phi^0
\end{pmatrix}.
\]
The components of the Higgs doublet are fixed by gauge invariance and charge conservation. In the fundamental representation of \(\mathrm{SU}(2)_L\), the generator \(T_3 = \frac{1}{2} \sigma_3\) has eigenvalues \(+\tfrac{1}{2}\) and \(-\tfrac{1}{2}\) acting on the upper and lower entries, respectively. With hypercharge \(Y = +\tfrac{1}{2}\), the electric charge of each component is given by \(Q = T_3 + Y\) \cite{Peskin}, resulting in \(Q = +1\) for the upper field and \(Q = 0\) for the lower one. The Higgs potential is given by:
\begin{equation}
\label{eqn:potencial}
V(\Phi)
~=~
-\,\mu^2\,\bigl(\Phi^\dagger\Phi\bigr)
~+~
\lambda\,\bigl(\Phi^\dagger \Phi\bigr)^2.
\end{equation}
For a stable vacuum, the quartic coupling \(\lambda\) must be positive. When \(\mu^2 > 0\), the Higgs field acquires a non-zero VEV, 
\[
\langle \Phi\rangle
=
\frac{1}{\sqrt{2}}
\begin{pmatrix}
0\\
v
\end{pmatrix},
\quad
v
=\sqrt{\frac{\mu^2}{\lambda}}.
\]

The spontaneous breaking of \(\mathrm{SU}(2)_L \times \mathrm{U}(1)_Y\) to \(\mathrm{U}(1)_{\mathrm{em}}\) results in the emergence of three Goldstone bosons which become the longitudinal components of the \(W^\pm\) and \(Z\) bosons. The remaining degree of freedom corresponds to the physical Higgs boson \(h\), with a mass given by:
\[
m_h^2 = 2\,\mu^2.
\]

The masses of the gauge bosons are generated through their interactions with the Higgs field. From the covariant derivative:
\[
D_\mu \Phi
=
\left(
\partial_\mu
+ i\,g_2\,\tau^a W^a_\mu
+ i\,g_1\,Y\,B_\mu
\right)\Phi,
\]
the term \((D_\mu\Phi)^\dagger(D^\mu \Phi)\) produces the masses:
\[
m_W = \frac{1}{2}g_2\,v,\quad
m_Z = \frac{1}{2}\sqrt{g_2^2 + g_1^2}\,v,
\quad
m_\gamma=0
\]
In a more explicit notation:
\begin{align}
W^\pm_\mu 
&= \frac{1}{\sqrt{2}}(W^1_\mu \mp i\,W^2_\mu),
\\
Z_\mu &= \frac{1}{\sqrt{g_2^2 + g_1^2}}(g_2\,W^3_\mu - g_1\,B_\mu),
\qquad
\\
A_\mu &= \frac{1}{\sqrt{g_2^2 + g_1^2}}(g_1\,W^3_\mu + g_2\,B_\mu)
\qquad
\end{align}
Thus, the \(W^\pm\) and \(Z\) bosons acquire mass, while the \(\gamma\) remains massless.

\subsection*{Yukawa Sector \texorpdfstring{$\boldsymbol{\mathcal{L}_{Yukawa}}$}{Lyuk}}
This sector is introducing the interactions between the Higgs field and the fermions, giving rise to fermion masses after SB.

The Yukawa sector is described by the Lagrangian:
\begin{align}
-\mathcal{L}_{Yukawa}
&=
\sum_{m,n=1}^{3}
\Bigl[
Y^u_{mn}\,\overline{Q'}_{Lm}\,\tilde{\Phi}\,u'_{Rn}
+
Y^d_{mn}\,\overline{Q'}_{Lm}\,\Phi\,d'_{Rn}
+
Y^l_{mn}\,\overline{L'}_{Lm}\,\Phi\,l'_{Rn}
\Bigr]
+h.c.,
\\
&\quad
\text{where }\tilde{\Phi} = i\,\sigma_2\,\Phi^*,\quad
m,n\ \text{are flavour indices.}
\nonumber
\end{align}

Here, \(\tilde{\Phi}\) is needed to ensure electric charge conservation in the up-type Yukawa interaction. The quarks and charged leptons obtain mass matrices:
\[
M^u
=
\frac{v}{\sqrt{2}}\,Y^u,\quad
M^d
=
\frac{v}{\sqrt{2}}\,Y^d,\quad
M^l
=
\frac{v}{\sqrt{2}}\,Y^l.
\]
These matrices can be diagonalised through unitary transformations, producing the physical masses of the fermions. For example, the up-type quark mass matrix is diagonalised as:
\[
\overline{M}^u = \mathrm{diag}(M_u,\,M_c,\,M_t),
\]
with similar expressions for the down-type quarks and charged leptons. 

\subsection{Flavour Diagonalisation and CKM}
The diagonalisation of the mass matrices produces the CKM, which describes the mixing between quark flavours in weak interactions. This matrix arises from the difference between the transformations that diagonalise the up-type and down-type quark mass matrices. In the SM, the Higgs boson does not mediate \ac{FCNCs} at tree-level, as the Yukawa couplings become flavour diagonal in the mass basis.

The mass terms for the quarks after EWSB are given by \cite{Schwartz:2014sze},
\[
\mathcal{L}_{\text{mass}} = -\frac{v}{\sqrt{2}} \left( \overline{d}_L Y_d d_R + \overline{u}_L Y_u u_R \right) + \text{h.c.},
\]
where \(Y_d\) and \(Y_u\) are the Yukawa matrices for the down-type and up-type quarks. These matrices can be diagonalised using the unitary matrices \(U_d\), \(U_u\), \(K_d\), and \(K_u\), 
\[
Y_d = U_d M_d K_d^\dagger, \quad Y_u = U_u M_u K_u^\dagger,
\]
where \(M_d\) and \(M_u\) are diagonal mass matrices. Then the CKM matrix \(V\) is,
\[
V = U_u^\dagger U_d,
\]
The CKM matrix can be parametrised by three mixing angles \(\theta_{12}\), \(\theta_{23}\), \(\theta_{13}\) and a complex phase \(\delta\) \cite{PhysRevLett.53.1802}:
\[
V = \begin{pmatrix}
c_{12}c_{13} & s_{12}c_{13} & s_{13}e^{-i\delta} \\
-s_{12}c_{23} - c_{12}s_{23}s_{13}e^{i\delta} & c_{12}c_{23} - s_{12}s_{23}s_{13}e^{i\delta} & s_{23}c_{13} \\
s_{12}s_{23} - c_{12}c_{23}s_{13}e^{i\delta} & -c_{12}s_{23} - s_{12}c_{23}s_{13}e^{i\delta} & c_{23}c_{13}
\end{pmatrix},
\]
where \(c_{ij} = \cos\theta_{ij}\) and \(s_{ij} = \sin\theta_{ij}\).

At this point it can be seen that there is a relatively weak flavour mixing as the CKM is nearly diagonal. The Wolfenstein parametrisation \cite{PhysRevLett.51.1945}, provides a useful approximation, 
\[
|V| \approx \begin{pmatrix}
1 - \frac{\lambda^2}{2} & \lambda & \lambda^3 \\
-\lambda & 1 - \frac{\lambda^2}{2} & \lambda^2 \\
\lambda^3 & \lambda^2 & 1
\end{pmatrix} + \mathcal{O}(\lambda^4),
\]
where \(\lambda = \sin\theta_{12} \approx 0.22\).

The unitarity of the CKM can be tested experimentally, and any deviation from it could indicate BSM physics.

\section{Motivations for Extending the Scalar Sector}
\label{sec:MotivationsBSM}

Extending the scalar sector of the SM can address several of its limitations:
\begin{itemize}
\item \textbf{Extra CP Violation:} Additional Higgs doublets can introduce new CP-violating phases \cite{PhysRevD.8.1226}, which are crucial to explain the matter-antimatter asymmetry. 
\item \textbf{Stronger Phase Transition:} A more complex Higgs sector can lead to a strongly first-order EW phase transition, enabling baryogenesis.
\item \textbf{Flavour Symmetries:} Introducing additional fields or mechanisms can explain the hierarchical pattern of fermion masses through SSB.
\item \textbf{Seesaw Mechanisms and Dark Matter:} Additional scalars can couple to heavy neutrinos or dark matter sectors, providing mechanisms for neutrino masses and dark matter candidates. \cite{PhysRevD.73.077301}
\end{itemize}

%==============================================================================
\section{Overview of BSM Models}

%==============================================================================
 The 2HDM extends the SM by adding a second scalar doublet \(\Phi_2\), with the same hypercharge as \(\Phi_1\). A theoretical motivation for introducing a second Higgs doublet is the possibility of spontaneous CP violation. Additionally, the model predicts a richer scalar spectrum, including charged scalars and additional neutral states, which give rise to new decay channels and distinctive collider signatures \cite{Gunion:1989we,PhysRevD.8.1226,Branco:2011iw}. However, the inclusion of additional scalar fields may also affect electroweak precision observables. In particular, the scalar potential should be constructed with care to avoid significant deviations from custodial symmetry, which ensures the tree-level relation \(\rho = 1\), with \(\rho\) defined as the ratio \(M_W^2 / (M_Z^2 \cos^2 \theta_W)\). This symmetry, often referred to as custodial isospin, protects the relation between the \(W\) and \(Z\) boson masses in the SM. Such deviations are tightly constrained by experimental data \cite{10.1093/oso/9780198503996.001.0001}.

 After the SSB, one obtains up to five physical Higgs-like states: \(h,\, H\) (CP-even), \(A\) (CP-odd), and \(H^\pm\) (charged scalars). These additional scalars can be lighter or heavier than the observed \(125\,\mathrm{GeV}\) Higgs, and can produce exotic decays, new loop effects in flavour observables, or new CP-violating phases. To avoid large tree-level FCNC, discrete symmetries or specific Yukawa textures are often imposed (Type I, II, X, Y, or the so-called Type-III with four-zero textures) \cite{HernandezSanchez:2012eg}. Possible solutions to SM deficiencies include the following: additional scalar fields can lead to a strong first order EW phase transition, which is an essential factor in some baryogenesis scenarios \cite{Ghosh_Mukherjee_Mukherjee_2025}; with carefully chosen textures, it is possible to incorporate realistic quark/lepton masses while controlling FCNC at tree-level, and certain anomalies in heavy flavour data might also be addressed \cite{Carrillo-Monteverde_Ramírez_Gómez-Ávila_López-Lozano_2025}; finally, the presence of \(H^\pm\) offers direct collider signatures, e.g.\ \(t \to H^+ b\), \(H^+ \to \tau^+ \nu,\, cb\), tested at the LHC \cite{Collaboration_2023, Kim_Lee_Sanyal_Song_Wang_2024}.

%==============================================================================
SM fermion masses extend over five orders of magnitude. The Yukawa couplings do not provide a reason for this vast range. A Flavon model suggests a Spontaneously Broken Flavour Symmetry, e.g.\ \(\mathrm{U}(1)_F\), under which, each fermion \(f\) is assigned an integer charge \(Q_f\). The Yukawa terms appear as higher dimensional operators suppressed by \(\bigl\langle \Phi_F \bigr\rangle / \Lambda\)~\cite{Babu2023}. One introduces a gauge singlet \(\Phi_F\), which is the so called Flavon, with a negative \(\mathrm{U}(1)_F\) charge. The effective operators for up type quarks look like
\[
\Bigl(\frac{\Phi_F}{\Lambda}\Bigr)^{Q_{Q_L} + Q_{u_R}}
\overline{Q}_L \tilde{\Phi}\,u_R,
\]
where \(\Lambda\) is some high flavour scale. A small ratio \(\epsilon = \frac{\langle \Phi_F\rangle}{\Lambda} < 1\) can then produce the observed hierarchy in \(m_u, m_c, m_t\). The remaining physical Flavon field can be mixed with the Higgs or induce exotic decays. If \(\langle \Phi_F\rangle\) is near the TeV domain, such effects might be accessible at collider experiments~\cite{Das_2023}. 

The phenomenological impact includes hierarchical fermion masses, as by powers of \(\epsilon\), large mass hierarchies arise naturally from integer charges. It also includes LFV, since the Flavon can couple off-diagonally to leptons, enabling processes like \(\tau\to \mu\gamma\) or \(h\to \tau\mu\) if there is a mixing with the SM Higgs~\cite{Li_Yue_2019}. Additionally, rare \emph{B}-decays might be impacted, as quark flavour transitions can receive contributions at loop level. This possibility can test the parameter space via meson observables~\cite{Datta_2024}.

We have established the essential features of the SM, the gauge group
\(\mathrm{SU}(3)_c\times \mathrm{SU}(2)_L\times \mathrm{U}(1)_Y\), the single Higgs doublet responsible for EWSB, and the Yukawa interactions that provide
fermions with mass. While robustly confirmed by experiments, including the discovery of the
\(\sim125\,\mathrm{GeV}\) Higgs boson \cite{Aad:2012tfa, cmsconference}, the SM fails to address several questions.

To confront these issues, many models extend the scalar sector. In this work, we use two distinct BSM frameworks: the 2HDM-III and a Flavon model via the FN mechanism, which we will analyse in the next chapter.

\chapter{Beyond Standard Models}
\label{chap:Beyond Standard Models}
\noindent\rule{\linewidth}{0.4pt}
% Explore various Higgs decay channels including ee, eµ, and muon-neutrino-cb.
\section{\label{SecII}Two-Higgs Doublet Model of Type III}

In this section, we present the theoretical framework, which conducts a thorough analysis of the Yukawa Lagrangian within the context of the 2HDM-III and subsequently derives the necessary Feynman rules for our calculations. Through this in-depth exploration, we construct a robust framework in order to facilitate our ensuing analysis of the charged Higgs boson production and decay at the LHC and HL-LHC.

\subsection{Scalar Potential}

The 2HDM-III incorporates an extra Higgs doublet in addition to what is in the SM. The two scalar doublets can be written as $\Phi_a^T=( \phi_{a}^{+},\phi_{a}^{0})$ for $a=1, 2$, and they have an associated hypercharge with a value of $+1$. During the SSB, the Higgs doublets acquire non-zero VEV, which can be expressed as
	
	\begin{center}
		$\braket{\Phi_a}=\frac{1}{\sqrt{2}}\left(\begin{array}{c}
			0\\
			\upsilon_a
		\end{array}\right)$.
	\end{center}

	The introduction of additional Higgs doublets engenders new interactions between these doublets and all different types of fermions ~\cite{Weinberg:1996kr}. Nonetheless, such interactions may lead to the occurrence of FCNCs at the tree-level. Given that empirical observations impose rigorous restrictions on FCNCs, mechanisms must be implemented to suppress their presence within the framework of the model ~\cite{papaqui}. An effective methodology for attaining this suppression involves the implementation of a specific four-zero texture in the Yukawa sector. This is a specific arrangement of zeros within the Yukawa matrices, which is crucial to define the mass and mixing angles of the fermions. This texture acts as a simplified theoretical construct that delineates the flavour interactions between fermions and the Higgs bosons. By employing a four-zero texture, we can effectively modulate the intensity of FCNCs while ensuring alignment with the experimental data. ~\cite{Hernandez-Sanchez:2020vax, Arroyo-Urena:2020mgg, Arroyo-Urena:2019qhl, Arroyo-Urena:2013cyf}
	
	The most general scalar potential that is invariant under the symmetry group $SU(2)_L \times U(1)_Y$ can be expressed as follows \cite{Gunion:2002zf}:  

 \begin{eqnarray}
V(\Phi_1,\Phi_2) &=& \mu_{1}^{2}(\Phi_{1}^{\dag}\Phi_{1}^{}) + \mu_{2}^{2}(\Phi_{2}^{\dag}\Phi_{2}^{}) - \mu_{12}^{2}(\Phi_{1}^{\dag}\Phi_{2}^{} + h.c.) \nonumber \\ 
&+& \frac{1}{2} \lambda_{1}(\Phi_{1}^{\dag}\Phi_{1}^{})^2 + \frac{1}{2} \lambda_{2}(\Phi_{2}^{\dag}\Phi_{2}^{})^2 \nonumber \\ 
&+& \lambda_{3}(\Phi_{1}^{\dag}\Phi_{1}^{})(\Phi_{2}^{\dag}\Phi_{2}^{}) + \lambda_{4}(\Phi_{1}^{\dag}\Phi_{2}^{})(\Phi_{2}^{\dag}\Phi_{1}^{}) \nonumber \\
&+& \frac{1}{2} \lambda_{5}(\Phi_{1}^{\dag}\Phi_{2}^{})^2  + \lambda_{6}(\Phi_{1}^{\dag}\Phi_{1}^{})(\Phi_{1}^{\dag}\Phi_{2}^{}) \nonumber \\
&+& \lambda_{7}(\Phi_{2}^{\dag}\Phi_{2}^{})(\Phi_{1}^{\dag}\Phi_{2}^{}) + h.c.
\label{potential2}
\end{eqnarray}

In order to prevent tree-level FCNC, a discrete symmetry is commonly imposed, under which the scalar fields transform as \(\Phi_1 \rightarrow \Phi_1\) and \(\Phi_2 \rightarrow -\Phi_2\)~\cite{PhysRevLett.37.657}. Such a symmetry forbids the \(\lambda_6\) and \(\lambda_7\) terms in the scalar potential, which break this discrete invariance. When this symmetry is exact, the potential is reduced by eliminating these terms. Furthermore, this discrete invariance ensures that only one Higgs doublet couples to each type of right handed fermion, avoiding flavour violation in the Yukawa sector. This symmetry also contributes to CP conservation, as long as all parameters remain real. As a result, the scalar potential simplifies to a CP conserving and flavour safe form, consistent with phenomenological constraints.

	In this study, we take all parameters in the scalar potential to be real, including the VEV's of the Higgs doublets. Under these conditions, the CP symmetry is conserved in this sector because there is no source of complex phases within the scalar potential itself. However, this does not imply global CP conservation across the model, as the CKM matrix in the fermion sector still introduces CP violation.

	Furthermore, the model includes several additional independent parameters, among which the mixing angle $\alpha$, emerges from diagonalising the CP-even scalar mass matrix and governs the transition from gauge eigenstates to mass eigenstates:
	
	\begin{eqnarray}
		H&=&\rm Re(\phi_1^0)\cos\alpha+\rm Re(\phi_2^0)\sin\alpha,\nonumber\\
		h&=&\rm Re(\phi_1^0)\sin\alpha+\rm Re (\phi_2^0)\cos\alpha,
		\label{eq:Hh_definition}
	\end{eqnarray}
	with 
	\begin{equation*}
		\tan2\alpha=\frac{2m_{12}}{m_{11}-m_{22}},
	\end{equation*}
	where $m_{11,\,12,\,22}$ correspond to the entries of the real component of the mass matrix $\mathbf{M}$ determining how the fields acquire their masses,
	\begin{center}
		$\rm Re(\mathbf{M})=\left(\begin{array}{cc}
			m_{11} & m_{12}\\
			m_{12} & m_{22}
		\end{array}\right)$,
		\par\end{center}
	where
	\begin{eqnarray}
		m_{11}&=&m_A^2 \sin^2\beta+v^2(\lambda_1\cos^2\beta+\lambda_{5}\sin^2\beta)\nonumber
		\\
		m_{12}&=&-m_A^2 \cos\beta\sin\beta+v^2(\lambda_{3}+\lambda_{4})\cos\beta\sin\beta\nonumber 
		\\
		m_{22}&=&m_A^2\cos^2\beta+v^2(\lambda_{2}\sin^2\beta+\lambda_{5}\cos^2\beta).
        \label{eq:mxx}
	\end{eqnarray}
	Eqs. \eqref{eq:Hh_definition} define the distinct scalar states in the model, which are one physical charged scalar boson, a pseudo Goldstone boson that is absorbed by the $W$ gauge fields, a separate CP-odd state, and another pseudo Goldstone boson corresponding to the $Z$ gauge boson. As detailed further in this section, \(h\) typically behaves as the SM-like Higgs boson, whereas \(H\) emerges as a heavier scalar state. In Eqs. \eqref{eq:mxx}, $m_A$ represents the mass of the CP-odd scalar, $A$.

	In addition to $\alpha$, there is a second mixing angle, $\beta$, which governs how the charged components of $\Phi_a$ are combined with the imaginary parts of the neutral components being crucial to distinguish the physical charged states from the Goldstone modes absorbed by the gauge bosons determining the final spectrum of charged scalar fields.
    
	\begin{eqnarray}
		G_W^{\pm}&=&\phi_1^{\pm}\cos\beta+\phi_2^{\pm}\sin\beta,\nonumber\\
		H^{\pm}&=&-\phi_1^{\pm}\sin\beta+\phi_2^{\pm}\cos\beta,\nonumber\\	
		G_Z&=&\rm Im(\phi_1^0)\cos\beta+\rm Im(\phi_2^0)\sin\beta,\nonumber\\
		A^0&=&-\rm Im(\phi_1^0)\sin\beta+\rm Im(\phi_2^0)\cos\beta.
	\end{eqnarray}
	
	The second mixing angle $\beta$ is conventionally defined through the VEV ratio, such than $\tan\beta=\upsilon_2/\upsilon_1$, playing a key role in determining how the Higgs doublets contribute to the masses of the fermions and gauge bosons.

	In this theoretical framework, the primary parameters of interest include the two mixing angles $\alpha$ and $\beta$, as well as the physical masses $M_{H^{\pm}}$, $M_h$, $M_H$ and $M_A$. The expressions for these masses in terms of the underlying parameters are given by
	
	\begin{eqnarray}
		M_{H^{\pm}}^2&=&\frac{\mu_{12}}{\sin\beta\cos\beta}\\&-&\frac{1}{2}v^2\Bigg(\lambda_4+\lambda_{5}+\cot\beta\lambda_{6}+\tan\beta\lambda_7\Bigg), \nonumber
		\\
		M_{h,H}^2&=&\frac{1}{2}\Bigg(m_{11}+m_{22}\mp\sqrt{(m_{11}-m_{22})^2+4m_{12}^2}\Bigg),  
		\\
		M_A^2&=&M_H^{\pm}+\frac{1}{2}v^2(\lambda_4-\lambda_{5}).
	\end{eqnarray}

\subsection{Yukawa Lagrangian}
	The Yukawa sector encapsulates the interactions between the two Higgs doublets and the fermion fields. It is described by the following Lagrangian\cite{HernandezSanchez:2012eg}:
	\begin{align}
\label{YukawaLagrangian} 
\mathcal{L}_Y = & -\left( Y_{1}^{u} \bar{Q}_{L} \tilde{\Phi}_{1} u_{R} + Y_{2}^{u} \bar{Q}_{L} \tilde{\Phi}_{2} u_{R} \right. \nonumber \\
& \left. + Y_{1}^{d} \bar{Q}_{L} \Phi_{1} d_{R} + Y_{2}^{d} \bar{Q}_{L} \Phi_{2} d_{R} \right. \nonumber \\
& \left. + Y_{1}^{l} \bar{L}_{L} \tilde{\Phi}_{1} l_{R} + Y_{2}^{l} \bar{L}_{L} \tilde{\Phi}_{2} l_{R} \right),
\end{align}

	here,  $\tilde{\Phi}_{a}$ with $(a=1, 2)$ is defined by $\tilde{\Phi}_{a} = i\sigma_2 \Phi_{a}^{*}$.  After that the EWSB takes place, the fermion mass matrices have the following form:
	
	\begin{equation}
		M_f = \frac{1}{\sqrt{2}} \left( v_1 Y_{1}^{f} + v_2 Y_{2}^{f} \right),\; f=u,d,\ell.	
	\end{equation}
	
	In this step, we take both Yukawa matrices to exhibit the previously described four-zero texture structure and we also assume that they are Hermitian to reduce the parameter space and contribute to maintaining consistency with the assumption of CP conservation.
	
	Following the diagonalisation process, we obtain
	\begin{eqnarray}
		\bar{M}_f = V_{fL}^{\dag} M_f V_{fR}&=& \tfrac{1}{\sqrt{2}} \left( v_1 \tilde{Y}_{1}^{f} + v_2 \tilde{Y}_{2}^{f} \right),\\
		\tilde{Y}_{a}^{f} &=& V_{fL}^{\dag} Y_{a}^{f} V_{fR}
	\end{eqnarray}
	from these relations, we can deduce the following expressions,
	\begin{equation}\label{RotateYukawas}
		\left[ \tilde{Y}_a^f \right]_{ij} = \frac{\sqrt{2}}{v_a}\delta_{ij}\bar{M}_{ij}^f-\frac{v_b}{v_a}\left[ \tilde{Y}_b^f \right]_{ij}
	\end{equation}
	where $f$ refers to massive fermions and the parameters $\left[\tilde{\chi}_a^f \right]_{ij}$ represent unknown dimensionless quantities in the model. It is important to mention that from Eq. \ref{RotateYukawas}, various types of interactions may be obtained. By selecting specific structures \cite{HernandezSanchez:2012eg}, the following models can be defined as:
	\begin{itemize}
		\item 2HDM-III I-like
		\begin{eqnarray}\label{I}
			\left[ \tilde{Y}_2^d \right]_{ij} &=& \frac{\sqrt{2}}{v\sin\beta}\delta_{ij}\bar{M}_{ij}^d-\cot\beta\left[ \tilde{Y}_1^d \right]_{ij}\nonumber\\
			\left[ \tilde{Y}_2^u \right]_{ij} &=& \frac{\sqrt{2}}{v\sin\beta}\delta_{ij}\bar{M}_{ij}^u-\cot\beta\left[ \tilde{Y}_1^u \right]_{ij}\nonumber\\
			\left[ \tilde{Y}_2^\ell \right]_{ij} &=& \left[ \tilde{Y}_1^d \right]_{ij}(d\to\ell).
		\end{eqnarray}  
		\item 2HDM-III II-like
		\begin{eqnarray}\label{II}
			\left[ \tilde{Y}_1^d \right]_{ij} &=& \frac{\sqrt{2}}{v\cos\beta}\delta_{ij}\bar{M}_{ij}^d-\tan\beta\left[ \tilde{Y}_2^d \right]_{ij}\nonumber\\
			\left[ \tilde{Y}_2^u \right]_{ij} &=& \frac{\sqrt{2}}{v\sin\beta}\delta_{ij}\bar{M}_{ij}^u-\cot\beta\left[ \tilde{Y}_1^u \right]_{ij}\nonumber\\
			\left[ \tilde{Y}_1^\ell \right]_{ij} &=& \left[ \tilde{Y}_1^d \right]_{ij}(d\to\ell).
		\end{eqnarray}  
		\item 2HDM-III Lepton Specific-like
		\begin{eqnarray}\label{LS}
			\left[ \tilde{Y}_2^d \right]_{ij} &=& \frac{\sqrt{2}}{v\sin\beta}\delta_{ij}\bar{M}_{ij}^d-\cot\beta\left[ \tilde{Y}_1^d \right]_{ij}\nonumber\\
			\left[ \tilde{Y}_2^u \right]_{ij} &=& \frac{\sqrt{2}}{v\sin\beta}\delta_{ij}\bar{M}_{ij}^u-\cot\beta\left[ \tilde{Y}_1^u \right]_{ij}\nonumber\\
			\left[ \tilde{Y}_1^\ell \right]_{ij} &=& \left[ \tilde{Y}_1^d \right]_{ij}(d\to\ell).
		\end{eqnarray}  
		\item 2HDM-III Flipped-like
		\begin{eqnarray}\label{F}
			\left[ \tilde{Y}_1^d \right]_{ij} &=& \frac{\sqrt{2}}{v\cos\beta}\delta_{ij}\bar{M}_{ij}^d-\tan\beta\left[ \tilde{Y}_2^d \right]_{ij}\nonumber\\
			\left[ \tilde{Y}_2^u \right]_{ij} &=& \frac{\sqrt{2}}{v\sin\beta}\delta_{ij}\bar{M}_{ij}^u-\cot\beta\left[ \tilde{Y}_1^u \right]_{ij}\nonumber\\
			\left[ \tilde{Y}_2^\ell \right]_{ij} &=& \left[ \tilde{Y}_2^d \right]_{ij}(d\to\ell).
		\end{eqnarray}  
		
	\end{itemize}
Where $v$ is defined as $v = \sqrt{v_1^2 + v_2^2}$, representing the combined VEVs of the two Higgs doublets. These models correspond to the following types of Higgs doublet couplings \cite{hunters, PhysRevD.94.055003}:
\begin{itemize}
    \item \textbf{Type I}: A single Higgs doublet provides mass to both up-type and down-type quarks.
    \item \textbf{Type II}: The neutral component of one Higgs doublet couples to up-type quarks, while the neutral component of the other doublet couples to down-type quarks.
    \item \textbf{Type III}: Both Higgs doublets can generate masses for up-type and down-type quarks simultaneously.
    \item \textbf{Type X (Lepton-Specific)}: The quark couplings are as in Type I, while the lepton couplings are as in Type II.
    \item \textbf{Type Y (Flipped)}: The quark couplings are as in Type II, while the lepton couplings are as in Type I.
\end{itemize}

\smallskip
\begin{table}[tbp]
\centering
{\small
\centering
\resizebox{13.5cm}{!} {
\begin{tabular}{|c|c|c|c|}
\hline
\centering
Type & Couples to up-type quarks & Couples to down-type quarks & Couples to charged leptons \tabularnewline
\hline
\hline
I & \(\phi_{2}\) & \(\phi_{2}\) & \(\phi_{2}\) \tabularnewline
\hline
II & \(\phi_{2}\) & \(\phi_{1}\) & \(\phi_{1}\) \tabularnewline
\hline
III & \(\phi_{1}\), \(\phi_{2}\) & \(\phi_{1}\), \(\phi_{2}\) & \(\phi_{1}\), \(\phi_{2}\) \tabularnewline
\hline
X & \(\phi_{2}\) & \(\phi_{2}\) & \(\phi_{1}\) \tabularnewline
\hline
Y & \(\phi_{2}\) & \(\phi_{1}\) & \(\phi_{2}\) \tabularnewline
\hline
\end{tabular}
}}
\caption{The fact that the up-type quark couples to the doublet \(\phi_{2}\) is by convention. Type III exhibits a flavour violation.
\label{tbt}}
\end{table}

	From Eqs. \eqref{YukawaLagrangian}-\eqref{RotateYukawas}, we obtain 
	\begin{equation}
		\mathcal{L}_Y^\phi=\phi\bar{f}_i(S_{ij}^{\phi}+iP_{ij}^\phi\gamma^5)f_j,
         \label{YukawaEffective}
	\end{equation}
	where $\phi=h,\,H,\,A$. The minus sign in Yukawa terms is absorbed into $S_{ij}^{\phi}$ and $P_{ij}^{\phi}$ in Eq.\eqref{YukawaEffective}, simplifying the notation.
	Those CP-conserving and CP-violating factors $S_{ij}^{\phi}$ and $P_{ij}^\phi$, respectively, incorporate flavour dynamics and are written as:
	\begin{eqnarray}
		S_{ij}^{\phi}&=&\frac{gm_f}{2M_W}c_f^\phi\delta_{ij}+d^\phi_f\left[ \tilde{Y}_a^f \right]_{ij},\nonumber\\
		P_{ij}^\phi&=&\frac{gm_f}{2M_W}e_f^\phi\delta_{ij}+g^\phi_f\left[ \tilde{Y}_a^f \right]_{ij}.\label{SijPij}
	\end{eqnarray}
	The coefficients $c_f^\phi,\,d_f^\phi,\,e_f^\phi,\,g_f^\phi$ depend on the presence of new physics in the Higgs sector. Within the SM $c_f^{\phi=h}=1$ and $d_f^{\phi=h}=e_f^{\phi=h}=g_f^{\phi=h}=0$, in contrast, within the 2HDM-III these coefficients, in the CP conserving case, are as shown in Table \ref{couplings}. In Eqs. \eqref{SijPij} we can choose $a$ for the Yukawa matrices according to \eqref{I}-\eqref{F}. Although we adopt a 2HDM-II like scenario within the broader 2HDM-III framework, for simplicity, we will refer to the setup simply as the 2HDM-III.  
	\begin{center}
		\begin{table}
			\begin{tabular}{|c|c|c|c|c|c|c|}
				\hline 
				Coefficient & $c_{f}^{h}$ & $c_{f}^{A}$ & $c_{f}^{H}$ & $d_{f}^{h}$ & $d_{f}^{A}$ & $d_{f}^{H}$\tabularnewline
				\hline 
				\hline 
				$d$-type & $-\frac{\sin\alpha}{\cos\beta}$ & $-\tan\beta$ & $\frac{\cos\alpha}{\sin\beta}$ & $\frac{\cos(\alpha-\beta)}{\cos\beta}$ & $\csc\beta$ & $\frac{\sin(\alpha-\beta)}{\cos\beta}$\tabularnewline
				\hline 
				$u$-type & $\frac{\cos\alpha}{\sin\beta}$ & $-\cot\beta$ & $\frac{\sin\alpha}{\sin\beta}$ & $-\frac{\cos(\alpha-\beta)}{\sin\beta}$ & $\sec\beta$ & $\frac{\sin(\alpha-\beta)}{\sin\beta}$\tabularnewline
				\hline 
				leptons $\ell$ & $-\frac{\sin\alpha}{\cos\beta}$ & $-\tan\beta$ & $\frac{\cos\alpha}{\sin\beta}$ & $\frac{\cos(\alpha-\beta)}{\cos\beta}$ & $\csc\beta$ & $\frac{\sin(\alpha-\beta)}{\cos\beta}$\tabularnewline
				\hline 
			\end{tabular}
			\caption{The coefficients for the $\phi$-Fermion couplings in the 2HDM-III with a $\mathcal{CP}$-conserving Higgs potential.}	\label{couplings}
			
		\end{table}
	\end{center}
\vspace{-8ex}
	
	The couplings of the charged scalar boson with quarks are given by

\begin{align}\label{YukChargedQuarks}
\mathcal{L}_Y^{H^{\pm}q_i q_j} &= \frac{\sqrt{2}}{v}\Bigg\{
\left[
\bar{d}_i\Bigl(m_u\cot\beta - \frac{v}{\sqrt{2}}\,g(\beta)
\left[\tilde{Y}_2^u\right]_{ij}\Bigr) u_j H^{-} V_{\rm CKM}^{ij*} \right. \nonumber\\[1mm]
&\quad \left. -\, \bar{u}_i\Bigl(m_d\tan\beta - \frac{v}{\sqrt{2}}\,f(\beta)
\left[\tilde{Y}_1^d\right]_{ij}\Bigr) d_j H^{+} V_{\rm CKM}^{ij}\right] P_R \nonumber\\[2mm]
&\quad -\left[
\bar{d}_i\Bigl(m_d\tan\beta - \frac{v}{\sqrt{2}}\,f(\beta)
\left[\tilde{Y}_1^d\right]_{ij}\Bigr) u_j H^{-} V_{\rm CKM}^{ij*} \right. \nonumber\\[1mm]
&\quad \left. +\, \bar{u}_i\Bigl(m_u\cot\beta - \frac{v}{\sqrt{2}}\,g(\beta)
\left[\tilde{Y}_2^u\right]_{ij}\Bigr) d_j H^{+} V_{\rm CKM}^{ij}\right] P_L
\Bigg\}.
\end{align}

where
	\begin{eqnarray}		\Big[\tilde{Y}_a^f\Big]_{ij}&=&\frac{\sqrt{m_{f_i}m_{f_j}}}{v}\chi_{ij}, (a=1,2),\\
		f(\beta)&=&\sqrt{1+\tan^2\beta},\\
		g(\beta)&=&\sqrt{1+\cot^2\beta},
	\end{eqnarray}
	while those couplings of the charged scalar boson with leptons are given by
\begin{equation}
\mathcal{L}_Y^{{H^{\pm}}\ell \nu_{{\ell}}} = \frac{\sqrt{2}m_{\ell_i}}{v}\bar{\nu}_{L}\Bigg(\tan\beta\frac{m_{\ell_i}}{m_{\ell_j}}\delta_{ij}-\frac{f(\beta)}{\sqrt{2}}\sqrt{\frac{m_{\ell_i}}{m_{\ell_j}}}\chi_{ij}^{\ell}\Bigg)\ell^-_{R}H^+  
+ H.c.
\end{equation}

	%%%%%%%%%%%%%%%%%%%%%%%%%%%%%%%%%%%%%%%%%%%%%%%%%%%
	\clearpage
	%%%%%%%%%%%%%%%%%%%
	\section{Constraints on the 2HDM-III Parameter Space}
	\label{SecIII}
	
	In order to obtain realistic predictions, we carry out a comprehensive analysis of multiple experimental constraints, which can be grouped into two main categories:
	\begin{itemize}
		\item Process induced for the neutral scalar (pseudo scalar) bosons $h,\,H,\,A$:
		\begin{itemize}
			\item LHC Higgs boson data \cite{CMS:2017con, ATLAS:2019pmk}.
			\item Neutral meson physics $B_{s}^0\to\mu^+\mu^-$ \cite{CMS:2022mgd}, $B_{d}^0\to\mu^+\mu^-$ \cite{CMS:2022mgd}.
			\item Lepton flavour violating decays $\ell_i\to \ell_j\ell_k\bar{\ell}_k$ and radiative decays $\ell_i\to \ell_j\gamma$ \cite{Workman:2022ynf}.
			\item Muon anomalous magnetic moment $a_\mu$\footnote{$a_\mu$ also receives contributions from the charged scalar boson, however, its contribution is subdominant.} \cite{Muong-2:2021ojo}.
			\item Double Higgs production cross-section $\sigma(pp\to H \to hh\to b\bar{b}\gamma\gamma)$ \cite{ATLAS:2021ifb}.
			\item Single scalar production $\sigma(gb\to \phi\to \tau\tau)$ ($\phi=H,\,A$) \cite{ATLAS-CONF, Sirunyan:2018zut}. 
		\end{itemize}
		\item Process induced for the charged scalar boson $H^{\pm}$:
		\begin{itemize}
			\item Limits on $\sigma(pp\to tbH^{\pm})\times \rm BR(H^{\pm}\to\tau^{\pm}\nu)$ \cite{ATLAS:2018gfm}.
			\item Limits on $\mathcal{BR}(t\to H^{\pm}b)\times \mathcal{BR}(H^{\pm}\to cb)$ \cite{ATLAS:2023bzb}.
			\item Radiative $b$ quark decay $b\to s\gamma$ \cite{Ciuchini:1997xe, Misiak:2017bgg}.
		\end{itemize} 
	\end{itemize}
	
	The free model parameters that directly influence our predictions can be summarised as follows:
	\begin{enumerate}
		\item The cosine of the difference of mixing angles: $\cos(\alpha-\beta)$.
		\item The ratio of the VEV's: $\tan\beta$.
		\item The masses of the additional bosons: $M_H$, $M_A$, $M_{H^{\pm}}$.
		\item The matrix elements $\chi_{cb}$, $\chi_{tb}$, $\chi_{\mu\mu}$, $\chi_{tt}$, $\chi_{bb}$. 
	\end{enumerate}
	There are several parameters $\chi_{ij}$ needed in the calculation of $\mathcal{BR}(H^+\to bc)$ and $\mathcal{BR}(t\to H^+b)$. Unless otherwise specified, all of these parameters are taken to be $\chi_{ij}=1$. Here the indices $i$ and $j$ denote fermions, and in general $i\neq j$. 

We also consider basic theoretical constraints, such as perturbativity, vacuum stability, tree-level unitarity~\cite{GOODSELL2019206}, and the oblique parameters (see Section~\ref{sec:oblique} for further discussion). Although these constraints are typically derived for $\mathbb{Z}_2$-symmetric models with $\lambda_6 = \lambda_7 = 0$, we adopt them in the Type-III case. Nevertheless, they do not impose significant restrictions on the explored parameter space, compared to the experimental ones.

\subsection{Constraint on \texorpdfstring{$\tan\beta$}{tan beta} and \texorpdfstring{$\cos(\alpha-\beta)$}{cos(alpha-beta)}}

	\subsection*{LHC Higgs boson data: Signal strength modifiers}\label{HiggsConstraints}
	For a decay $S\to X$ (where $X$ denotes a specific final state), or a production process $\sigma(pp\to S)$, the signal strength is defined by the following parameterisation \cite{Aad2016}
	\begin{equation}\label{muX}
		\mathcal{\mu}_{X}=\frac{\sigma(pp\to h)\cdot\mathcal{BR}(h\to X)}{\sigma(pp\to h^{\text{SM}})\cdot\mathcal{BR}(h^{\text{SM}}\to X)},
	\end{equation}
	here, $\sigma(pp\to S)$ denotes the production cross-section of $S$, where $S=h,\,h^{\text{SM}}$; in this notation, $h$ represents the SM-like Higgs boson coming from an extension of the SM, whereas $h^{\text{SM}}$ is the SM Higgs boson. Furthermore,  $\mathcal{BR}(S\to X)$ is the \ac{BR}, i.e., the probability of the decay $S\to X$, with $X=b\bar{b},\;\tau^-\tau^+,\;\mu^-\mu^+,\;WW^*,\;ZZ^*,\;\gamma\gamma$ \cite{ATLAS-CONF-2021-053,CMS-PAS-HIG-19-005}.  
	%%%%%%%%%%%%%%%%%%%%%%%%%%%%%%%%%%%%%%%%%%%%%%%%%%%%%%%%%%%%%%%%
	\subsection*{Neutral meson physics}\label{NeutralMesonsConstraints}
	%%%%%%%%%%%%%%%%%%%%%%%%%%%%%%%%%%%%%%%%%%%%%%%%%%%%%%%%%%%%%%%%
	To complement the constraints provided by LHC Higgs boson data, we also examine LFV processes. These are mediated by the neutral scalar bosons $H$ and $h$, a pseudoscalar $A$, (In principle, each of these can induce FCNC at tree-level), and a charged scalar boson $H^{\pm}$. These observables are:
    \begin{itemize}
  \item The decay $B_s^0 \to \mu^+\mu^-$.
  \item The muon anomalous magnetic moment $a_{\mu}$.
  \item Radiative processes $\ell_i \to \ell_j \gamma$ 
    ($\ell_i = \tau,\mu$; $\ell_j = \mu,e$; $i \neq j$).
  \item $\ell_i \to \ell_j \ell_k \bar{\ell}_k$ 
    ($k = \mu,e$) and the flavour changing decay 
    of the Higgs boson, $h \to \ell_i \ell_j$.
\end{itemize}

\subsubsection{Decays \texorpdfstring{$B_{s}^0 \to \mu^+ \mu^-$, $B_{d}^0 \to \mu^+ \mu^-$ and $B \to B_s^0, B_d^0$}{Bs and Bd decays}}

\begin{figure}[!ht]
		\centering
		\includegraphics[width=7cm]{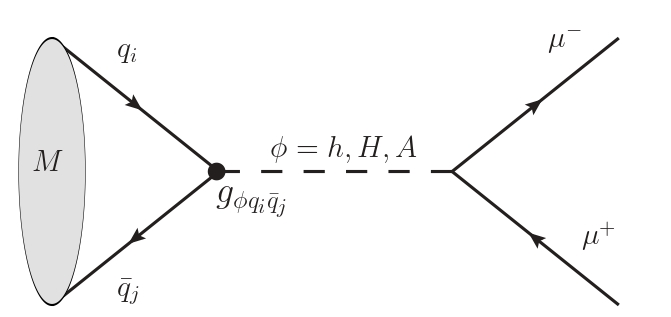}
		\caption{Generic Feynman diagram for the decays of a neutral meson $M$ (such as $B^0$, $K^0$, or $D^0$) into $\mu^+ \mu^-$. The black circle indicates a Flavour-changing vertex in the quark sector.}\label{FDMmumu} 
	\end{figure}

In the SM, neutral mesons that decay into muons are strongly suppressed. This suppression arises due to three primary reasons: (i) these processes proceed via loop level diagrams, which are inherently weaker than tree-level processes; (ii) helicity suppression further reduces the interaction probability; and (iii) certain CKM matrix elements are very small, thereby lowering the decay rate even more. As a result, the corresponding BRs are extremely small. Although other decay channels such as those involving electrons or $\tau$ leptons, can occur in principle, they are heavily suppressed in one case and challenging to reconstruct in the other, respectively. Fig. \ref{FDMmumu} presents a representative Feynman diagram for this process.
	
	The decay of $B_{s,d}^0$ mesons into a $\mu^+\mu^-$ pair is both compelling and highly constraining, given its sensitivity to BSM physics. Within the theoretical framework of the SM, the BRs are \cite{Beneke:2019slt}: 
\[
\mathcal{BR}(B_s^0\to \mu^+\mu^-) = (3.66\pm 0.14)\times 10^{-9}
\]
and 
\[
\mathcal{BR}(B_d^0\to \mu^+\mu^-) = (1.03\pm 0.05)\times 10^{-10}.
\]

According to measurements by the CMS collaboration \cite{2023137955}, the current experimental value is, 
\[
\mathcal{BR}(B_s^0\to \mu^+\mu^-) = \left(4.02^{+0.40}_{-0.38}(\mathrm{stat})^{+0.28}_{-0.23}(\mathrm{syst})^{+0.18}_{-0.15}(\mathrm{ext})\right)\times 10^{-9},
\]

where \textit{stat} denotes the statistical uncertainty, 
\textit{syst} the experimental systematic uncertainty, 
and \textit{ext} the external uncertainty from the
branching fractions \(\mathcal{B}(B^{+}\!\to J/\psi K^{+})\),
\(\mathcal{B}(B_s^{0}\!\to J/\psi\phi)\), and from the
fragmentation fraction ratio of the probabilities for a \(b\) quark to hadronise into
a \(B_s^{0}\) or \(B^{+}\) meson. While

\[
\mathcal{BR}(B_d^0\to \mu^+\mu^-) < 1.9\times 10^{-10}
\]
at $95\%$ C.L. \cite{CMS:2022mgd}. In this context of the 2HDM-III, these decays, $B_{s,d}^0\to\mu^+\mu^-$, can be mediated at tree-level by the SM-like Higgs boson, a heavy scalar $H$, or the pseudoscalar $A$. The relevant Feynman diagram for these decays is depicted in Fig.~\ref{FDMmumu}. In the case of $B_s^0\to \mu^+\mu^-$ (or $B_d^0\to \mu^+\mu^-$), the quark indices are $q_i=s,\,\bar{q}_j=\bar{b}$ (or $q_i=d,\,\bar{q}_j=\bar{b}$).
	
	The effective Hamiltonian governing the $B_s^0\to \mu^+\mu^-$ transition is,
	\begin{align}
\mathcal{H}_{\rm eff}^{B_{s}^0\to \mu^+\mu^-}=-\frac{G_F^2m_W^2}{\pi^2}\Big(C_A^{bs}\mathcal{O}_A^{bs}+C_S^{bs}\mathcal{O}_S^{bs}+C_P^{bs}\mathcal{O}_P^{bs} \notag \\
+C_A^{\prime bs}\mathcal{O}_A^{\prime bs}+C_S^{\prime bs}\mathcal{O}_S^{\prime bs}+C_P^{\prime bs}\mathcal{O}_P^{\prime bs}\Big)+h.c.,
\end{align}

where \( C_i^{bs} \) and \( C_i^{\prime bs} \) (with \( i=A,S,P \)) are the Wilson coefficients encoding the short distance contributions.	The Wilson operators $\mathcal{O}_i^{bs}$ describe the low energy effective interactions between the quark and lepton fields, they take the form,
	\begin{eqnarray}\label{WilsonOperators}
		\mathcal{O}_A^{bs}&=&\left(\bar{b}\gamma_\mu P_Ls\right)\left(\mu^+\gamma_\mu\gamma_5\mu^-\right),\nonumber \\
		\mathcal{O}_S^{bs}&=&\left(\bar{b} P_Ls\right)\left(\mu^+\mu^-\right),\\
		\mathcal{O}_P^{bs}&=&\left(\bar{b} P_Ls\right)\left(\mu^+\gamma_5\mu^-\right)\nonumber.
	\end{eqnarray}
	The primed operators are obtained by interchanging $P_L\leftrightarrows P_R$.
	The BR for this decay is then given by 

	\begin{align}\label{BRMlili}
		\begin{array}{ccl}
			\mathcal{BR}\left(M\rightarrow\ell^{+}\ell^{-}\right) & = & \frac{G_{F}^{4}m_{W}^{4}}{8\pi^{5}}\sqrt{1-4\frac{m_{\ell}^{2}}{m_{M}^{2}}}m_{M}f_{M}^{2}m_{\ell}^{2}\tau_{M}\\
			& \times & \left[\left|\frac{m_{M}^{2}\left(C_{P}^{ij}-C_{P}^{\prime ij}\right)}{2\left(m_{i}+m_{j}\right)m_{\ell}}-C_{A}^{\rm SM}\right|^{2}+\left|\frac{m_{M}^{2}\left(C_{S}^{ij}-C_{S}^{\prime ij}\right)}{2\left(m_{i}+m_{j}\right)m_{\ell}}\right|^{2}\left(1-4\frac{m_{\ell}^{2}}{m_{M}^{2}}\right)\right],
		\end{array}
	\end{align}

	here, $\ell^{+(-)}=\mu^{+(-)}$ and $i(j)=s(\bar{b})$. The parameter $m_M=m_{B_s^0}=5.36692$ GeV denotes the $B_s^0$ meson mass, while $f_M=f_{B_s^0}=0.2303$ GeV is the $B_s^0$ meson decay constant. The lifetime of the $B_s^0$ meson is given by $\tau_M=\tau_{B_s^0}=2.311\times 10^{12}$ GeV$^{-1}$ \cite{Bauer_2016}. The Fermi constant is represented by $G_F$, and the SM contribution at one loop, $C_A^{\rm SM}$, is given by,

	\begin{equation}
		C_A^{\rm SM}=-V_{tb}^* V_{ts} Y\Bigg(\frac{m_t^2}{m_W^2}\Bigg)-V_{cb}^* V_{cs} Y\Bigg(\frac{m_c^2}{m_W^2}\Bigg),
	\end{equation}
	where $Y$ is written as $Y=\eta_Y Y_0$ to include NLO QCD effects through $\eta_Y=1.0113$ \cite{Buras:2012ru}. The loop Inami-Lim function \cite{Inami:1980fz} is given by
	\begin{equation}
		Y_0(x)=\frac{x}{8}\Big[\frac{4-x}{1-x}+\frac{3x}{(1-x)^2}\ln(x)\Big].	
	\end{equation}
Finally, the form factors are 
	\begin{eqnarray}\label{FormFactors}
	C_{S}^{ij}&=&\frac{\pi^{2}}{2G_{F}^{2}m_{W}^{2}}\underset{\phi=h,H}{\sum}\frac{2g_{\phi\ell^+\ell^-}g_{\phi ij}}{M_{\phi}^{2}},\nonumber\\
	C_{P}^{ij}&=&\frac{\pi^{2}}{2G_{F}^{2}m_{W}^{2}}\frac{2g_{\phi\ell^+\ell^-}g_{\phi ij}}{M_{A}^{2}},\\
	C_{S}^{\prime ij}&=&C_{S}^{ij}\left(g_{\phi ij}\leftrightarrows g_{\phi ji}\right),\nonumber\\
	C_{P}^{\prime ij}&=&C_{S}^{ij}\left(g_{\phi ij}\leftrightarrows g_{\phi ji}\right)\nonumber.
\end{eqnarray}

The symbol \( \leftrightarrows \) indicates that the couplings \( g_{\phi ij} \) and \( g_{\phi ji} \) must be exchanged in the expression. This substitution accounts for the contribution where the external fermions are reversed, corresponding to a change in the flavour indices of the scalar or pseudoscalar interaction vertex. 
To obtain the corresponding $\mathcal{BR}(B_d^0\to \mu^+\mu^-)$, we make the following substitutions in the BR expression Eq. \eqref{BRMlili} the replacements $m_{B_s^0}\to m_{B_d^0}=5.27966$ GeV, $f_{B_s^0}\to f_{B_d^0}=0.190$, $\tau_{B_s^0}\to\tau_{B_d^0}=2.312\times 10^{12}$ GeV; In Eqs. \eqref{FormFactors}  $g_{\phi \bar{b}s}\to g_{\phi \bar{b}d}$ and in Eqs. \eqref{WilsonOperators} $s\to d$.
	
	%\subsubsection*{$\underline{K_L\to \mu^+\mu^-\,\rm and\,\bar{D}^0\to \mu^+\mu^-}$}
	%As far as the decays $K_L\to \mu^+\mu^-$ and $\bar{D}^0\to \mu^+\mu^-$ are concerned, the BR and the Wilson coefficients can be obtained via the necessary replacements 
	%%%%%%%%%%%%%%%%%%%%%%%%%%%%%%%%%%%%%%%%%%%%%%%%%%%%%%%%%%%%%%%%
	\subsection*{Lepton Flavour violating processes}\label{LFVconstraints}
	%%%%%%%%%%%%%%%%%%%%%%%%%%%%%%%%%%%%%%%%%%%%%%%%%%%%%%%%%%%%%%%%
\subsubsection{Muon anomalous magnetic moment \texorpdfstring{$a_\mu$}{a\_mu}}

\begin{figure}[!htb]
		\centering
		\includegraphics[width=9cm]{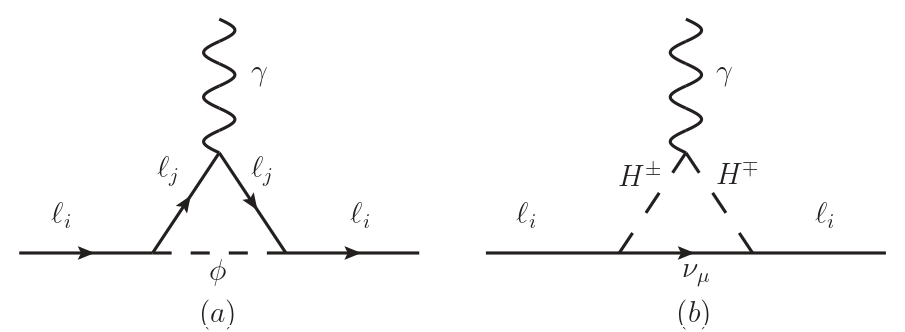}
         \\
         \includegraphics[width=4.8cm]{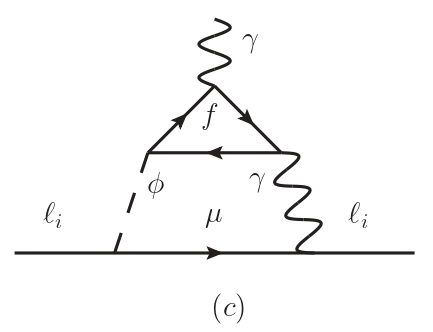}
		\caption{Feynman diagrams contributing to $a_\mu$. Here $\phi$ represents a CP-even scalar, CP-odd scalar, or the SM-like Higgs boson. $H^{\pm}$ denotes charged scalar bosons. In these diagrams $\ell_i=\mu$.}\label{FDmuonAMDM}
	\end{figure}

Considering the persistent discrepancy between the experimental measurement of $a_\mu$ and its theoretical prediction within the SM, we use this observable to constrain the relevant free parameters of the model. The Feynman diagrams contributing to $a_\mu$ are shown in Fig. \ref{FDmuonAMDM} \azul{~\cite{Ilisie:2015tra}}.
	The one-loop contributions are given by
	\begin{equation}
		\delta a_\mu^\phi=\frac{G_F m_\mu^2}{4\pi^2 \sqrt{2}}\sum_{\ell_i} g_{\mu\ell_i}^2 R_{\phi}F_{\phi}(R_{\phi}),	
	\end{equation}
	with
	\begin{eqnarray}
		F_{h,\,H}&=&\int_{0}^{1}dx\frac{x^2(2-x)}{R_{h,\,H}x^2-x+1},\\
		F_A&=&\int_{0}^{1}dx\frac{-x^3}{R_Ax^2-x+1},\\
		F_H^{\pm}&=&\int_{0}^{1}dx\frac{-x^2(1-x)}{R_{H^\pm}x^2+(1-R_{H^\pm})x}.
	\end{eqnarray}
	where $R_{\phi}=m^2_\mu/M^2_{\phi}$ ($\phi=h,\,H,\,A,\,H^{\pm}$). The dominant two-loop effect arises from a diagram in which $A$ circulates inside the loop kind box, and it is given by 
	\begin{equation}
		\delta a_\mu^{2-loops}=\frac{\alpha^2}{8\pi^2 s_W^2}\frac{m_\mu^2 g_{A\mu\mu}}{m_W^2}\sum_{f}N_c^f Q_f^2 R_A \bar{F}_A g_{Af_i\bar{f}_j},
	\end{equation}
	where
	\begin{equation}
		\bar{F}_A=\int_{0}^{1}dy\frac{\log\Big(\frac{R_A}{y(1-y)}\Big)}{R_A-y(1-R_A)},
	\end{equation}

    here, \( g_{Af_i\bar{f}_j} \) represents the Yukawa coupling of the CP-odd scalar \( A \) to fermions \( f_i \) and \( \bar{f}_j \). Only diagonal terms are considered, i.e., \( i = j = \mu \).

	\subsubsection{\texorpdfstring{$\ell_i \to \ell_j \gamma$}{li -> lj gamma} decays}

\begin{figure}[!htb]
		\centering
		\includegraphics[width=6.5cm]{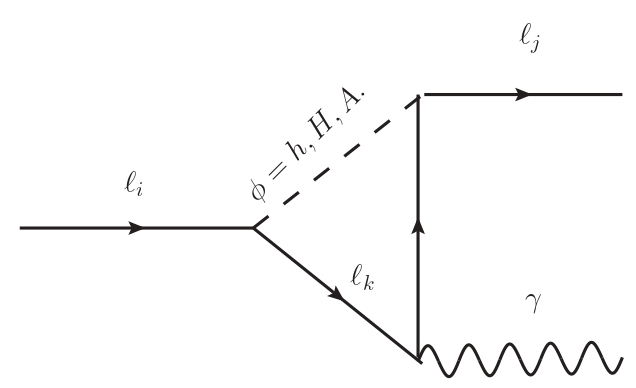}
		\caption{Feynman diagram for process $\ell_i\to\ell_j\gamma$ at one-loop level induced by $\phi=h,\,H,\,A$.}\label{FDliljgamma}
	\end{figure}

The effective Lagrangian describing $\ell_i\to\ell_j\gamma$ processes is 
	\begin{equation}\label{EffLagrangian}
		\mathcal{L}_{\text{eff}}=C_L Q_{L\gamma} C_R Q_{R\gamma}+h.c.,
	\end{equation}
	the relevant dim-5 electromagnetic penguin operators are
	\begin{equation}\label{dim5Op}
		Q_{L\gamma,\,R\gamma}=\frac{e}{8\pi^2}(\bar{\ell}_j\sigma^{\alpha\beta}P_{L,\,R}\ell_i)F_{\alpha\beta},
	\end{equation}
	here $F_{\alpha\beta}$ denotes the electromagnetic field strength tensor. The Feynman diagram for the process $\ell_i\to\ell_j\gamma$ is shown in Fig. \ref{FDliljgamma}.

	The Wilson coefficients $C_{L,\,R}$ receive contributions at one-loop order and also significant contributions from the Barr-Zee diagrams at two-loops. For the specific case in which  $\ell_i=\tau$ and $\ell_j=\mu$, we adopt the approximations $g_{\phi\mu\mu}\ll g_{\phi\tau\tau}$ and $m_\mu\ll m_{\tau}\ll m_\phi$. Based on these assumptions, the one-loop Wilson coefficients $C_{L,\,R}$ simplify as follows \cite{Harnik:2012pb, Blankenburg:2012ex}
	
    \begin{align}\label{WilsonCoe1loop}
    C_L^{1-loop} &\simeq \sum_{\phi}\frac{g_{\phi\tau\tau}g_{\phi\tau\mu}}{12m_\phi^2}\Bigg(-4+3\log\frac{m_\phi^2}{m_{\tau}^2}\Bigg),\notag \\
    C_R^{1-loop} &\simeq\sum_{\phi}\frac{g_{\phi\tau\tau}g_{\phi\tau\mu}}{12m_\phi^2}\Bigg(-4+3\log\frac{m_\phi^2}{m_{\tau}^2}\Bigg).
    \end{align}

	The numerical expressions for the 2-loop contributions are given by
	
	\begin{align}\label{WilsonCoe2loop}
		C_L^{2-loops}&=\sum_{\phi=h,\,H,\,A}g_{\phi\tau\mu}^*(-0.082g_{\phi tt}+0.11)/(m_\phi\text{GeV})^2, \nonumber \\
		C_R^{2-loops}&=C_L^{2-loops}(g_{\phi\tau\mu}^*\to g_{\phi\tau\mu}).
	\end{align}
	%where $Y_{tt}=\bar{m}_t/v=0.67$, with $\bar{m}_t\simeq 164 \text{GeV}$.
	
	The rate for $\tau\to\mu\gamma$ is
	
	\begin{equation}\label{ratetaumugamma}
		\Gamma(\tau\to\mu\gamma)=\frac{\alpha m_{\tau}^2}{64\pi^4}(|C_L|^2+|C_R|^2).
	\end{equation}
	
	To obtain the corresponding width decays for the processes $\mu\to e\gamma$ and $\tau\to e\gamma$, we make the substitutions $\tau\to \mu,\, \mu\to e$ in the first case, and $\mu\to e$ in the second one, applying them consistently from Eqs. \eqref{EffLagrangian} to \eqref{ratetaumugamma}.
	
\subsubsection{\texorpdfstring{$\ell_i\to\ell_j\ell_k\bar{\ell}_k$}{li -> lj lk lbar_k decays}}

\begin{figure}[H]
		\centering
		\includegraphics[width=4.1cm]{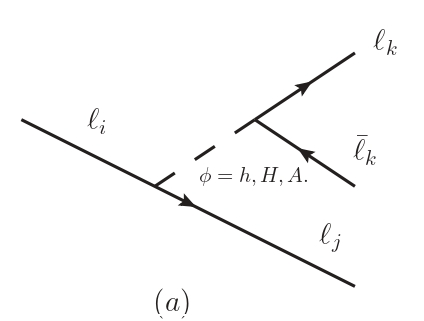}
		\includegraphics[width=4.1cm]{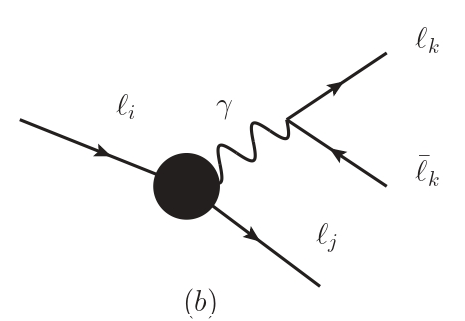}
		\caption{Feynman diagrams contributing to the process $\ell_i\to \ell_j\ell_k\bar{\ell}_k$. Panel  (a) shows the tree-level and (b) one-loop level, where the black circle denotes a loop of the type as Feynman diagram of Fig. \ref{FDliljgamma}.}\label{FDliljlklk}
	\end{figure}

Within the 2HDM-III framework, these types of decays can proceed at tree-level via the exchange of $h,\,H,\,A$, as shown in Fig. \ref{FDliljlklk}. Nonetheless, the process is suppressed by the LFV Yukawa couplings $Y_{\ell_i\ell_j}$ and by the Flavour-conserving couplings $Y_{\ell_k\ell_k}$. There are also higher order contributions at one-loop and two-loops level. The partial width for the corresponding flavour violating decay is
	\begin{equation}
		\Gamma(\tau\to3\mu)\sim \frac{\alpha m_{\tau}^5}{6(2\pi)^5}\Big|\log\frac{m_{\mu}^2}{m_{\tau}^2}-\frac{11}{4} \Big|(|C_L|^2+|C_R|^2),
	\end{equation}  
here any terms suppressed by the muon mass are neglected. The Wilson coefficients $C_{L,\,R}$ are given in Eqs. \eqref{WilsonCoe1loop}-\eqref{WilsonCoe2loop}. 
	
\subsubsection{\texorpdfstring{$h\to \ell_i \ell_j$}{h -> li lj decays}}\label{hlilj}

\begin{figure}[!htb]
		\centering
		\includegraphics[width=5.5cm]{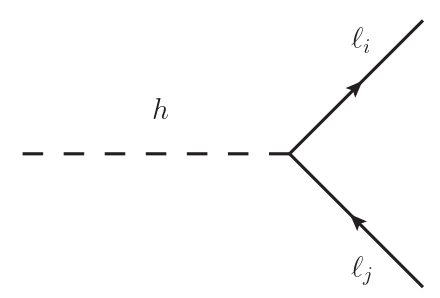}
		\caption{Feynman diagram for the tree-level decay $h\to\ell_i\ell_j$.}\label{FDhlilj}
	\end{figure}

The LFV processes $h\to \ell_i\ell_j$ ($\ell_{i,\,j}=\ell_{i,\,j}^-\ell_{i,\,j}^+$) where $\ell_i\ell_j=e\mu,\,e\tau,\,\tau\mu$ may occur at tree-level in various SM extensions, as shown in Fig. \ref{FDhlilj}.
	
	The relevant interactions can be derived from the Yukawa Lagrangian
	\begin{equation}
		\mathcal{L}_Y\supset -Y_{ij}\bar{\ell}_L^i \ell_R^jh+h.c.
	\end{equation}	 
	The total decay width for $h\to \bar{f}_i f_j$ is given by
	\begin{align}
		\Gamma(h\to \bar{f}_i f_j)=\frac{N_c g^2_{h \bar{f}_i f_j }m_h}{128\pi}\Bigg[ 4-(\sqrt{\tau_{f_i}} 
        +\sqrt{\tau_{f_j}})^2  \Bigg]^{3/2}\sqrt{4-(\sqrt{\tau_{f_i}}-\sqrt{\tau_{f_j}})^2},
	\end{align}
	where $g_{h \bar{f}_i f_j}$ is the coupling associated with $h\bar{f}_i f_j$ arising from an SM extension, $Nc=3\,(1)$ is the colour factor for quarks (leptons), $m_h$ denotes the Higgs boson mass and $\tau_i=4m_i^2/m_h^2$. 
	
\begin{figure}[!htb]
		\centering
		\includegraphics[width=9.5cm]{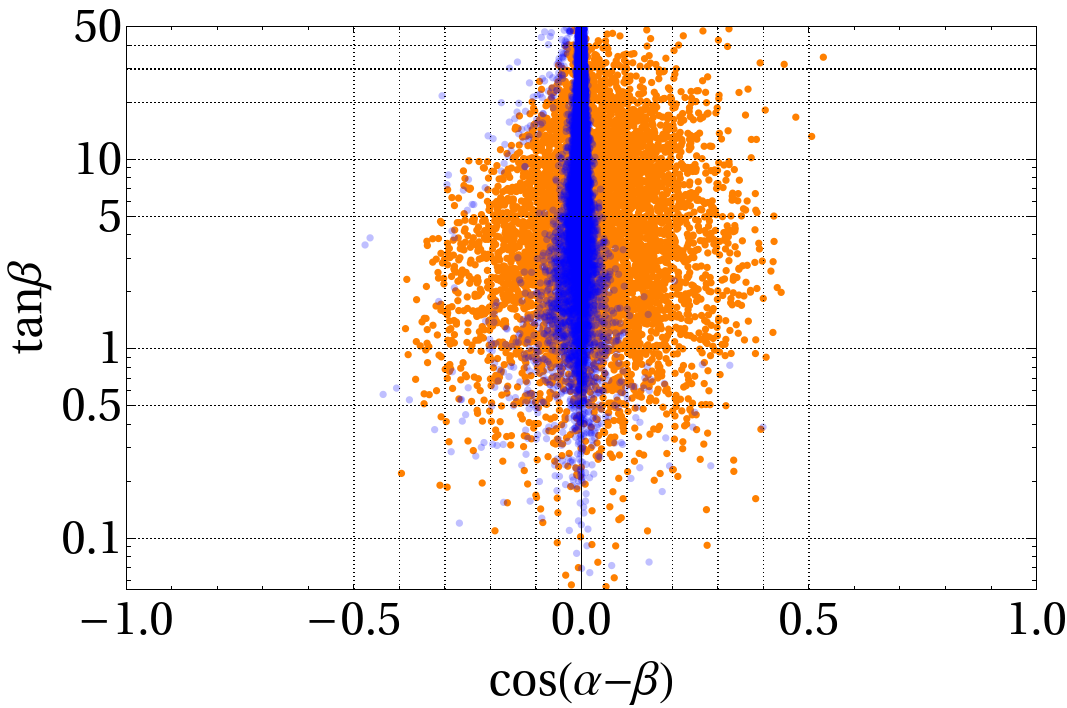}
		\caption{In the $\cos(\alpha-\beta)-\tan\beta$ plane, the blue points represent parameter values allowed by all signal strength modifiers $\mathcal{\mu}_{X}$, whereas the orange points indicate those consistent with the LFV processes. The dataset was generated using the \texttt{SpaceMath} framework. }\label{HixBosonData}
	\end{figure}

	Having presented all the analytical expressions needed to calculate the BRs for the observables described, we proceed to evaluate all processes discussed in Sec. \ref{HiggsConstraints} with the \texttt{Mathematica} package called \texttt{SpaceMath}\footnote{This software implements all the experimental constraints considered in this research.} \cite{Arroyo-Urena:2020qup}, in order to find the permissible values of the free parameters in our analysis. 
	In Fig. \ref{HixBosonData}, we present the plane $\cos(\alpha-\beta)-\tan\beta$, where each point corresponds to a parameter combination allowed by experimental constraints. The orange points satisfy the measured (or upper limits) BRs of $B_{s,d}\to\mu^+\mu^-$ and the LFV processes described above. Meanwhile, the blue points accomodate the LHC Higgs boson data. By considering all the observables jointly, it is possible to identify a parameter region that is consistent with every constraint, thus enabling the exploration of different scenarios to predict further observables. As an illustration, two such realistic parameter scenarios are:
	\begin{itemize}
		\item $-0.04\lesssim \cos(\alpha-\beta)\lesssim 0.025$ for $1\lesssim \tan\beta \lesssim 10$,
		\item  $-0.01\lesssim \cos(\alpha-\beta)\lesssim 0.01$ for $1\lesssim \tan\beta \lesssim 50$.
	\end{itemize}
	
These bounds extend to the decoupling limit \texorpdfstring{$\alpha-\beta\to \pi/2$}{alpha-beta -> pi/2}, which suppresses flavour-violating decays of the SM-like Higgs boson. Simultaneously, having \texorpdfstring{$\cos(\alpha-\beta)\sim 0$}{cos(alpha-beta) ~ 0} can enhance such processes in decays of the heavy Higgs boson $H$.

	Further details on the individual allowed parameter values for each observable are included in Appx.~\ref{V}. 
	%%%%%%%%%%%%%%%%%%%%%%%%%%%%%%%%%%%%%%%%%%%%%%%%%%%%%%%%%%%%%%%%
	\subsection{Constraint on \texorpdfstring{$M_H$, $M_A$, and $M_{H^\pm}$}{MH, MA, MHpm}}
\label{subsubsection:constraintoncharguedscalar}

	%%%%%%%%%%%%%%%%%%%%%%%%%%%%%%%%%%%%%%%%%%%%%%%%%%%%%%%%%%%%%%%%
\subsubsection{Collider Constraints} 
\begin{figure}[!htb]
		\center{\includegraphics[scale=0.8]{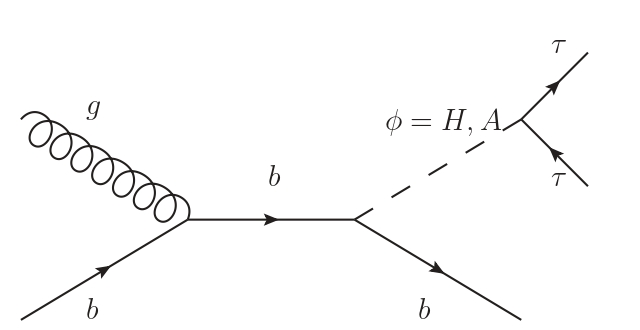}}
		\caption{Feynman diagram for the production of $\phi$ in association with a bottom quark at the LHC, followed by its decay into a $\tau\tau$ pair.  \label{FeynmanDiagram2}}
	\end{figure}

The ATLAS and CMS collaborations have conducted several searches to look for additional neutral Higgs bosons in different channels. Among these is the di-tau channel $gb\to\phi\to\tau\tau$, with $\phi=A,\,H$. The corresponding Feynman diagram is shown in Figure \ref{FeynmanDiagram2}. Although no evidence of extra Higgs bosons was found, upper limits were imposed on the production cross-section $\sigma(gb\to\phi)$ multiplied by $\mathcal{BR}(\phi\to\tau\tau)$.
	
\begin{figure}[!htb]
    \centering
    \begin{subfigure}{0.45\textwidth}
        \centering
        \includegraphics[scale=0.38]{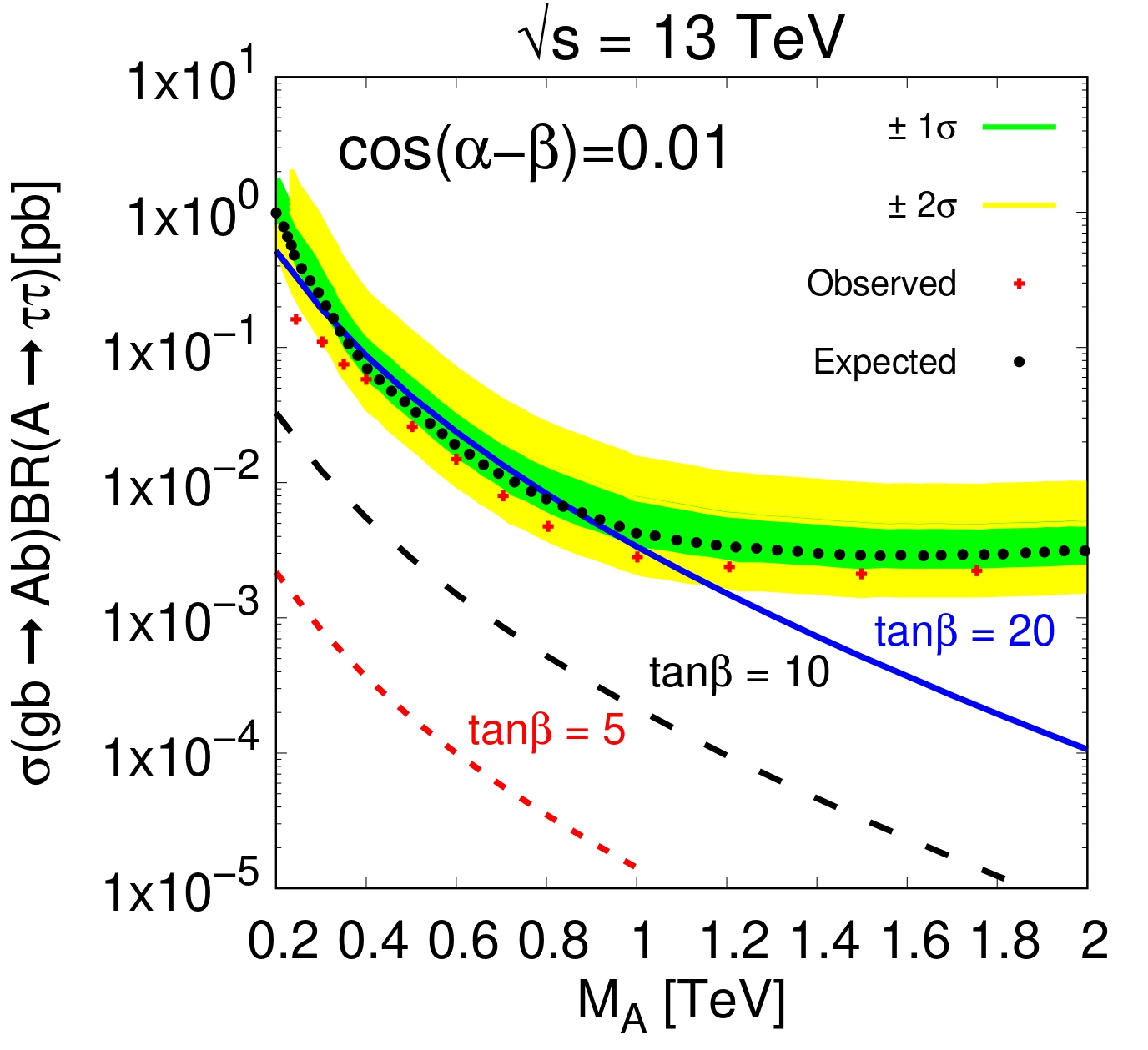}
        \caption{}\label{XS_A_timesBRHtautau}
    \end{subfigure}
    \\ % this will fill space between the two subfigures if necessary
    \begin{subfigure}{0.45\textwidth}
        \centering
        \includegraphics[scale=0.38]{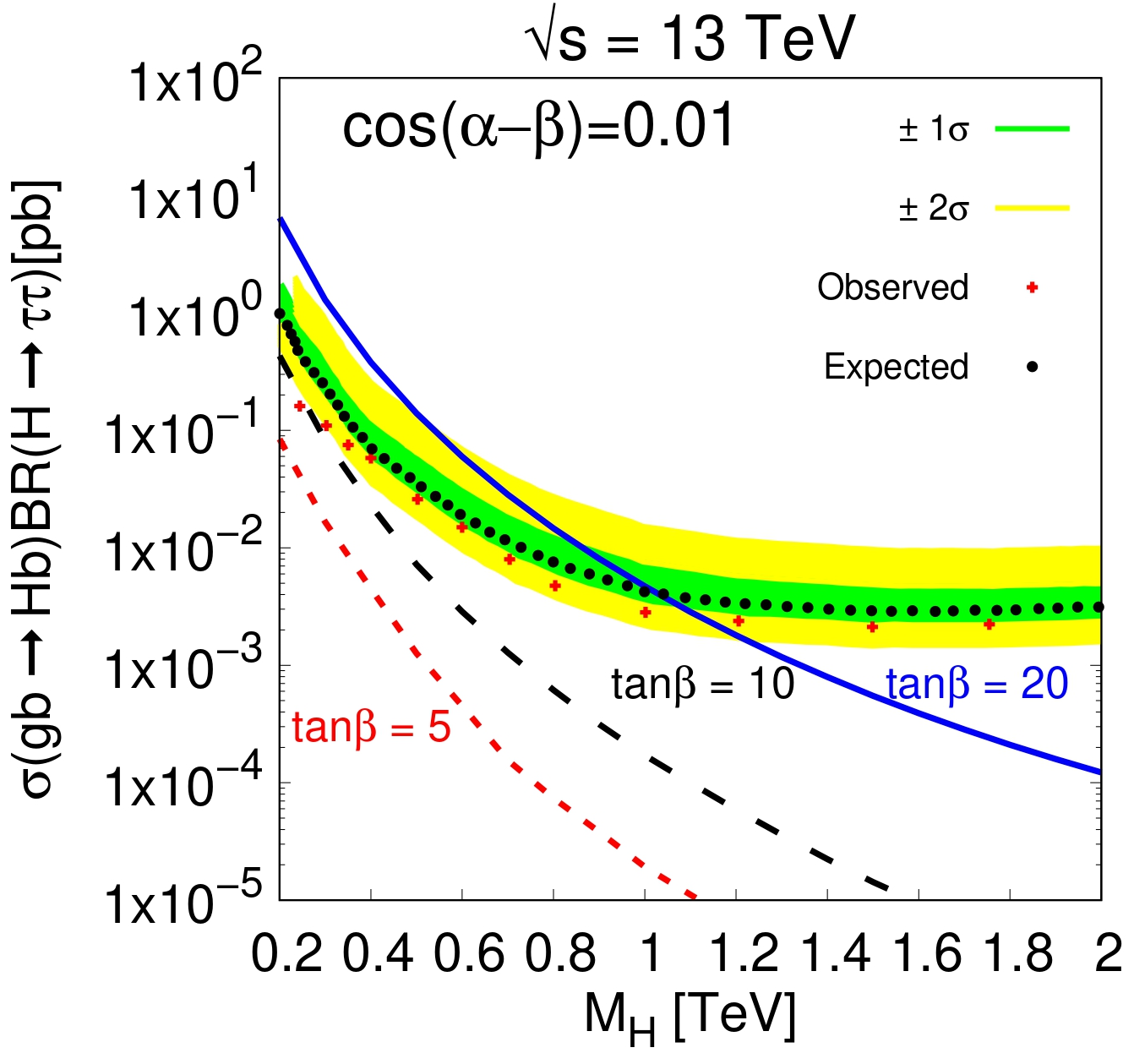}
        \caption{}\label{XS_H_timesBRHtautau}
    \end{subfigure}
    \caption{Observed and expected 95\% CL upper limits on the production cross-section multiplied by the di-tau BR for a scalar boson produced in association with a bottom quark. The results are shown as a function of (a) $M_A$ and (b) $M_H$, where we consider $tan_{\beta}=$ 5, 10, 20 and $cos_{\alpha\beta}=0.01$.}
    \label{constMasses}
\end{figure}

	Fig. \ref{XS_A_timesBRHtautau} shows the production cross-section $\sigma(gb\to Ab)\times\mathcal{BR}(A\to\tau\tau)$ as a function of $M_A$. The curves are shown for representative values of $\tan{\beta}=5,\,10,\,20$ and $\cos(\alpha-\beta)=0.01$. Meanwhile, Fig. \ref{XS_H_timesBRHtautau} presents the analogous results for $\phi=H$. In both plots, the black points correspond to the expected 95$\%$ CL upper limits, whereas the red crosses indicate the observed limits at the same confidence level. Additionally, the green (yellow) shaded bands illustrate the $\pm 1 \sigma$ ($\pm 2 \sigma$) intervals around the expected limits, providing a visual measure of the experimental and theoretical uncertainties.

\vspace{-4ex}
	
\begin{figure}[!htb]
		\center{\includegraphics[scale=0.35]{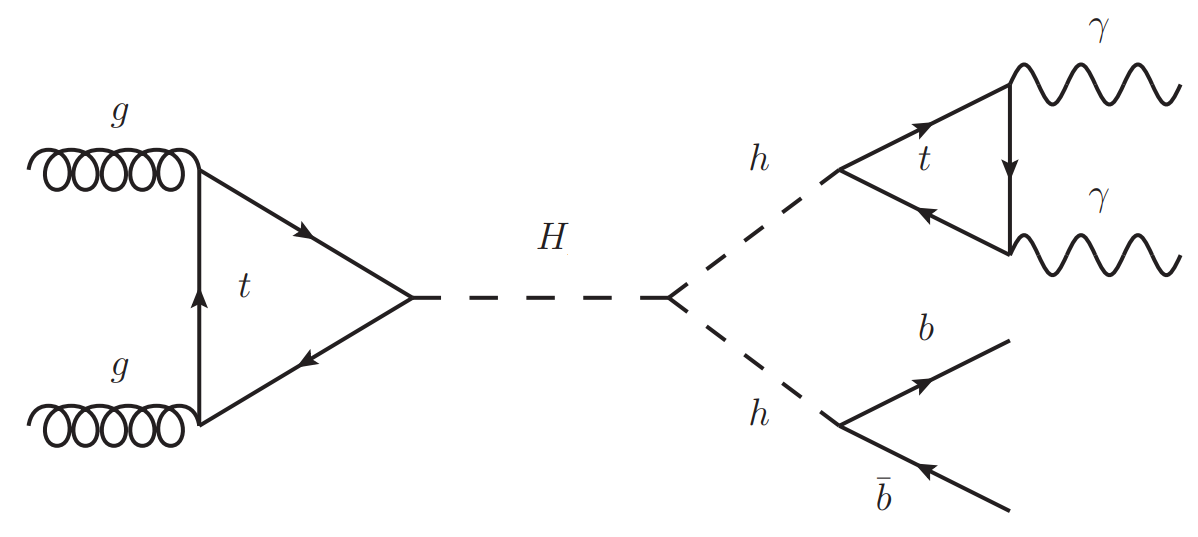}}
		\caption{Feynman diagram illustrating the production process of $H$ and its subsequent decay into a pair of Higgs bosons $hh$.\label{FDHhh}}
	\end{figure}

	Moreover, from Fig. \ref{XS_A_timesBRHtautau}, it is observed that $M_A\lesssim 1$ TeV ($M_A\lesssim 1.2$ TeV) is excluded at 1$\sigma$ (2$\sigma$) CL for $\tan{\beta}= 20$. On the other hand, for lower values of $\tan{\beta}\lesssim 5,\,10$, the upper limit on $\sigma(gb\to \phi b)\times\mathcal{BR}(\phi\to\tau\tau)$ is readily satisfied, indicating that such scenarios are not significantly constrained by the current experimental data. Similarly, from Fig. \ref{XS_H_timesBRHtautau}, it is found that $M_H\lesssim 1.1$ TeV ($M_H\lesssim$ 1.3 GeV) is excluded at 1$\sigma$ (2$\sigma$) for $\tan{\beta}= 20$. For $\tan\beta\lesssim 15$, a wide range of masses is allowed, particularly for $M_A>300$.  Additionally, another process used to constrain the mass of the heavy scalar $M_H$ involves its decay into a pair of Higgs bosons, which subsequently decay into $b\bar{b}$ and $\gamma\gamma$, i.e., $pp\to H\to hh, h\to b\bar{b}, h\to \gamma\gamma$, as shown in Fig. \ref{FDHhh}. We focus on this final state because both ATLAS and CMS have performed dedicated searches in this channel \cite{ATLAS:2021ifb}. The $\gamma\gamma$ final state provides excellent mass resolution, allowing the comparison of our results with the existing experimental constraints in a well studied channel.
This decay, $h \to \gamma\gamma$, receives loop-level contributions from the $W^\pm$ bosons and the top quark, as in the Standard Model, and an additional contribution from the charged scalar boson $H^\pm$ present in the 2HDM-III. The partial width can be computed using the effective vertex approximation \cite{Gunion:1989we},
\begin{equation}
\Gamma(h \to \gamma\gamma) = \frac{\alpha^2 m_h^3}{1024\pi^3 m_W^2} \left| A^{h\gamma\gamma} \right|^2,
\end{equation}
where the total amplitude includes contributions from all particles running in the loop,
\begin{equation}
A^{h\gamma\gamma} = \sum_{s} A_s^{h\gamma\gamma}(\tau_s),
\end{equation}
\vspace*{-4ex}
with $s$ denoting the spin of the loop particle. For the scalar contribution, $s=0$, and $\tau_0 = \tau_{H^\pm} = 4 m_{H^\pm}^2 / m_h^2$. The amplitude $A_0^{h\gamma\gamma}$ is then given by
\begin{equation}
A_0^{h\gamma\gamma}(\tau_{H^\pm}) = \frac{m_W\, g_{hH^+H^-}}{m_{H^\pm}^2} F_0(\tau_{H^\pm}), \tag{16}
\end{equation}
where $g_{hH^+H^-}$ is the trilinear coupling between the Higgs boson $h$ and the charged Higgs bosons. The loop functions $F_0(x)$ and $f(x)$ are defined as
\begin{equation}
F_0(x) = x\left[1 - x f(x)\right],
\end{equation}
\begin{equation}
f(x) = \left\{
\begin{array}{ll}
\left[\arcsin\left(\frac{1}{\sqrt{x}}\right)\right]^2, & \quad x \geq 1, \\
-\frac{1}{4} \left[ \log\left( \frac{1 + \sqrt{1 - x}}{1 - \sqrt{1 - x}} \right) - i\pi \right]^2, & \quad x < 1.
\end{array}
\right.
\end{equation}

The vertex used in the simulation for the interaction between the $H$ and the top quarks follows the effective Lagrangian
\[
\mathcal{L}_{Y}^{\phi}= \phi\,\bar{f}_{i}\!\left(S^{\phi}_{ij}+iP^{\phi}_{ij}\gamma^{5}\right)\!f_{j},
\]
where $S^\phi_{ij}$ and $P^\phi_{ij}$ denote the scalar and pseudoscalar couplings, respectively. In the CP-conserving case, the pseudoscalar term vanishes for $H$, so $P^H_{tt}=0$.

For the scalar component, the relevant term is
\[
S^{H}_{tt}= \frac{g\,m_{t}}{2M_{W}}\,c^{H}_{u}+d^{H}_{u}\,[\tilde{Y}_{a}^{u}]_{tt},
\qquad \text{with } \frac{g\,m_{t}}{2M_{W}} = \frac{m_t}{v}.
\]
Using the expressions $c^H_u = \sin\alpha/\sin\beta$ and $d^H_u = \sin(\alpha-\beta)/\sin\beta$, along with the Yukawa texture $[\tilde{Y}_{a}^{u}]_{tt}= (m_t/v)\,\chi_{tt}$, the coupling becomes
\begin{equation}
S^{H}_{tt}= \frac{m_{t}}{v}\left[
\frac{\sin\alpha}{\sin\beta}
+\frac{\sin(\alpha-\beta)}{\sin\beta}\,\chi_{tt}\right],
\label{eq:HttScalarCoupling}
\end{equation}
defining a correction factor
\[
\kappa_{Htt}= \frac{\sin\alpha}{\sin\beta}
              +\frac{\sin(\alpha-\beta)}{\sin\beta}\,\chi_{tt}.
\]
Thus, the SM-like coupling $(m_t/v)\,\bar{t}th$ is replaced by $(m_t/v)\,\kappa_{Htt}\,\bar{t}tH$.

\begin{figure}[!htb]
		\centering
		{\includegraphics[scale=0.3]{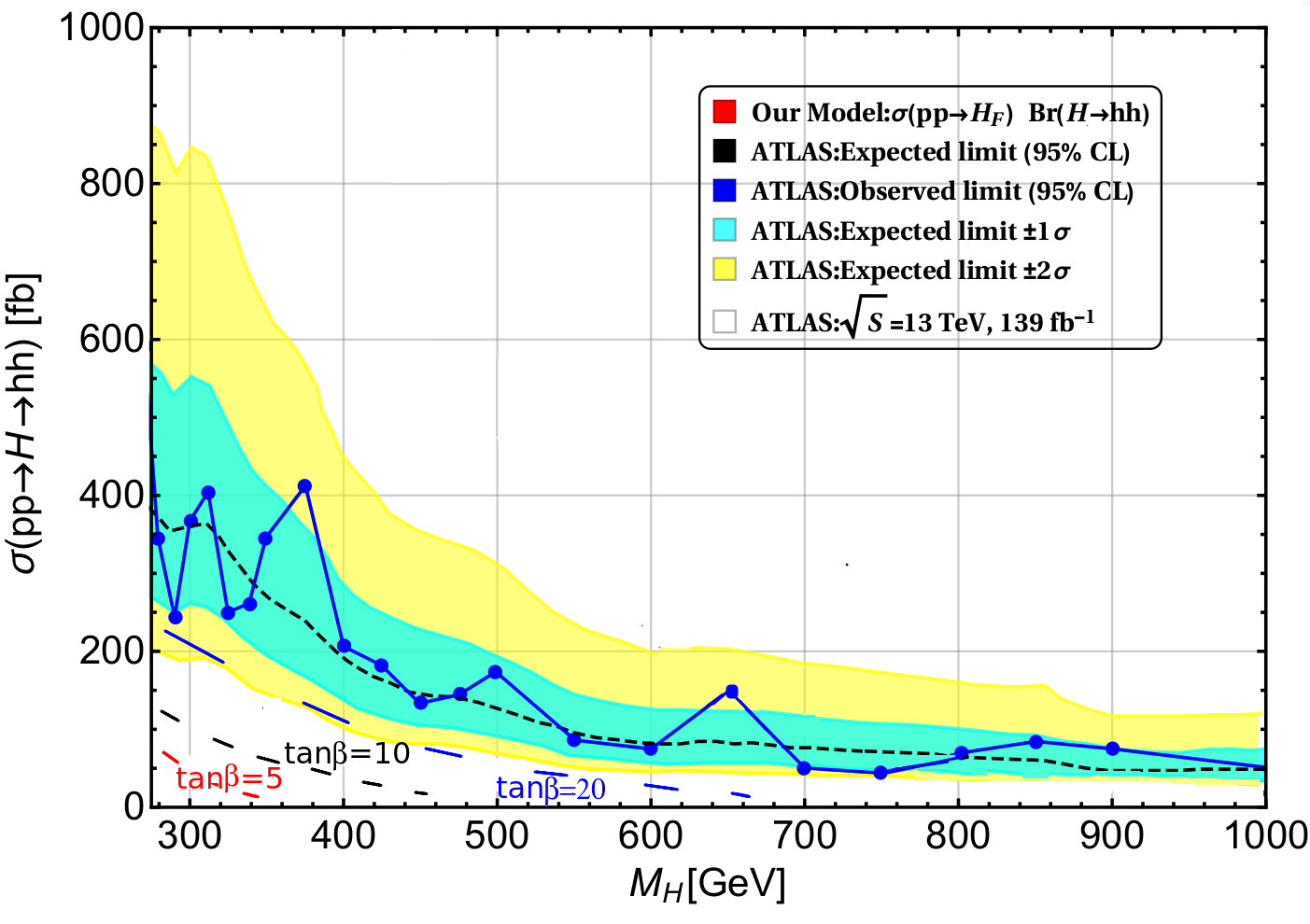}}
		\caption{The 95$\%$ CL upper limits on the production cross-section multiplied by the di-Higgs BR are presented for a scalar boson produced in $pp$ collisions as a function of $M_H$. We observed and expected limits are shown, considering $\tan{\beta}=$ 5, 10, 20 and $\cos(\alpha-\beta)=0.01$.}\label{Hhh}
	\end{figure}

		The cross-section $\sigma(pp\to H\to hh, h\to b\bar{b}, h\to \gamma\gamma)$ as a function of $M_H$ is shown in Fig. \ref{Hhh}, where the upper limits at 1$\sigma$ (green band) and 2$\sigma$ (yellow band) are displayed alongside the expected limit (dashed black line) and the observed limit (blue line). These results, reported by the ATLAS collaboration \cite{ATLAS:2021ifb}, are compared with the theoretical predictions of the 2HDM-III for $\tan\beta=5,\,10,\,20$ and $\cos(\alpha-\beta)=0.01$. We note that the aforementioned limits on $M_H$ are readily satisfied within the 2HDM-III framework for values of $\tan\beta\lesssim 18$. 
        The $H\rightarrow VV, (V=WZ)$ channel is also examined \cite{ATLAS:2020tlo,CMS:2019bnu}. However, the constraints derivated from this channel are not particularly stringent in the current model due to the proportionality of the coupling $g_{HVV}$ to $\cos(\alpha-\beta)$. Additionally, the $t\bar{t}$ channel is analysed and its predictions are compared with experimental results from a model independent analysis ~\cite{ATLAS:2024jja}. The cross-section predicted by the 2HDM-III is found to be heavily suppressed, being 4 to 8 orders of magnitude smaller, as it scales inversely with $\tan^4 \beta$.
		 
\subsubsection{\texorpdfstring{$b \to s \gamma$}{b->s gamma} decay}

The detection of the charged scalar boson $H^{\pm}$ would provide a clear signature of new physics. Constraints on its mass $M_{H^{\pm}}$ have been obtained from collider searches focusing on the production of $H^{\pm}$ and its subsequent decay into a $\tau^{\pm}\nu_\tau$ pair~\cite{ATLAS:2018gfm}. However, we find that such processes are not particularly effective in  constraining the charged scalar mass $M_{H^{\pm}}$ within the framework of the 2HDM-III. In contrast, the decay $b\to s\gamma$, imposes stringent lower limits on $M_{H^\pm}$ due to the contributions from the charged scalar boson~\cite{Ciuchini:1997xe, Misiak:2017bgg}.
	In Fig.~\ref{mCH}(a)-(b), a scatter plot in the $M_{H^\pm}$-$\tan\beta$ plane is presented, where red and blue points represent regions allowed by the ratio $2.77\times 10^{-3}<R_{\rm quark}<3.67\times 10^{-3}$ ~\cite{Misiak:2017bgg}, defined as,
	\begin{equation}
		R_{\rm quark}=\frac{\Gamma(b\to X_s\gamma)}{\Gamma(b\to X_ce\nu_e)},
		\label{eq:rquark}
	\end{equation} 
here, $X_s$ and $X_c$ denote inclusive hadronic final states containing a strange and charm quark, respectively.

\begin{figure}
    \centering
    \includegraphics[width=0.5\linewidth]{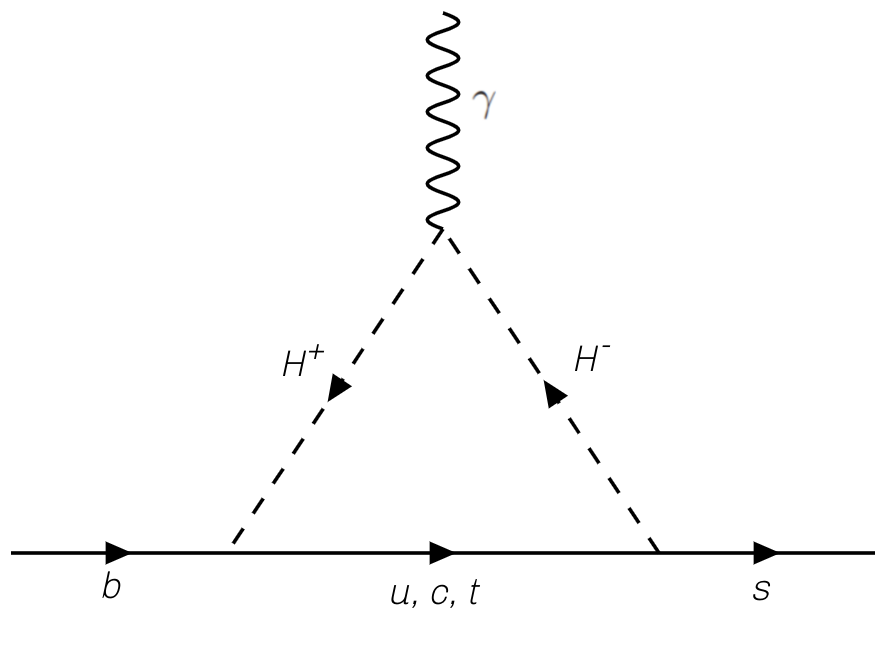}
    \caption{Radiative decay $b \to s \gamma$ mediated by charged Higgs bosons. The loop involves up-type quarks ($u$, $c$, $t$), with the dominant contribution coming from the top quark. The photon is emitted from the charged scalar line.}
    \label{Hbs}
\end{figure}
	
\begin{figure}[!htb]
    \centering
    \begin{subfigure}{0.45\textwidth}
        \centering
        \includegraphics[scale=0.4]{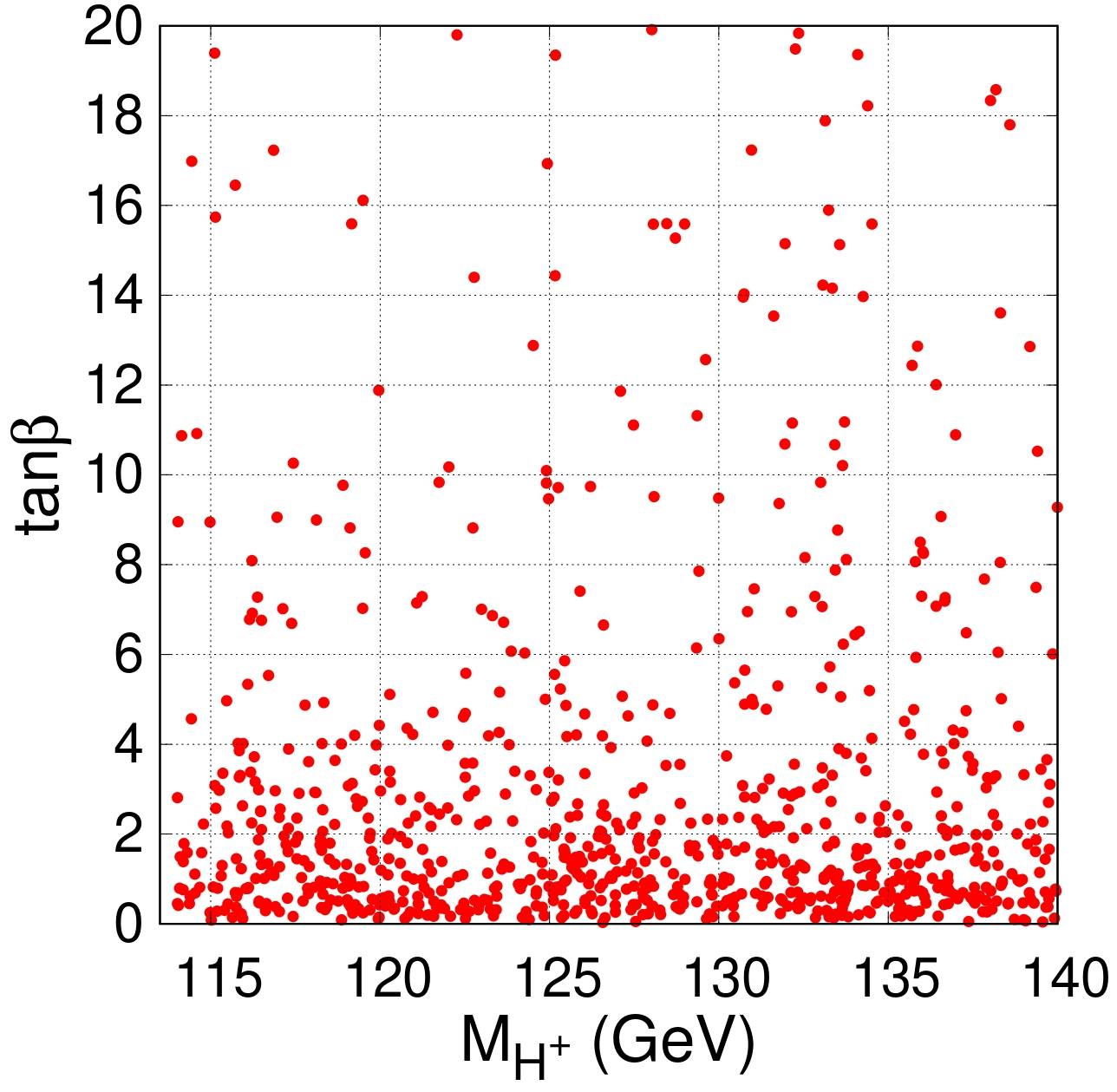}
        \caption{} % optional subcaption
    \end{subfigure}
    \\ % this will fill space between the two subfigures if necessary
    \begin{subfigure}{0.45\textwidth}
        \centering
        \includegraphics[scale=0.4]{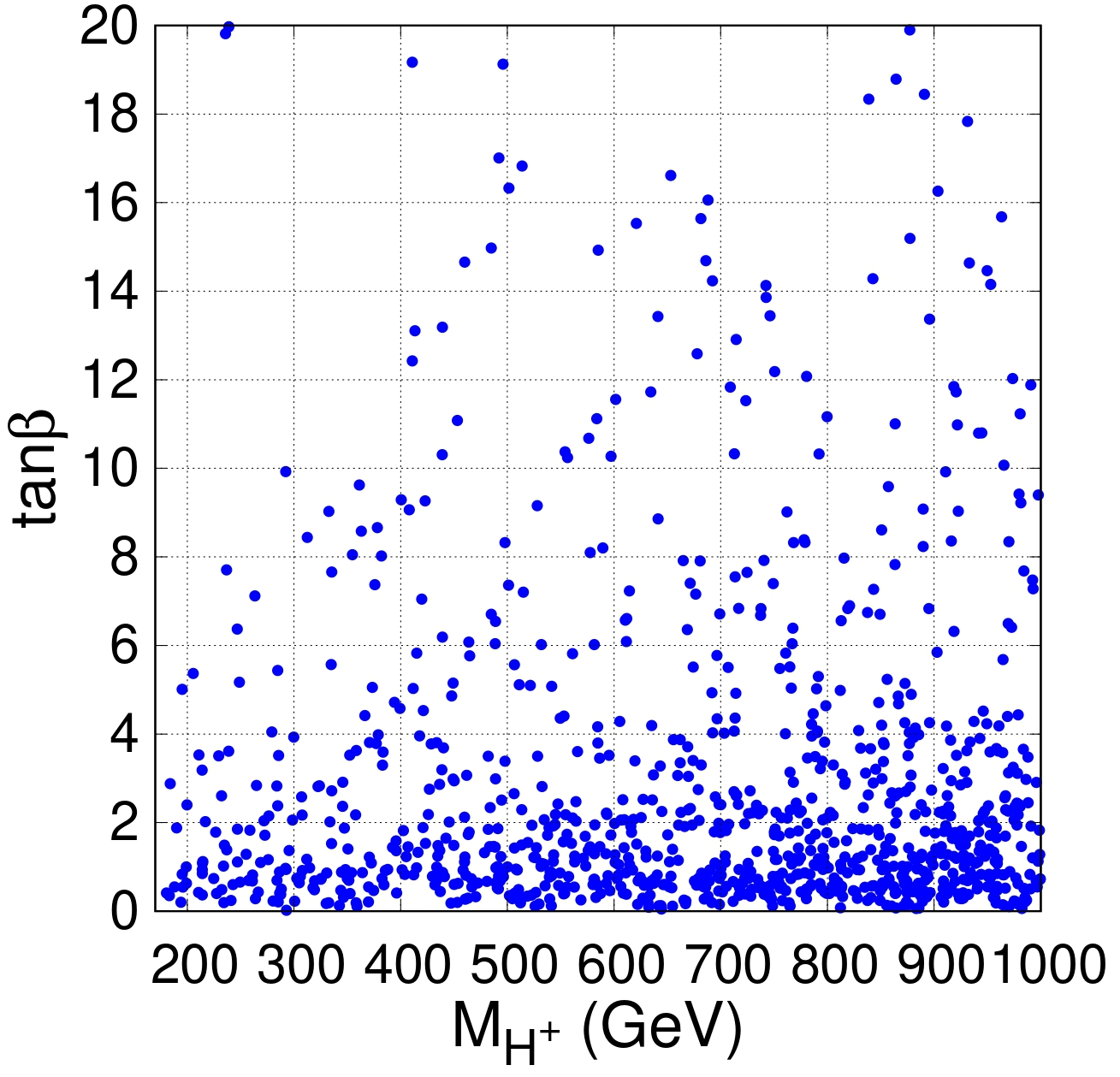}
        \caption{} % optional subcaption
    \end{subfigure}
    \caption{Scatter plot in the \(M_{H^{\pm}}-\tan\beta\) plane. (a) Red points stand for these regions allowed by \(R_{\rm quark}\) for the mass interval \(114\leq M_{H^{\pm}}\leq 140\) GeV. (b) The same as in (a) but for the mass range \(m_t+m_b\leq M_{H^{\pm}}\leq 1000\). The \(R_{\rm quark}\) is defined in Eq. \eqref{eq:rquark}.}\label{mCH}
\end{figure}

	A preference for low $\tan\beta$ is observed in both mass ranges, $114\leq M_{H^\pm}\leq 140$ GeV and $m_t+m_b\leq M_{H^\pm}\leq 1000$ GeV. This preference arises from the structure of the contributions in the 2HDM-III, where the couplings $g_{\bar{u}dH^{-}}\sim 1/\tan\beta(1-\chi_{tt}/\sqrt{2})$ and $g_{\bar{d}uH^{+}}\sim \tan\beta(1-\chi_{bb}/\sqrt{2})$ indicate the large values of $\tan\beta$ could lead to significant contributions to $b\to s\gamma$ potentially conflicting with experimental constraints.  The partial width is proportional to the squared amplitude, $\Gamma(b \to s\gamma) \propto |C_7^{(R)}|^2 + |C_7^{(L)}|^2$, where $C_7^{(R)}$ and $C_7^{(L)}$ are the Wilson coefficients of the magnetic dipole operators contributing to the effective Hamiltonian. The contribution to $C_7^{(L)}$ ~\cite{PhysRevD.87.094031}, takes the form 
\[
C_7^{(L)} = \frac{v^2}{\lambda_t m_b} \sum_j \frac{\Gamma_{u_j d_2}^{RL H^+ *} \Gamma_{u_j d_3}^{LR H^+}}{m_{u_j}} C_{7,XY}^0(y_j)
+ \frac{v^2}{\lambda_t} \sum_j \frac{\Gamma_{u_j d_2}^{RL H^+ *} \Gamma_{u_j d_3}^{LR H^+}}{m_{u_j}^2} C_{7,YY}^0(y_j),
\]
where the charged Higgs couplings are given by
\[
\Gamma_{u_f d_i}^{LR H^+} = \sum_j \sin\beta\, V_{fj} \left( \frac{m_{d_j}}{v_d} \delta_{ji} - \epsilon_{ji}^d \tan\beta \right), \]
\[\Gamma_{d_f u_i}^{LR H^+} = \sum_j \cos\beta\, V_{jf}^* \left( \frac{m_{u_j}}{v_u} \delta_{ji} - \epsilon_{ji}^u \tan\beta \right).
\].

\begin{figure}[!htb]
		\centering
		\includegraphics[scale=0.25]{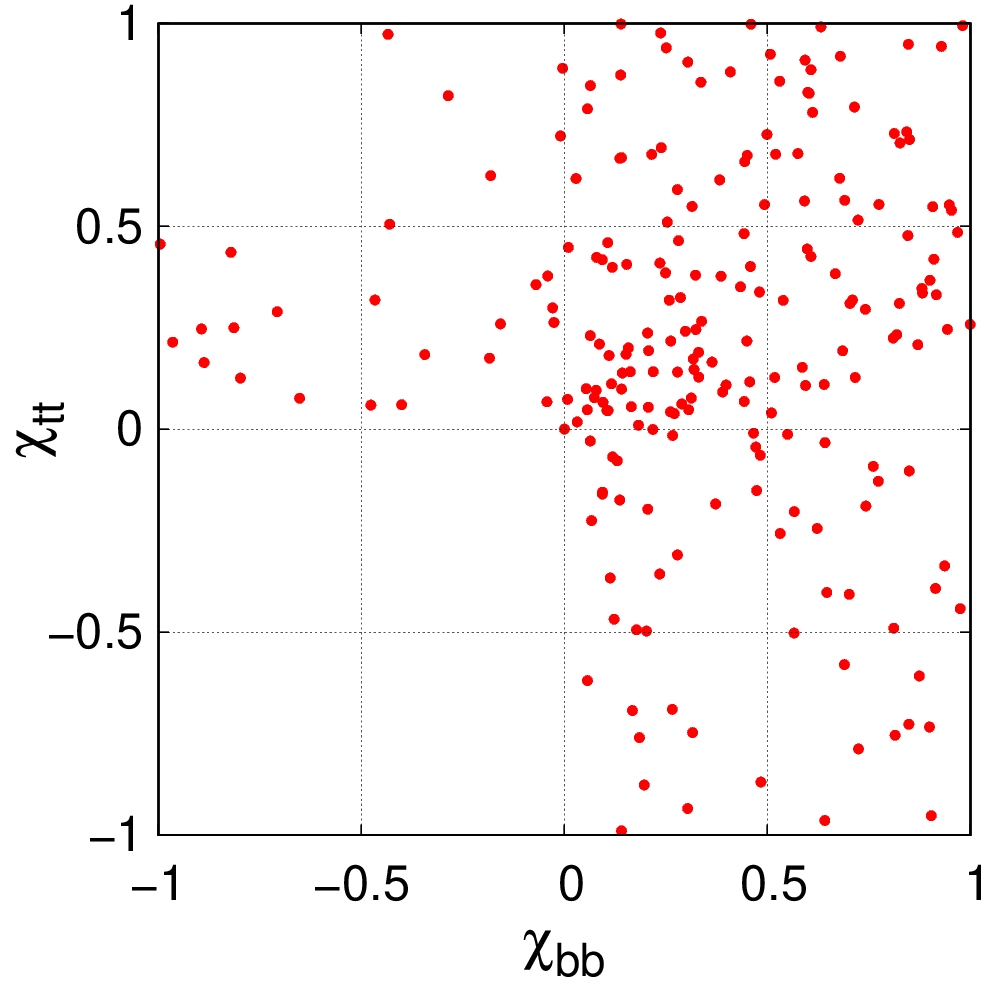}
		\caption{Scatter plot in the $\chi_{bb}-\chi_{tt}$ plane. Red points represent parameter combinations allowed by allowed by $R_{\rm quark}$ as defined in Eq. \eqref{eq:rquark}.}\label{chitt-chibb}
	\end{figure}

Here, $\lambda_t = V_{tb} V_{ts}^*$. The functions $C_{7,XY}^0(y)$ and $C_{7,YY}^0(y)$ are defined in~\cite{PhysRevD.58.074004}, with $y_j = m_{u_j}^2 / m_{H^+}^2$. The corresponding contribution to $C_7^{(R)}$ is suppressed by light quark masses and flavour-changing couplings.
 The lower mass range, $114\leq M_{H^\pm}\leq 140$, is particularly motivated by the recent excess of events reported by the ATLAS collaboration \cite{ATLAS:2023bzb}. It is worth noting that the parameters $\chi_{tt}$ and $\chi_{bb}$ play a crucial role in attenuating the magnitude of the $g_{\bar{d}uH^{+}}$ and $g_{\bar{u}dH^{-}}$ couplings. This attenuation allows for lighter charged scalar masses in the 2HDM-III compared to the 2HDM-I and -II, where the couplings are fixed as $g_{\bar{u}dH^{-}}^{\rm I}=g_{\bar{d}uH^{+}}^{\rm I}= 1/\tan\beta$, $g_{\bar{u}dH^{-}}^{\rm II}= 1/\tan\beta$ and $g_{\bar{d}uH^{+}}^{\rm II}= \tan\beta$. 
		In Fig. \ref{chitt-chibb}, the $\chi_{bb}-\chi_{tt}$ plane is displayed, with red points representing parameter combinations satisfying the experimental constraints on $R_{\rm quark}$ as defined in Eq. \eqref{eq:rquark}. 		
		Meanwhile, the scanned parameters are summarised in Tab. \ref{scan2}.
	\begin{table}
		
		\begin{centering}
			
			\begin{tabular}{|c|c|}
				\hline 
				Parameter & Scanned range\tabularnewline
				\hline 
				\hline 
				$\tan\beta$ & $[0.1,20]$\tabularnewline
				\hline 
				$\chi_{tt}$ & $[-1,1]$\tabularnewline
				\hline 
				$\chi_{bb}$ & $[-1,1]$\tabularnewline
				\hline 
				$M_{H^{\pm}}$ & $[114,140]$ GeV\tabularnewline
				\hline 
			\end{tabular}
	\caption{Scanned range of the parameters. The values of $\chi_{ij}$ are set to 1 for fermions $i$ and $j$ not included in the Table (in general $i\neq j$). }\label{scan2}		\par\end{centering}
	\end{table}
	
\begin{figure}[htbp]
    \centering
    % Primera fila: (a) y (b) lado a lado
    \begin{subfigure}{0.45\textwidth}
        \centering
        \includegraphics[scale=0.31]{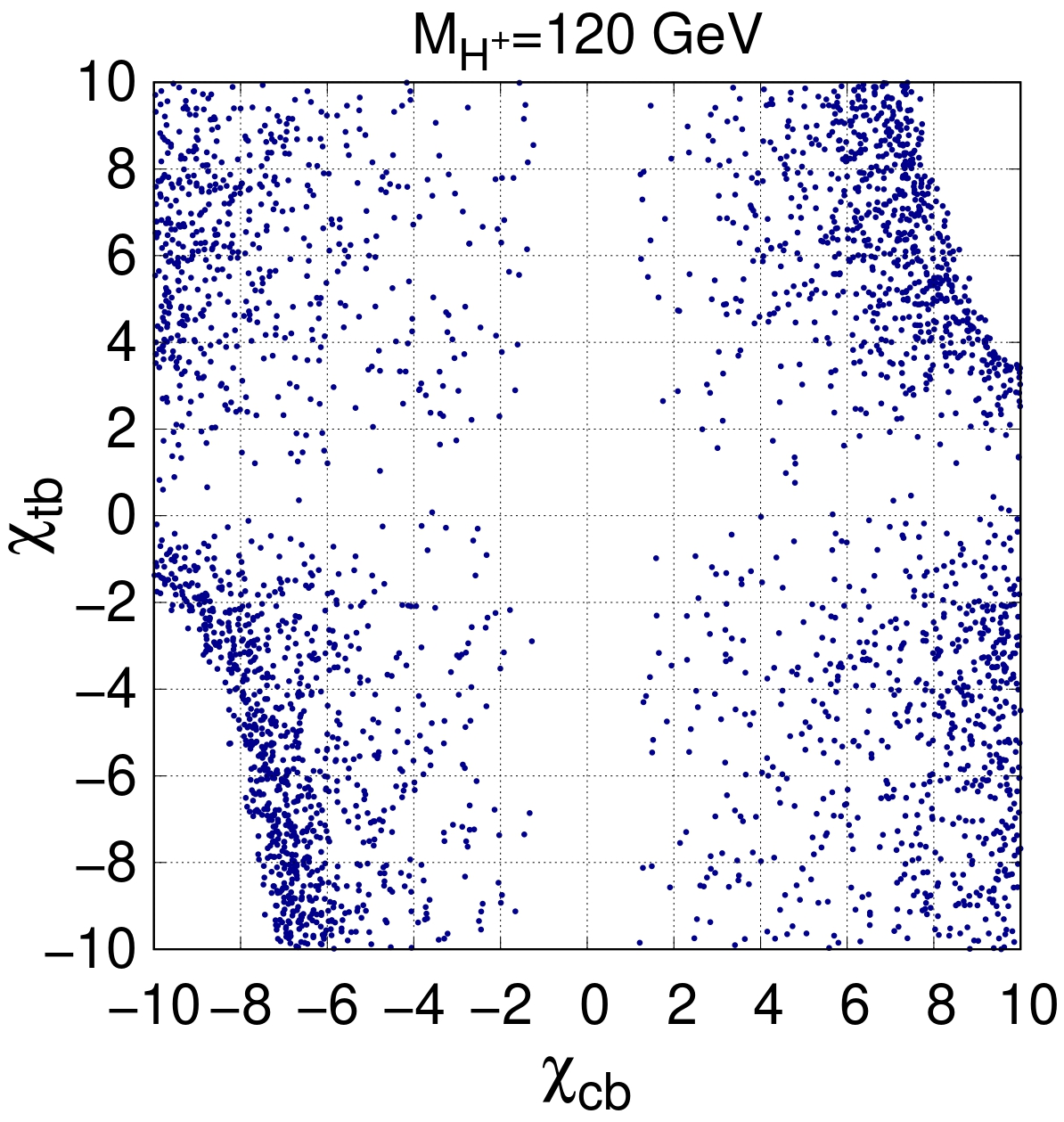}
        \caption{}
    \end{subfigure}
    \hfill
    \begin{subfigure}{0.45\textwidth}
        \centering
        \includegraphics[scale=0.31]{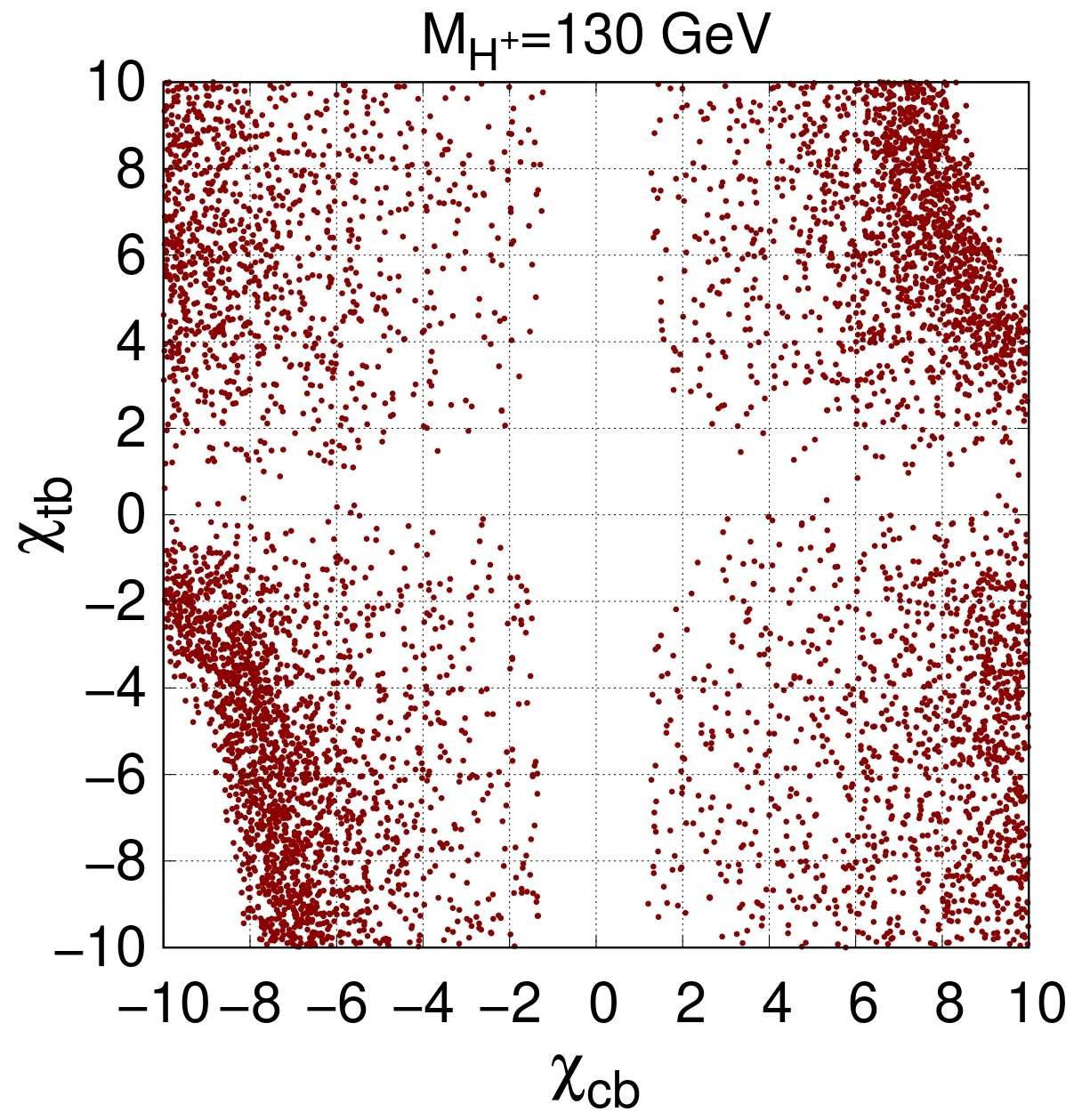}
        \caption{}
    \end{subfigure}

    \vspace{1em}

    % Segunda fila: (c) centrada
    \begin{subfigure}{0.45\textwidth}
        \centering
        \includegraphics[scale=0.31]{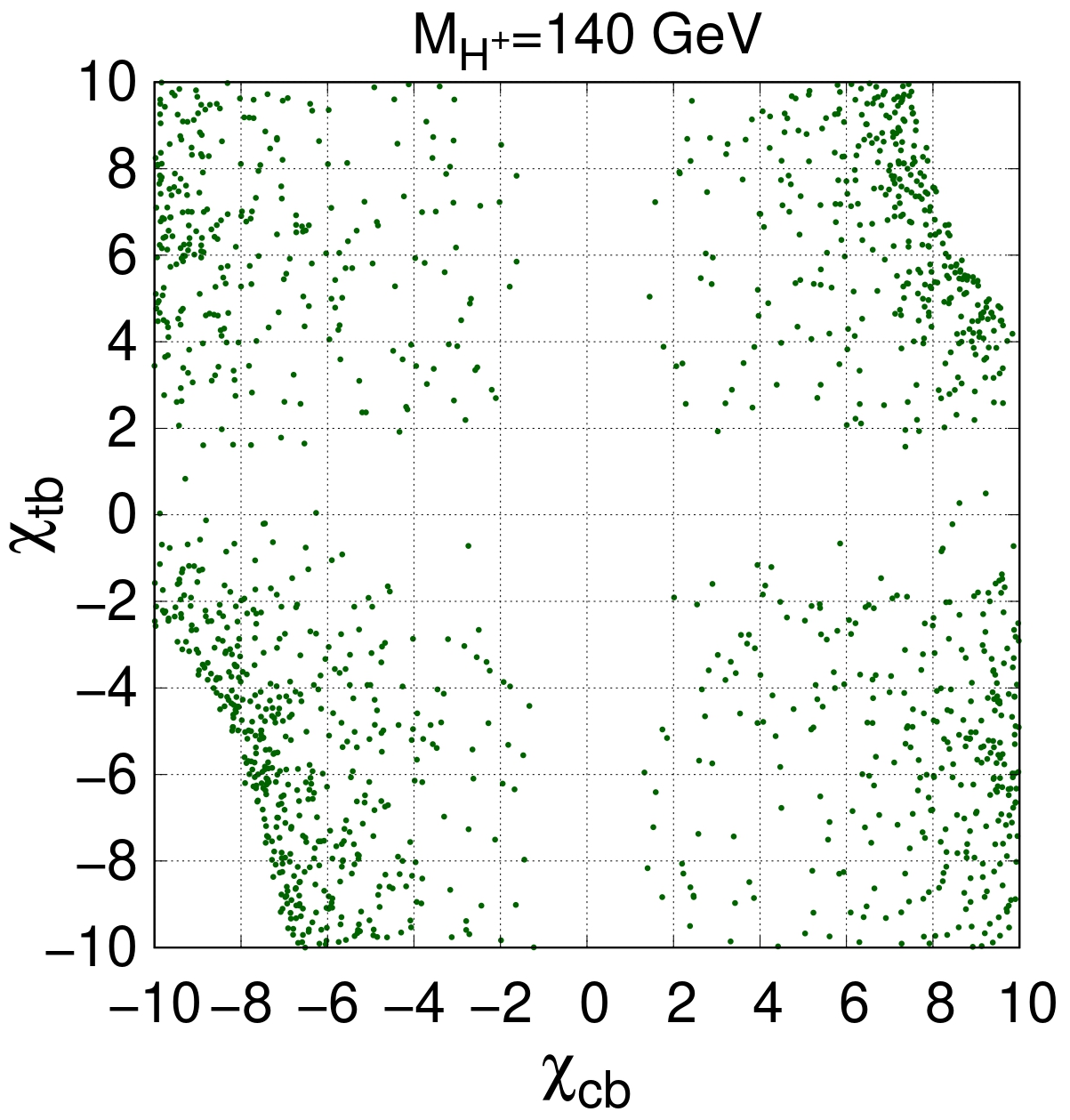}
        \caption{}
    \end{subfigure}

    \caption{$\chi_{cb}$--$\chi_{tb}$ plane. The coloured points indicate the values of $\chi$ parameters that accommodate the current excess of events reported by the ATLAS collaboration. (a) Blue points represent parameter values that explain the excess for $M_{H^{\pm}}=120$ GeV. (b) and (c) show the same as in (a) but for $M_{H^{\pm}}=130$ GeV and $M_{H^{\pm}}=140$ GeV, respectively.}
    \label{excessMHp}
\end{figure}

	%Because we are interested in analysing a regimen of low charged scalar masses 
	Given that this analysis focuses on light charged scalar masses
	($M_{H^{\pm}}<m_t+m_b$), it is important to highlight that that the model under consideration could potentially explain the recent slight excess in data, with a significance of around $3\sigma$, reported by the ATLAS collaboration for $M_{H^{\pm}}=130$ \cite{ATLAS:2023bzb}. This analysis used a data set of collisions $pp$ collected at a centre of mass energy of $\sqrt{s}=13$ TeV, corresponding to an integrated luminosity of 139 fb$^{-1}$. The search targeted a data sample enriched in top quark pair production, where one top quark decays into a leptonically decaying $W$ boson and a bottom quark, while the other top quark may decay into a $H^{\pm}$ boson and a bottom quark. The model independent exclusion at the 95\% CL on the product of BRs, $\mathcal{BR}=\mathcal{BR}(t\to H^{\pm}b)\times \mathcal{BR}(H^{\pm}\to cb)$ was reported as a function of $M_{H^{\pm}}$ ~\cite{ATLAS:2023bzb}..
	In Fig. \ref{excessMHp} we present the $\chi_{cb}-\chi_{tb}$ plane for $0.1<\tan\beta<20$, $\chi_{ij}=1$, $-10<\chi_{\mu\mu}<10$. Blue, red, and green points correspond to parameter values that can accommodate the current excess for $M_{H^{\pm}}=120\,(130,\,140)$ GeV, respectively.
	
We highlight two key characteristics:
	\begin{enumerate}
		\item The density of red points is significantly higher compared to blue and green points. This is because the red region corresponds to parameter values associated with the excess for $M_{H^{\pm}}=130$ GeV, which represents the largest excess reported by the ATLAS collaboration,. In contrast, the excess for $M_{H^{\pm}}=140$ GeV is less pronounced, as illustrated in Fig. 8 of Ref. \cite{ATLAS:2023bzb}.  
		\item Based on our scan over the model parameter space, as shown in Table \ref{scan1}, the parameters $\chi_{tb}$ and $\chi_{cb}$ play a crucial role in accommodating the current excess due to their high sensitivity. This is a distinctive feature of the 2HDM-III, as the 2HDM-I, II, Lepton Specific and Flipped, which lack these parameters. It is worth noting that several parameters $\chi_{ij}$, associated with fermions $i$ and $j$, influence the total width decay of the charged scalar boson $H^{\pm}$. In our analysis, we have set $\chi_{ij}=1$ for simplicity.    
	\end{enumerate}
	\begin{table}
		
		\begin{centering}
			
			\begin{tabular}{|c|c|}
				\hline 
				Parameter & Scanned range\tabularnewline
				\hline 
				\hline 
				$\tan\beta$ & $[0.1,20]$\tabularnewline
				\hline 
				$\chi_{tb}$ & $[-10,10]$\tabularnewline
				\hline 
				$\chi_{cb}$ & $[-10,10]$\tabularnewline
				\hline 
				$\chi_{\mu\mu}$ & $[-10,10]$\tabularnewline
				\hline 
				$M_{H^{\pm}}$ & $[114,140]$ GeV\tabularnewline
				\hline 
			\end{tabular}
\caption{Scanned range of the parameters. The values of $\chi_{ij}$ are set to 1 for fermions $i$ and $j$ not included in the Table (in general $i\neq j$). }\label{scan1}			\par\end{centering}
	\end{table}

\subsubsection{Oblique parameters}
\label{sec:oblique}
We require that the Peskin-Takeunchi parameters $S$, $T$ and $U$ lie within the experimental results \cite{ParticleDataGroup:2024cfk}, 
	\begin{itemize}
		\item $S=-0.04\pm 0.1$,
		\item $T=0.01\pm 0.12$,
		\item $U=-0.01\pm 0.09$.   
	\end{itemize}
        These oblique parameters quantify the effects of new physics on electroweak precision observables by modifying the propagation of the gauge bosons. The parameter $S$ reflects new contributions to neutral current processes via the $Z$ boson self energy; $T$ measures the breaking of custodial symmetry through differences in the $W$ and $Z$ propagators at zero momentum; and $U$ accounts for momentum dependent corrections to charged current processes involving the $W$ boson. These parameters are defined with the SM contributions subtracted and offer a model independent framework for constraining extended scalar sectors \cite{GRIMUS200881}. The expressions for $S$, $T$, and $U$ are provided in Appx.~\ref{ObParam}. For a scenario with $\cos(\alpha-\beta)=0.01$ and $0.1 \leq \tan\beta \leq 15$, the oblique parameters impose stringent constraints on the mass differences among the scalars predicted by the model.
  
\begin{figure}[!htb]
  	\centering
  	{\includegraphics[scale=0.31]{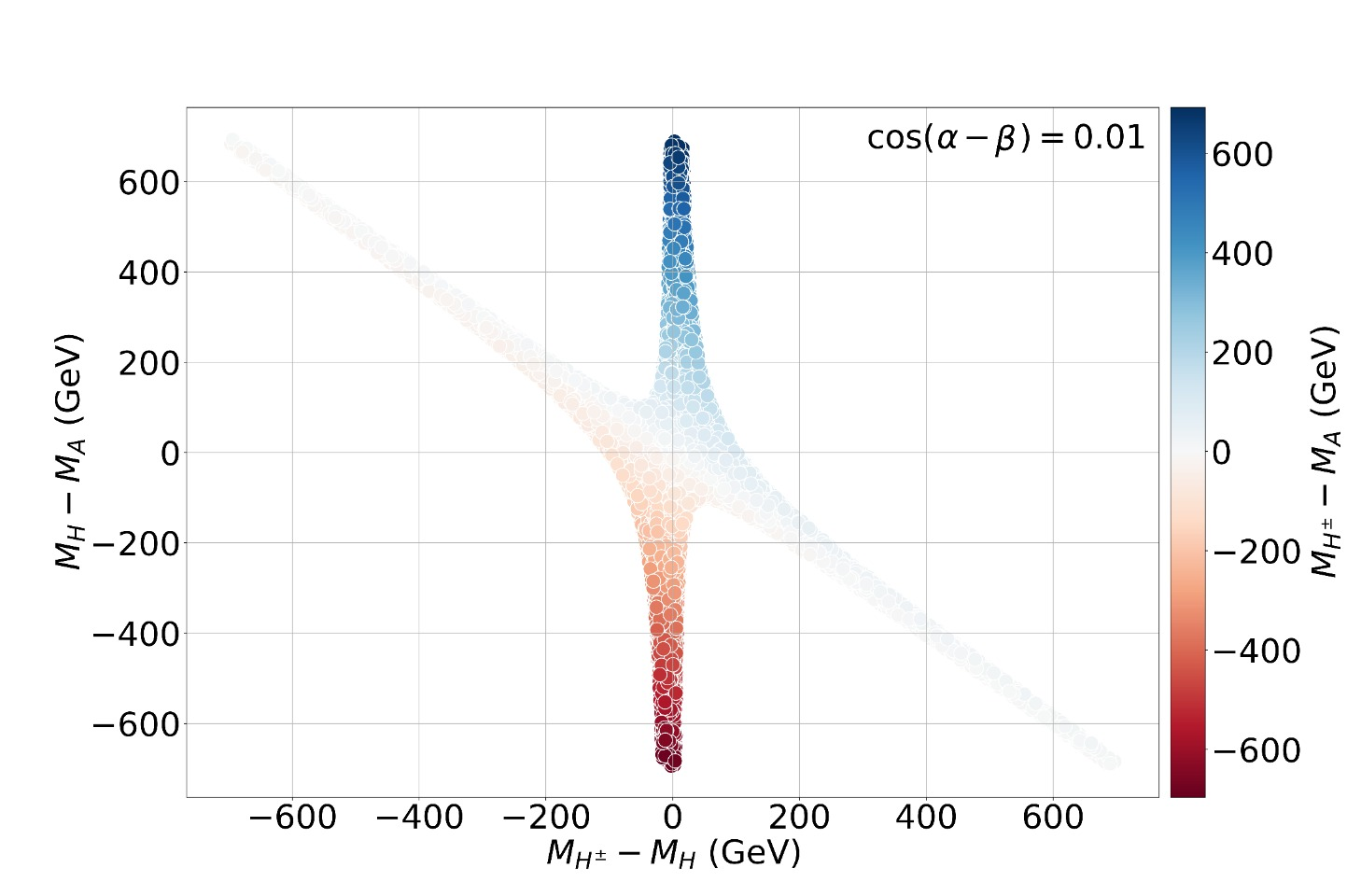}}
  	\caption{Allowed points (simultaneously) for the difference of scalar masses by oblique parameters.}\label{OP}
  \end{figure}

  We present in Fig. \ref{OP} a scattering plot where the points correspond to values of $M_H$, $M_A$ and $M_{H^+}$ that simultaneously satisfy the constraints imposed by the oblique parameters $U$, $S$, $T$. We observe that the dominant contribution to the production cross-section $\sigma(pp\to H^- H^+)$ arises from the on-shell decay  $H\to H^+ H^-$. This could lead to a suppression of the signal cross-section if the masses $M_{H^\pm}$, $M_H$, $M_A$ are nearly degenerate. To avoid this, we analyse scenarios where $M_{H^\pm}-M_H\leq -250$ GeV. Specifically, we consider three mass values for the neutral scalar boson $M_H=500,\,800,\,1000$ GeV. For the case $M_H=500$ GeV, the charged scalar mass is constrained to $M_{H^\pm}\leq 250$ GeV corresponding to $M_H-M_A\approx 250$ GeV and $M_{H^\pm}\approx M_A$. Although the oblique parameters $U$, $S$, and $T$ impose stringent constraints on the mass differences, the scenarios proposed below allow us to evade their constraints, as well as restrictions from collider searches and LFV processes.

 In summary, based on our systematic exploration of the 2HDM-III parameter space, we propose three phenomenologically viable scenarios for collider studies, which are outlined in Table \ref{scenarios}. These scenarios will serve as benchmarks for the simulations discussed in the subsequent section.
		 \begin{table}[!htb]
		  	\begin{centering}
		 		\begin{tabular}{ccccccc}
		 			\hline 
		 			Scenario & $\tan\beta$&$\chi_{tt}$ & $\chi_{bc}$ & $\chi_{\mu\mu}$ & $\cos(\alpha-\beta)$ & $M_{H}$(GeV)\tabularnewline
		 			\hline 
		 			\hline 
		 			$S1$ & 1 &0.1& 10 & 10 & 0.01 & 500, 800, 1000\tabularnewline
		 			\hline 
		 			$S2$ & 5 &0.1& 5 & 5 & 0.01 & 500, 800, 1000\tabularnewline
		 			\hline 
		 			$S3$ & 10 &0.1& 1 & 1 & 0.01 & 500, 800, 1000\tabularnewline
		 			\hline 
		 		\end{tabular}
		 		\par\end{centering}
                \caption{Benchmark scenarios to be used in the subsequent calculations.}\label{scenarios}
		 \end{table}
	 
		 	Through these simulations, we seek to explore the phenomenological consequences of the model and identify potential signatures that could be detected in forthcoming collider experiments.

\section{\label{Flavon Singlet Model}Flavon Model}

In addition to the 2HDM-III, we explore the Flavon Singlet Model or \ac{FNSM}, a different extension of the SM that introduces a complex singlet $S_F$, referred to here as the Flavon. In this FNSM, the VEV of the complex singlet  \( \langle S_F \rangle = v_s/\sqrt{2} \) spontaneously breaks the global $U(1)_F$ Flavour symmetry. As a result, the physical spectrum includes a CP-even Flavon, $H_F$, and a CP-odd Flavon, $A_F$, alongside the SM-like Higgs boson, $h$.                                                                                                                              
In this section we present an analysis of the scalar and Yukawa sectors within the FNSM framework. We derive constraints from perturbativity, unitarity, and vacuum stability in a similar manner to the 2HDM-III treatment. We then incorporate experimental bounds on Higgs observables, including signal strengths, LFV processes, and rare meson decays, to highlight the viable parameter space of the FNSM.

	\subsection{Scalar Potential} 
	The scalar sector of the FNSM extends the SM by introducing the singlet complex FN scalar $S_F$, which can be expressed as,
	%%%%%%%%%%%
	%%%%%%%%%%%
	%%%%%%%%%%%
	\begin{eqnarray} 
		%	& \Phi = \left( \begin{array}{  c} 0 \\ \frac{  v + \phi^0}{\sqrt 2}\\
			%	\end{array}  \right), \label{dec_doublets}&\\ 
		& S_F = \frac {(v_s + S_R + i S_I )}{ \sqrt 2 }   , \label{dec_Singlet} &
	\end{eqnarray}
	%%%%%%%%%%%%%%%%%%%
	%%%%%%%%%%%
	%%%%%%%%%%%
	where $v_s$ denotes the VEV of the FN singlet. The scalar potential remains invariant under the $U(1)_F$ Flavour symmetry associated with the FN mechanism. Under this Flavour symmetry, the $S_F$ transforms as $S_F \to e^{i\alpha} S_F$, while the SM Higgs doublet \(\Phi=\left(\frac{v+\phi^0}{\sqrt{2}}, \phi^+\right)^T\) remains unchanged, i.e., \(\Phi \to \Phi\), it is important to note the convention used here differs from the definition above. The scalar potential allows a complex VEV of the form \((S_F)_0 = \frac{v_s}{\sqrt{2}} e^{i\alpha}\). However, in this work, we restrict our analysis to the CP-conserving scenario by setting \(\alpha = 0\). In this case, the CP-conserving Higgs potential takes the following form:
	%%%%%%%%%%%
	%%%%%%%%%%%
	%%%%%%%%%%%
\begin{align}
\label{potential}
V_0 
&= -\frac{1}{2} m_1^2\,\Phi^\dagger \Phi 
   \;-\;\frac{1}{2} m_2^2\,S_F^*S_F
   \;+\;\frac{1}{2}\,\lambda_1 \bigl(\Phi^\dagger \Phi\bigr)^2
   \nonumber \\
&\quad +\;\lambda_2\,\bigl(S_F^*S_F\bigr)^2 
       \;+\;\lambda_{3}\,\bigl(\Phi^\dagger \Phi\bigr)\bigl(S_F^* S_F\bigr).
\end{align}
After that the VEVs of the spin-0 fields $(\Phi, S_F)$ spontaneously break the $U(1)_F$ Flavour symmetry, a massless Goldstone boson appears in the physical spectrum. We introduce a soft $U(1)_F$ breaking term into the scalar potential to give it a mass,
	%%%%%%%%%%%
	%%%%%%%%%%%
	%%%%%%%%%%%
	\begin{eqnarray}
		V_{\rm soft} = -\frac{m_3^2}{2} \left (S_F^{2} + S_F^{*2} \right).  
	\end{eqnarray}
	Thus, the full scalar potential is expressed as,
	\begin{eqnarray}
		V = V_0 + V_{\rm soft}. 
	\end{eqnarray}
	%%%%%%%%%%%
	%%%%%%%%%%%
	%%%%%%%%%%%
	The parameter $\lambda_3$ in Eq. \eqref{potential} enables the mixing between the
	Flavon and Higgs fields once the $U(1)_F$ Flavour and EW symmetries are 
	spontaneously broken, thereby determining the masses of these two states. In contrast, the soft $U(1)_F$ breaking term $V_{\rm soft}$ is responsible for generating the pseudoscalar Flavon $(S_I)$ mass. Upon minimising $V$, one obtains the following relations among its parameters
	\begin{eqnarray} 
		m_{1} ^2  &=&  v^2 \lambda_1 + v_s^2 \lambda_{3},   \\
		m_{2} ^2 &=& -2 m^2_{3} + 2 v_s^2 \lambda_2 + v^2 \lambda_{3}.
	\end{eqnarray}
	%%%%%%%%%%%
	%%%%%%%%%%%
	%%%%%%%%%%% 
	Since all the parameters of the scalar potential are taken to be real, the imaginary and real components remain separate. Consequently, the CP-even mass matrix can be expressed in the $(\phi_0, S_R)$ basis as
	%%%%%%%%%%%
	%%%%%%%%%%%
	%%%%%%%%%%% 
	\begin{equation} 
		M^2_S =
		\left( \begin{array}{cc} 
			\lambda_1 v^2      &  \lambda_{3} v v_s \\
			\lambda_{3}v v_s   &  2 \lambda_2 v_s^2
		\end{array}  \right),
	\end{equation} 
	%%%%%%%%%%%
	%%%%%%%%%%%
	%%%%%%%%%%%
	whose mass eigenstates are derived via the $2\times 2$ rotation,
	%%%%%%%%%%%
	%%%%%%%%%%%
	%%%%%%%%%%%
	\begin{eqnarray} 
		\phi^0   &=&\cos\alpha \  h + \sin\alpha  \  H_F,   \\
		S_R    &=& -\sin\alpha \  h + \cos\alpha \  H_F, 
	\end{eqnarray}
	%%%%%%%%%%%
	%%%%%%%%%%%
	%%%%%%%%%%% 
	where $\alpha$ denotes the mixing angle. We identify $h$ with the SM-like Higgs boson with a mass $M_h$ = 125.5 GeV. While the other CP-even eigenstate $H_F$ is associated with the Flavon. The CP-odd Flavon is denoted as $A_F \equiv S_I$ and acquires its mass through the soft breaking term $V_{\rm {soft}}$ such that $M^2_{A_F} = 2m_3^2$. It is assumed that both the CP-even $H_F$ and the CP-odd $A_F$ are heavier than $h$.
	In this analysis, we treat the physical masses $M_{S}\, (S=h,\,H_F,\,A_F)$ and the mixing angle $\alpha$ as free parameters. Their relations with the quartic couplings of the scalar potential in Eq.~(\ref{potential}) can be expressed as,
	%%%%%%%%%%%
	%%%%%%%%%%%
	%%%%%%%%%%%
	%%%%%%%%%%%%
	\begin{eqnarray}\label{eq:relate}
		\lambda_1&=& \frac{ \cos\alpha^2 M_h^2+\sin\alpha^2 M_{{H_F}}^2}{v^2},\nn\\
		\lambda_2&=& \frac{M_{{A_F}}^2+{\cos\alpha }^2 M_{{H_F}}^2+{\sin\alpha }^2 M_h^2}{2 v_s^2},\\
		\lambda_3&=& \frac{ \cos\alpha \, \sin\alpha }{ v v_s} \, ( M_{{H_F}}^2 -  M_h^2).\nn
	\end{eqnarray} 
	%%%%%%%%%%%%%%
	%%%%%%%%%%%
	%%%%%%%%%%%
	%%%%%%%%%%%
	
	%%%%%%%%%%%%%%%%%%%
	\subsection{Yukawa Lagrangian} 
	%%%%%%%%%%%%%%%%%%%%%%
	The effective $U(1)_{F}$ invariant Yukawa Lagrangian, which yields the Yukawa couplings once the $U(1)_{F}$ Flavour symmetry is spontaneously broken, is given by \cite{ Froggatt:1978nt}
	%%%%%%%%%%%
	%%%%%%%%%%%
	%%%%%%%%%%%
	\begin{align} 
		\mathcal{L}_ Y &= \rho^d_{ ij } \left( \frac{S_F}{\Lambda} 
		\right)^{q_{ ij }^d}  \bar{Q}_i d_j  \tilde \Phi 
		+ \rho^u_{ ij } \left(\frac{S_F}{\Lambda }\right)^{q_{ij }^u}\bar{Q}_i u_j 
		\Phi \nonumber\\&+ \rho^\ell_{ij}\left(\frac{S_F}{\Lambda}\right)^{q_{ij}^\ell}
		\bar{L}_i \ell_j \Phi  + \rm h.c.,
		\label{eq:fermlag}  
	\end{align} 
	%%%%%%%%%%%
	%%%%%%%%%%%
	%%%%%%%%%%%
	In this expression, $\rho^{f}_{ij}$ ($f=u,\,d,\,\ell$) are dimensionless parameters, typically of order $\mathcal{O}(1)$. The quantities $q_{ij}^f$ denote the Abelian charges that are assigned in order to reproduce the observed fermion masses, and $\Lambda$ represent the ultraviolet mass scale. In order to derive the Yukawa coupling from the Lagrangian in \eqref{eq:fermlag}, it is necessary to spontaneously break both the $U(1)_{F}$ Flavour symmetry and the EW symmetry. After this symmetry breaking, the resulting $S f_i\bar{f}_i$ interactions emerge as it is shown in Table \ref{Diagonal_couplings}. It is important to note that to avoid significant deviations from the SM couplings, the conditions $v_s\approx\Lambda$ and $\cos\alpha\approx -1$ must be satisfied.
	\begin{table}

		\begin{centering}
			\begin{tabular}{c c}
				\hline 
				Vertex $SXX$ & Coupling\tabularnewline
				\hline 
				\hline 
				$hf_i\bar{f}_i$ & $\frac{v_{s}m_{f}}{v\Lambda^{2}}\Big(v\sin\alpha-v_{s}\cos\alpha\Big)$\tabularnewline
				\hline 
				$hZZ$ & $g\frac{m_{Z}}{c_{W}}\cos\alpha$\tabularnewline
				\hline 
				$hWW$ & $gm_{W}\cos\alpha$\tabularnewline
				\hline 
				$H_{F}f_i\bar{f}_i$ & $-\frac{v_{s}m_{f}}{v\Lambda^{2}}\Big(v_{s}\sin\alpha+v\cos\alpha\Big)$\tabularnewline
				\hline 
				$H_{F}ZZ$ & $g\frac{m_{Z}}{c_{W}}\sin\alpha$\tabularnewline
				\hline 
				$H_{F}WW$ & $gm_{W}\sin\alpha$\tabularnewline
				\hline 
				$A_{F}f_i\bar{f}_i$ & $-\frac{v_{s}m_{f}}{\Lambda^{2}}$\tabularnewline
				\hline 
				$A_{F}ZZ$ & $0$\tabularnewline
				\hline 
				$A_{F}WW$ & $0$\tabularnewline
				\hline 
			\end{tabular}
			\par\end{centering}
            \caption{Diagonal $S XX$ interactions, ($S=h,\,H_F,\,A_F$).}\label{Diagonal_couplings}
	\end{table}	
    
	Meanwhile, to induce non-diagonal $S f_i\bar{f}_j$ interactions, we proceed as follows. In the unitary gauge, one cab perform a first-order expansion of the neutral component of the neutral component of the heavy Flavon field $S_F$ around its $v_s$. 
	%%%%%%%%%%%
	%%%%%%%%%%%
	%%%%%%%%%%%
	\begin{align} 
		\Bigg(\frac{S_F}{\Lambda_F}\Bigg)^{q_{ ij }} &=\left(\frac{v_s+S_R+iS_I}{  \sqrt 2\Lambda_F}  \right)^{q_{ ij }} \nonumber\\&
		\simeq \left(\frac{v_s}{ \sqrt 2\Lambda_F}  \right)^{q_{ ij }} \left[1+q_{ ij }\left(\frac{S_R+iS_I}{v_s}\right)\right],
	\end{align}
	%%%%%%%%%%%
	%%%%%%%%%%%
	%%%%%%%%%%% 
    This expansion leads to the following interaction Lagrangian upon replacing the mass eigenstates,%%%%%%%%%%%
	%%%%%%%%%%%
	%%%%%%%%%%%
	\begin{eqnarray} \label{YukaLagrangian} 
		\mathcal {L}  _Y &=& \frac 1  v [\bar{U}  M^u U+\bar{D}  M^d D+\bar{L} M^ \ell L](c_ \alpha h+s_ \alpha H_F) \nonumber\\
		&+&\frac{v }{ \sqrt 2 v_s } [\bar{U}_i\tilde Z_{ij}^u U_j+\bar{D}_i\tilde Z_{ij}^d D_j+\bar{L}_i\tilde Z_{ij}^ \ell  L_j]\nonumber\\&\times&
		(-\sin\alpha h+\cos\alpha H_F+iA_F)+ \rm h.c.,  
	\end{eqnarray} 
	%%%%%%%%%%%
	%%%%%%%%%%%
	%%%%%%%%%%%
	Here, $M^f$ denotes the diagonal fermion mass matrix. We encapsulate the Higgs-Flavon couplings in the matrices $\tilde{Z}_{ij}^f=U_L^f Z_{ij}^f U_L^{f\dagger}$. In the flavour basis, the $Z_{ij}^f$ matrix elements are given by
	%%%%%%%%%%%
	%%%%%%%%%%%
	%%%%%%%%%%%
	\begin{equation}
		Z_{ij}^f= \rho_{ij}^f \left(\frac{v_s}{\sqrt 2\Lambda_F}
		\right)^{q_{ij}^f}q_{ij}^f,
	\end{equation}
	%%%%%%%%%%%
	%%%%%%%%%%%
	%%%%%%%%%%%
	which, in general, remains non-diagonal even after the mass matrices are diagonalised, thereby giving rise to flavour-violating interactions.

    Finally, in addition to the standard Yukawa couplings, we also require the $H_Fhf_i\bar{f}_j$ couplings for our calculations. In the FNSM, these interactions are described by
	\begin{equation}
		H_Fhf\bar{f}=\frac{m_f v_s}{\sqrt{2}\Lambda^2}(1-2\cos^2\alpha).
        \label{vertfnsm}
	\end{equation} 
    As a particular case, we explore the scenario where $f = b$. This choice is motivated by experimental reports on Higgs pair searches \cite{ATLAS:2023qzf, ATLAS:2024lsk, ATLAS:2023gzn, ATLAS:2024yuv, ATLAS:2023elc}.

	\section{Constraints on the FNSM Parameter Space}
	\label{se:conts}
	%%%%%%%%%%%%%%%%%%%%%%%

	To compute a realistic numerical analysis of the signals proposed in this project, namely 
    \[
    pp\to H_F\to hb\bar{b}\quad \text{(with } h\to b\bar{b},\, \gamma\gamma\text{)},
    \]  
    it is necessary to constrain the free FNSM parameters involved in the forthcoming calculations. These free parameters are:
	\begin{itemize}
		\item The mixing angle $\alpha$ of the real components of the doublet $\Phi$ and the FN singlet $S$.
		\item FN singlet VEV $v_s$.
		\item The ultraviolet mass scale $\Lambda$.
		\item CP-even scalar mass $M_{H_F}$.
	\end{itemize} 
	
    These parameters can be constrained by several theoretical requirements, such as absolute vacuum stability, triviality, perturbativity, and unitarity of scattering matrices and by various experimental data, mainly the upper limits on the production cross-sections of additional Higgs states from LHC measurements. We also consider bounds from \ac{LFVp}, such as $L_i\to \ell_j\ell_k\bar{\ell}_k$ and $\ell_i\to \ell_j\gamma$, as well as measurements of $\mathcal{BR}(B_{s,\,d}^0\to\mu^+\mu^-)$. Finally, the anomalous magnetic moments of the muon and electron $\Delta a_{\mu}$ and $\Delta a_{e}$, respectively, are also taken into account.
	
	\subsection{Theoretical Constraints}
	\subsubsection{Stability of the Scalar Potential}
	It is essential to ensure that the scalar potential in Eq.~\eqref{potential} is bounded from below, meaning that it does not approach negative infinity in any direction of the field space ($h, H_F, A_F$) at large field values. At high field strengths, the quartic terms dominate over the quadratic ones. Therefore, the conditions for absolute stability read \cite{Khan:2014kba},
    \[
    \Bigl(\lambda_1,\, \lambda_2,\, \lambda_3 + \sqrt{2 \lambda_1 \lambda_2}\Bigr) > 0.
    \]
    The quartic couplings are evaluated at the scale $\Lambda$ using the Renormalisation Group Evolution (RGE) equations. If the scalar potential in Eq.~(\ref{potential}) exhibits a metastable EW vacuum, these conditions must be appropriately modified \cite{Khan:2014kba}. In order to constrain the scalar field masses $M_{\phi}$, the VEV of the complex singlet $v_s$, and the mixing angle $\alpha$, one can use Eqs.~\eqref{eq:relate} to translate the stability limits into constraints on the model parameters.

	\subsubsection{Perturbativity and Unitarity Constraints}\label{theoretical_constraints}
	
	%%%%%%%%%%%%%%%%%%%%%%%%%%%%%%%%%%%%%%%%%%%%
	%\begin{figure}[!t]
	%	\begin{center}
		%		\includegraphics[scale=0.155]{LamSPlot.png}
		%		\includegraphics[scale=0.155]{Lam11Plot.png}
		%		\includegraphics[scale=0.155]{LamUPlot.png}
		%	\end{center}
	%	\caption{ In the first two plots we show the perturbative bounds on the quartic couplings $\lambda_{2,3}$ while the third plot shows the stringent unitary bounds on $\lambda_U$.}
	%	\label{PLimit}
	%\end{figure}
	%%%%%%%%%%%%%%%%%%%%%%%%%%%%%%%%%%%%%%%%%%%%%%%%%%%
	
	%%%%%%%%%%%%%%%
	The upper limits provided in Eq.~\eqref{UL4pi} are essential to ensure that the radiatively corrected scalar potential of the FNSM remains perturbative at all energy scales ~\cite{Arroyo-Urena:2022oft},
\vspace{-2ex}
	%%%%%%
	%%%%%%%%%%%
	%%%%%%%%%%%
	%%%%%%%%%%%
	\begin{equation}\label{UL4pi}
		\mid \lambda_1, \lambda_2, \lambda_3\mid \leq 4 \pi.
	\end{equation}
	%%%%%%%
	%%%%%%%%%%%
	%%%%%%%%%%%
	%%%%%%%%%%%
	%%%%%%
	
	The quartic couplings emerging from the scalar potential are subject to stringent constraints arising from the unitarity of the $S$-matrix. In practice, even at large field values, one can derive the $S$-matrix by analysing various $2\to2$ scattering processes involving interactions between pseudo scalar bosons, scalars, and gauge bosons—denoted here as $(P)S$, $V$, and $(P)S-V$ interactions. Unitarity requires that all eigenvalues of the $S$-matrix remain below $8\pi$ \cite{Cynolter:2004cq,Khan:2014kba}. In the context of the FNSM, and by employing the equivalence theorem, the resulting unitary bounds can be expressed as,	\begin{eqnarray}\label{UnitaryBounds}
		\lambda_1 \leq 16 \pi  \quad {\rm and} \quad \Big| {\lambda_1}+{\lambda_2} \pm \sqrt{ ({\lambda}_1-{\lambda_2})^2+(2/3 \lambda_3)^2}\Big|\equiv |\lambda_U^{\pm}| \leq 16/3 \pi.
	\end{eqnarray}
	%%%%%%%%%%%
	%%%%%%%%%%%
	%%%%%%%%%%%
	%%%%%%%%%%%%% 
	By using the Eqs.~\eqref{eq:relate}, \eqref{UL4pi} and \eqref{UnitaryBounds}, we can derive constraints on the scalar singlet VEV $v_s$, the masses of the heavy Higgs bosons $M_{H_F}$ and $M_{A_F}$, as well as on the mixing angle $\alpha$.

\begin{figure}[htb!]
	\begin{center}
		\includegraphics[scale=0.31]{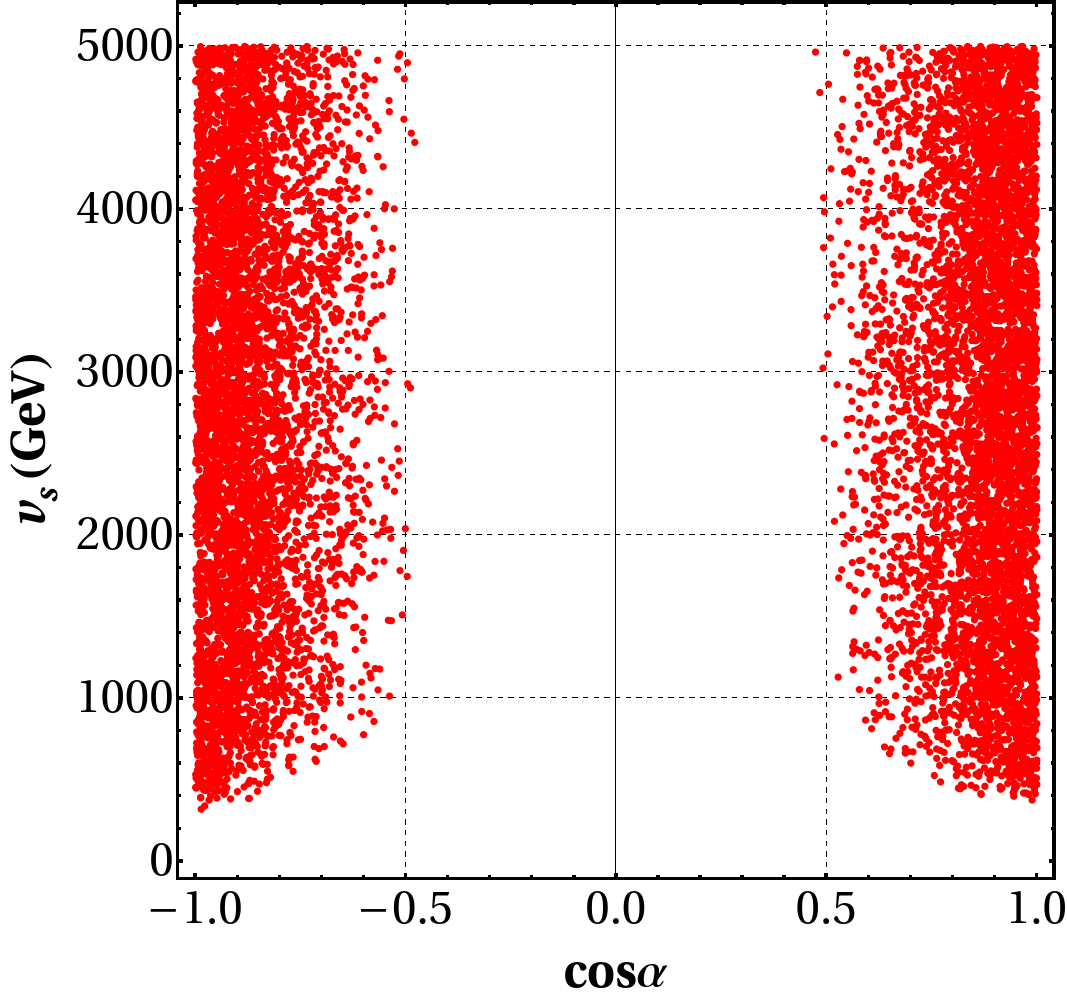}
	\end{center}
	\caption{The VEV of the FN singlet $v_s$ as a function of the cosine of the mixing angle $\alpha$. The red points denote the region of parameter space that is consistent with all the theoretical constraints discussed in the text.}
	\label{PLimit}
\end{figure}

	Figure~\ref{PLimit} displays the $\cos\alpha$–$v_s$ plane, where each point represents a parameter combination that satisfies all the theoretical constraints, including perturbativity and the unitarity conditions imposed on the $S$-matrix. For this analysis, we generated a set of random points that comply with the relations given in Eqs.~\eqref{eq:relate}, \eqref{UL4pi}, and \eqref{UnitaryBounds}. The ranges of the scanned parameters to achieve that proposal are summarised in Table~\ref{RPTC}.

	\begin{table}

		\begin{centering}
			\begin{tabular}{|c|c|}
				\hline 
				Parameter & Interval\tabularnewline
				\hline 
				\hline 
				$\cos\alpha$ & {[}-1, 1{]}\tabularnewline
				\hline 
				$M_{H_{F}}$ & 800-1500 GeV\tabularnewline
				\hline 
				$M_{A_{F}}$ & 800-1500 GeV\tabularnewline
				\hline 
				$v_{s}$ & $v$-5000 GeV\tabularnewline
				\hline 
			\end{tabular}
			\par\end{centering}
            \caption{Scanned parameters for theoretical viability in the FNSM.}
            \label{RPTC}
	\end{table}
	
	We find that the upper bound $|\lambda_U^+| \leq \frac{16\pi}{3}$ represents the most stringent constraint on the scalar quartic couplings. This bound translates into a lower limit on the scalar singlet VEV, namely $v_s \geq (276,\,345,\,415,\,519)$ GeV for specific choices of the heavy Higgs masses $M_{H_F} = M_{A_F} = (800,\,1000,\,1200,\,1500)$ GeV and for $\cos\alpha = -0.995$. It should be noted that our analysis is performed only at the EW scale, and a detailed Renormalisation Group Evolution (RGE) analysis is beyond the scope of this thesis. Moreover, we observe a clustering of allowed points near $\cos\alpha \approx 1$ and $\cos\alpha \approx -1$, this is expected, since $\alpha$ must remain close to zero in order to avoid large deviations from the SM $hf\bar{f}$ coupling.

    It is also important to emphasise that the range of masses considered in Table~\ref{RPTC} was chosen based on a previous study \cite{Arroyo-Urena:2022oft}, in which it was shown that $M_{H_F} > 800$ GeV is necessary to alleviate the constraints imposed by the cross-section limits on the process $pp \to S \to hh$ reported by the ATLAS Collaboration \cite{ATLAS:2021ifb}, where $S$ represents a resonant pseudo-scalar particle.

	\subsection{Experimental Constraints}
	\subsubsection{LHC Higgs boson data}
	
	We complement the theoretical constraints by incorporating experimental measurements from Higgs boson physics \cite{ParticleDataGroup:2024cfk}. In particular we consider the \emph{signal strengths} defined as
    \begin{equation}
	\mathcal{R}_{X} = \frac{\sigma(pp \to h) \cdot \mathcal{BR}(h \to X)}{\sigma(pp \to h^{\text{SM}}) \cdot \mathcal{BR}(h^{\text{SM}} \to X)},
    \end{equation}
    where $\sigma(pp \to H_i)$ denotes the production cross-section of $H_i$ (with $H_i = h,\, h^{\text{SM}}$); here, $h$ represents the SM-like Higgs boson arising in the extended model, while $h^{\text{SM}}$ is the SM Higgs boson. The BR $\mathcal{BR}(H_i \to X)$ corresponds to the decay of $H_i$ into final states $X$, with $X$ taking values such as $b\bar{b}$, $c\bar{c}$, $\tau^-\tau^+$, $\mu^-\mu^+$, $WW^*$, $ZZ^*$, or $\gamma\gamma$.

    \subsubsection{Lepton Flavour Violating Processes}

    In addition, we examine several LFVp to impose further constraints on the parameters pertinent to our subsequent calculations. Specifically, we utilise (i) the upper limits on $\mathcal{BR}(L_i \to \ell_j \ell_k \bar{\ell}_k)$ and $\mathcal{BR}(\ell_i \to \ell_j \gamma)$, and (ii) the measurements of $\mathcal{BR}(B_{s,\,d}^0 \to \mu^+\mu^-)$ together with the anomalous magnetic moment of the muon, $\Delta a_{\mu}$. To facilitate the exploration of the FNSM parameter space, the model is implemented also in the \texttt{Mathematica} package \texttt{SpaceMath}.

\begin{figure}[htb!]
		\begin{center}
			\includegraphics[scale=0.4]{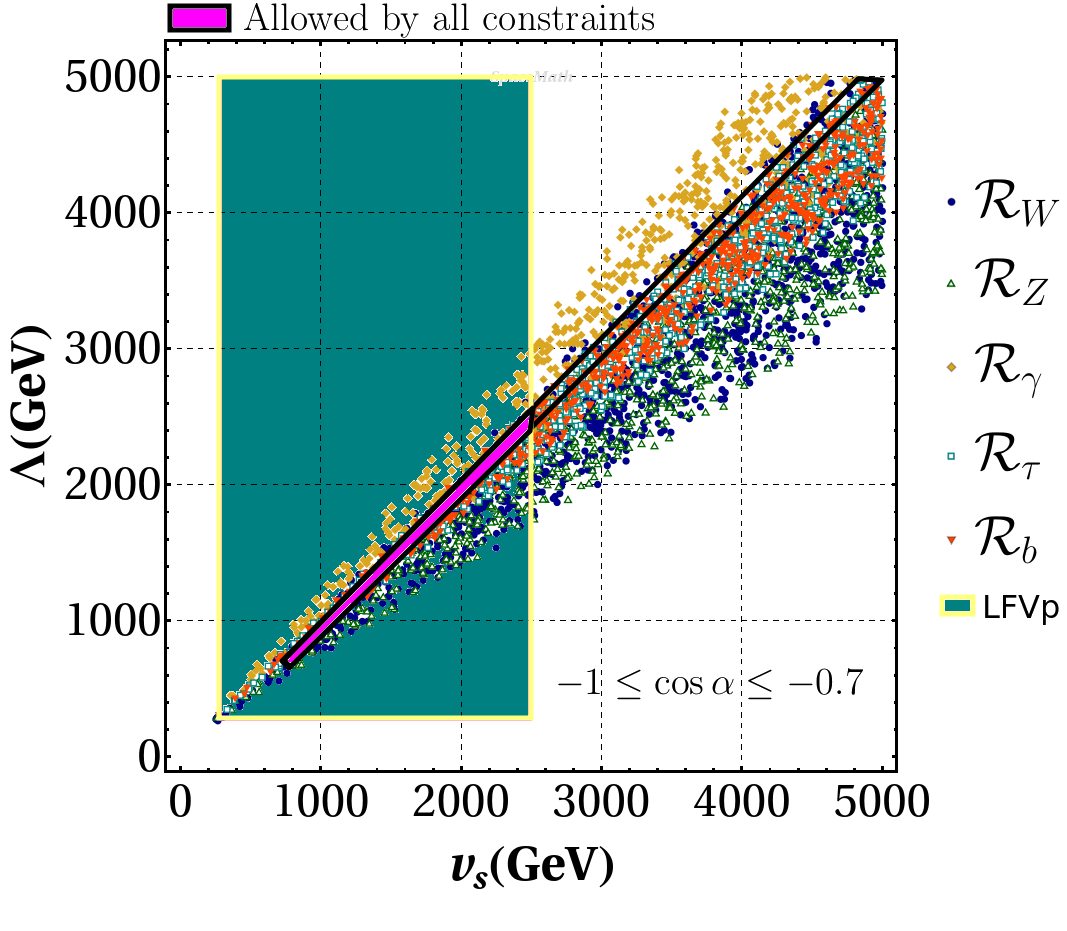}
		\end{center}
		\caption{This plot shows the ultraviolet mass scale $\Lambda$ as a function of the VEV of the complex singlet $v_s$. The area enclosed by the solid black lines corresponds to the intersection of all constraints, while the different points represent the individual $\mathcal{R}_X$ measurements. The green rectangle indicates the parameter space allowed by the LFVp constraints.}
		\label{fig:RXs}
	\end{figure}

    In Fig.~\ref{fig:RXs}, the ultraviolet mass scale $\Lambda$ is presented as a function of the VEV of the complex singlet $v_s$. Different symbols in this plot denote the individual signal strengths $\mathcal{R}_X$: blue circles represent $\mathcal{R}_W$, green triangles correspond to $\mathcal{R}_Z$, yellow diamonds indicate $\mathcal{R}_\gamma$, green squares denote $\mathcal{R}_\tau$, orange triangles correspond to $\mathcal{R}_b$, and green rectangles represent additional channels. The overlapping region common to all these individual measurements is highlighted by the area enclosed within solid black lines. Additionally, the green rectangle marks the region permitted by the LFVp constraints, while the cyan area reflects the intersection of all the theoretical and experimental constraints discussed.

    Motivated by the analysis presented in Sec.~\ref{theoretical_constraints}, we once again perform a scan over the model parameters involved in the evaluation of $\mathcal{R}_X$, as summarised in Table~\ref{Rs_Constraints}. It is noteworthy that $v_s \approx \Lambda$ is expected since the $hf\bar{f}$ coupling scales as $v_s^2/\Lambda^2$ when $\cos\alpha \approx -1$, as detailed in Table~\ref{Diagonal_couplings}. We have also explored the allowed ranges for $\mathcal{R}_{\mu}$ and $\mathcal{R}_c$ (although these results are not depicted in Fig.~\ref{fig:RXs}). However, these observables do not impose particularly stringent constraints on the FNSM.

We explored the theoretical frameworks of 2HDM-III as well as FNSM and reviewed the experimental and theoretical constraints. In the following chapter, we introduce our data analysis methodology by describing how simulated collider events are processed and subsequently analysed using \acp{BDT}. This Machine Learning based approach is aimed at optimising the signal versus background discrimination and in that way, refining our predictions for the potential discovery of new scalar particles.

	\begin{table}

		\begin{centering}
			\begin{tabular}{|c|c|}
				\hline 
				Parameter & Interval\tabularnewline
				\hline 
				\hline 
				$\cos\alpha$ & {[}-1, -0.7{]}\tabularnewline
				\hline 
				$\Lambda$ & $v$-5000 GeV\tabularnewline
				\hline 
				$v_{s}$ & $v$-5000 GeV\tabularnewline
				\hline 
			\end{tabular}
			\par\end{centering}
            \caption{Scanned parameters for $\mathcal{R}_X$ evaluation in the FNSM.}
            \label{Rs_Constraints}  
	\end{table}

	%%%%%%%%%%%%%%%%%%%%%%%%%%%%%%%%%%%%%%%%%%%%%%%%%%%%%%%%%%%%%%%%%%%%%%%%%%%%%%%%%%%%%%%%%%%%%%%%%%%%%%%%%%%%%%%%%%%%%%%%%%%%%%%%%%%%%%%%%%%%%%%%%%%%%%%%%%%%%%%%%%%%%%%%%%%%%%%%%%%%%%%%%%%%%%%%%%%%%

\chapter{Methodology BDT Data Analysis}
\label{chap:methodology}

\noindent\rule{\linewidth}{0.4pt}

In the study of BSMs, simulations are an essential tool to explore phenomenology and test theoretical models. This chapter describes the methodology used to process the simulated data, extract observables information and train \acp{BDT} to separate signal from background in the phenomenological analysis of BSM physics. \ac{BDT} is a Machine Learning model that applies a series of conditions on the input variables to progressively divide a dataset into smaller groups. In this approach, the decision trees are designed to correct the misclassifications made in previous steps \cite{Hastie:2009itz, doi:10.1142/9789811234033_0002}.

Monte Carlo event generators, such as MadGraph5 \cite{MadGraphNLO}  produce output files in the “.root” format, containing simulated collider events with detailed kinematic information. We have to distinguish the “.root” file extension from the CERN developed software framework called ROOT.
The ROOT framework is widely used in high energy physics to read, process and analyse “.root” files. However, these data files can be very large and unwieldy. While ROOT or other analysis tools can be employed to extract kinematic observables, the internal structure of .root files is not directly readable. 

To address this challenge and get direct information from .root files, it is a common practice to convert .root files into the LHC Olympics (LHCO) format \cite{Kasieczka_2021}. The LHCO format was originally introduced in the ‘LHC Olympics', a series of community challenges around the start of LHC operations where researchers were trying to reconstruct signals from ‘black box’ detector level events. This format is well known for being lightweight and text based, while preserving essential information for phenomenological studies such as the particle types, pseudorapidity, azimuthal angle, transverse momentum, and invariant mass. As described in the manual~\cite{lhco_manual}, each event in that LHCO file is structured as a block of rows.

For conveniently reading and processing these LHCO files, there is a Python module called $LHCO\_reader$ \cite{Fowlie2015LHCO}, which analyses detector level events stored in LHCO format. Nevertheless, $LHCO\_reader$ does not perform the final analysis or extract physical observables, rather, it provides an interface to read the LHCO file data quickly and easily into Python structures.

The Python \cite{python3} based framework presented here, called PrakritiMLPrep \cite{prakritimlprep}, is designed to process the large volumes of simulation data by extracting primitive and derived physical observables from the events. These computed observables are then used to classify the events, enabling the creation of datasets that can be used to train Machine Learning models and thereby improving signal versus background discrimination. This approach typically achieves a higher statistical significance than obtaining it by empirical cuts in the observables. 

By converting .root files to LHCO format, we can significantly reduce data size while retaining relevant kinematic details needed for Machine Learning training and analysis. 

Its main functionalities include user-friendly interaction with LHCO files, complete with basic .py examples to guide users in reading LHCO files and defining custom observables. This makes it easy to extend the analysis to specific decay modes or processes. Additionally, it supports data preparation for Machine Learning, focusing on the creation of datasets that blend simulated signal and background events for training models such as \acp{BDT}. It also features a flexible architecture with a modular design, allowing users to add more observables.

\subsection{The LHCO File Format}
The LHCO format is a simple, text based format used to show collider event data. It consists of a series of block events where each block contains rows. Every row corresponds to a physics object, such as a lepton, photon, jet, or \ac{MET}.

\subsection{Column Definitions and Structure}
Each row in an LHCO file contains several columns. The Table~\ref{tab:lhco_columns} summarises the standard column definitions.

\subsection{Example for Reading a LHCO}

In the following example (Table~\ref{tab:lhco_event_example}), we show how to read an LHCO file.  Despite the apparent simplicity of the LHCO format, each event block contains kinematic variables and object classifications necessary for a robust analysis. The code example below walks through these steps one at a time so we can clearly see how it works.

\begin{table}
\centering
\scriptsize
\renewcommand{\arraystretch}{1.3} % Increase row height
\resizebox{1\textwidth}{!}{%
\begin{tabular}{|c|m{2.8cm}|m{8cm}|}
\hline
\textbf{Column \#} & \textbf{Name/Type} & \textbf{Description} \\
\hline
1 & Counter & Sequential index labelling the object in the event. \\
\hline
2 & Type & Object type indicator: 0 = photon, 1 = electron, 2 = muon, 3 = hadronically decaying tau, 4 = jet, 6 = MET. \\
\hline
3 & $\eta$ (Pseudorapidity) & The pseudorapidity, defined as $\eta=-\ln\!\bigl[\tan\!\bigl(\frac{\theta}{2}\bigr)\bigr]$, where $\theta$ is the polar angle relative to the beam axis. For massless objects, $\eta$ is the rapidity $y$. \\
\hline
4 & $\phi$ (Azimuthal Angle) & The azimuthal angle around the beam axis in radians. \\
\hline
5 & $p_T$ (Transverse Momentum) & The momentum component perpendicular to the beam direction, measured in GeV. \\
\hline
6 & Invariant Mass ($jmass$) & For jets, the invariant mass calculated from the energy deposits within the jet. \\
\hline
7 & $ntrk$ (Number of Tracks) & Number of tracks associated with the object. For leptons, this number is multiplied by the charge (e.g., -1 for a muon, +1 for a positron). \\
\hline
8 & $btag$ & B-tagging information: values 1 or 2 indicate different levels of heavy-flavour tagging for jets; for other objects (e.g.\ muons), this column can carry special meaning such as the index of the closest jet. \\
\hline
9 & Had/EM Ratio & The ratio of hadronic to electromagnetic energy deposition in the associated calorimeter cells. Typically, jets yield values greater than 1, electrons or photons yield values much less than 1. For muons this field is formatted as \texttt{xxx.yy}, where \texttt{xxx} is the isolation $p_T^{iso}$ (the sum of transverse momenta in a cone of radius $R=0.4$, excluding the muon) and \texttt{yy} is the energy ratio in a surrounding grid. \\
\hline
10 and 11 & Dummy & Reserved for future use, currently set to zero. \\
\hline
\end{tabular}%
}
\caption{LHCO File Column Definitions.}
\label{tab:lhco_columns}
\end{table}

\begin{table}
\centering
\scriptsize
\resizebox{1\textwidth}{!}{%
\begin{tabular}{|c|c|c|c|c|c|c|c|c|c|}
\hline
\textbf{1} & \textbf{2} & \textbf{3} & \textbf{4} & \textbf{5} & \textbf{6} & \textbf{7} & \textbf{8} & \textbf{9} & \textbf{10 \& 11} \\
\hline
\textbf{Counter} & \textbf{Type} & \textbf{$\eta$} & \textbf{$\phi$} & \textbf{$p_T$} & \textbf{Jmass} & \textbf{Ntrk} & \textbf{Btag} & \textbf{Had/em} & \textbf{Dummies} \\
\hline
\multicolumn{10}{|c|}{\textbf{Event Block 0}} \\
\hline
0 &  & 0 & 0 &  &  &  &  &  &  \\
1 & 2 & -1.925 & -1.406 & 138.29 & 0.11 & 1.0 & 6.0 & 138.00 & 0.0 \\
2 & 4 & -0.773 & 2.467 & 58.03 & 11.73 & 7.0 & 0.0 & 0.27 & 0.0 \\
3 & 4 & 2.136 & 0.930 & 46.42 & 8.49 & 7.0 & 0.0 & 2.86 & 0.0 \\
4 & 4 & -2.052 & 0.987 & 45.30 & 2.94 & 3.0 & 0.0 & 0.00 & 0.0 \\
5 & 4 & 0.275 & 2.311 & 30.45 & 6.27 & 6.0 & 0.0 & 0.00 & 0.0 \\
6 & 4 & -0.142 & -1.995 & 29.20 & 7.34 & 5.0 & 0.0 & 0.42 & 0.0 \\
7 & 4 & 1.537 & 0.003 & 23.67 & 11.55 & 5.0 & 0.0 & 4.10 & 0.0 \\
8 & 6 & 0.000 & 2.020 & 28.44 & 0.00 & 0.0 & 0.0 & 0.00 & 0.0 \\
\hline
\multicolumn{10}{|c|}{\textbf{Event Block 1}} \\
\hline
0 &  & 1 & 0 &  &  &  &  &  & \\
1 & 1 & 1.037 & 2.787 & 47.10 & 5.21 & 7.0 & 0.0 & 3.62 & 0.0 \\
2 & 1 & 0.049 & 2.905 & 32.17 & 5.99 & -7.0 & 0.0 & 0.00 & 0.0 \\
3 & 4 & -0.647 & -2.051 & 24.05 & 3.49 & 3.0 & 0.0 & 999.90 & 0.0 \\
4 & 6 & 0.000 & -0.081 & 74.31 & 0.00 & 0.0 & 0.0 & 0.00 & 0.0 \\
\hline
\multicolumn{10}{|c|}{\textbf{Event Block 2}} \\
\hline
0 &  & 2 & 0 &  &  &  &  &  & \\
1 & 2 & -1.649 & 0.979 & 232.50 & 0.11 & 1.0 & 5.0 & 237.03 & 0.0 \\
2 & 4 & -2.088 & -2.437 & 172.57 & 17.59 & 15.0 & 0.0 & 0.07 & 0.0 \\
3 & 4 & -1.500 & -2.354 & 49.42 & 7.43 & 10.0 & 0.0 & 1.92 & 0.0 \\
4 & 4 & 0.018 & -2.715 & 44.89 & 4.55 & 4.0 & 0.0 & 0.00 & 0.0 \\
5 & 4 & -1.119 & 0.429 & 30.85 & 6.95 & 7.0 & 0.0 & 0.25 & 0.0 \\
6 & 4 & 2.757 & -0.564 & 23.64 & 5.09 & 0.0 & 0.0 & 999.90 & 0.0 \\
7 & 6 & 0.000 & -1.166 & 45.86 & 0.00 & 0.0 & 0.0 & 0.00 & 0.0 \\
\hline
\end{tabular}%
}
\caption{Detailed example of LHCO event data formatted for multiple event blocks.}
\label{tab:lhco_event_example}
\end{table}

In this event blocks, each row corresponds to a distinct object recorded during a simulated proton–proton collision. The structure follows a standardised layout:

\subsection*{Event Block 0}

\begin{itemize}
    \item \textbf{Row 0 (Event Header):}  
    The first row in the block (with counter 0). It contains the event number and trigger information. In this case, all entries are 0, indicating that no additional details are provided. This row is primarily used for identification purposes and can generally be ignored during analysis. However, it is useful to identify as a block in our Python framework.

    \item \textbf{Row 1 (Lepton Entry):}  
    The second row is labelled with counter 1 and has a type value of 2, which indicates that the object is a lepton (specifically, a muon). The kinematic properties are as follows:
    \begin{itemize}
        \item \textbf{Pseudorapidity ($\eta$):} -1.925, which describes the particle's angle relative to the beam axis.
        \item \textbf{Azimuthal Angle ($\phi$):} -1.406 radians, specifying the orientation around the beam.
        \item \textbf{Transverse Momentum ($p_T$):} 138.29 GeV, representing the momentum perpendicular to the beam axis.
        \item \textbf{Invariant Mass (Jmass):} 0.11 GeV. For leptons, the invariant mass is typically very small.
    \end{itemize}
    The seventh column (Ntrk), is given as 1.0. According to the LHCO conventions, for leptons this number is multiplied by the charge. A positive value here is indicating that the lepton is positively charged (i.e., a $\mu^+$). The eighth column, (Btag) shows a value of 6.0, which, although primarily used for jets, is repurposed in some analyses to encode additional identification or isolation information for leptons. Finally, the ninth column (Had/em) is 138.00; for muons, this field is formatted as \texttt{xxx.yy}, where the integer part represents the isolation transverse momentum ($p_T^{iso}$, measured in GeV within a cone of $R=0.4$ around the muon, excluding the muon itself) and the decimal part gives a percentage ratio of the surrounding transverse energy. Here, the large value emphasises the isolation characteristic of this object.

    \item \textbf{Rows 2 to 7 (Jet Entries):}  
    Rows 2 through 7 correspond to jets (objects with type 4). For example, considering Row 2:
    \begin{itemize}
        \item \textbf{Pseudorapidity ($\eta$):} -0.773
        \item \textbf{Azimuthal Angle ($\phi$):} 2.467 radians
        \item \textbf{Transverse Momentum ($p_T$):} 58.03 GeV
        \item \textbf{Invariant Mass (Jmass):} 11.73 GeV, which is the mass computed from the energy and momentum constituents of the jet.
        \item \textbf{Number of Tracks (Ntrk):} 7.0, indicating the jet is composed of several charged tracks.
        \item \textbf{Btag:} 0.0, meaning the jet was not tagged as containing a b-quark.
        \item \textbf{Had/em Ratio:} 0.27, suggesting that the energy deposited in the hadronic calorimeter is relatively low compared to that in the electromagnetic calorimeter. 
    \end{itemize}
    The remaining jet rows (Rows 3 through 7) are structured similarly, with variations in the kinematic quantities reflecting the different characteristics of each jet.

    \item \textbf{Row 8 (MET):}  
    The final row in the block is marked with a type value of 6, which denotes the MET. In this row:
    \begin{itemize}
        \item \textbf{$p_T$ (MET):} 28.44 GeV, representing the imbalance of transverse energy in the event.
        \item All other quantities (invariant mass, number of tracks, b-tag, etc.) are set to 0, as MET does not correspond to a reconstructed particle.
    \end{itemize}
\end{itemize}

\subsection*{Event Block 1}

\begin{itemize}
    \item \textbf{Row 0 (Event Header):}\\
    This row is marked by a counter value of 0 and serves as the event header. In this block, the header shows a value of 1 in the third column, which can be interpreted as the event label. 

    \item \textbf{Row 1 (First Lepton):}\\
    The row with counter 1 contains an object of \emph{Type 1}, which corresponds to an electron. Its kinematic and other properties are:
    \begin{itemize}
        \item \textbf{Pseudorapidity ($\eta$):} 1.037, which quantifies the angle relative to the beam axis.
        \item \textbf{Azimuthal Angle ($\phi$):} 2.787 radians, indicating the particle's orientation in the transverse plane.
        \item \textbf{Transverse Momentum ($p_T$):} 47.10 GeV.
        \item \textbf{Invariant Mass (Jmass):} 5.21 GeV.
    \end{itemize}
    Ntrk is recorded as 7.0. This value is multiplied by the particle's charge; therefore, a positive value indicates a positively charged electron (i.e., a positron). The b-tag column is 0.0 since b-tagging is not applicable to electrons. The Had/em ratio is given as 3.62, which in the context of leptons, is interpreted using a special format where the integer part corresponds to the isolation transverse momentum ($p_T^{iso}$) and the decimal part to the relative energy in a surrounding grid. 
    
    \item \textbf{Row 2 (Second Lepton):}\\
    This row also represents an electron (Type 1). Its kinematic values are:
    \begin{itemize}
        \item \textbf{Pseudorapidity ($\eta$):} 0.049.
        \item \textbf{Azimuthal Angle ($\phi$):} 2.905 radians.
        \item \textbf{Transverse Momentum ($p_T$):} 32.17 GeV.
        \item \textbf{Invariant Mass (Jmass):} 5.99 GeV.
    \end{itemize}
    Here, Ntrk is listed as \textbf{-7.0}. The negative sign indicates that this electron is negatively charged. As in Row 1, the b-tag value is 0.0 and the Had/em ratio is 0.00.

    \item \textbf{Row 3 (Jet):}\\
    Row 3 contains an object with \emph{Type 4}, corresponding to a jet. Its properties include:
    \begin{itemize}
        \item \textbf{Pseudorapidity ($\eta$):} -0.647.
        \item \textbf{Azimuthal Angle ($\phi$):} -2.051 radians.
        \item \textbf{Transverse Momentum ($p_T$):} 24.05 GeV.
        \item \textbf{Invariant Mass (Jmass):} 3.49 GeV.
    \end{itemize}
It is associated with 3 tracks, and the b-tag value is 0.0, indicating no heavy-flavour tagging. The Had/em ratio is 999.90, an unusually high value that likely serves as a flag to indicate a special jet characteristic or a reconstruction anomaly.
    \item \textbf{Row 4 (Missing Transverse Energy – MET):}\\
    The final row in this block is marked by type 6, denoting MET. For this entry:
 \begin{itemize}
        \item \textbf{Azimuthal Angle ($\phi$):} -0.081 radians.
        \item \textbf{Transverse Momentum ($p_T$):} 74.31 GeV.
    \end{itemize}
\end{itemize}

\subsection*{Event Block 2}

In this block, the structure is as follows:

\begin{itemize}
  \item \textbf{Row 0 (Event Header):}  
    3 provides the event number.

  \item \textbf{Row 1 (Lepton Entry):}  
    Type 2 value, indicating that the object is a lepton. The kinematic and identification information is:
    \begin{itemize}
      \item \textbf{Pseudorapidity ($\eta$):} -1.649.
      \item \textbf{Azimuthal Angle ($\phi$):} 0.979 radians.
      \item \textbf{Transverse Momentum ($p_T$):} 232.50 GeV.
      \item \textbf{Invariant Mass (Jmass):} 0.11 GeV.
    \end{itemize}
    Here Ntrk is 1.0, indicating that the lepton is positively charged (i.e., a positron). 
    Row 6 and Row 7 have a similar interpretation as the last block analysed previously.
       \end{itemize}

\subsection{Setup and Usage of Our Python Framework}
\label{sec:setup}

The framework requires a fully functional \verb|Python 3| environment and depends on the \verb|pandas| library for data manipulation. These dependencies must be installed to ensure proper functionality.

\subsection*{Obtaining the Project}
The PrakritiMLPrep framework can be cloned from this repository:
\begin{verbatim}
git clone https://gitlab.com/ed_in/prakritimlprep.git
\end{verbatim}

\subsection*{Running the Tool}

To run the tool, it is necessary to run it as: 
\begin{verbatim}
python PrakritiMLPrep.py $Signal_File.lhco $Background_File.lhco
\end{verbatim}
Here, \verb|$Signal_File.lhco| is the LHCO file containing the signal events and the LHCO file with the background events, is \verb|$Background_File.lhco|. The tool will then begin processing these files and generate the output dataset for further Machine Learning analysis.

\subsection*{Output Details}
After executing the command, the directory \texttt{\$Signal\_File\_\allowbreak\$Background\_File} is created, containing the dataset file \url{training-$Signal_File.csv}. This .csv file includes
both, the signal and background information, randomly mixed and formatted to be
ready for training Machine Learning models. The structure of the .csv file is
as follows:

\begin{itemize}
    \item The first column, \verb|New_ID|, is used to uniquely identify each row (or event).
    \item The subsequent columns contain the calculated observables contained in the \verb|Modules| directory. By convention, the raw quantities measured by the detector (the \emph{Primitive Variables}) are labeled with the prefix \verb|PRI_|, while the computed quantities (the \emph{Derived Variables}) are labeled with the prefix \verb|DER_|.
    \item The final column indicates whether the row corresponds to a signal or a background event.
\end{itemize}

\subsection{Adding New Observables}

The code is designed using a modular approach which allows, new physical observables  be easily added by creating new .py modules within the \verb|Modules/| directory.

 Each observable is computed in a separate .py file, which allows to add or modify observables without affecting the core functionality.

To include a new observable in the framework, one begins by creating a dedicated .py file (for example, \verb|DER_NewObservable.py|) within the \verb|Modules/| directory. Next, a function is defined in this file to accept the raw event data, calculate the required observable, and record the resulting values as an additional column in a .csv output.

As a basic example, the Code Block~\ref{code:lepton-processing} shows the Python code used to process LHCO data for lepton entries. This code extracts \emph{primitive variables} and computes \emph{derived variables}  for pairs of leptons.

\begin{minted}[
    linenos,
    frame=single,
    bgcolor=bgcolor,
    fontsize=\small,
    breaklines=true
]{python}
import csv
import sys
from math import cosh, cos, sqrt

def process_data(input_file, output_file):
    # List to store results for each lepton pair.
    results = []
    
    # Read all lines from the input LHCO file.
    with open(input_file, 'r') as file:
        lines = file.readlines()
        i = 0
        
        # Process each event.
        while i < len(lines):
            line = lines[i].strip()
            
            # Identify the beginning of a new event with a header row (starting with '0').
            if line.startswith('0'):
                leptons = []  # List to store lepton data for this event.
                i += 1
                
                # Read subsequent lines until the next event header is encountered.
                while i < len(lines):
                    line = lines[i].strip()
                    words = line.split()
                    
                    # If a new event header is reached and lepton data exists, break.
                    if line.startswith('0') and leptons:
                        break
                    
                    # Process rows representing electrons (type '1') or muons (type '2').
                    if len(words) >= 8 and words[1] in ['1', '2']:
                        # Append a tuple: (object type, pT, eta, phi)
                        leptons.append((words[1], float(words[4]), float(words[2]), float(words[3])))
                    i += 1
                
                # If at least two leptons are found, compute kinematic variables for every pair.
                if len(leptons) >= 2:
                    for j in range(len(leptons)):
                        for k in range(j+1, len(leptons)):
                            type1, pt1, eta1, phi1 = leptons[j]
                            type2, pt2, eta2, phi2 = leptons[k]
                            
                            # Calculate differences in pseudorapidity and azimuthal angle.
                            delta_eta = abs(eta1 - eta2)
                            delta_phi = abs(phi1 - phi2)
                            
                            # Compute the angular separation (Delta R) in (eta, phi) space.
                            delta_r = sqrt(delta_eta**2 + delta_phi**2)
                            
                            # Compute the invariant mass using the formula for massless leptons:
                            # m = sqrt(2 * pt1 * pt2 * (cosh(delta_eta) - cos(delta_phi)))
                            invariant_mass = sqrt(2 * pt1 * pt2 * (cosh(delta_eta) - cos(delta_phi)))
                            
                            # Save the computed variables for this lepton pair.
                            results.append({
                                'pt1': pt1,
                                'eta1': eta1,
                                'phi1': phi1,
                                'pt2': pt2,
                                'eta2': eta2,
                                'phi2': phi2,
                                'delta_eta': delta_eta,
                                'delta_r': delta_r,
                                'invariant_mass': invariant_mass
                            })
            else:
                i += 1
    
    # Write the results to a .csv file with extended columns.
    with open(output_file, 'w', newline='') as csvfile:
        fieldnames = ['pt1', 'eta1', 'phi1', 'pt2', 'eta2', 'phi2', 'delta_eta', 'delta_r', 'invariant_mass']
        writer = csv.DictWriter(csvfile, fieldnames=fieldnames)
        writer.writeheader()
        for res in results:
            writer.writerow(res)
\end{minted}
\codecaption{Example of Python code for processing lepton data.}
\label{code:lepton-processing}

\subsection*{Explanation of the Main Code Steps}

\begin{enumerate}
    \item \textbf{File Reading and Event Loop:}\\
    The function \texttt{process\_data} opens the input LHCO file and reads all lines. It then iterates over these lines to detect the beginning of each event, which is indicated by a row starting with \texttt{'0'}.

    \item \textbf{Extracting Lepton Data:}\\
    For each event, the code initializes a list \texttt{leptons} and processes subsequent rows. If the row represents an electron (Type \texttt{`1'}) or a muon (Type \texttt{`2'}), the code extracts the transverse momentum (\texttt{pT}), pseudorapidity (\texttt{eta}), and azimuthal angle (\texttt{phi}), storing these values as a tuple.

    \item \textbf{Computing Derived Variables:}\\
    Once at least two leptons have been collected for an event, the code computes kinematic variables for each unique pair:
    \begin{itemize}
        \item \textbf{$\Delta\eta$}: The absolute difference in pseudorapidity.
        \item \textbf{$\Delta\phi$}: The absolute difference in azimuthal angle.
        \item \textbf{$\Delta R$}: The angular separation computed as 
        \[
        \Delta R = \sqrt{(\Delta\eta)^2 + (\Delta\phi)^2}.
        \]
        \item \textbf{Invariant Mass} \cite{OpenDataEducation2025}:  
        \[
        m = \sqrt{2\, p_{T1}\, p_{T2}\,\left[\cosh(\Delta\eta) - \cos(\Delta\phi)\right]},
        \]
        which is standard for massless leptons.
    \end{itemize}
    These derived quantities are stored in a list of dictionaries.

    \item \textbf{Output Generation:}\\
Finally, the results are written to a .csv file. The .csv file includes a header and rows for each lepton pair, with the following fields: \texttt{pt1}, \texttt{eta1}, \texttt{phi1}, \texttt{pt2}, \texttt{eta2}, \texttt{phi2}, \texttt{delta\_eta}, \texttt{delta\_r}, and \texttt{invariant\_mass}. In this example, the main column of interest is the invariant mass, column number 9. 

This codes can be tested individually by running:

\begin{verbatim}
python \$Observable.py \$LHCOFile.lhco \$CSVOutputfile.csv
\end{verbatim}

Where \texttt{\$Observable.py} refers to the Python script itself. \texttt{\$LHCOFile.lhco} is the LHCO event file (whether signal or background), while \texttt{\$}\url{CSVOutputfile.csv} is the user specified output filename where the computed observable will be stored.

\end{enumerate}

\subsection{BDT Training}

\subsubsection{Introduction to BDT Training}

A decision tree is a flowchart structure that begins with a root node (representing the entire dataset) and subsequently splits into internal nodes, like branches, based on threshold values of various features. In our case these features are the kinematic observables. At each branching, events that satisfy a split criterion (e.g. “Observable X is above some cutoff”) move along one branch, while the others follow a different and separated path. This process continues down multiple levels ending in terminal nodes, as “leafs”, where events are assigned a predicted class. Because of each threshold split attempts to make the classes as pure as possible, in our case, the signal vs. the background, the entire tree maps out a series of observable based decisions that aim to classify new events accurately.

\acp{BDT} are a form of ensemble Machine Learning algorithm that combine many decision trees to improve classification performance \cite{breiman1984classification}. By utilising a boosting technique, which is a method where several simple models such as decision trees, are trained one after the other, the trees are built sequentially. Each tree is focusing on the mistakes of the previous one, thereby reducing the overall error. This ensemble approach yields a powerful discriminator that is capable of separating signal which are our desired events from background, with higher accuracy than any single tree alone.

In \ac{HEP} analyses, \acp{BDT} have become an important tool. Instead of applying a series of one dimensional cuts, a \ac{BDT} learns an optimal combination of many observables at once, improving sensitivity to new phenomena. Modern implementations, such as \ac{XGBoost} \cite{xgboost}, further enhance performance and efficiency. XGBoost uses gradient boosting, which is a boosting technique using the gradients of a loss function to guide the training of each new learner, ensuring that each step is aiming to minimise the general error. Also XGBoost works with built in regularisation, internal methods to control the model complexity, such as, penalising overly deep or complex trees. This helps to prevent overfitting by ensuring that the ensemble does not become too finely tuned to the training data.
Actually, XGBoost gained significant attention in HEP following its successful application in the Higgs Boson Machine Learning Challenge (2014) \cite{kaggle_Higgs}.

\subsubsection{Dataset Preparation}

Ensuring a balanced number of signal and background events is crucial to avoid systematic error in the training process causing that the model could be performing well on the training data but poorly on new, unseen data, leading to inaccurate predictions or unfair decisions. If one class dominates, the \acp{BDT} might learn to simply label all events as belonging to that class.

Before training, we also screened the input observables for redundancy. For this purpose, a Pearson correlation matrix was computed on the training sample, and whenever a pair of variables exhibited a very strong linear correlation ($|\rho|>0.95$), the second member of the pair in the correlation matrix was discarded. By eliminating such highly correlated features, we mitigate multicollinearity and improve control of overfitting.

\subsubsection{Feature Importance and Cuts}
One advantage of using \acp{BDT} is the ability to assess the feature importance of each observable. The \ac{BDT} calculates the contribution of each one to the classification task, thereby providing a hierarchy of observables that are discriminate signal from background.

For example, key observables identified can include:
\begin{itemize}
    \item $p_{T}(b)$: Transverse momentum of the $b$-jet.
    \item MET: Missing transverse energy.
    \item $p_{T}(c)$: Transverse momentum of the $c$-jet.
    \item $M_{T}(\mu)$: Transverse mass of the muon.
    \item $M_{\mathrm{inv}}(bc)$: Invariant mass of the $b$ and $c$-jet pair.
\end{itemize}

These observables could then be examined to see if simple cut selections could isolate the signal. However, rather than applying multiple cuts individually, the \ac{BDT} can create a combined discriminant (referred to as the \texttt{xgb} output when we are using XGBoost, as aforementioned) that encapsulates the discriminating feature of all observables simultaneously. This composite variable enables a more precise selection by applying a single cut on the \ac{BDT} output score.

\subsubsection{Hyperparameter Tuning and Optimisation}
Training a \ac{BDT} involves setting several hyperparameters, which are parameters that control aspects such as model complexity and learning efficiency, and are set before the learning procces begins. The ones that we primarily use are:

\begin{itemize}
    \item \textbf{Number of trees:} The total number of decision trees in the ensemble.
    \item \textbf{Maximum depth per tree:} The maximum number of splits in each tree, controlling the complexity of the model. Deeper trees can capture more detail but may also lead to overfitting.
    \item \textbf{Learning rate:} A scaling factor that determines how much each new tree contributes to correcting the errors of previous trees. A smaller learning rate generally requires more trees but can lead to have a better performance.
    \item \textbf{Subsampling ratio:} The fraction of the training data used to build each tree. 
\end{itemize}

\subsubsection{Performance Evaluation}
The performance of the \ac{BDT} is primarily evaluated using the \ac{ROC} curve, which plots the true positive rate (signal efficiency) against the false positive rate (background contamination) for different thresholds on the \ac{BDT} output. The \ac{AUC} provides a single measure of classifier performance, where an AUC of 1 corresponds to perfect classification and 0.5 corresponds to random guessing. \cite{FAWCETT2006861}

We also perform checks for overtraining by comparing the \ac{BDT} output distributions between the training and testing sets. The \ac{KS} test \cite{Ross2014Probability}, is used to quantify the similarity between these distributions. A KS statistic close to zero indicates that the cumulative distributions of the two samples being compared are very similar. In the context of model generalisation, this indicates that the model predictions on the training data are very similar to its predictions on new, unseen data suggesting that the model has not overfitted to the training set and is performing well on new data \cite{Coadou2022}.

\subsubsection{XGB Output Variable and Final Significance Estimation}
A cut on the \texttt{xgb} variable is then applied to separate signal from background events. The optimal cut is determined by maximising the \textit{Signal Significance}, defined as \cite{Logic1718}

\begin{equation}
\label{eq:signalsignificance}
Z = \frac{S}{\sqrt{S + B}},
\end{equation}

where $S$ and $B$ are the number of signal and background events passing the cut, respectively, and they are defined as, 

\[
S = L \times \sigma_S,
\quad
B = L \times \sigma_B.
\]

Incorporating the integrated luminosity, cross-sections, and BRs allows the estimation of the discovery potential under realistic experimental conditions. Thus, the \ac{BDT} training process involves preparing and balancing a representative dataset, selecting key observables and understanding their relative importance, adjusting hyperparameters to optimise performance while avoiding overfitting, evaluating the model using ROC curves, KS tests, and significance calculations, and using the combined \texttt{xgb} variable to perform the final event selection.

This approach shows that the \ac{BDT} is not only capable of effectively distinguishing signals from background, but also provides a robust tool to estimate the potential discovery of new physics processes.

In the following chapter, we are applying our previous methodology to perform a detailed collider analyses within the frameworks of the 2HDM-III and the FNSM focusing specifically on the identification of relevant collider signatures.

\chapter{Collider Analysis}
\label{chap:collyder_analysis}
\noindent\rule{\linewidth}{0.4pt}

In this chapter, we run a detailed collider analysis within the two different BSM frameworks that we are working on. We examine the decay processes and production mechanisms pertinent to these models, emphasising the identification of key signals and their differentiation from the SM background. Furthermore, we utilise multivariate analysis techniques to enhance the discrimination power and improve the sensitivity for a potential experimental discovery of the processes. The chapter thus sets the stage for a systematic exploration of the collider signatures associated with our proposed scenarios.

\section{Collider Analysis for the 2HDM-III}
\begin{figure}[!h]
	\centering
	\begin{subfigure}[b]{0.5\textwidth}
		\includegraphics[scale=0.35]{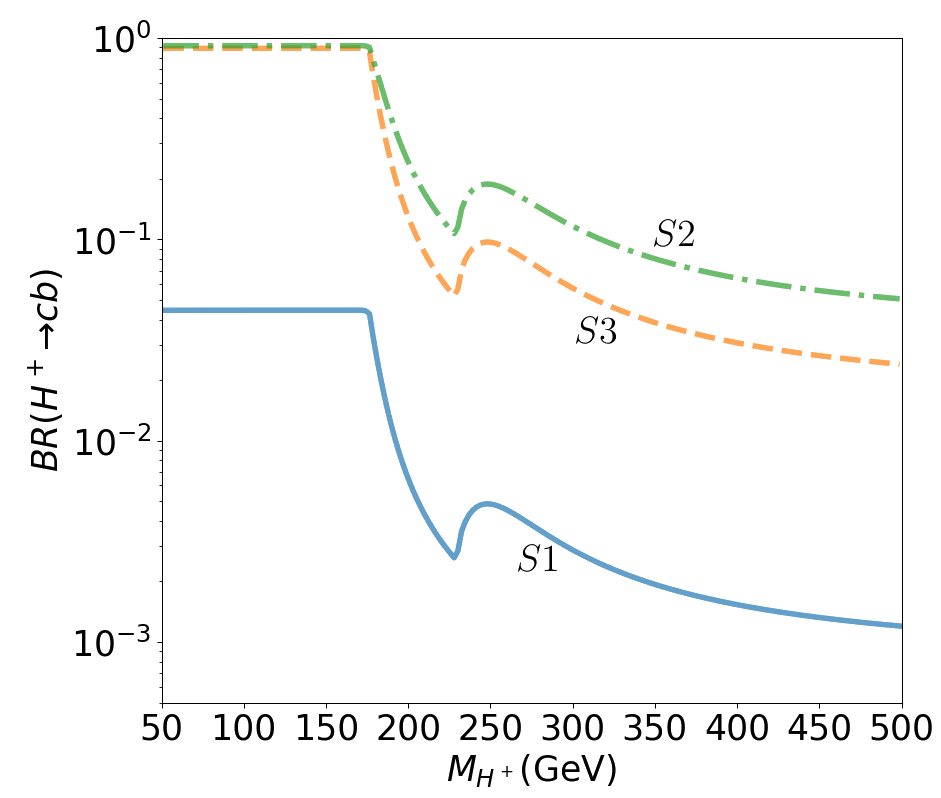}
		\caption{}
		\label{fig:BRHp-cb}
        
	\end{subfigure}%
    \\
	\begin{subfigure}[b]{0.5\textwidth}
		\includegraphics[scale=0.35]{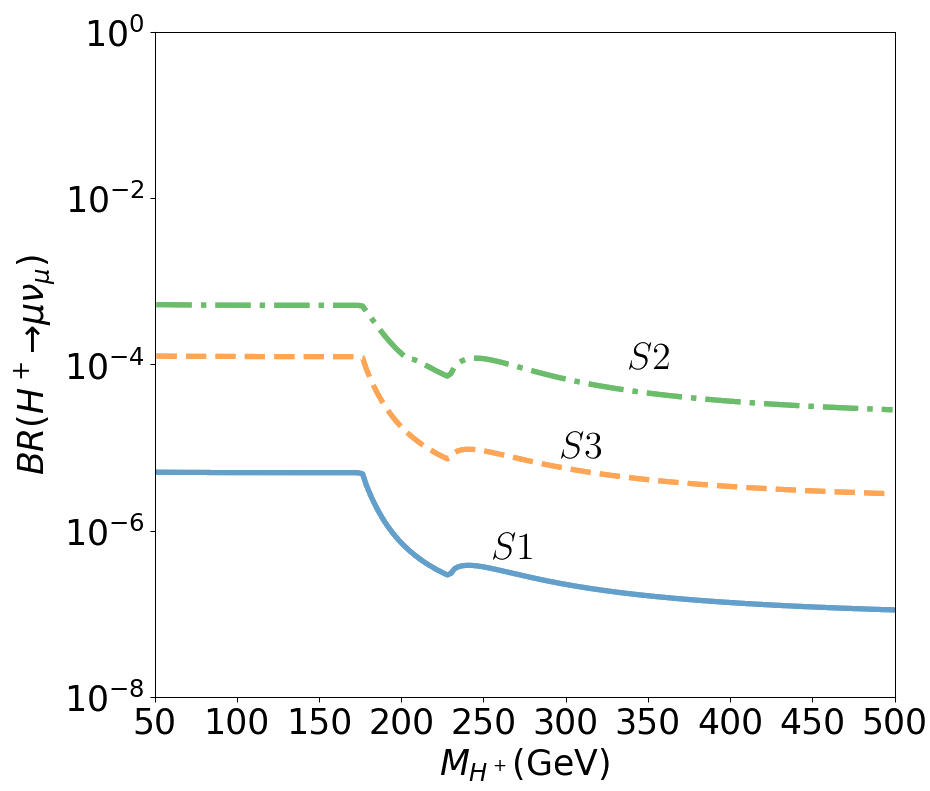}
		\caption{}
		\label{fig:BRHp-munu}
	\end{subfigure}
	\caption{Branching Ratios (a) $BR(H^+\to cb)$, (b) $BR(H^+\to \mu\nu)$, as a function of the charged scalar mass $M_{H^+}$.}
	\label{BRs}
\end{figure}

We present the analytical expressions required to compute the decay widths for the processes $H^{\pm} \to u_i d_j$ and $H^{\pm} \to \ell^{\pm}\nu_\ell$. These formulas are essential for describing the underlying dynamics of these decays and are given by,

\begin{eqnarray*}
	\Gamma(H^{\pm}\to u_{i}d_{j})&=&\frac{3G_{F}M_{H^{\pm}}}{\sqrt{2}4\pi}\Big(m_{u_{i}}^{2}|Y_{ij}|^{2}+m_{d_{j}}^{2}|X_{ij}|^{2}\Big),\\ \nonumber
	\Gamma(H^{\pm}\to \ell^{\pm}\nu_{\ell})&=&\frac{G_{F}M_{H^{\pm}}}{\sqrt{2}4\pi}\Big(m_{\ell}^{2}|Z_{ij}^\ell|^{2}\Big)\nonumber,
	\end{eqnarray*}
where 
\begin{eqnarray*}
    X_{ij} &=& \sum_{l=1}^{3}(V_{\rm CKM})_{il}\left[t_\beta \frac{m_{d_l}}{m_{d_j}}\delta_{l j}-\frac{\sqrt{1+t^2_\beta}}{\sqrt{2}}\sqrt{\frac{m_{d_l}}{m_{d_j}}}\chi_{lj}^d \right],\\
    Y_{ij} &=& \sum_{l=1}^{3}\left[\frac{1}{t_\beta}\delta_{il}-\frac{\sqrt{1+\frac{1}{t^2_\beta}}}{\sqrt{2}}\sqrt{\frac{m_{u_l}}{m_{u_j}}}\chi_{lj}^u\right](V_{\rm CKM})_{lj},\\
    Z_{ij}^\ell &=& \left[t_\beta \frac{m_{\ell_i}}{m_{\ell_j}}\delta_{i j}-\frac{\sqrt{1+t_\beta^2}}{\sqrt{2}}\sqrt{\frac{m_{\ell_i}}{m_{\ell_j}}}\chi_{ij}^\ell \right],
\end{eqnarray*}

where $t_\beta \equiv \tan\beta$. 
In Fig.~\ref{BRs}, we present the BRs for the corresponding decay channels.
$
H^{+}\to cb
$ and $
H^{+}\to \mu \nu_\mu,
$ plotted as functions of the charged scalar mass $M_{H^+}$, with the analysis carried out for the scenarios \textbf{S1}, \textbf{S2}, and \textbf{S3}, which are defined in Table~\ref{BRs}.

	\subsection{Signal and Background}
    We examine the signal and background processes arising from $pp$ collisions at the LHC, carefully assessing their contributions to the observed event rates.
	
\begin{figure}[!htb]
			\centering
			{\includegraphics[scale=0.34]{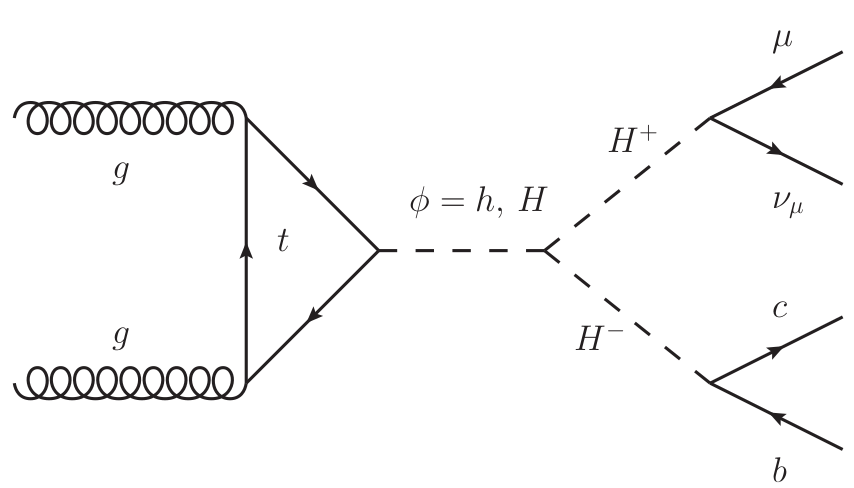}
            \\
                \includegraphics[scale=0.35]{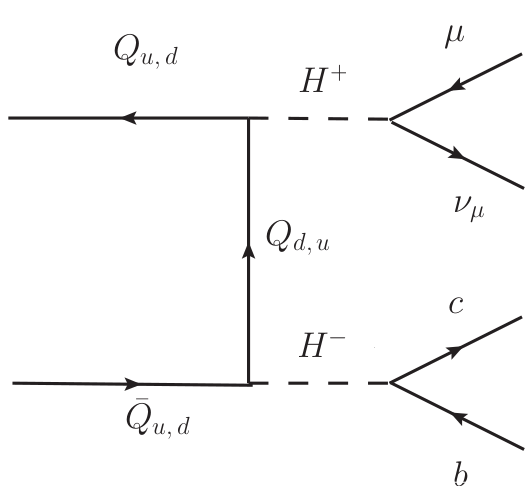}}
			\caption{Feynman diagrams of the production cross-section of the signal $pp\to H^+H^-\to \mu\nu_\mu cb$.}\label{FDsignal}
		\end{figure}

	\begin{itemize}
		\item \textbf{SIGNAL:} We focus on the search for a specific final state, $cb\mu\nu_{\mu}$, which arises from the pair production of charged Higgs bosons in proton-proton collisions. This process can be expressed as $pp \to H^- H^+ \to \mu^+\nu_{\mu}\bar{c}b + \mu^-\bar{\nu}_{\mu}c\bar{b}$, and we denote this final state as $\mu\nu_\mu cb$. For our analysis, we assume a $b$-tagging efficiency of $\epsilon_b = 80\%$, a $c$-jet mis-tagging rate of $\epsilon_c = 10\%$, and a misidentification rate for other jets of $\epsilon_j = 1\%$. The dominant contributions to this final state within the framework of the 2HDM-III are illustrated in Fig.~\ref{FDsignal}.
        \newpage

		\item \textbf{BACKGROUND:} The dominant SM background comes from the final state $bj\ell\nu_\ell$, which is produced by

		\begin{itemize}
			\item $Wjj+Wb\bar{b}$,
			\item $tb+tj$,
			\item $t\bar{t}$.
		\end{itemize}
	\end{itemize}

    In the case of the $t\bar{t}$ background process, either one of the two leptons is missed in semi-leptonic top quark decays, or two of the four jets are not reconstructed when one of the top quarks undergoes a semi-leptonic decay. The numerical values of the cross-sections and BRs for the signal, corresponding to $M_H = 500,\, 800,\, 1000$ GeV in scenario \textbf{S2}, are summarised in Table~\ref{XSsignal}. Meanwhile, the cross-sections of the dominant SM background processes are presented in Table~\ref{XSSMBGD}.

\begin{figure}[!h]
		\centering
		{\includegraphics[scale=0.28]{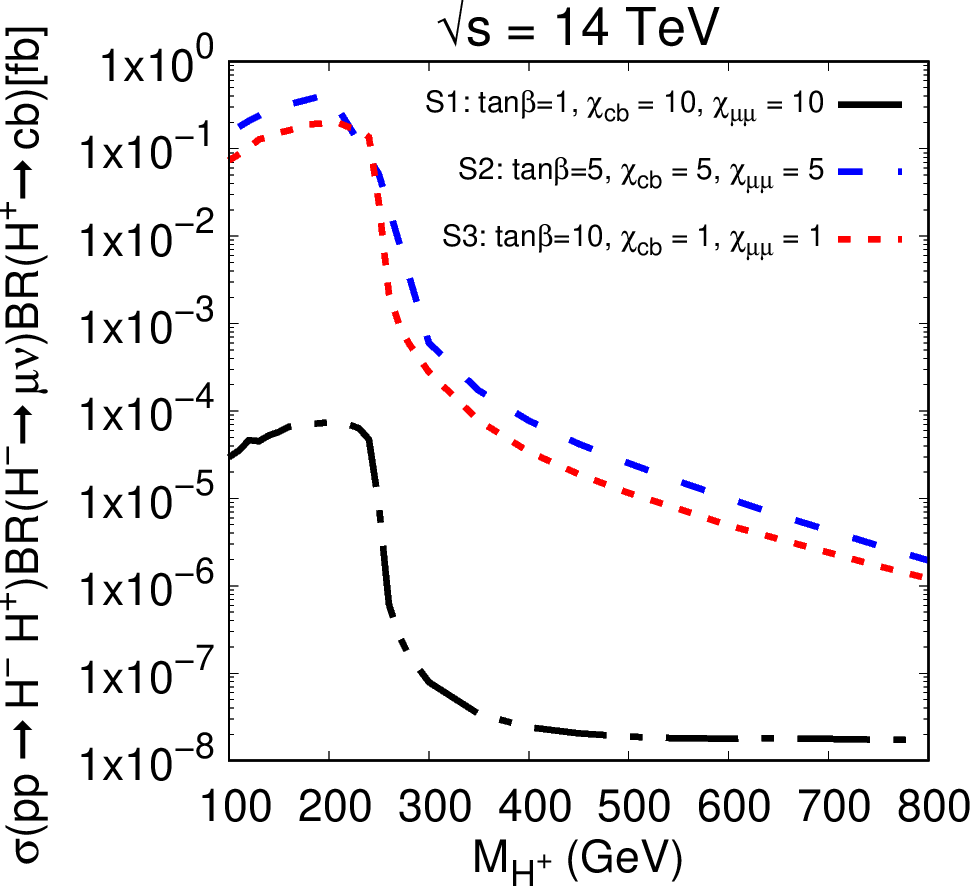}}
		\caption{Production cross-section of a charged Higgs pair with their subsequent decays into $\mu\nu_{\mu} cb$ for $M_H=500$ GeV. }\label{Brs}
	\end{figure}

\begin{figure}[!htb]
		\centering
		{\includegraphics[scale=0.32]{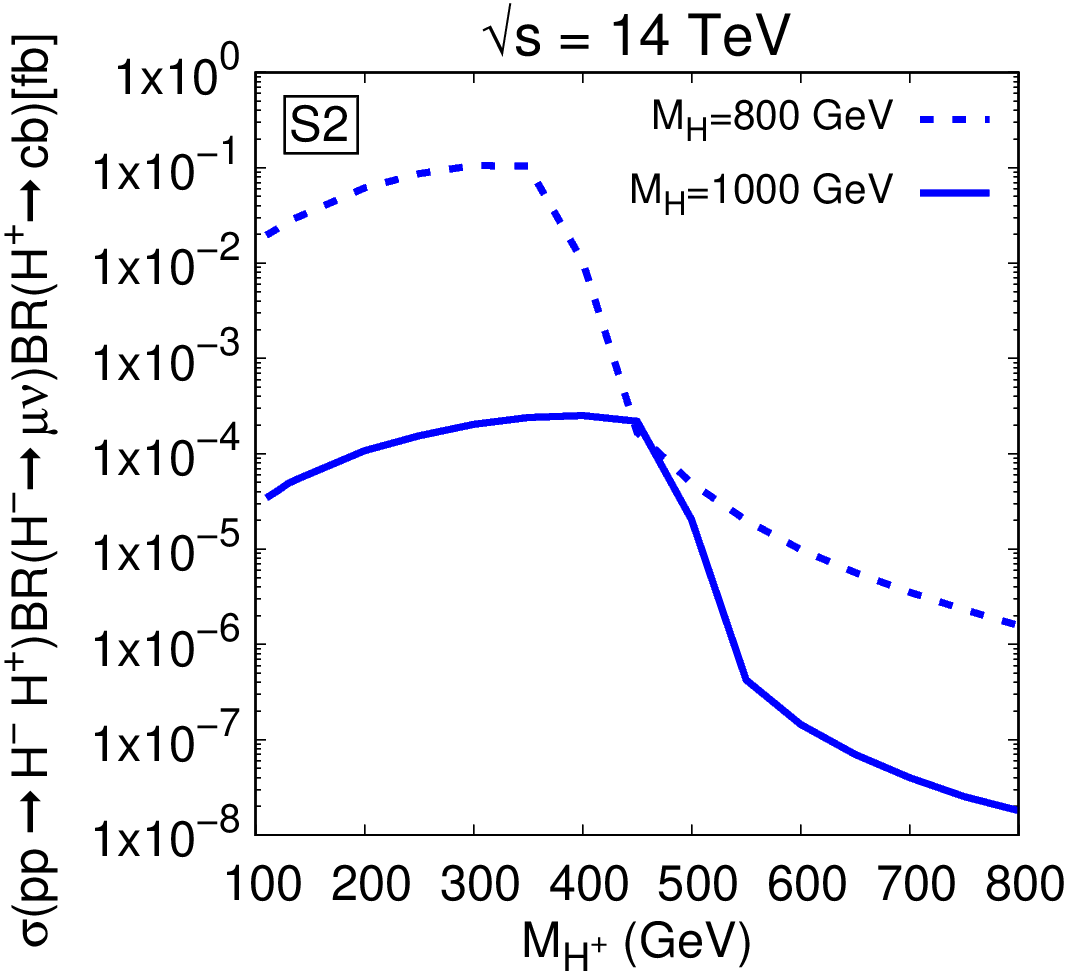}}
		\caption{Production cross-section for a pair of charged Higgs bosons followed by their subsequent decays into $\mu\nu_{\mu} cb$, for $M_H = 800,\, 1000$ GeV in scenario \textbf{S2}.}\label{XS8001000}
	\end{figure}

\begin{table}[!htb]
\centering

\resizebox{\textwidth}{!}{%
\begin{tabular}{|c|c|c|ccc|c|c|}
\hline
\multirow{2}{*}{$M_{H}$ (GeV)} & \multirow{2}{*}{$M_{H^{\pm}}$ (GeV)} 
& \multirow{2}{*}{$\sigma(gg \to H)$} 
& \multicolumn{3}{c|}{BRs} 
& \multirow{2}{*}{$\sigma\bigl(pp \to \mu^{+}\nu_{\mu}b\bar{c}\bigr)$} 
& \multirow{2}{*}{Events (3000 fb$^{-1}$)} \\
\cline{4-6}
& & 
& $H \to H^{-}H^{+}$ 
& $H^{-}\to b\bar{c}$ 
& $H^{+}\to \mu^{+}\nu_{\mu}$ 
& & \\
\hline
\multirow{2}{*}{500} 
& 100 
& 1200 fb 
& 0.32 
& 0.944 
& $5.27\times10^{-4}$ 
& 0.189 fb 
& 567 
\\
& 150 
& 1200 fb 
& 0.54 
& 0.943 
& $5.26\times10^{-4}$ 
& 0.324 fb 
& 972 
\\
\hline
\multirow{2}{*}{800} 
& 100 
& 151 fb 
& 0.27 
& 0.944 
& $5.27\times10^{-4}$ 
& 0.02 fb 
& 59 
\\
& 150 
& 151 fb 
& 0.40 
& 0.943 
& $5.26\times10^{-4}$ 
& 0.03 fb 
& 100 
\\
\hline
\multirow{2}{*}{1000} 
& 100 
& 42 fb 
& $1.67\times10^{-3}$ 
& 0.944 
& $5.27\times10^{-4}$ 
& $3.5\times10^{-5}$ fb 
& 0.1 
\\
& 150 
& 42 fb 
& $2.83\times10^{-3}$ 
& 0.943 
& $5.26\times10^{-4}$ 
& $5.9\times10^{-5}$ fb 
& 0.2 
\\
\hline
\end{tabular}%
}
\caption{Leading contributions to the cross-section and BRs for scenario \textbf{S2}.}
\label{XSsignal}
\end{table}
	%%%%%%%%%%%%%%%%%%%%%%%%%%%%%%%%%%%%%%%%%%%%%%%%%%%%%%%%%%%%%%%%%%%%%%%%%%%%%%%%%%%%%%%%%%%%%%%%%%%%%%%%
%	\end{widetext}	
%	
		\begin{table}[!htb]
		
		\begin{centering}
			\begin{tabular}{|c|c|c|}
				\hline 
				SM backgrounds & Cross-section {[}fb{]}&Events (3000 fb$^{-1}$)\tabularnewline
				\hline 
				\hline 
				$pp\to Wjj+Wb\bar{b}\,(W\to\ell\nu_{\ell})$ & $3745960$&$\mathcal{O}(10^{10})$\tabularnewline
				\hline 
				$pp\to tb+tj\,(t\to\ell\nu_{\ell}b)$ & $1734$&$\mathcal{O}(10^{6})$\tabularnewline
				\hline 
				$pp\to t\bar{t}\,(t\to\ell\nu_{\ell}b,\,t\to q_{i}q_{j}b)$ & $431001$&$\mathcal{O}(10^{9})$\tabularnewline
				\hline 
			\end{tabular}
			\par\end{centering}
            \caption{Cross-section of the dominant SM background processes.}\label{XSSMBGD}
	\end{table}
	Figure~\ref{Brs} provides an overview of the production cross-section as a function of $M_{H^{\pm}}$ for the scenarios \textbf{S1}, \textbf{S2}, \textbf{S3}, with $M_H = 500$\,GeV. In contrast, Fig.~\ref{XS8001000} illustrates the corresponding plane for $M_H = 800,\,1000$\,GeV, focusing exclusively on scenario \textbf{S2}.

	For the three scenarios, we can observe that the cross-sections are significantly higher when the charged Higgs boson masses lie in the range of $100 \leq M_{H^{\pm}} \leq 250$ GeV. This is due to the dominant contribution from the neutral heavy Higgs boson ($M_H = 500\,\rm GeV$), which is capable of producing two real charged Higgs bosons in this mass range. As $M_{H^\pm}$ increases beyond $250\,\rm GeV$, the mediation of the heavy Higgs boson becomes virtual, leading to a suppression of the cross-section with increasing $M_{H^\pm}$. For $M_{H^\pm} \gtrsim 400$\,GeV, the cross-section tends to stabilise, particularly in scenario \textbf{S1}, where a plateau emerges. This behaviour may be attributed to the increasing relevance of alternative decay channels, which dominate the total rate once the neutral Higgs bosons are produced off-shell. In contrast, scenarios \textbf{S2} and \textbf{S3} do not exhibit a similar plateau, likely due to their specific parameter choices, which prevent the cross-section from levelling off at higher masses.

From Tables \ref{XSsignal} and \ref{XSSMBGD}, we can note that for $M_{H^{\pm}} = 100\,\rm GeV$ and $M_H = 500\,\rm GeV$, the signal cross-section exceeds the corresponding background production cross-section by up to seven orders of magnitude. Assuming an integrated luminosity of $3000\,\rm fb^{-1}$, this corresponds to a difference of approximately $10^8$ events. However, for $M_H = 1000\,\rm GeV$ no events are expected to be produced even at the maximum integrated luminosity achievable at HL-LHC. This effectively rules out the possibility of observing a massive scalar in this mass range for the given luminosity.

In contrast, for $M_H = 800\,\rm GeV$, although the number of expected signal events is small, our analysis suggests that the HL-LHC may still have the potential to see the proposed signal. This possibility will be explored further in this work.
	
    In our computational framework we begin by implementing the full model using \texttt{FeynRules} \cite{Alloul:2013bka} within \texttt{MadGraph5} \cite{MadGraphNLO}, employing the NNPDF2.3LO \cite{BALL2013244} Parton Distribution Functions (PDFs). Subsequently, the model is interfaced with \texttt{Pythia8} \cite{Sjostrand:2008vc} for parton showering and hadronisation and with \texttt{Delphes 3} \cite{delphes} for detector simulations. For jet reconstruction, we use the \texttt{FastJet} package \cite{Cacciari:2011ma} in conjunction with the \texttt{anti-$k_t$} algorithm \cite{Cacciari:2008gp}.

	\subsection{Signal Significance}

    Traditional strategies based on kinematic cuts applied to observables often reject a significant number of background events. However, these cuts also discard a substantial fraction of signal events reducing the overall sensitivity. To address this limitation, we employ a Multivariate Analysis (MVA) approach, which provides a more efficient way to discriminate between signal and background events. Specifically, as mentioned in Chapter ~\ref{chap:methodology}, we compute a \ac{BDT} \cite{Woodruff2018} training it on the set of variables listed in Table~\ref{VarBDT} of App.~\ref{VI}. 

    The relevant hyperparameters for the \ac{BDT} are chosen as follows: number of trees (\texttt{NTree}) is set to 50, the maximum depth of each decision tree (\texttt{MaxDepth}) is set to 5, and the maximum number of leaves per tree (\texttt{MaxLeaves}) is set to 8. All other parameters are left at their default values.
	
\begin{figure}[!h]
		\centering
		\includegraphics[scale=0.5]{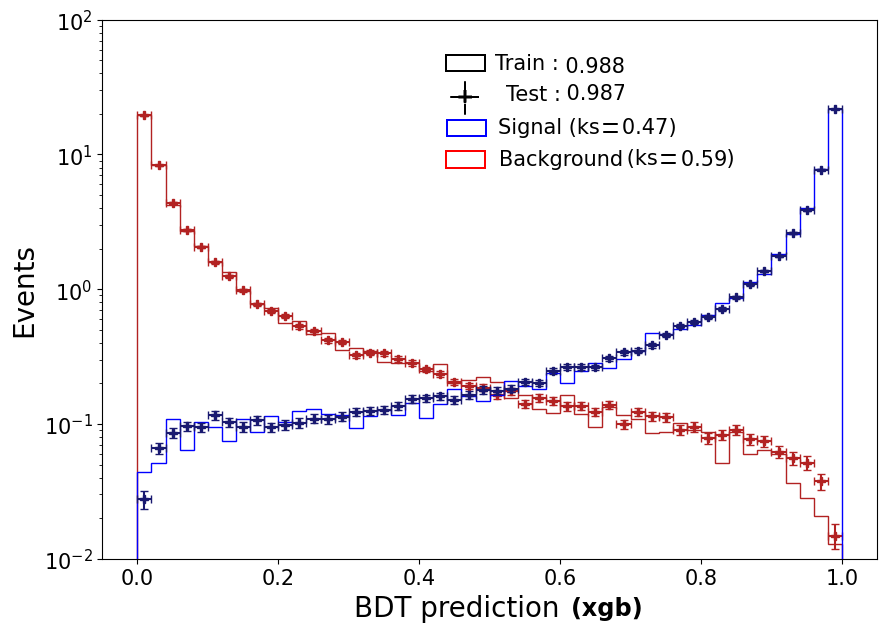}
		\caption{Plot of the discriminant for signal and background data.}\label{Discriminat}
	\end{figure}

    In Fig.~\ref{Discriminat}, we present the discriminant distributions for both the signal and background processes. The goodness of fit is evaluated using the KS test. We find that the KS value lies within the permissible range of [0,1], with values of 0.47 for the signal and 0.59 for the background, indicating a reasonable agreement between the observed and expected distributions.

Once the classifier was trained, it outputs the single variable denoted as $\textbf{xgb}$, which effectively separates signal events from background events, as illustrated in Fig.~\ref{Discriminat}. Subsequently, the signal-to-background ratio is optimised based on this discriminant variable.
	
Our analysis reveals that the most discriminative observables are as follows: (i) the transverse momentum of the $b$-jet, (ii) the missing transverse energy $\slashed{E}_T$, (iii) the transverse momentum of the $c$-jet, (iv) the transverse mass of the muon $M_T[\mu]$, and (v) the invariant mass of one of the charged Higgs bosons decaying into a $bc$-jet pair, $M_{\text{inv}}[bc]$. For the scenario \textbf{S2} with $M_{H^{\pm}} = 110\,\text{GeV}$ and $M_H = 500\,\text{GeV}$, the distributions of these observables are presented in Fig.~\ref{distributions}.

A notable feature emerges from the distributions of $M_T[\mu]$ and $M_{\text{inv}}[bc]$, a clear distinction between the behaviour of the signal and background processes. Specifically, both observables exhibit a resonant peak around $M_{H^{\pm}} = 110\,\text{GeV}$, which corresponds to the presence of the charged scalar bosons. One of these bosons decays into a $bc$-jet pair, while the other decays into $\mu + \slashed{E}_T$. These features represent the most distinctive signatures of our signal. Based on these observations, one could directly impose cuts on these variables as follows,
\begin{enumerate}
	\item $M_{H^{\pm}}-20<M_T[\mu]<M_{H^{\pm}}+20$ GeV,
	\item $M_{H^{\pm}}-30<M_{inv}[bc]<M_{H^{\pm}}+20$ GeV,
	\item $P_T[b]>50$ GeV,
	\item $\rm MET>20$ GeV,
	\item $P_T[j_1]>30$ GeV.
\end{enumerate}    
With the aforementioned kinematic cuts (which we rigorously verify using the \texttt{MadAnalysis5} package \cite{Conte:2012fm}), a good \textit{Signal Significance} can be achieved. Nevertheless, we instead choose to impose cuts on the variable $\textbf{xgb}$ as detailed below.
\begin{figure}[!htb]
    \centering
    \begin{subfigure}[b]{0.5\textwidth}
        \includegraphics[scale=0.6]{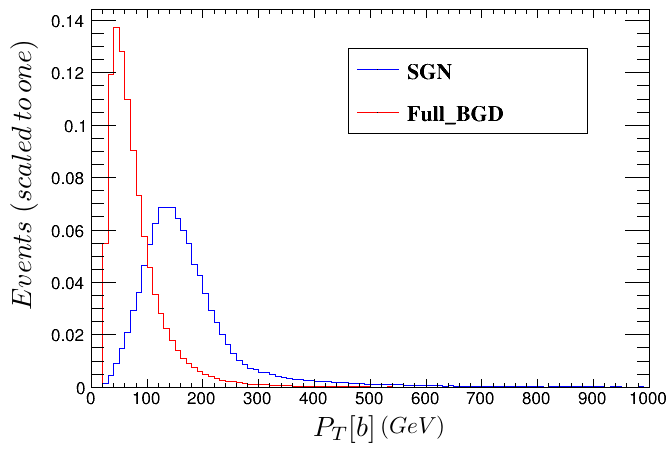}
        \caption{Transverse moment of the $b$-jet}
        \label{fig:a}
    \end{subfigure}%
    \\
    \begin{subfigure}[b]{0.5\textwidth}
        \includegraphics[scale=0.6]{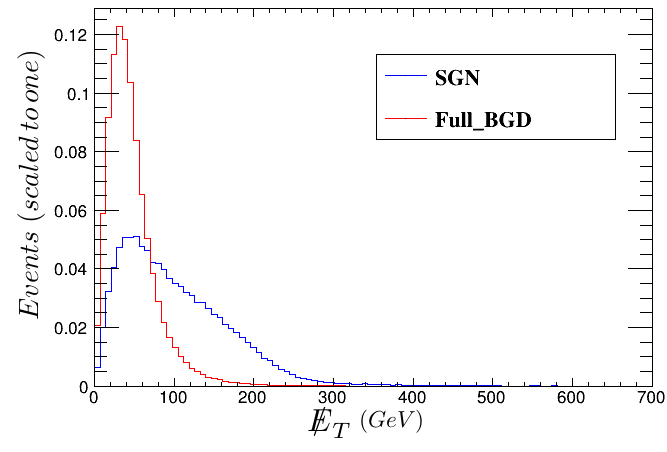}
        \caption{Missing energy transverse $\slashed{E}_T$}
        \label{fig:b}
    \end{subfigure}
    \\
    \begin{subfigure}[b]{0.5\textwidth}
        \includegraphics[scale=0.6]{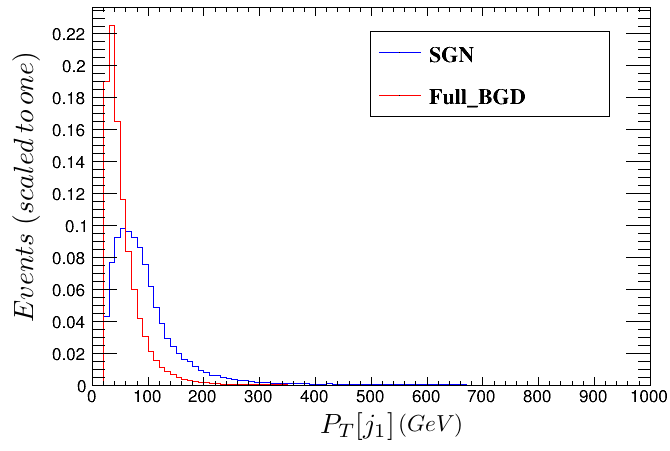}
        \caption{Transverse moment of the $c$-jet.}
        \label{fig:c}
    \end{subfigure}
    \\
    \begin{subfigure}[b]{0.5\textwidth}
        
    \end{subfigure}
\end{figure}

\begin{figure}[!htb]
    \centering
    \begin{subfigure}[b]{0.5\textwidth}
    \includegraphics[scale=0.6]{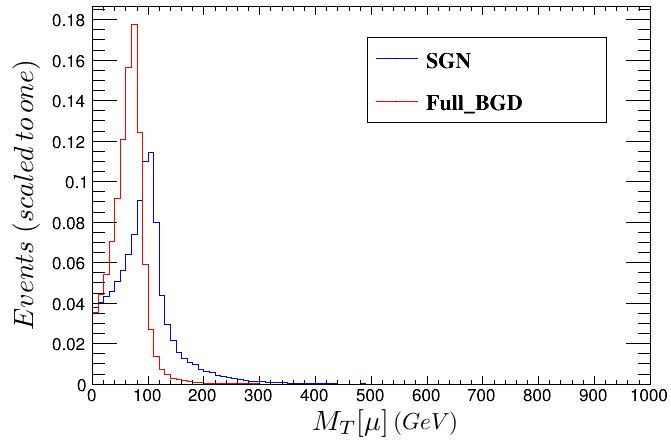}
            \captionsetup{labelformat=empty} % Disable automatic label

        \caption{\hspace{2.5em}(d) Transverse mass of the muon $M_{T}[\mu]$.}
        \label{fig:d}
        \end{subfigure}
    \\
    \begin{subfigure}[b]{0.5\textwidth}
    \includegraphics[scale=0.36]{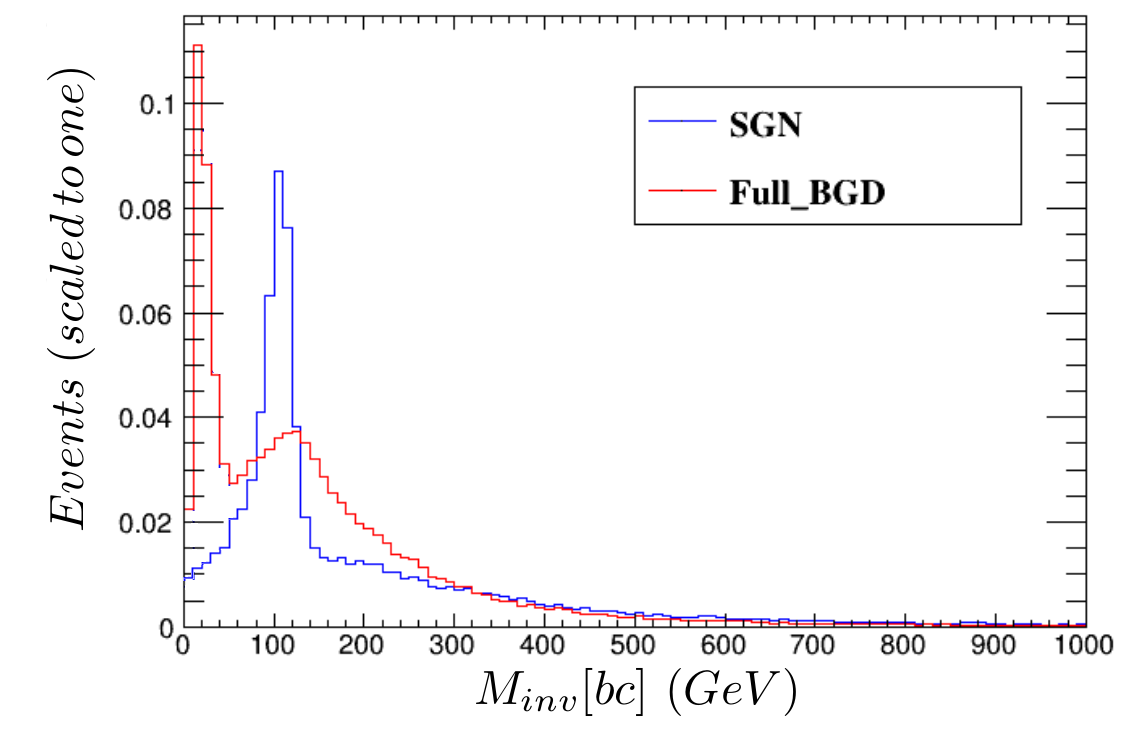}       
        \captionsetup{labelformat=empty} % Disable automatic label
        \caption{\hspace{3em}(e)~\hspace{-0.6em}Invariant mass of one of the \phantom{Invariant} ~\hspace{0.1em}charged Higgs bosons decaying into \phantom{Invariant} ~\hspace{0.1em}a $bc$-jet pair.}
        \label{fig:e}
    \end{subfigure}
    \caption{Kinematic distributions for the signal and SM background processes are shown for the following observables: (a) transverse momentum of the $b$-jet, (b) missing transverse energy $\slashed{E}_T$, (c) transverse momentum of the $c$-jet, (d) transverse mass of the muon $M_T[\mu]$ and (e) invariant mass of one of the charged Higgs bosons decaying into a $bc$-jet pair, $M_{\text{inv}}[bc]$. In all plots the signal is labeled as \texttt{SGN}, while the total background is labeled as \texttt{Full$\_$BGD}.}
    \label{distributions}
\end{figure}
\clearpage
Our procedure involves maximising the \textit{Signal Significance}, defined as the ratio given in Eq.~\eqref{eq:signalsignificance}. This maximum significance is achieved by applying a cut on the \ac{BDT} output, requiring \textbf{xgb} > 0.95, as clearly illustrated in Fig.~\ref{Discriminat}.

\begin{figure}[H]
    \centering
    \begin{subfigure}[b]{0.3\textwidth}
        \includegraphics[scale=0.29]{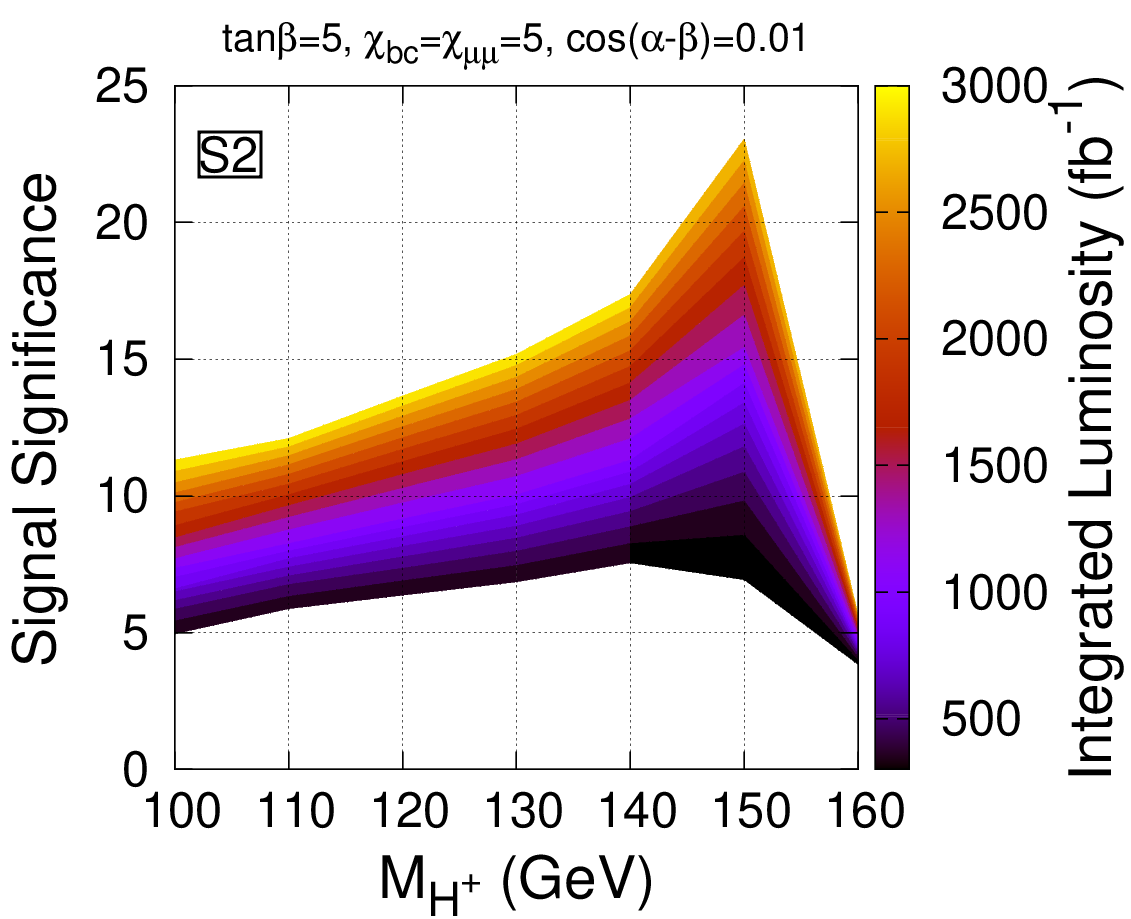}
        \caption{\textbf{S2}, $M_H=500$ GeV}
        \label{fig:b}
    \end{subfigure}%
    \\
    \begin{subfigure}[b]{0.3\textwidth}
        \includegraphics[scale=0.29]{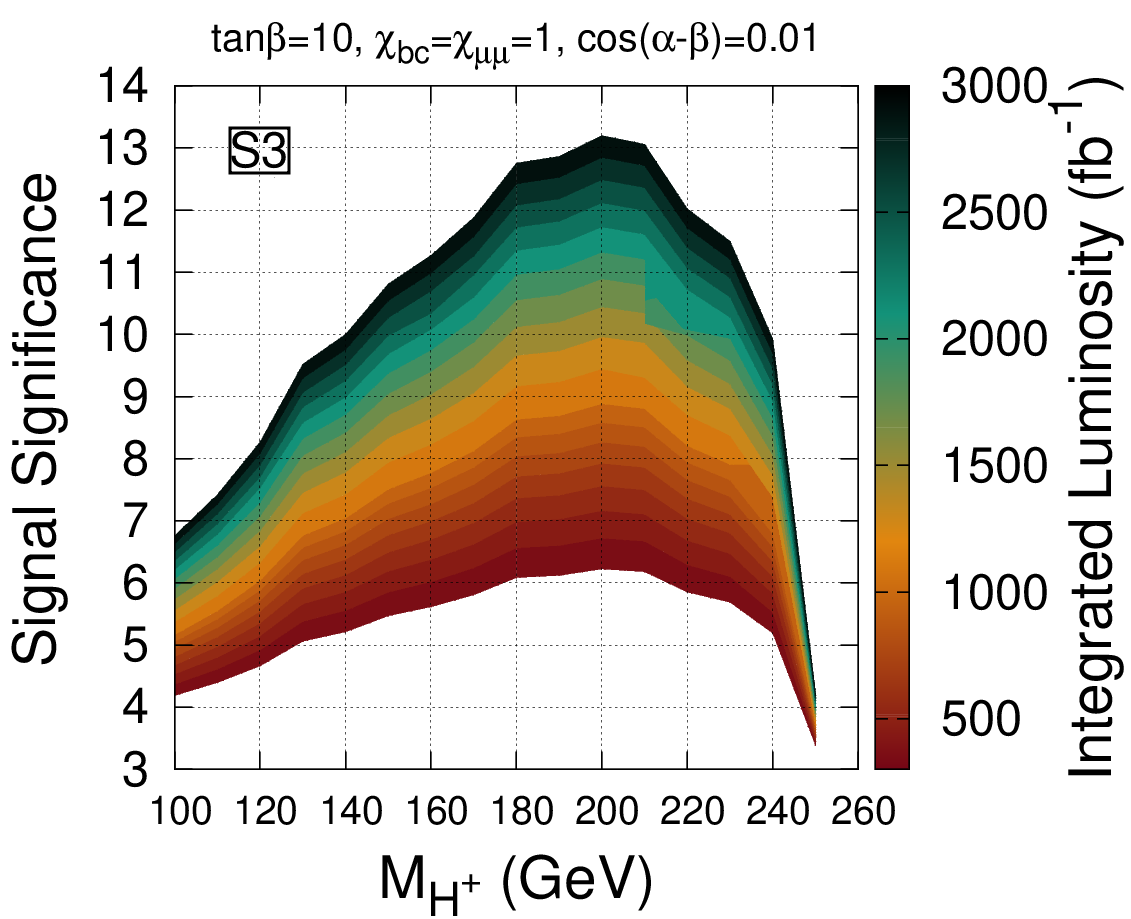}
        \caption{\textbf{S3}, $M_H=500$ GeV}
        \label{fig:c}
    \end{subfigure}
    \\
    \begin{subfigure}[b]{0.3\textwidth}
        \includegraphics[scale=0.29]{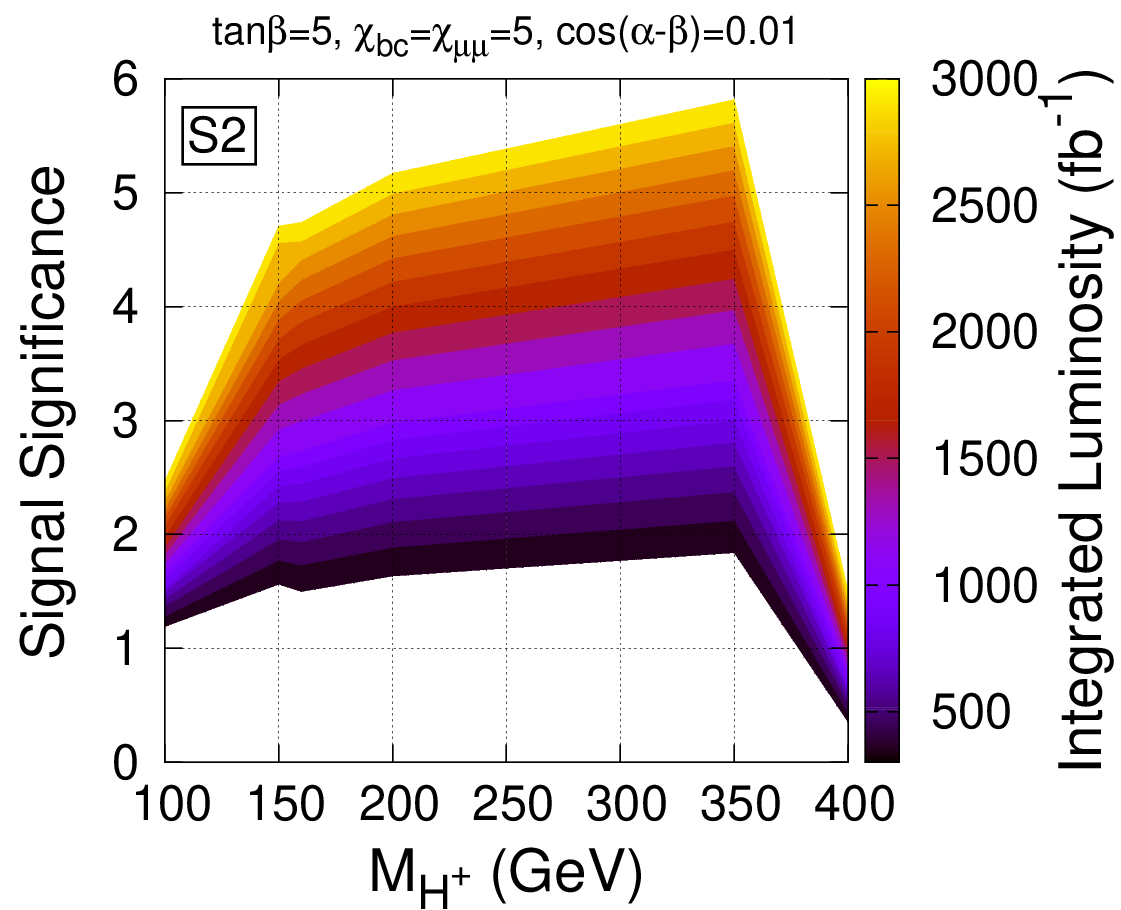}
        \caption{\textbf{S2}, $M_H=800$ GeV}
        \label{fig:d}
    \end{subfigure}
    \caption{\textit{Signal Significance} as a function of the charged Higgs boson mass $M_{H^{\pm}}$ and the integrated luminosity for the following scenarios: (a) \textbf{S2}, (b) \textbf{S3} with $M_H = 500\,\text{GeV}$, and (c) \textbf{S2} with $M_H = 800\,\text{GeV}$.}
    \label{SignalSignificance}
\end{figure}

Figure~\ref{SignalSignificance} presents contour plots of the \textit{Signal Significance} as a function of the charged Higgs boson mass $M_{H^{\pm}}$ and the integrated luminosity for scenarios \textbf{S2}\footnote{We explore masses up to $M_{H^{\pm}} = 200\,\text{GeV}$. However, the \textit{Signal Significance} remains negligible in this range.} and \textbf{S3}. We find that scenario \textbf{S1} poses significant experimental challenges as can be seen in Fig.~\ref{BRs}, which is why it has been excluded from our analysis.

	We find that scenario \textbf{S2} is the most promising, predicting a statistical significance of $5\sigma$ in the mass range $100 \leq M_{H^\pm} \leq 160\,\text{GeV}$ for an integrated luminosity of $250$--$300\,\text{fb}^{-1}$. This scenario could potentially be explored at the LHC and subsequently tested with greater precision at the HL-LHC. 

    Regarding scenario \textbf{S3}, it also holds potential for experimental investigation, with the possibility of exploring charged Higgs boson masses up to $M_{H^\pm} \sim 250\,\text{GeV}$. We anticipate achieving a statistical significance of $5\sigma$ for $100 \lesssim M_{H^\pm} \lesssim 250\,\text{GeV}$ once an integrated luminosity in the range $220$--$1000\,\text{fb}^{-1}$ is accumulated.

	%%%%%%%%%%%%%%%%%%%%%%%%%%%%%%%%%%
	%%%%%%%%%%%%%%%%%%%%%%%%%%%%%%%%%
	%%%%%%%%%%%%%%%%%%%%%%%%%%%%%%%%%%

\section{Collider Analysis for the FNSM}

	\label{se:col_an}
	
\begin{figure}[!hbt]
		\includegraphics[width=10cm]{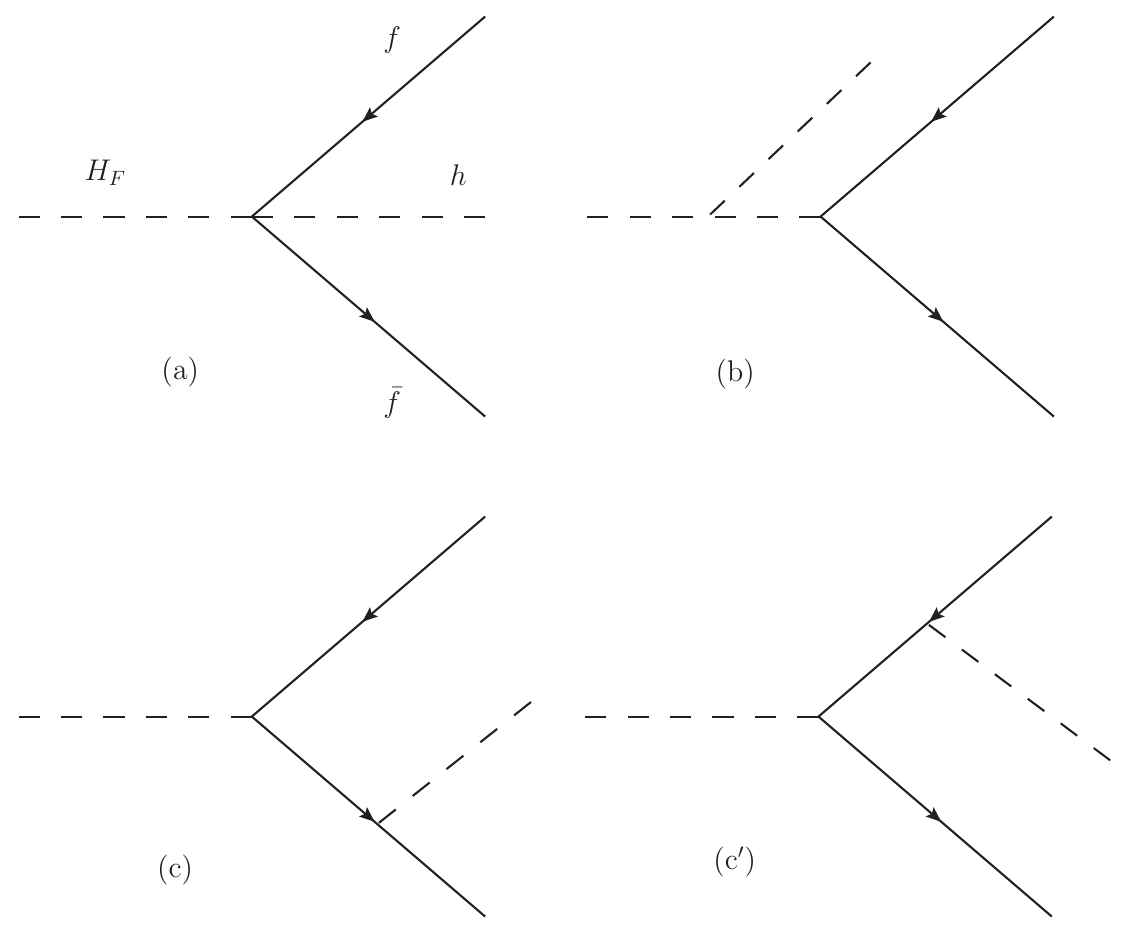}
		\caption{Feynman diagrams inducing the $H_F\to \bar{f}fh$ decay in the FNSM. Diagram (a) shows an effective vertex corresponding to a dimension-5 operator suppressed by $1/\Lambda$. Such interactions are non renormalisable and are treated within the effective field theory framework, where higher dimensional operators are systematically incorporated and suppressed by powers of the new physics scale $\Lambda$, as shown in Eq.~\eqref{vertfnsm}. This ensures the consistency of the model at energies below $\Lambda$.}
\label{Htohff}
	\end{figure}

Having completed the collider analysis for the 2HDM, we proceed to perform an analogous analysis within the FNSM framework. We first present the decay width $\Gamma(H_F \to h\bar{f}f)$ for the three-body process. The study of such processes is of particular interest, as they can exhibit a sizeable BR. The Feynman diagrams contributing to these reactions are shown in Fig.~\ref{Htohff} ~\cite{Burgess:2020eft}.

	The decay width can be written as follows
	\begin{eqnarray}
		\Gamma(H_F\to h\bar{f}f)&=&\frac{M_{H_F}}{256 \pi
			^3} \int  dx_a\int dx_b |\overline {\cal M} |^2,
	\end{eqnarray}
	where the average square amplitude is given by
	\begin{align}
	|\overline {\cal M}|^2 
	&= \frac{1}{2\left(x_a+x_b+x_h-2\right)^2}
	\left(x_a+x_b+x_h-4 x_t-1\right)
	\left(\left(x_a+x_b+x_h-2\right)C_a+C_b\right)^2 \nonumber\\[2mm]
	&\quad +\frac{2}{\left(x_a-1\right)^2\left(x_b-1\right)^2}\Bigg(
	\left(x_a-1\right)\left(x_b-1\right)\left(x_a-x_b\right)^2 \nonumber\\[1mm]
	&\quad \quad -16\left(x_a+x_b-2\right)^2 x_t^2 \nonumber\\[1mm]
	&\quad \quad +4\left(x_a+x_b-2\right)
	\left(2-3x_b+x_a\left(4 x_b-3\right)\right)x_t \nonumber\\[1mm]
	&\quad \quad +x_h\Bigl(4\left(x_a+x_b-2\right)^2x_t-\left(x_a-x_b\right)^2\Bigr)
	\Bigg)C_c^2 \nonumber\\[2mm]
	&\quad -\frac{4 \sqrt{x_t}}{\left(x_a-1\right)\left(x_b-1\right)}
	\Bigl(x_a^2+2 \left(3 x_b+x_h-4 x_t-3\right) x_a+x_b^2-4 x_h \nonumber\\[1mm]
	&\quad \quad +2 x_b\left(x_h-4 x_t-3\right)+16x_t+4\Bigr)C_a C_c \nonumber\\[2mm]
	&\quad -\frac{4 \sqrt{x_t}}{\left(x_a-1\right)\left(x_b-1\right)\left(x_a+x_b+x_h-2\right)} \nonumber\\[1mm]
	&\quad \quad \times \Biggl(x_a^2+2\left(3 x_b+x_h-4 x_t-3\right)x_a
	+x_b^2-4 x_h \nonumber\\[1mm]
	&\quad \quad \quad +2 x_b\left(x_h-4 x_t-3\right)+16x_t+4\Biggr)C_b C_c.
\end{align}

with $x_a=(m_a/M_{H_F})^2$. The factors $C_a=g_{H_F hff}$, $C_b=g_{H_F h h}g_{h ff}/m_h^2$, 
and $C_c=g_{H_F ff}g_{hff}/m_h$ represent the coupling constants involved in the Feynman 
diagrams of Fig. \ref{Htohff}.

Finally, the integration domain is given by
\begin{equation}
	2 \sqrt{x_t}\leq x_a\leq 1-x_h-2 \sqrt{x_t x_h},
\end{equation}
\begin{equation}
	x_b \gtreqqless\frac{2 (1-x_h+2x_t)+x_a \left(x_a+x_h-2 x_t-3\right)\mp\sqrt{x_a^2-4 x_t}\sqrt{\left(x_a+x_h-1\right)^2-4 x_h x_t}}{2 \left(1-x_a+x_t,\right)}.
\end{equation}

	Meanwhile, the production cross-section of the heavy CP-even Flavon $H_F$ (or the pseudo scalar $A_F$, for that matter) depends primarily on the $g_{H_F t\bar{t}}= \frac{ c_\alpha v + s_\alpha
		v_s}{v_s}\, \frac{y_t}{\sqrt{2}}$ ($g_{A_F t\bar{t}}=\frac{v}{v_s}\, \frac{y_t}{\sqrt{2}} $) coupling.  
	The corresponding term in the effective Lagrangian reads \cite{Plehn:2009nd}:
	%%%%%%%%%%%
	%%%%%%%%%%%
	%%%%%%%%%%%
	\begin{eqnarray}
		\mathcal{L}_{\rm eff}&=&\frac{1}{v} \, g_{hgg} \, h \, G_{\mu\nu} G^{\mu\nu},\\
		g_{Sgg} &=& -i \, \frac{\alpha_S}{8\pi}\, \tau (1+(1-\tau)\,f(\tau))~~~~~{\rm with}~~\tau = \frac{4 M_t^2}{M_h^2},\\
		f(\tau)&=& \begin{cases} 
			(\sin^{-1}\sqrt{ \frac{1}{\tau} })^2, \quad\quad\quad\quad\quad\quad \tau\geq 1,\\ 
			-\frac{1}{4}[\ln\frac{1+\sqrt{ 1-\tau}}{1-\sqrt{1-\tau}}-i\pi]^2\quad\quad\quad \tau<1.
		\end{cases}
	\end{eqnarray}
	%%%%%%%%%%%
	%%%%%%%%%%%
	%%%%%%%%%%%
	In FNSM, the $ggh$, $gg H_{F}$ and $gg A_{F}$ couplings are given, respectively, by:
	\begin{eqnarray}
		g_{hgg}&=&\left(\frac{c_\alpha v_s - s_\alpha v }{v_s}\right)\, g_{Sgg},\nonumber\\
		g_{H_F gg}&=&\left (\frac{c_\alpha v +s_\alpha v_s }{v_s}\right)\, g_{S gg},\nonumber \\
		g_{A_F gg}&=&\frac{v}{v_s}\,(-i \,\alpha_S/\pi)\, \tau \,f(\tau),
	\end{eqnarray}
	The scalar $A_F$ is CP-odd, while $h$ and $H_F$ are CP-even scalars. When the couplings involving left and right-handed fields are expressed in terms of Dirac fields, the Hermitian part of the coupling in Eq.~\eqref{eq:fermlag} introduces an imaginary factor $i = \sqrt{-1}$ for $h$ and $H_F$ and a $\gamma_5$ coupling for $A_F$. Consequently, the result of the top quark loop integral differs for $h$, $H_F$, and $A_F$.\cite{Pak:2011hs,LHCHiggsCrossSectionWorkingGroup:2016ypw}. It is {to be} noted that for $M_{H_F, A_F}>2\, M_t$, $f(\tau)=-\frac{1}{4}[\ln\frac{1+\sqrt{ 1-\tau}}{1-\sqrt{1-\tau}}-i\pi]^2$.
	
Following a similar procedure as described previously for the 2HDM-III framework, we first use {\tt FeynRules} \cite{Alloul:2013bka} to derive the FNSM model in our computational framework and generate the corresponding UFO files for {\tt MadGraph5} \cite{Alwall:2014hca}. We compute the production cross-sections for the aforementioned production and decay processes using the resulting particle spectrum.

   Background events within the SM were generated using the \texttt{$\rm MadGraph\_aMC@NLO$} framework \cite{Alwall:2014hca}. Simulations of parton showering and hadronisation were subsequently performed using \texttt{Pythia-8} \cite{Sjostrand:2014zea}. For this analysis, the default ATLAS card provided in the \texttt{Delphes-3.4.2} package was employed as well as to emulate the detector response  \cite{deFavereau:2013fsa}. We consider the Leading Order cross-sections calculated by \texttt{$\rm MadGraph\_aMC@NLO$} for both signal and background processes.

	Having outlined the methodology for evaluating collider observables, we now define in Table~\ref{Scenarios} three different scenarios to be analysed in the following analysis. For practicality, we label these scenarios in the same way as for the 2HDM-III collider analysis: \textbf{S1}, \textbf{S2}, and \textbf{S3}.

\begin{table}[!htb]

		\begin{centering}
			\begin{tabular}{|c|c|c|c|}
				\hline 
				Parameter & \textbf{S1} & \textbf{S2} & \textbf{S3}\tabularnewline
				\hline 
				\hline 
				$\cos\alpha$ & $0.995$ & $0.995$ & $0.995$\tabularnewline
				\hline 
				$\Lambda$ & $1$ TeV & $1.5$ TeV & $2.5$ TeV\tabularnewline
				\hline 
				$v_{s}$ & $1$ TeV & $1.5$ TeV & $2.5$ TeV\tabularnewline
				\hline 
			\end{tabular}
			\par\end{centering}
\caption{Scenarios (\textbf{S1}, \textbf{S2}, \textbf{S3}) used in the calculations.}\label{Scenarios}
	\end{table}
	
	In Table~\ref{XS-signal} we present the numerical cross-sections of the proposed signal, along with the corresponding number of events generated for the scenarios $\textbf{S1}$, $\textbf{S2}$ and $\textbf{S3}$.
	\begin{table}[!htb]

		\begin{centering}
			\begin{tabular}{|c|c|c|c|}
				\hline 
				Scenario & $M_{H_{F}}$(GeV) & $\sigma(pp\to H_{F}\to hb\bar{b})$(fb) & Events $(\mathcal{L}_{{\rm int}}=300$fb$^{-1})$\tabularnewline
				\hline 
				\hline 
				$\textbf{S1}$ & $(800,\,900,\,1000)$ & $(6.8,\,3.3,\,1.7)$ & $(2040,\,990,\,510)$\tabularnewline
				\hline 
				$\textbf{S2}$ & $(800,\,900,\,1000)$ & $(4.3,\,2.3,\,1.3)$ & $(1290,\,690,\,390)$\tabularnewline
				\hline 
				$\textbf{S3}$ & $(800,\,900,\,1000)$ & $(3.8,\,2.1,\,1.1)$ & $(1140,\,630,\,330)$\tabularnewline
				\hline 
			\end{tabular}
			\par\end{centering}
            \caption{Cross-section of the signal for scenarios $\textbf{S1},\,\textbf{S2},\,\textbf{S3}$.}\label{XS-signal}
	\end{table}
	
\begin{figure}[!htb]
		\begin{center}
			\includegraphics[scale=0.28]{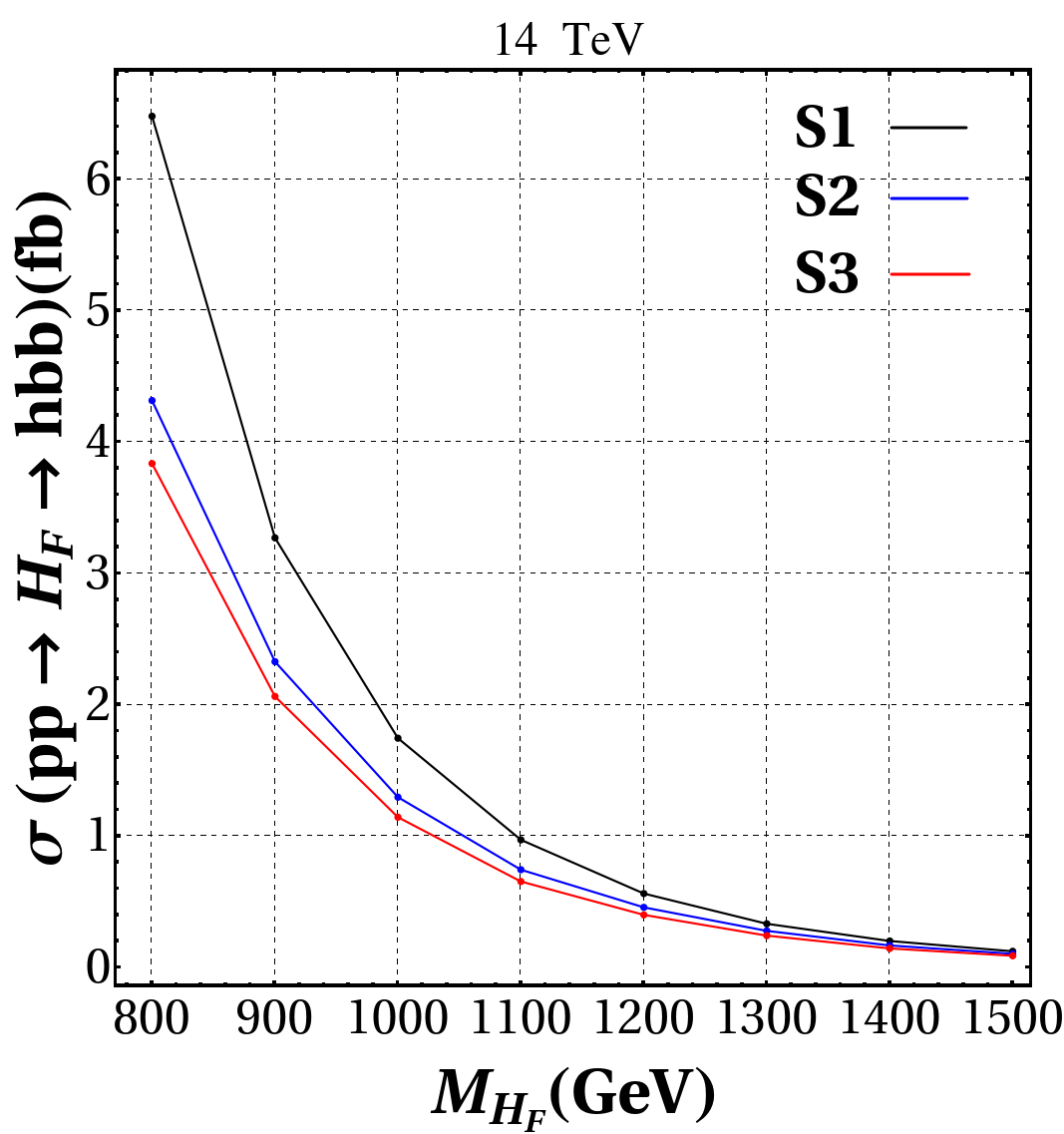}
		\end{center}
		\caption{Production cross-section of the signal $pp\to H_F\to h b\bar{b}$. The centre-of-mass energy was set to 14 TeV.}
		\label{XS1}
	\end{figure}

Meanwhile, Fig.~\ref{XS1} provides an overview of the production cross-section for the signal process $pp \to H_F \to h b\bar{b}$, considering the three scenarios $\textbf{S1}$, $\textbf{S2}$, and $\textbf{S3}$.
%	We observe that scenario $\textbf{S1}$ ($\textbf{S2},\,\textbf{S3}$), for $M_{H_F}=800$ GeV and an integrated luminosity ($\mathcal{L}_{\rm{int}}$) of $300$ fb$^{-1}$, predicts $1935$ ($1290,\,1140$) signal events ($hb\bar{b}$).   
	
	On the other hand, we analyse two specific decay channels of the Higgs boson: (i) $pp \to H_F \to h b \bar{b} (h \to b \bar{b})$ and (ii) $pp \to H_F \to h b \bar{b} (h \to \gamma \gamma)$.

    The identification of $b$-jets produced from the fragmentation and hadronisation of bottom quarks plays a crucial role in distinguishing the signal from background processes, which typically involve gluons, light-flavour jets ($u$, $d$, $s$) and charm-quark fragmentation. To address this challenge we employ the \texttt{FastJet} package~\cite{Cacciari:2011ma} (via \texttt{MadAnalysis}~\cite{Conte:2012fm}) and utilise the anti-$k_T$ algorithm~\cite{Cacciari:2008gp} for jet reconstruction. Additionally, we incorporate a $b$-tagging efficiency of $\epsilon_b = 90\%$. The probabilities of misidentifying a $c$-jet or any other light jet ($j$) as a $b$-jet are set to $\epsilon_c = 5\%$ and $\epsilon_j = 1\%$, respectively~\cite{ATLAS:2023gog}.

	\subsubsection{$pp \to H_F\to h b \bar{b}(h\to b \bar{b})$}
	The SM background processes for the $h\to b \bar{b}$ channel are given by
	\begin{itemize}
		\item $pp\to t\bar{t},\,(t\to W^+ \bar{b},\,W^+\to c\bar{b},\,\bar{t}\to W^- b,\,W^+\to b\bar{c}$),
		\item $pp\to Wh,\, (W\to cb,\, h\to b\bar{b}$),
		\item $pp\to ZZ,\, (Z\to b\bar{b},\, Z\to b\bar{b}$),
		\item $pp\to Zh,\, (Z\to b\bar{b},\, h\to b\bar{b}$),
		\item $pp\to b\bar{b}jj$, where $j$ denotes non-bottom-quark jets. 
	\end{itemize} 
{The numerical cross-section of the SM background processes is presented in Table~\ref{XSbgd_4b}.}
	
	\begin{table}

		\begin{centering}
			\begin{tabular}{cc}
				\hline 
				SM background process & Cross-section (fb)\tabularnewline
				\hline 
				\hline 
				$pp\to t\bar{t}$ & $26910$\tabularnewline
				\hline 
				$pp\to Wh$ & $0.5463$\tabularnewline
				\hline 
				$pp\to ZZ$ & $231.5$\tabularnewline
				\hline 
				$pp\to Zh$ & $79.75$\tabularnewline
				\hline 
				$pp\to\bar{b}bjj$ & $5.441\times10^{8}$\tabularnewline
				\hline 
			\end{tabular}
			\par\end{centering}
    		\caption{Cross-section of the SM background processes for the \(h \to b\bar{b}\) channel.}\label{XSbgd_4b}
	\end{table}

    For this channel, the $b$-jets originating from the primary vertex are expected to exhibit high transverse momentum, denoted as $p_T(b_1, b_2)$. In contrast, the $b$-jets produced via the decay $h \to b\bar{b}$ possess lower transverse momentum, $p_T(b_3, b_4)$, compared to the primary $b$-jets. A key feature of our signal, and its most distinctive signature, is the resonant effect arising from the decay chain $H_F \to h b\bar{b} \,(h \to b\bar{b}) \to b\bar{b}b\bar{b}$.
	
	In Fig.~\ref{Minv4b} we present the invariant mass $M_{\rm inv}(b_1b_2b_3b_4)$ for the scenario \textbf{S1} with $M_{H_F} = 800\,\text{GeV}$.

    \newpage

    \begin{figure}[!htb]
    \centering
    %--- Subfigure 1
    \begin{subfigure}[b]{0.58\textwidth}
        \centering
        \includegraphics[width=\textwidth]{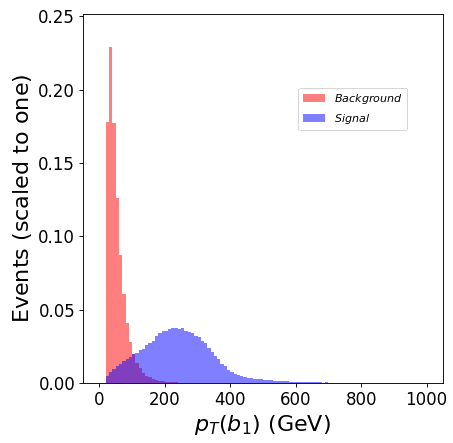}
        \caption{$p_T(b_1)$}
        \label{fig:ptb1}
    \end{subfigure}
    \\
    %--- Subfigure 2
    \begin{subfigure}[b]{0.58\textwidth}
        \centering
        \includegraphics[width=\textwidth]{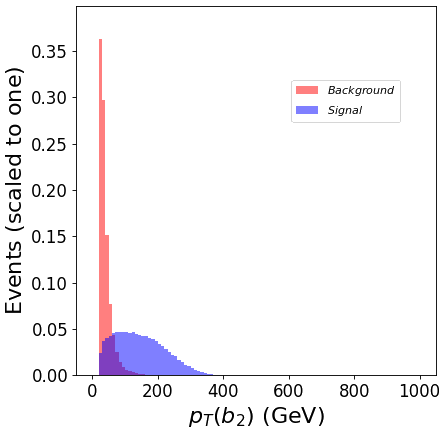}
        \caption{$p_T(b_2)$}
        \label{fig:ptb2}
    \end{subfigure}
\end{figure}
\begin{figure}
    %--- Subfigure 3
    \begin{subfigure}[b]{0.58\textwidth}
        \centering
        \includegraphics[width=\textwidth]{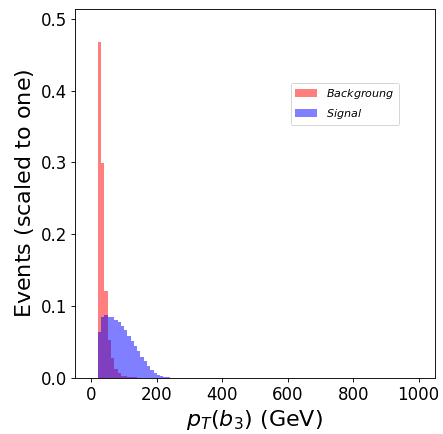}
        \captionsetup{labelformat=empty}
        \caption{(c) $p_T(b_3)$}
        \label{fig:ptb3}
    \end{subfigure}
    \\
    %--- Subfigure 4
    \begin{subfigure}[b]{0.58\textwidth}
        \centering
        \includegraphics[width=\textwidth]{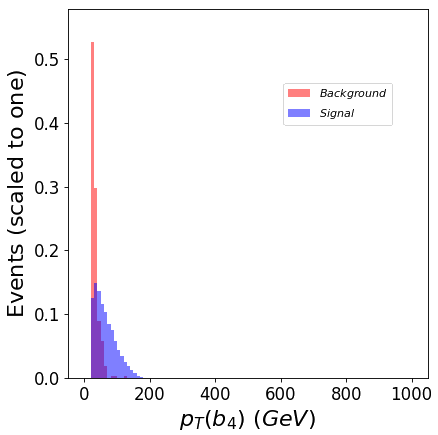}
        \captionsetup{labelformat=empty}
        \caption{(d) $p_T(b_4)$}
        \label{fig:ptb4}
    \end{subfigure}
    \caption{$b$-jet transverse momentum normalised distributions for the signal and total background following the application of the acceptance cuts.}
    \label{distributions_bb-channel}
\end{figure}

    \begin{figure}[H]
    \centering
    \includegraphics[width=9cm]{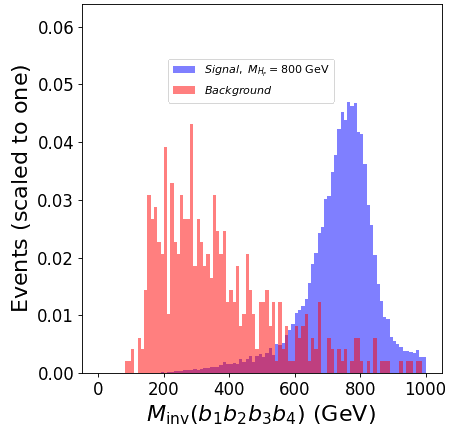}
    \caption{Normalised distribution of the reconstructed invariant mass $M_{\rm inv}(b_1b_2b_3b_4)$ for the signal and background processes.}\label{Minv4b}
    \end{figure}
	
	Meanwhile, Fig.~\ref{distributions_bb-channel} displays the $p_T(b_i)$ distributions ($i = 1, 2, 3, 4$) for both the signal and the SM background processes. The subscript $1$ ($4$) corresponds to the $b$-jet with the highest (lowest) transverse momentum, while the subscripts $2$ and $3$ represent the second and third most energetic $b$-jets, respectively.

	From Fig.~\ref{distributions_bb-channel}, we observe that $p_T(b_1, b_2)$ is higher than $p_T(b_3, b_4)$, as the former originates from the primary vertex, while the latter arises from the decay of the Higgs boson. Both $p_T(b_i)$ and $M_{\rm inv}(b_1b_2b_3b_4)$ are the most critical variables for isolating the signal from the background.
	
	\subsubsection{$pp \to H_F\to h b \bar{b}(h\to \gamma \gamma)$}
	Regarding the di-photon channel, the SM background processes are as follows,
	\begin{itemize}
		\item $pp\to h t\bar{t},\,(h\to\gamma\gamma,\,t\to W^+ \bar{b},\,W^+\to \ell^+ \nu_\ell,\,\bar{t}\to W^- b,\,W^+\to \ell^- \bar{\nu}_\ell$),
		\item $pp\to t\bar{t}\gamma\gamma,\,(t\to W^+ \bar{b},\,W^+\to \ell^+ \nu_\ell,\,\bar{t}\to W^- b,\,W^+\to \ell^- \bar{\nu}_\ell$),
		\item $pp\to Wh,\, (W\to cb,\, h\to \gamma\gamma$),
		\item $pp\to Zh,\, (Z\to b\bar{b},\, h\to \gamma\gamma$),
		\item $pp\to hjj,\,  (h\to \gamma\gamma$),
		\item $pp\to \gamma\gamma jj$
		\item $pp\to \gamma\gamma b\bar{b}$.
	\end{itemize} 
The numerical cross-section of the SM background processes is presented in Table~\ref{XSbgd_2b2g}.

\begin{table}

	\begin{centering}
		\begin{tabular}{cc}
			\hline 
			SM background process & Cross-section (fb)\tabularnewline
			\hline 
			\hline 
			$pp\to ht\bar{t}$ & $3.2\times10^{-2}$\tabularnewline
			\hline 
			$pp\to t\bar{t}\gamma\gamma$ & $0.57$\tabularnewline
			\hline 
			$pp\to Wh$ & $9.7\times10^{-4}$\tabularnewline
			\hline 
			$pp\to Zh$ & $0.14$\tabularnewline
			\hline 
			$pp\to hjj$ & $13.62$\tabularnewline
			\hline 
			$pp\to\gamma\gamma jj$ & $1.1\times10^{5}$\tabularnewline
			\hline 
			$pp\to b\bar{b}\gamma\gamma$ & $5113$\tabularnewline
			\hline 
		\end{tabular}
		\par\end{centering}
	\caption{Cross-section of the SM background processes for the \(h \to \gamma\gamma\) channel.}\label{XSbgd_2b2g}
\end{table}

In this case we observe a scenario similar to the previous channel with respect to the resonant effect arising from the decay $H_F \to b\bar{b}\gamma\gamma$. Figure~\ref{Minv2b2gam} displays the invariant mass distribution $M_{\rm inv}(bb\gamma\gamma)$ for $M_{H_F} = 900\,\text{GeV}$.
	As in the previous channel, we also present in Fig. \ref{fig:distributions_gammagamma-channel} the the $p_T(b_i)$, $p_T(\gamma_i)$ ($i=1,\,2$).

	%\subsubsection{$pp \to H_F\to h b \bar{b}(h\to W W^*)$}
	
	%\subsubsection{$pp \to H_F\to h b \bar{b}(h\to Z Z^*)$}
	
	%\subsubsection{$pp \to H_F\to h b \bar{b}(h\to \tau^-\tau^+)$}
	
	\subsection{Multivariate Analysis}
		Following the kinematic analysis, we found that most of the observables used to distinguish the signal from the background exhibit relatively weak discriminating power. Consequently, we perform the final candidate selection using MVA discriminators, combining these observables into the single, more powerful classifier. As for the 2HDM-III, we employ a \ac{BDT} algorithm implemented via the XGBoost library.

        \newpage

\begin{figure}[!htb]
    \centering
    % (a)
    \begin{subfigure}[b]{0.6\textwidth}
        \centering
        \includegraphics[width=\textwidth]{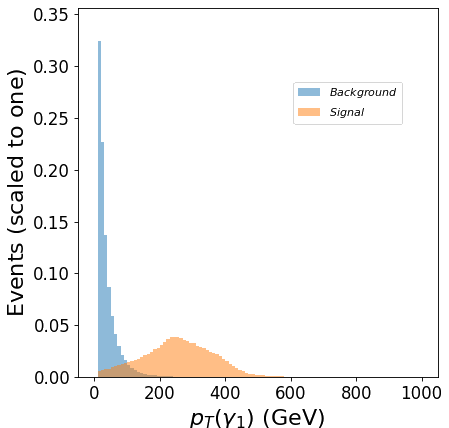}
        \caption{$p_T(\gamma_1)$}
        \label{fig:pt-gamma1}
    \end{subfigure}
    \\
    % (b)
    \begin{subfigure}[b]{0.6\textwidth}
        \centering
        \includegraphics[width=\textwidth]{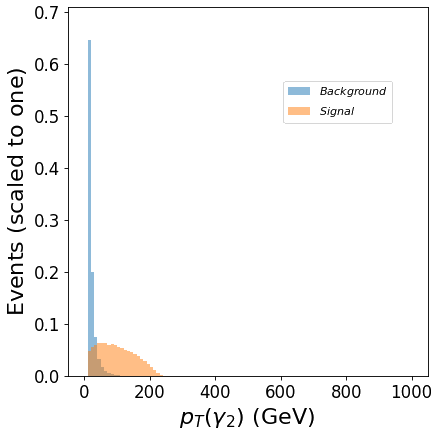}
        \caption{$p_T(\gamma_2)$}
        \label{fig:pt-gamma2}
    \end{subfigure}
\end{figure}

\begin{figure}[!htb]
    \centering
    % (c)
    \begin{subfigure}[b]{0.6\textwidth}
        \centering
        \includegraphics[width=\textwidth]{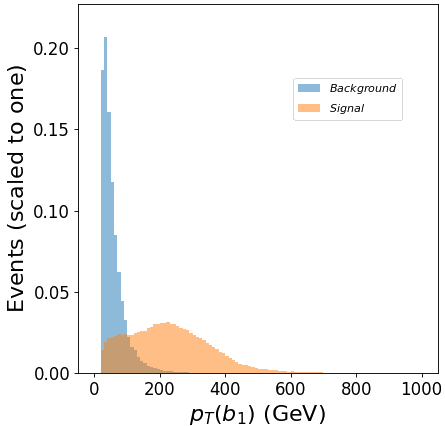}
        \captionsetup{labelformat=empty}
        \caption{(c) $p_T(b_1)$}
        \label{fig:pt-b1}
    \end{subfigure}
    \\
    % (d)
    \begin{subfigure}[b]{0.6\textwidth}
        \centering
        \includegraphics[width=\textwidth]{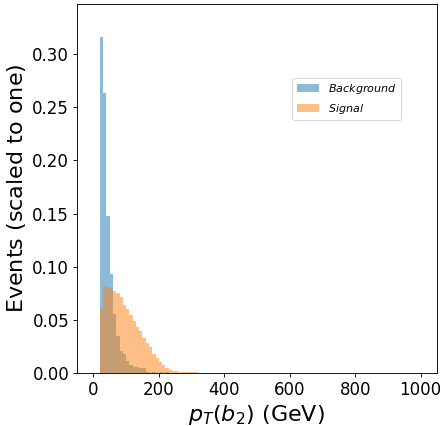}
        \captionsetup{labelformat=empty}
        \caption{(d) $p_T(b_2)$}
        \label{fig:pt-b2}
    \end{subfigure}
    \caption{Normalised $p_T$ distributions for photons ($\gamma$) and $b$-jets, comparing signal and total background after acceptance cuts.}
    \label{fig:distributions_gammagamma-channel}
\end{figure}
\newpage

%The BDT classifiers are trained using variables related to the kinematics of final and intermediate state particles, including the transverse momentum ($p_T$) and rapidity of the daughter $b$-jets (or photons, in the case $h\to \gamma \gamma$), as well as the $p_T$ of Flavon and Higgs particles. 

For this framework, we train the \ac{BDT} classifiers using variables associated with the kinematics of both final-state and intermediate-state particles, including the following:\\
\\
\textbf{Case 1:} Four $b$-jets $(b_1, b_2, b_3, b_4)$:
\begin{itemize}
    \item Transverse momentum ($p_T$):  $p_T(b_1)$, $p_T(b_2)$, $p_T(b_3)$, $p_T(b_4)$
    \item Rapidity ($\eta$): $\eta(b_1)$, $\eta(b_2)$, $\eta(b_3)$, $\eta(b_4)$
    \item Combined transverse momentum: $p_T(b_1b_2b_3b_4)$
\end{itemize}
\textbf{Case 2:} Two $b$-jets $(b_1, b_2)$ and two photons ($\gamma_1$, $\gamma_2$):
\begin{itemize}
    \item Transverse momentum ($p_T$): $p_T(b_1)$, $p_T(b_2)$, $p_T(\gamma_1)$, $p_T(\gamma_2)$ 
    \item Rapidity ($\eta$):  $\eta(b_1)$, $\eta(b_2)$, $\eta(\gamma_1)$, $\eta(\gamma_2)$ 
    \item Invariant mass of the photons: $M(\gamma_1 \gamma_2)$ 
    \item Combined transverse momentum: $p_T(b_1b_2\gamma_1\gamma_2)$
\end{itemize}

\begin{figure}[!htb]
	\includegraphics[width=9cm]{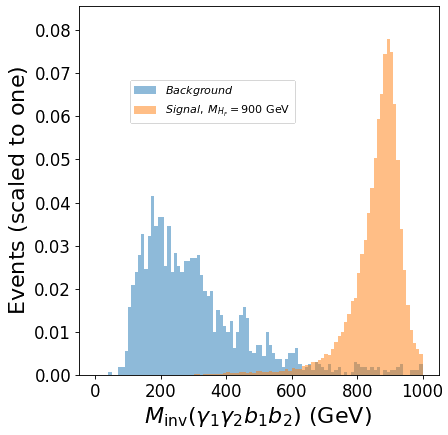}
	\caption{Normalised distribution of the reconstructed invariant mass $M_{\rm inv}(\gamma_1\gamma_2 b_1b_2)$ for the signal and background processes.}\label{Minv2b2gam}
\end{figure}
%The BDT training is performed using the MC-simulated samples. 
The training of the \ac{BDT} classifier is carried out using Monte Carlo simulated samples. The dataset comprises 200,000 signal events and an equivalent number of background events, distributed across the various background processes described earlier, each contributing 200,000 events. The data is partitioned such that 70\% of the total events are allocated to the training dataset, while the remaining 30\% are reserved for testing. To optimise the performance of the \ac{BDT} model, this time we employ \textit{HyperOpt}~\cite{bergstra2013hyperparameter,Hiper:2013} for hyperparameter tuning. 
\textit{HyperOpt} library employs a Bayesian optimisation strategy to maximise classification performance. Rather than exhaustively testing every possible combination, this library treats the performance metric, such as accuracy, as a function of the hyperparameters. It builds a probabilistic model that estimates how changes in hyperparameters affect the performance. 

This iterative process continues until it finds a set of hyperparameters that maximise the classifier's performance on validation data. This optimisation process helps prevent overfitting by finding the optimal balance between the model complexity and generalisation. Regularisation parameters in XGBoost also reduce overfitting by penalising overly complex trees.
As an example of the resulting performance, Fig.~\ref{fig:Roc_hbb2} presents the ROC curve for the $h \to b\bar{b}$ channel. Moreover, in Fig.~\ref{fig:Roc_hbb} we show the relationship between the 1$/$Background Acceptance and the Signal Acceptance for the same decay mode.

\begin{figure}
    \centering
    \begin{subfigure}[b]{0.4\textwidth}
        \includegraphics[scale=0.37]{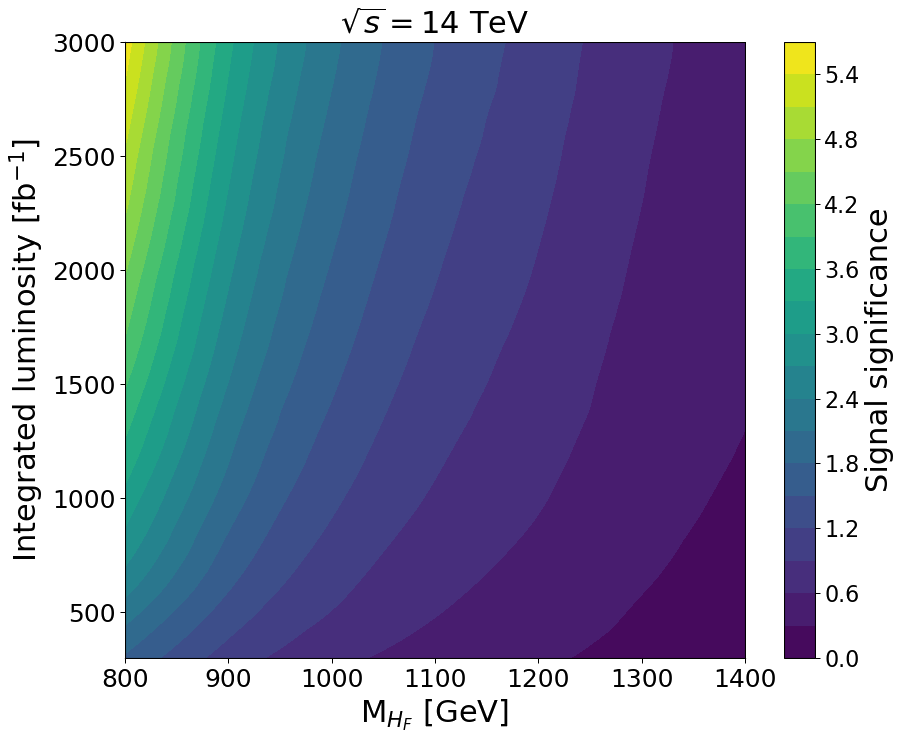}
        \caption{$\textbf{S1}$}
        \label{fig:esc1}
    \end{subfigure}%
    \\
    \begin{subfigure}[b]{0.4\textwidth}
        \includegraphics[scale=0.37]{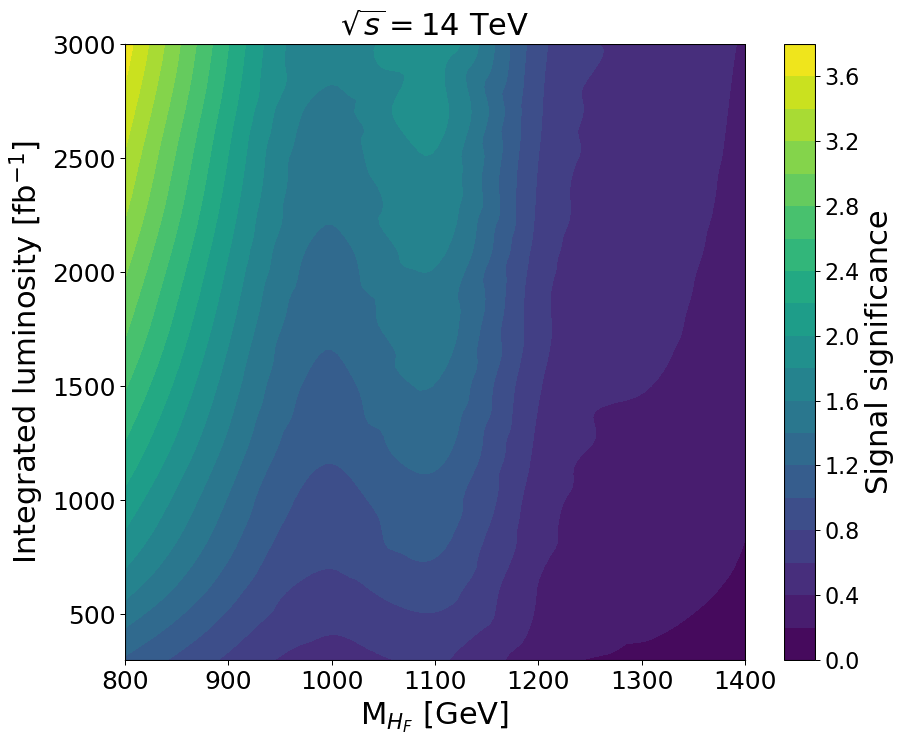}
        \caption{$\textbf{S2}$}
        \label{fig:esc2}
    \end{subfigure}%
    \\
    \begin{subfigure}[b]{0.4\textwidth}
        \includegraphics[scale=0.37]{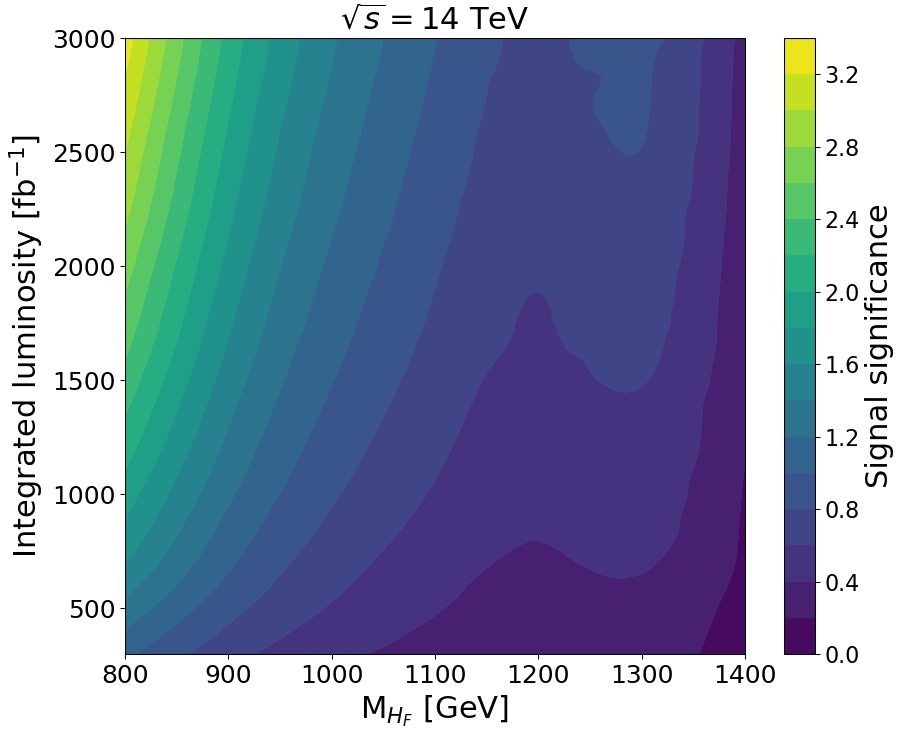}
        \caption{$\textbf{S3}$}
        \label{fig:esc3}
    \end{subfigure}
    \caption{Density plot showing the \textit{Signal Significance} for the $h\to b\bar{b}$ channel as a function of the integrated luminosity and the Flavon mass $M_{H_F}$, the subplots correspond to the following benchmark scenarios: (a) $\textbf{S1}$, (b) $\textbf{S2}$, and (c) $\textbf{S3}$.}
    \label{fig:Hbbh_SigplotS1}
\end{figure}

The expected number of candidates is determined based on the integrated luminosity and the corresponding cross-sections, and the signal and background samples are scaled accordingly.  To maximise the \textit{Signal Significance}, Eq.~\eqref{eq:signalsignificance}, the \ac{BDT} selection is optimised independently for each channel. Figure~\ref{fig:Hbbh_SigplotS1} illustrates the \textit{Signal Significance} (for the three scenarios) as a function of luminosity and the Flavon mass $M_{H_F}$ for the process $pp \to H_F \to h b \bar{b} (h \to b \bar{b})$. 

        For scenario $\textbf{S1}$, we observe strong sensitivity to various choices of $M_{H_F}$ at a 14 TeV $pp$ collider with an integrated luminosity of 3000 fb$^{-1}$. Specifically, in the range $800 < M_{H_F} < 950$ GeV, a \textit{Signal Significance} between $3$-$5.6\sigma$ is achieved. Consequently, if nature favours scenario $\textbf{S1}$, the Flavon particle could potentially be discovered at the HL-LHC. In contrast, scenarios $\textbf{S2}$ and $\textbf{S3}$ exhibit a depletion in significance around $M_{H_F} = 1.2$ TeV, followed by an increase near $M_{H_F} = 1.3$ TeV. This behaviour can be attributed to the slightly enhanced performance of the \ac{BDT} classifier for $M_{H_F} = 1.3$ TeV compared to $1.2$ TeV.

	Fig.~\ref{fig:Haabb_SigplotS1} illustrates the \textit{Signal Significance} as a function of luminosity and the Flavon mass $M_{H_F}$ for the process $pp \to H_F \to h b \bar{b} (h \to \gamma \gamma)$. A key distinction in this channel is its reduced sensitivity compared to the earlier case, which is primarily attributed to the Branching Ratio $\mathcal{BR}(h \to \gamma \gamma) \sim 10^{-3}$. This suppression factor of $10^{-3}$ significantly diminishes the observable signal.

    Despite this limitation, the prospect of a future 100 TeV collider \cite{HL-lhc2} offers renewed optimism. Our calculations suggest that, with an integrated luminosity of 30 ab$^{-1}$, the sensitivity could achieve up to $5\sigma$, making this channel a promising avenue for exploration in next generation experiments.

        In Table $\text{\ref{tab:Signal-significance_StragCUTS}}$
	we show the \textit{Signal Significance} values for the $h\to b\bar{b}$ channel,
	where we have taken into account the basic cut, included in the HL-LHC card~\cite{HL-LHC_card}, along with the following kinematic cuts applied for $M_{H_F}=800$ GeV:
	\begin{enumerate}
		\item $p_{T}(b_1b_2b_3b_4)>80,\,60,\,50,\,25$ GeV,
		\item $650<M_{{\rm inv}}(b_{1}b_{2}b_{3}b_{4})<900$ GeV,
		\item Combined transverse momentum $p_{T}(b_{1}b_{2}b_{3}b_{4})>550$ GeV.
	\end{enumerate}

We can notice that the use of \ac{BDT} substantially improves the isolation and identification of the proposed signals, resulting in a more effective and accurate analysis.

    \begin{table}

		\begin{centering}
			\begin{tabular}{|c|c|}
				\hline 
				$M_{H_{F}}$(GeV) & \textit{Signal Significance}\tabularnewline
				\hline 
				\hline 
				800 & 1.75 $\sigma$\tabularnewline
				\hline 
				900 & 0.87$\sigma$\tabularnewline
				\hline 
				1000 & 0.4872$\sigma$\tabularnewline
				\hline 
			\end{tabular}
			\par\end{centering}
            \caption{The \textit{Signal Significance} for the decay channel $h\to b\bar{b}$ evaluated at $\mathcal{{L}_{{\rm int}}}$=3000 
			fb$^{-1}.$\label{tab:Signal-significance_StragCUTS} }
	\end{table}

\begin{figure}[H]
    \centering
    % Row 1: 800 and 900 GeV
    \begin{subfigure}[b]{0.43\textwidth}
        \includegraphics[width=\textwidth]{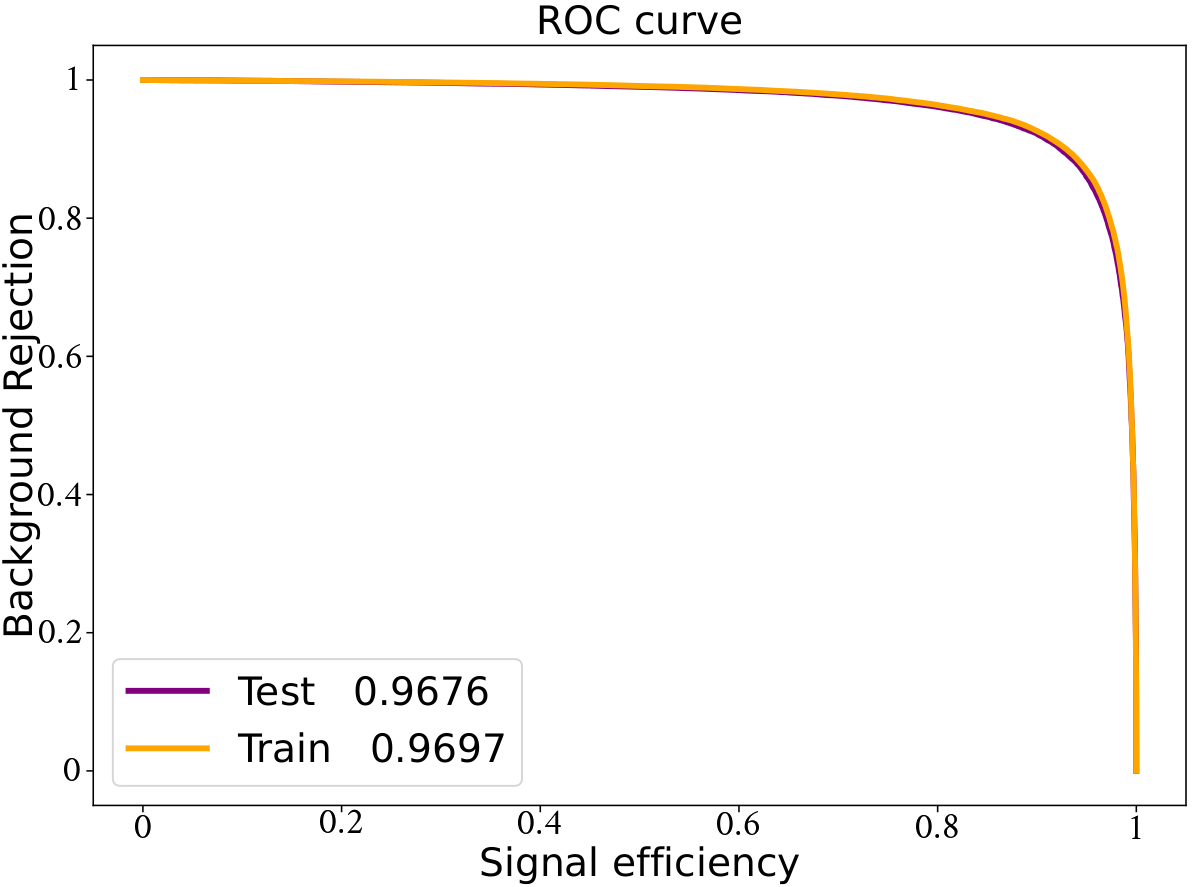}
        \caption{$M_{H_F} = 800$ GeV}
        \label{fig:roc-800}
    \end{subfigure}
    \hfill
    \begin{subfigure}[b]{0.43\textwidth}
        \includegraphics[width=\textwidth]{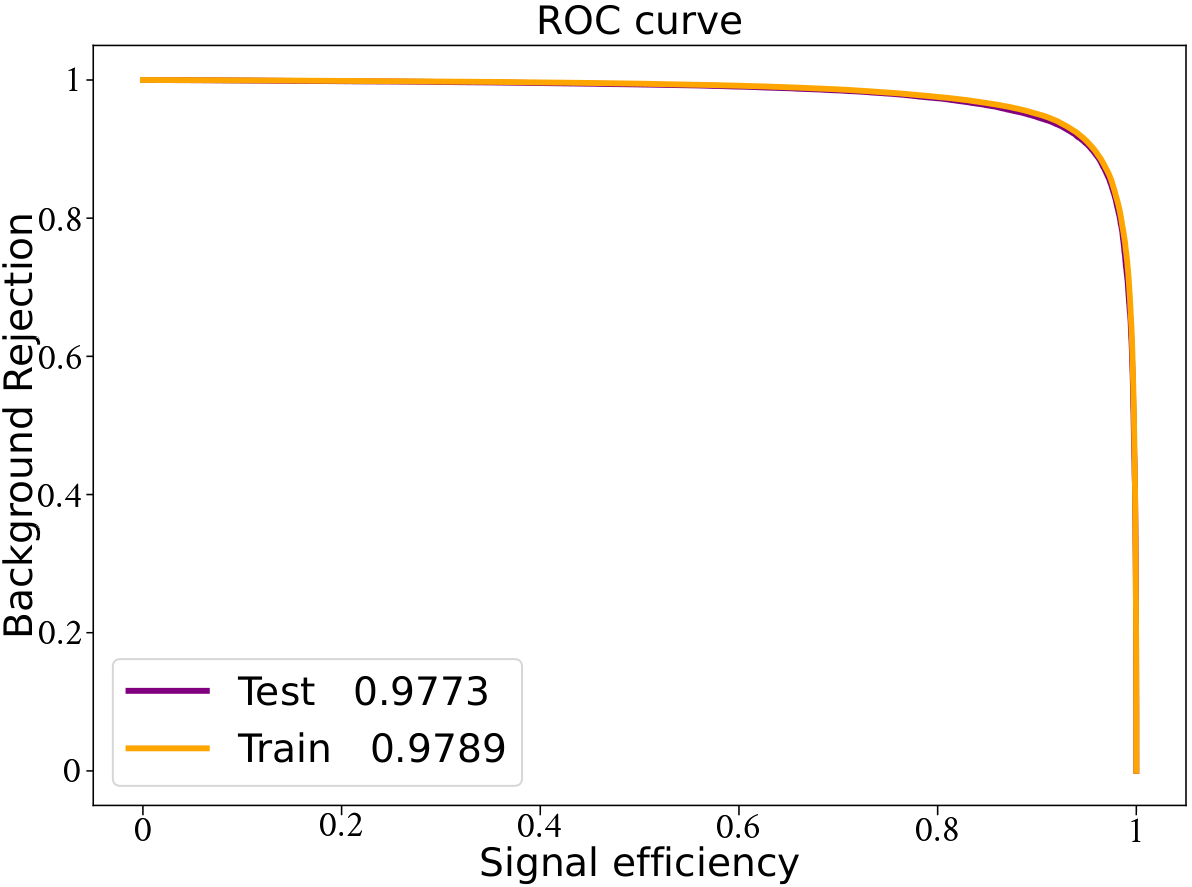}
        \caption{$M_{H_F} = 900$ GeV}
        \label{fig:roc-900}
    \end{subfigure}
    
    \vspace{0.5em}
    
    % Row 2: 1000 and 1100 GeV
    \begin{subfigure}[b]{0.43\textwidth}
        \includegraphics[width=\textwidth]{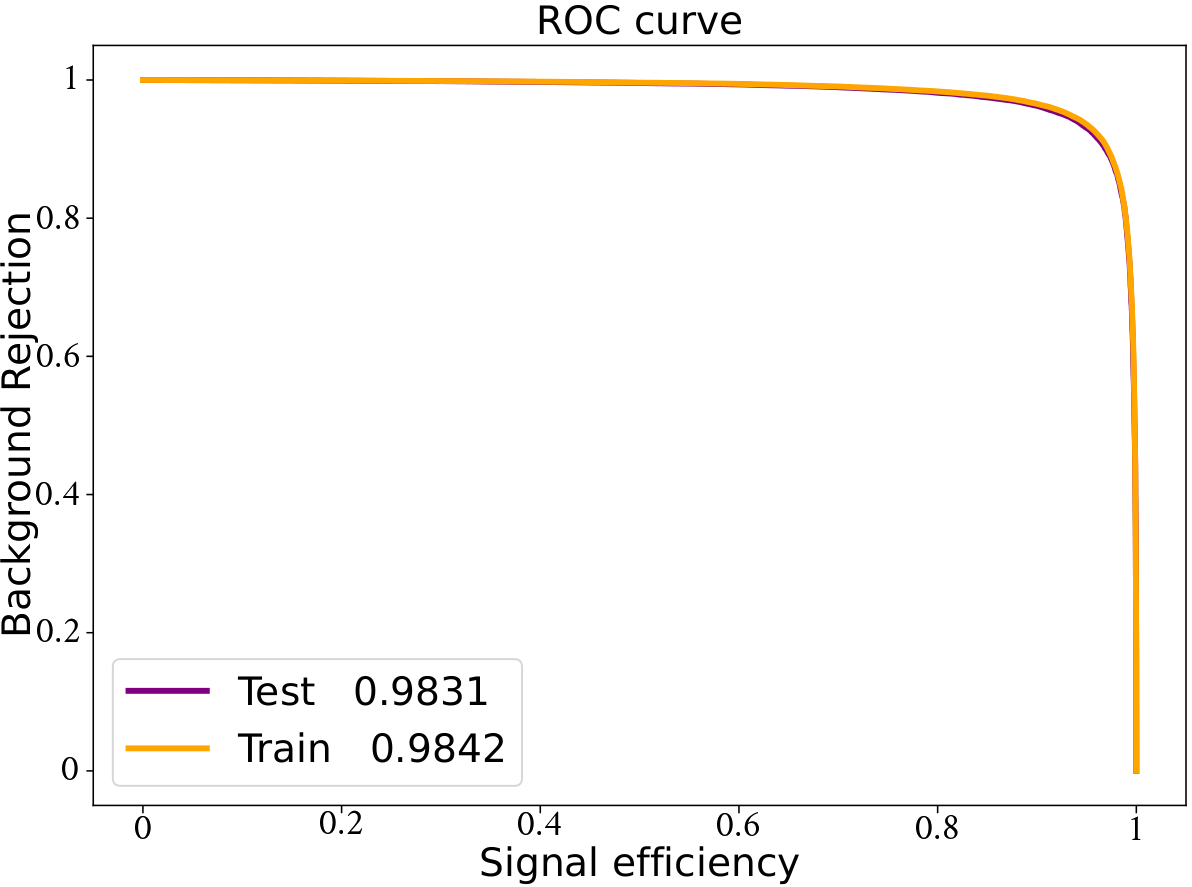}
        \caption{$M_{H_F} = 1000$ GeV}
        \label{fig:roc-1000}
    \end{subfigure}
    \hfill
    \begin{subfigure}[b]{0.43\textwidth}
        \includegraphics[width=\textwidth]{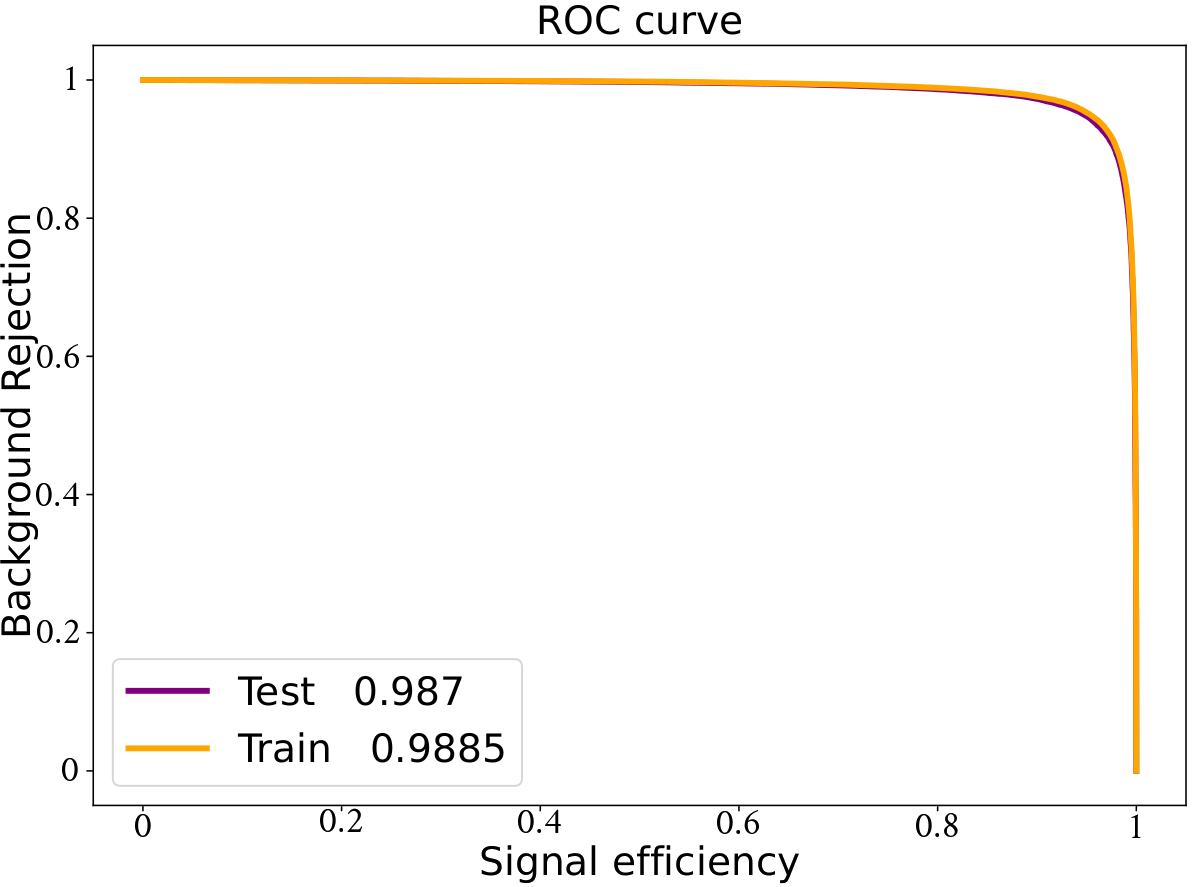}
        \caption{$M_{H_F} = 1100$ GeV}
        \label{fig:roc-1100}
    \end{subfigure}
    
    \vspace{0.5em}
    
    % Row 3: 1200 and 1300 GeV
    \begin{subfigure}[b]{0.43\textwidth}
        \includegraphics[width=\textwidth]{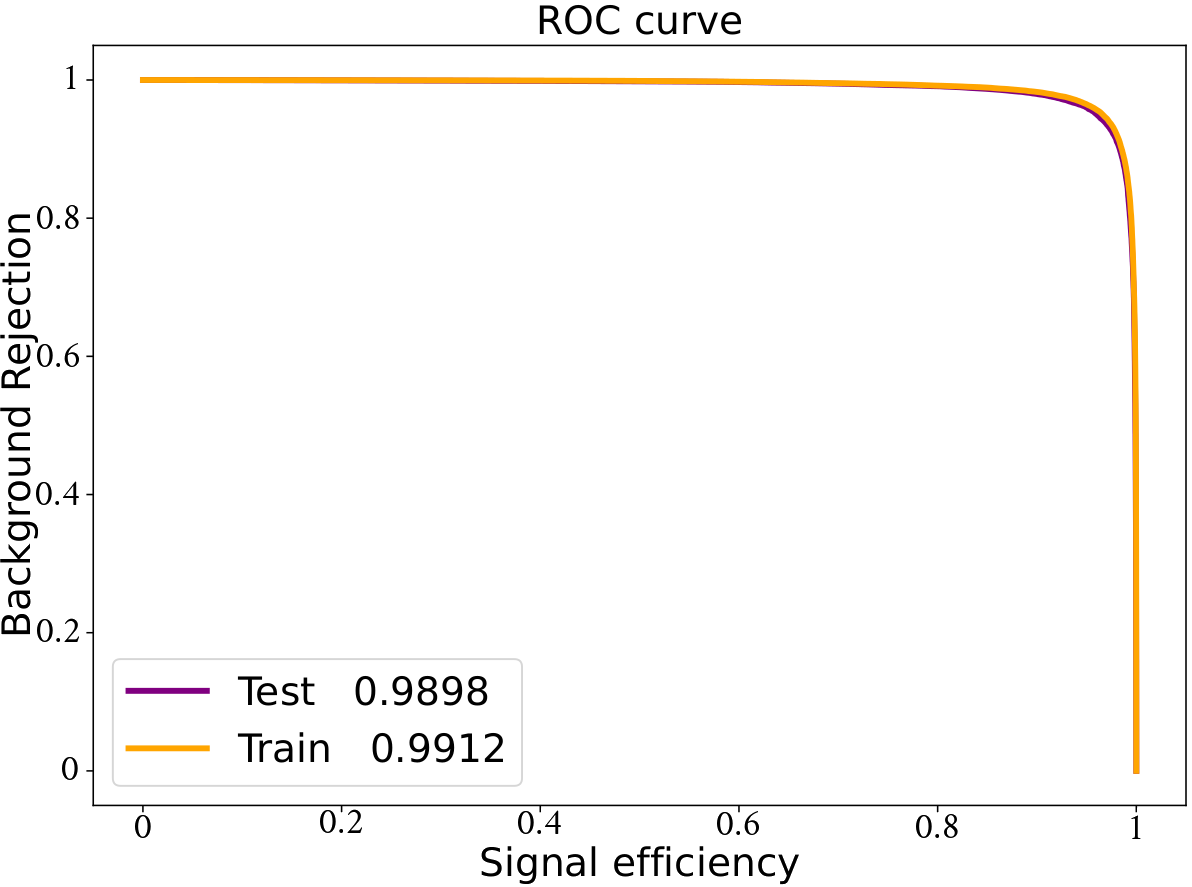}
        \caption{$M_{H_F} = 1200$ GeV}
        \label{fig:roc-1200}
    \end{subfigure}
    \hfill
    \begin{subfigure}[b]{0.43\textwidth}
        \includegraphics[width=\textwidth]{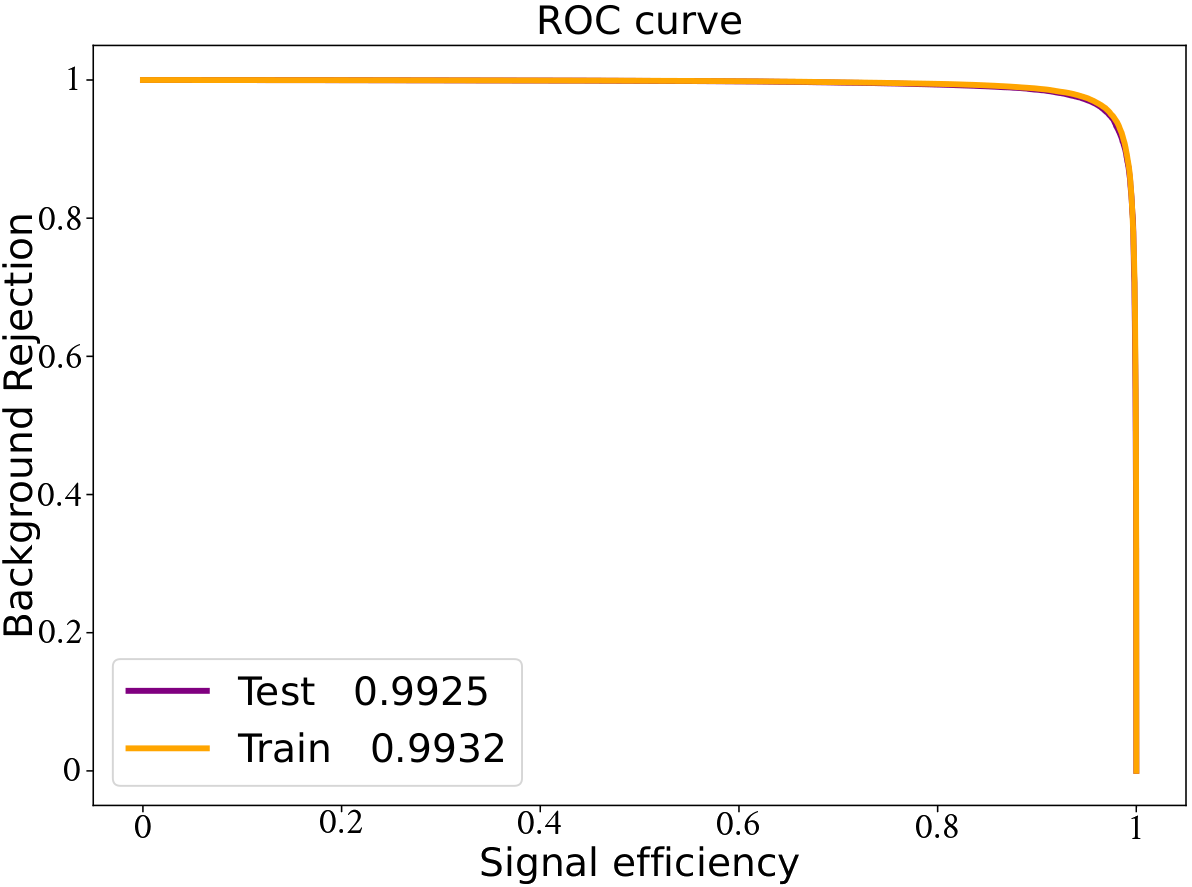}
        \caption{$M_{H_F} = 1300$ GeV}
        \label{fig:roc-1300}
    \end{subfigure}
    
    \vspace{0.5em}
    
    % Row 4: 1400 GeV (centered)
    \begin{subfigure}[b]{0.43\textwidth}
        \centering
        \includegraphics[width=\textwidth]{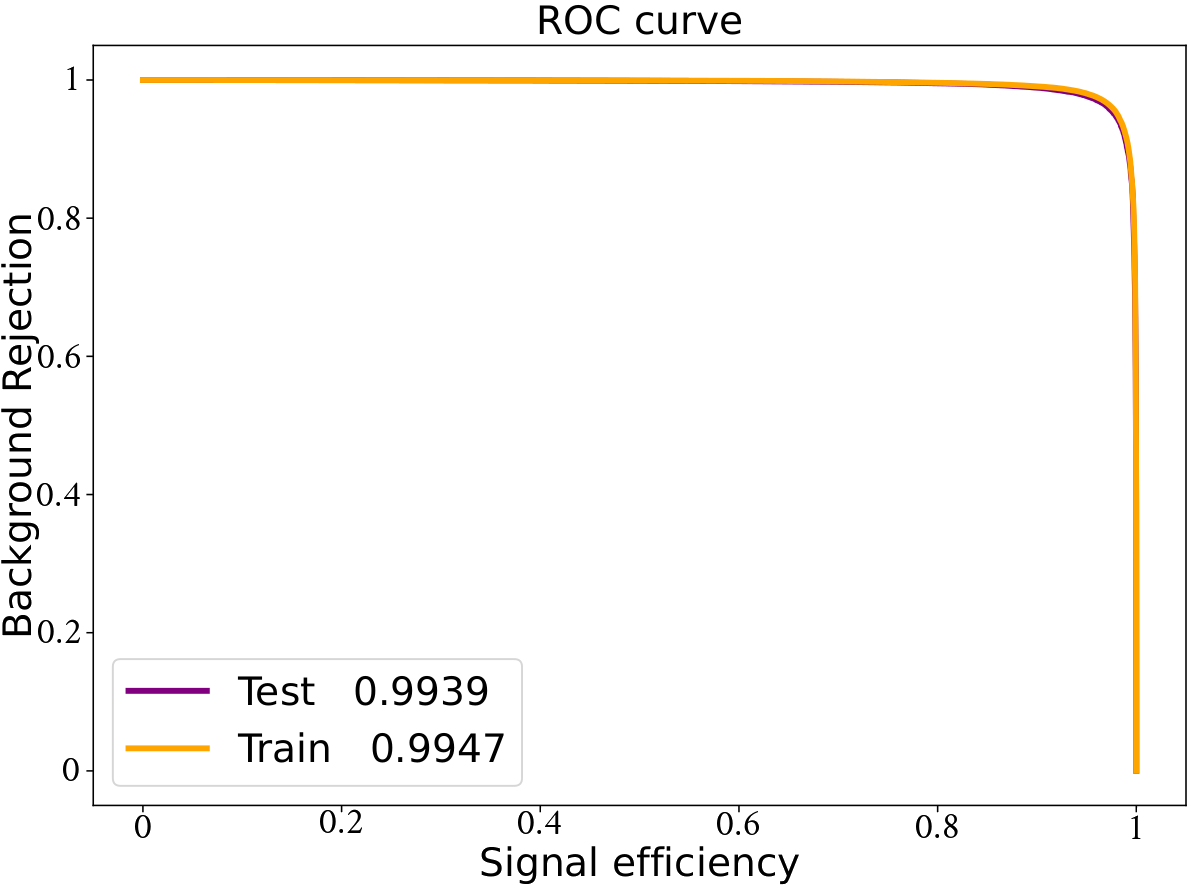}
        \caption{$M_{H_F} = 1400$ GeV}
        \label{fig:roc-1400}
    \end{subfigure}
    
    \caption{ROC curve for $h \to b\bar{b}$ channel showing the model's ability to predict training data (used to fit the model) and independent test data (evaluating performance on new, unseen data). Each plot corresponds to a different Flavon mass as indicated.}
    \label{fig:Roc_hbb2}
\end{figure}

\begin{figure}[H]
    \centering
    % Row 1: 800 and 900 GeV
    \begin{subfigure}[b]{0.43\textwidth}
        \includegraphics[width=\textwidth]{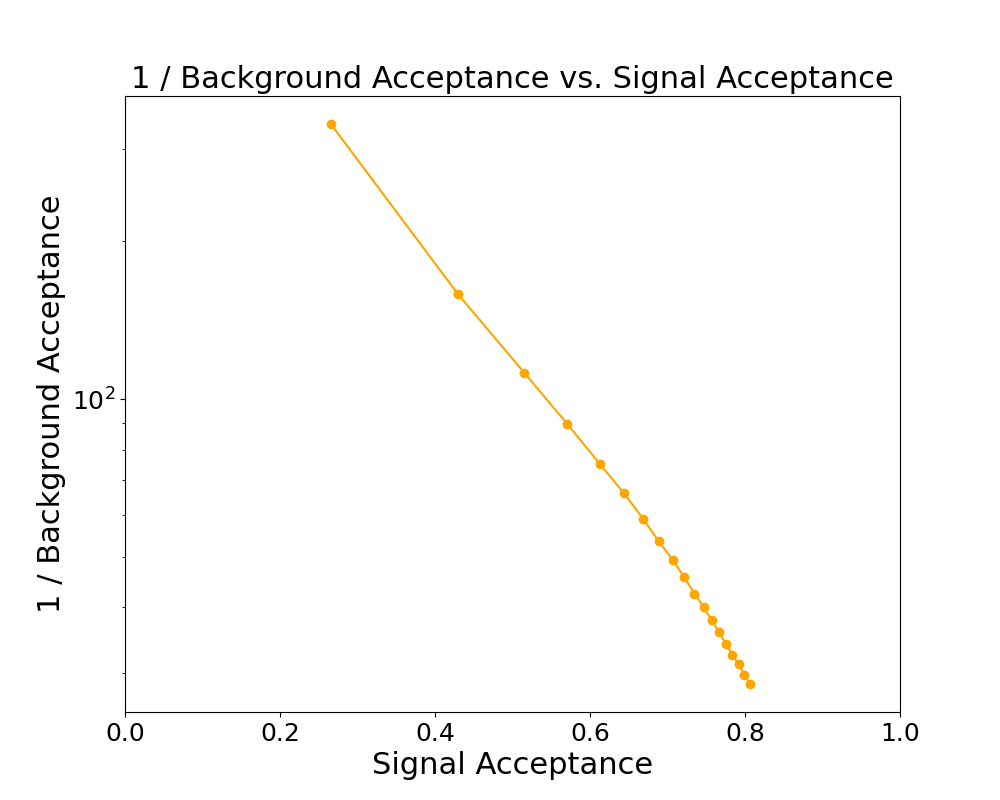}
        \caption{$M_{H_F} = 800$ GeV}
        \label{fig:Accep_MassVal800}
    \end{subfigure}
    \hfill
    \begin{subfigure}[b]{0.43\textwidth}
        \includegraphics[width=\textwidth]{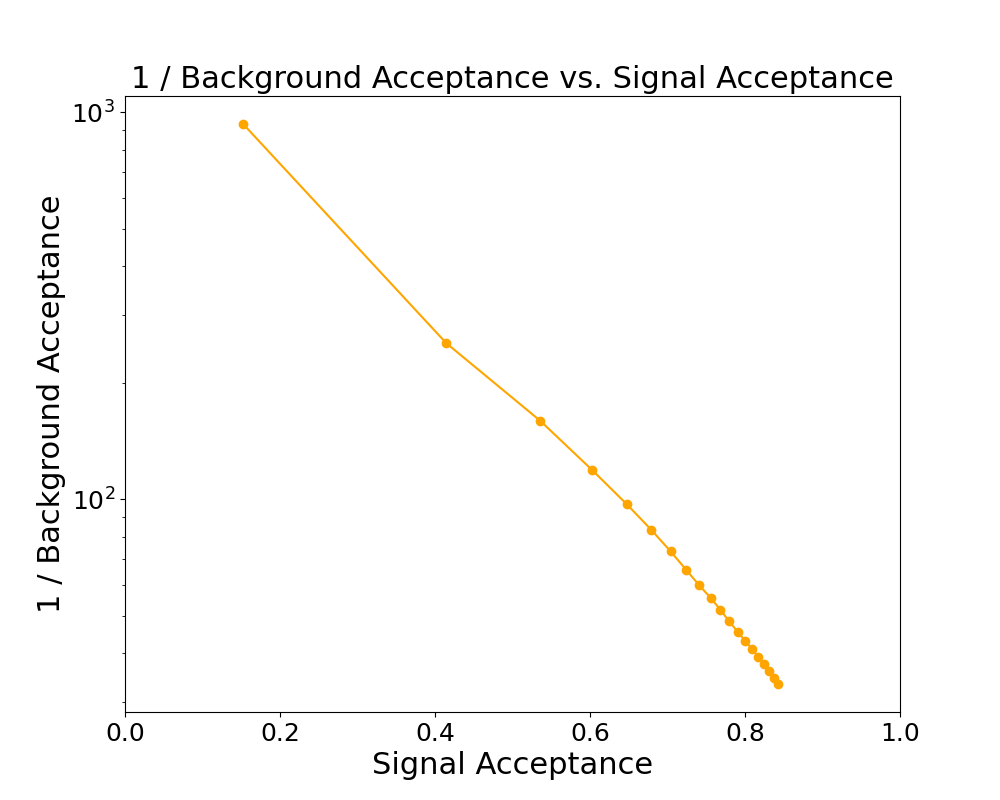}
        \caption{$M_{H_F} = 900$ GeV}
        \label{fig:Accep_MassVal900}
    \end{subfigure}
    
    \vspace{0.5em}
    
    % Row 2: 1000 and 1100 GeV
    \begin{subfigure}[b]{0.43\textwidth}
        \includegraphics[width=\textwidth]{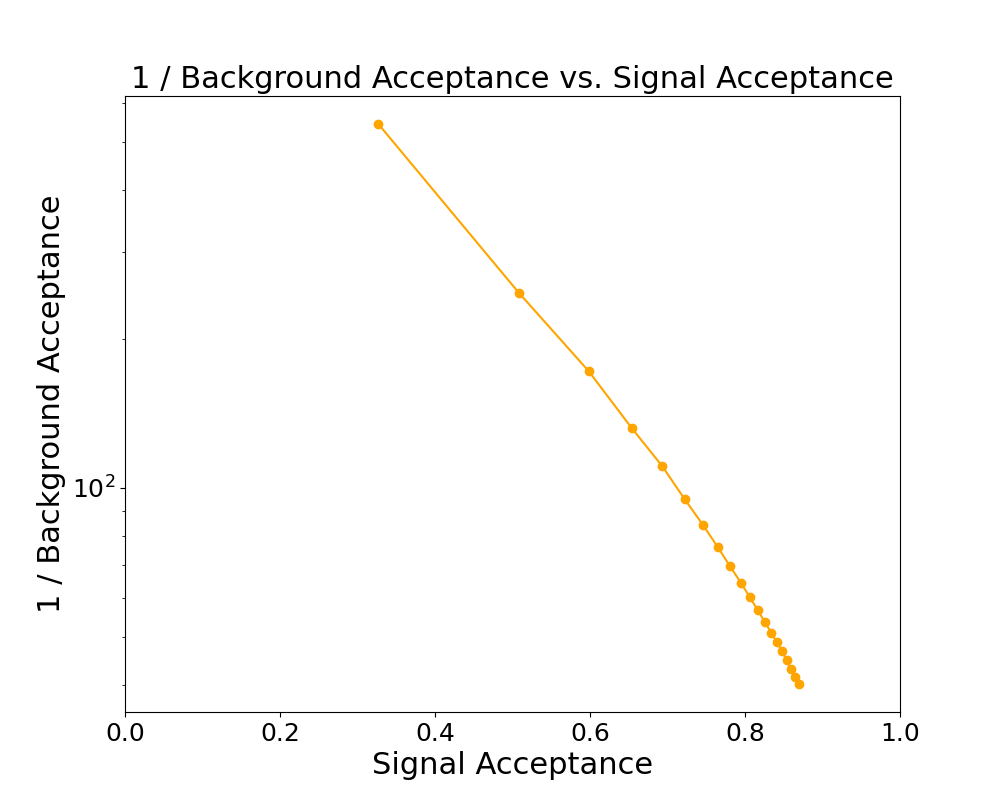}
        \caption{$M_{H_F} = 1000$ GeV}
        \label{fig:Accep_MassVal1000}
    \end{subfigure}
    \hfill
    \begin{subfigure}[b]{0.43\textwidth}
        \includegraphics[width=\textwidth]{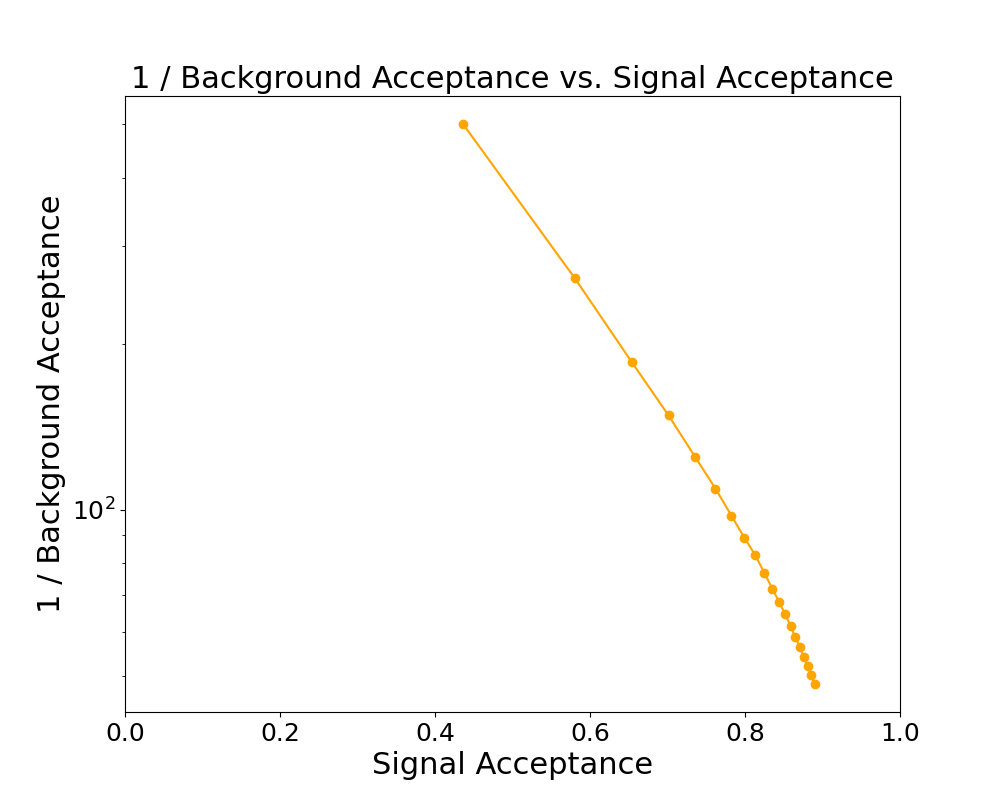}
        \caption{$M_{H_F} = 1100$ GeV}
        \label{fig:Accep_MassVal1100}
    \end{subfigure}
    
    \vspace{0.5em}
    
    % Row 3: 1200 and 1300 GeV
    \begin{subfigure}[b]{0.43\textwidth}
        \includegraphics[width=\textwidth]{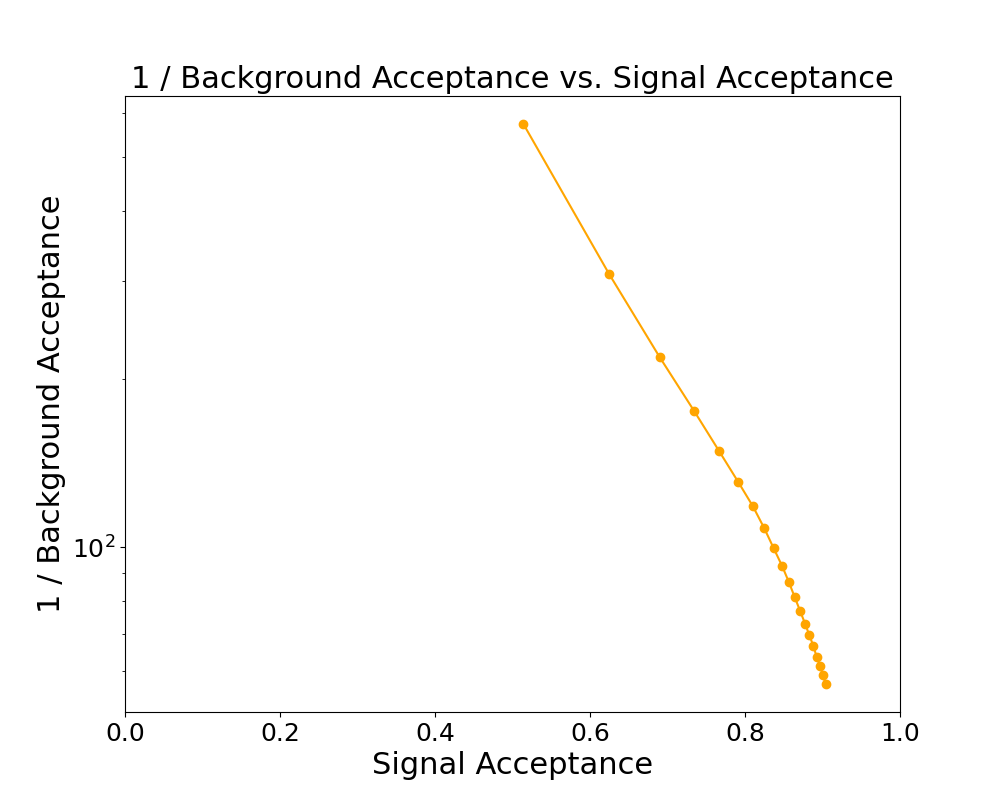}
        \caption{$M_{H_F} = 1200$ GeV}
        \label{fig:Accep_MassVal1200}
    \end{subfigure}
    \hfill
    \begin{subfigure}[b]{0.43\textwidth}
        \includegraphics[width=\textwidth]{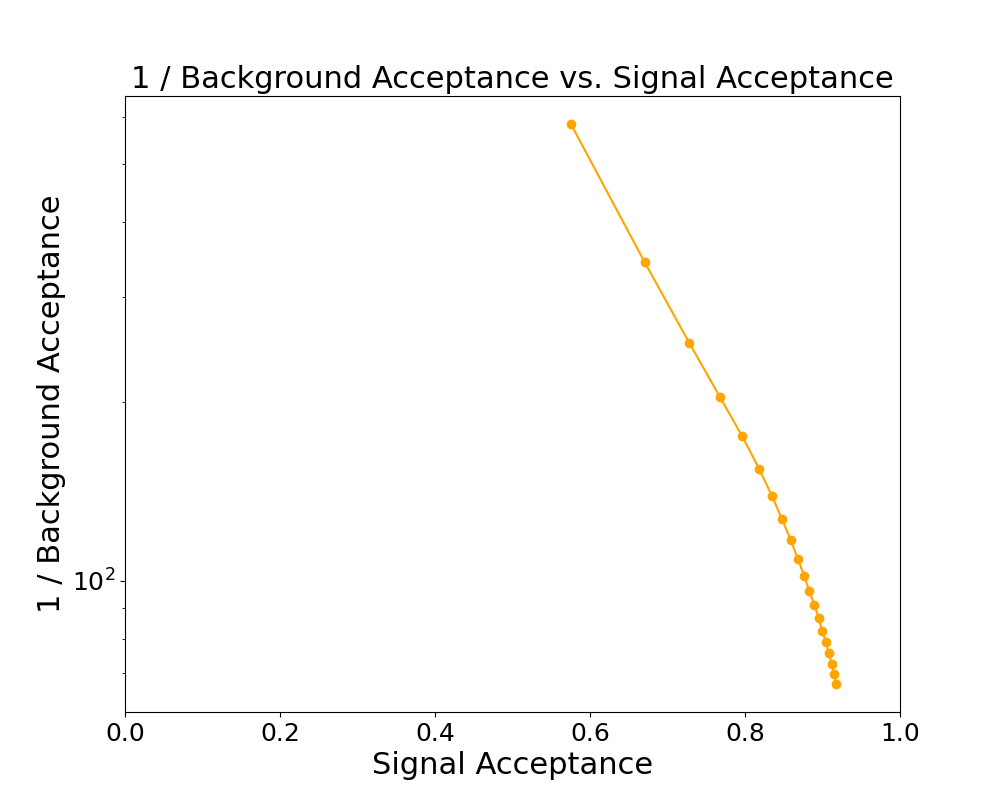}
        \caption{$M_{H_F} = 1300$ GeV}
        \label{fig:Accep_MassVal1300}
    \end{subfigure}
    
    \vspace{0.5em}
    
    % Row 4: 1400 GeV (centered)
    \begin{subfigure}[b]{0.43\textwidth}
        \centering
        \includegraphics[width=\textwidth]{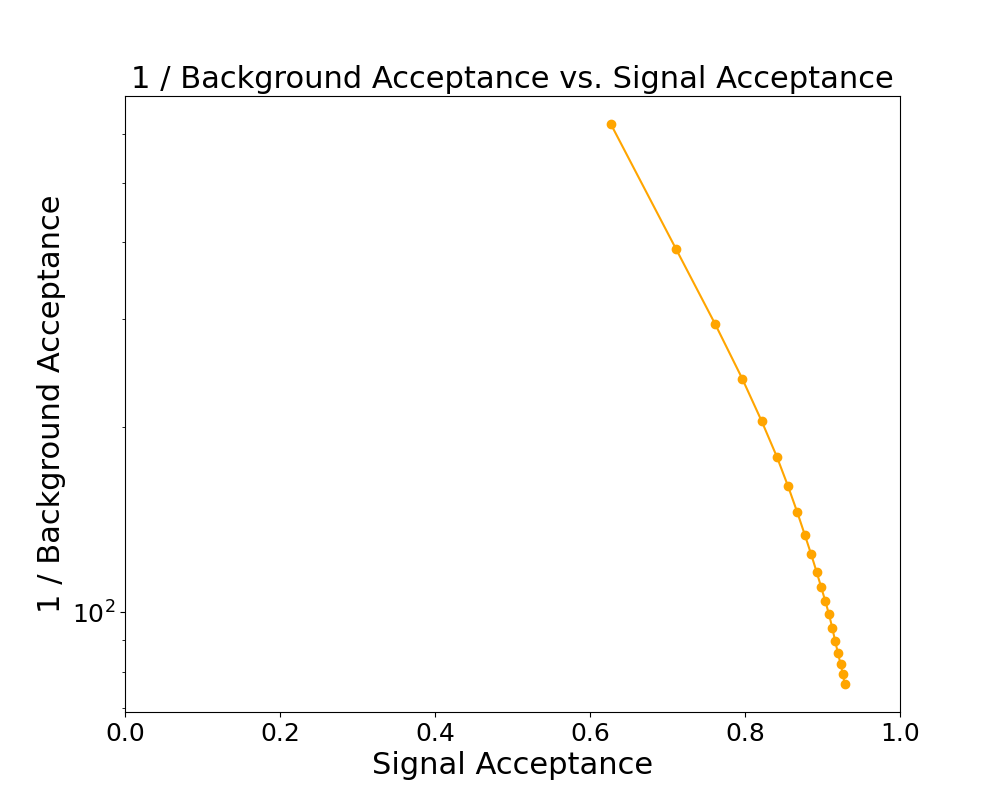}
        \caption{$M_{H_F} = 1400$ GeV}
        \label{fig:Accep_MassVal1400}
    \end{subfigure}
    
    \caption{1/Background Acceptance against the Signal Acceptance for the $h \to b\bar{b}$ channel. Each plot corresponds to a different Flavon mass as indicated.}
    \label{fig:Roc_hbb}
\end{figure}

\begin{figure}[H]
    \centering
    \begin{subfigure}[b]{0.4\textwidth}
        \includegraphics[scale=0.37]{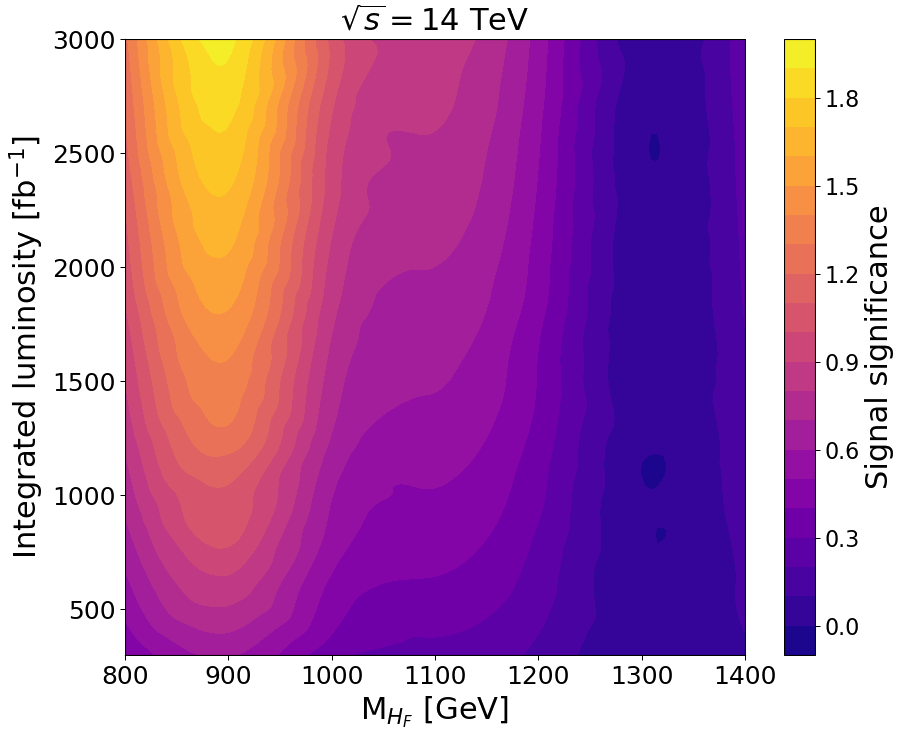}
        \caption{$\textbf{S1}$}
        \label{fig:esc1-aabb}
    \end{subfigure}%
    \\
    \begin{subfigure}[b]{0.4\textwidth}
        \includegraphics[scale=0.37]{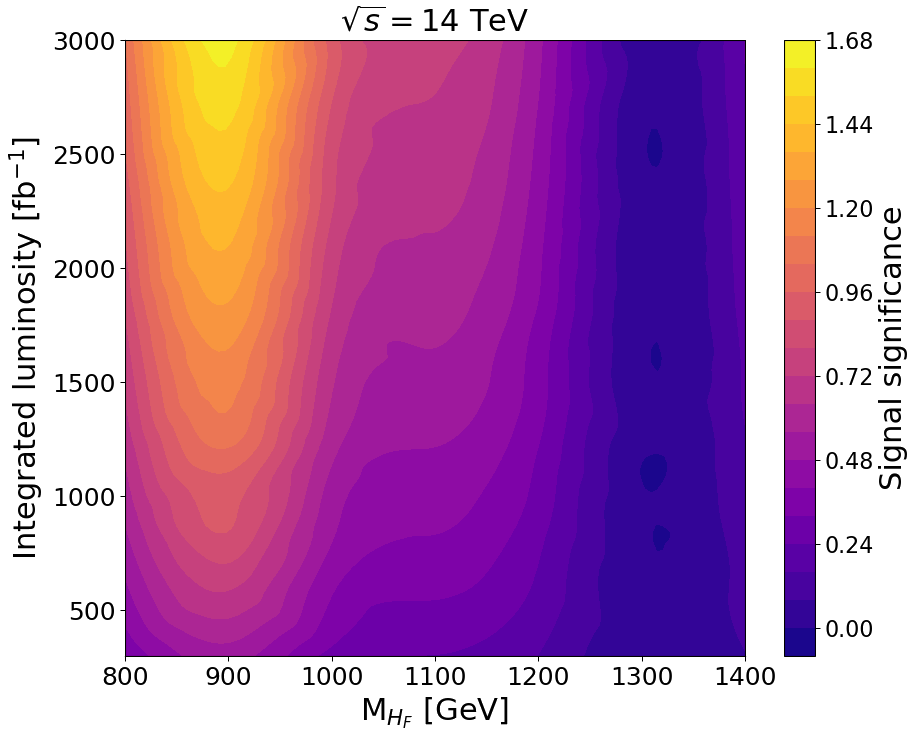}
        \caption{$\textbf{S2}$}
        \label{fig:esc2-aabb}
    \end{subfigure}%
    \\
    \begin{subfigure}[b]{0.4\textwidth}
        \includegraphics[scale=0.37]{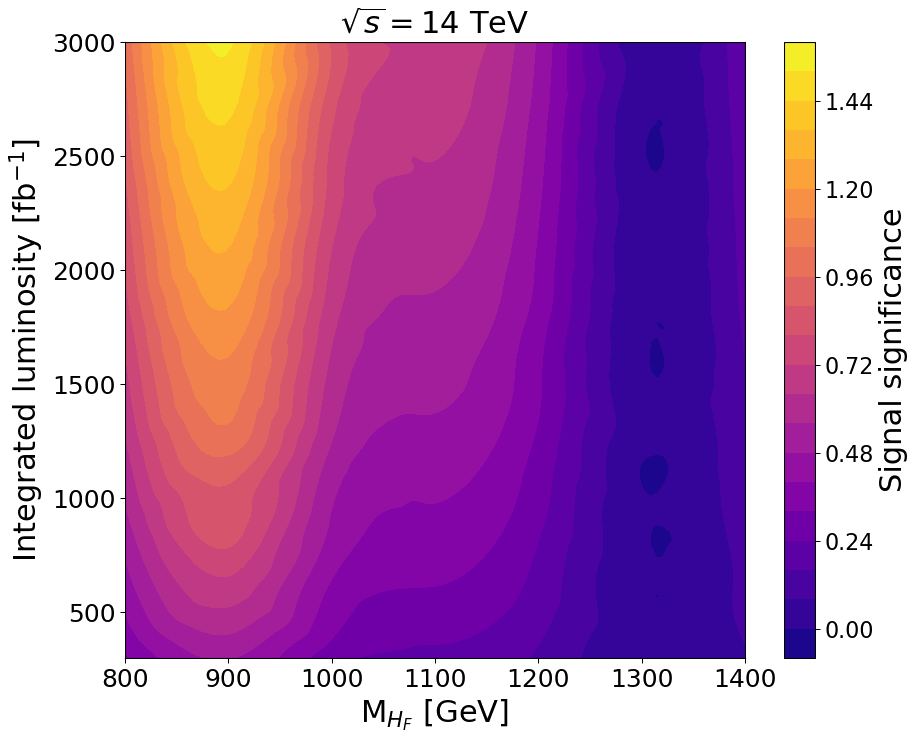}
        \caption{$\textbf{S3}$}
        \label{fig:esc3-aabb}
    \end{subfigure}
     \caption{Density plot displaying the \textit{Signal Significance} as a function of the integrated luminosity and the Flavon mass $M_{H_F}$ for the $h \to \gamma\gamma$ decay channel, the subplots correspond to the following benchmark scenarios: (a) $\textbf{S1}$, (b) $\textbf{S2}$, and (c) $\textbf{S3}$.}
    \label{fig:Haabb_SigplotS1}
\end{figure}

\subsection{Lepton-Flavour-Violating Decays \texorpdfstring{$h \to e\mu$}{}}
\label{sec:LFVemu}

Finally, we turn our attention to a distinctive signature of LFV decay of the SM-like Higgs boson, focusing on the channel \(h \to e\mu\). In the scenarios we analyse, such a process is enabled by the extended scalar sector and associated Yukawa couplings, which offer a direct probe of flavour dynamics within the FNSM framework.

\begin{figure}[H]
    \centering
    \includegraphics[width=0.4\textwidth]{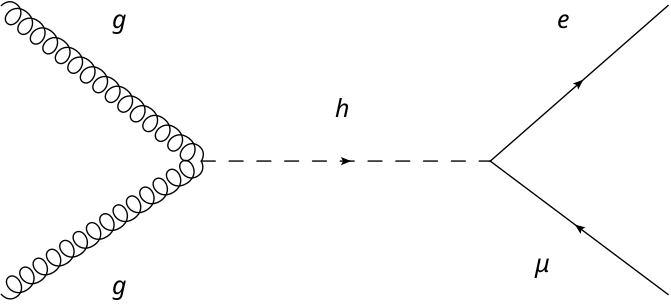}
    \caption{Feynman diagram of the LFV process \(h \to e\mu\).}
    \label{fig:feynmanemu}
\end{figure}

The signal process is
\[
pp \;\to\; h \;\to\; e\,\mu,
\]
As illustrated in Fig.~\ref{fig:feynmanemu}, where \(h\) is the Higgs-like scalar but now assumed to have LFV couplings. As in the previous channels, our study targets the LHC with a centre-of-mass energy of \(14\)\,TeV (and potentially the HL-LHC). 

The principal SM background processes arise primarily from:

\begin{itemize}
    \item \textbf{Drell-Yan Production:} 
    \begin{itemize}
        \item \( Z/\gamma^* \to \tau^+\tau^- \to e\nu_e\bar{\nu}_\tau + \mu\nu_\mu\bar{\nu}_\tau \)
    \end{itemize}

    \item \textbf{Top Quark Pair Production (\( t\bar{t} \)):}
    \begin{itemize}
        \item \( p p \to t\bar{t} \), with each top quark decaying as \( t \to bW^+ \) and \( \bar{t} \to \bar{b}W^- \), followed by \( W^+ \to e^+\nu_e \) and \( W^- \to \mu^-\bar{\nu}_\mu \)
    \end{itemize}

    \item \textbf{Diboson Production:}
    \begin{itemize}
        \item \( p p \to W^+W^- \), with \( W^+ \to e^+\nu_e \) and \( W^- \to \mu^-\bar{\nu}_\mu \)
        \item \( p p \to ZZ \), with each \( Z \) decaying into \( Z \to \mu^+\mu^- \)
        \item \( p p \to W^+Z \), with \( W^+ \to e^+\nu_e \) and \( Z \to \mu^+\mu^- \)
    \end{itemize}
\end{itemize}

Although these SM processes can lead to final states with different flavour leptons, they lack an explicit LFV vertex within a single decay. Then in the SM with massive neutrinos, those processes could appear. However, the BR is extremely suppressed, making them practically undetectable. Consequently, their kinematic distributions, especially around a resonance in the \(e\mu\) invariant mass, differ from a genuine two-body decay \(h \to e\mu\).

\subsubsection{Multivariate Analysis}

Also for this decay, we generate both signal and background samples at parton level using \texttt{MadGraph5\_aMC@NLO}, interfaced with \texttt{FeynRules} for the model implementation. Parton shower and hadronisation are simulated with \texttt{Pythia8}, and a fast detector simulation is performed via \texttt{Delphes3}, using an ATLAS-like parametrisation (consistent with the previous section). For the present analysis, we select a balanced dataset of 200,000 signal events and an equivalent number of background events.

To improve the separation of signal from background, we employ a Multivariate Analysis based on a \ac{BDT}, similarly to the methodology applied in the previous sections. Here, we implemented XGBoost, optimising the hyperparameters through \texttt{Optuna}~\cite{10.1145/3292500.3330701}. Here, we choose \texttt{Optuna} due to its more efficient exploration of large hyperparameter spaces and increased robustness against local minima compared to earlier Bayesian optimisation tools. This iterative approach enabled us to determine an optimal set of parameters, such as the number of trees (\texttt{NTree}) is set to 82, 
the maximum depth of each decision tree (\texttt{MaxDepth}) is set to 3, 
and the maximum number of leaves per tree (\texttt{MaxLeaves}) is set to 32. All other parameters are left at their default values.

\begin{figure}
    \centering
    \includegraphics[width=0.85\textwidth]{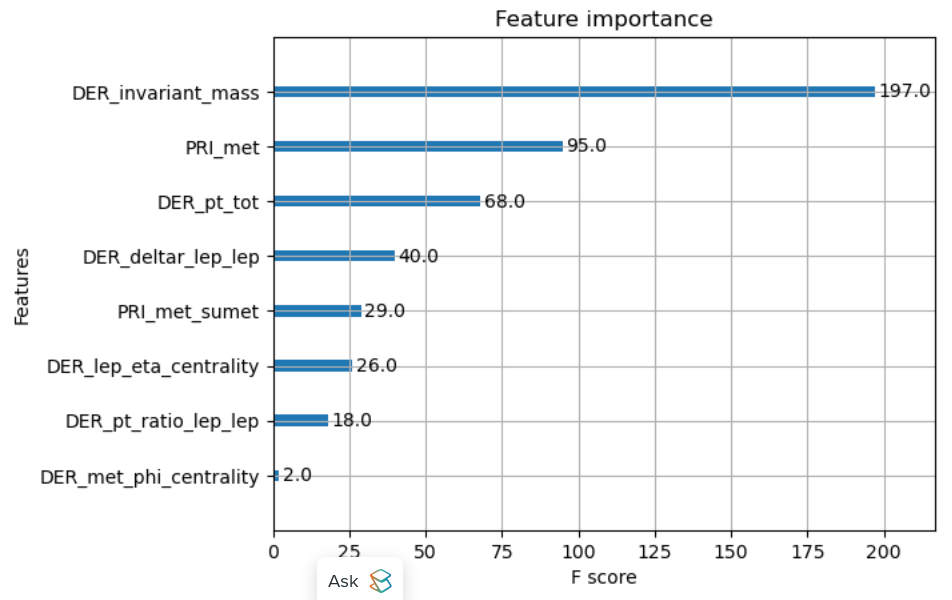}
    \caption{Feature importances from the XGBoost classifier, showing which variables contribute most strongly to separate signal from background for \(h \to e\mu\). (Correlated features were removed prior to training, as discussed in the dataset preparation section of the previous chapter.)}
    \label{fig:FeatureImportance_emu}
\end{figure}

After training, the \ac{BDT} outputs a single discriminant variable (\(\textbf{xgb}\)), which we can cut on in order to isolate a high purity signal region. Figure~\ref{fig:FeatureImportance_emu} shows the feature importances for the chosen input variables, illustrating that the invariant mass \(\mathrm{DER\_invariant\_mass}\) (closely related to \(M_{e\mu}\)) emerges as the strongest discriminator, followed by other kinematic observables such as \(\mathrm{PRI\_met}\) and total transverse momentum \(\mathrm{DER\_pt\_tot}\).

We define the \textit{Signal Significance} according to the usual approximation from Eq.~\eqref{eq:signalsignificance}. By scanning over different values of the model parameters (such as \(\cos\alpha\), \(v_s\), and LFV couplings like \(Z_{e\mu}\)), as well as varying the assumed integrated luminosity, we can identify regions in which the LFV decay \(h \to e\mu\)  that could be observed with evidence-level (\(3\sigma\)) or discovery-level (\(5\sigma\)) significance.

We consider the three benchmark scenarios (labelled \textbf{S4}, \textbf{S5}, and \textbf{S6}), corresponding to different choices of \((\cos\alpha,\;v_s,\;Z_{e\mu})\). We summarise below their cross-sections and the maximal significance obtained at an integrated luminosity of \(\mathcal{L} = 3000\,\mathrm{fb}^{-1}\):

\begin{table}[h]
\centering
\begin{tabular}{@{}lccccr@{}}
\toprule
Scenario & \( \cos\alpha \) & \( v_s \) (GeV) & \( Z_{e\mu} \) & \( \sigma \) (pb) & Luminosity (\( \text{fb}^{-1} \)) \\ \midrule
\textbf{S4}        & 0.991           & 350            & 0.0025         & 0.002395         & 3000 \\
\textbf{S5}       & -0.8949         & 2198.396       & 0.005601       & 0.003759         & 1800 \\
\textbf{S6}       & -0.9548         & 2461.078       & 0.009781       & 0.004404         & 1300 \\ \bottomrule
\end{tabular}
\caption*{These parameters correspond to a significance of $\sim 
5 \sigma$.}
\label{tab:HL3}

\end{table}

\begin{table}[h]
\centering

\begin{tabular}{@{}lccccr@{}}
\toprule
Scenario & \( \cos\alpha \) & \( v_s \) (GeV) & \( Z_{e\mu} \) & \( \sigma \) (pb) & Luminosity (\( \text{fb}^{-1} \)) \\ \midrule
\textbf{S4}        & 0.991           & 350            & 0.0025         & 0.002395         & 1500 \\
\textbf{S5}       & -0.8949         & 2198.396       & 0.005601       & 0.003759         & 700 \\
\textbf{S6}       & -0.9548         & 2461.078       & 0.009781       & 0.004404         & 500 \\ \bottomrule
\end{tabular}
\caption*{These parameters correspond to a significance of $\sim 3\sigma$.}
\caption{Parameters for scenarios \textbf{S4}, \textbf{S5} and \textbf{S6}}

\end{table}

\begin{figure}
    \centering    \includegraphics[width=0.53\textwidth]{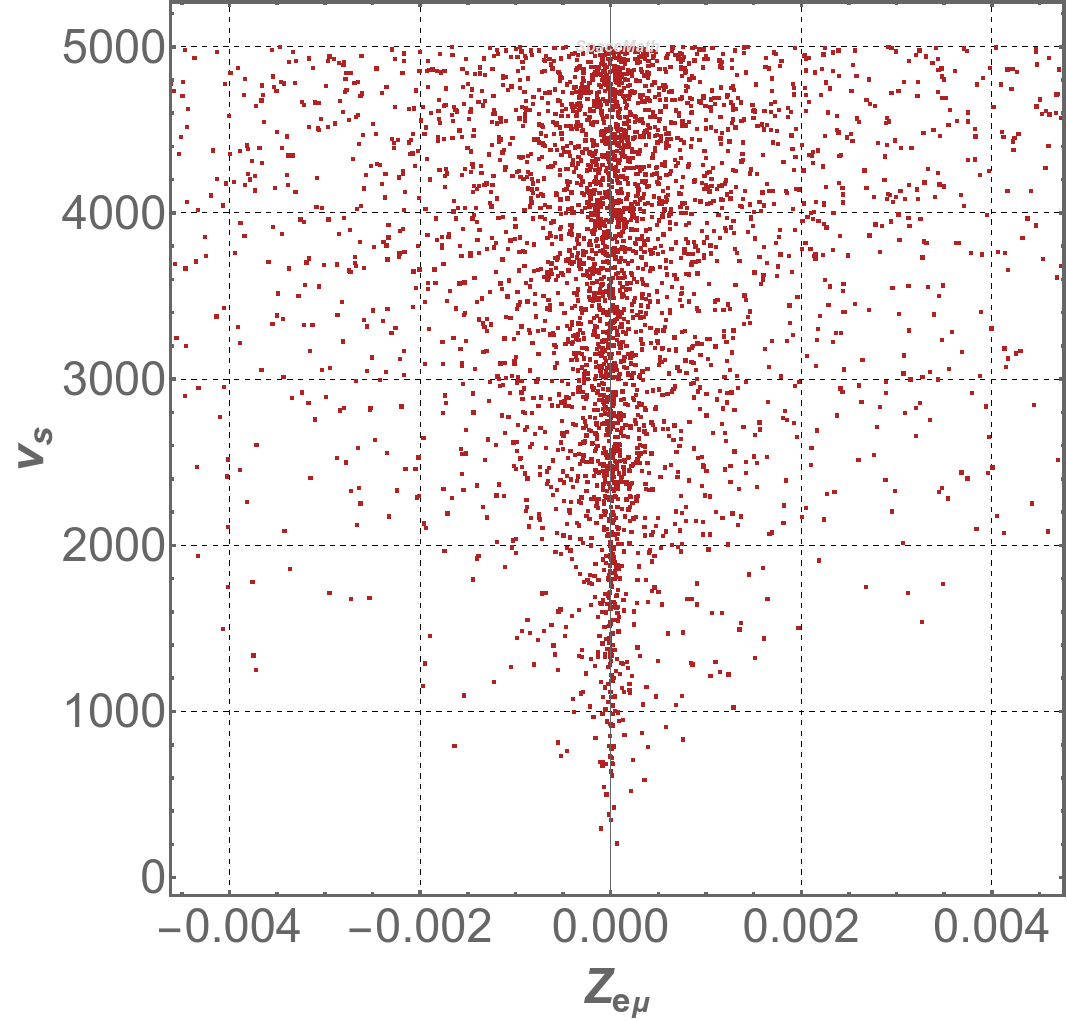}
    \caption{Parameter space in the plane $(v_s, Z_{e\mu})$ for $h \to e\mu$. Since this decay has not been observed, no signal strength is measured for this channel. Instead, only upper limits on the BR are available. In this analysis, we impose the experimental constraint $\mathcal{B}(h \to e\mu) < 6.2 \times 10^{-5}$.}
    \label{fig:parameterspaceemu}
\end{figure}

In Figure~\ref{fig:parameterspaceemu} we show an example of distribution in the \((v_s,\,Z_{e\mu})\) plane ~\cite{Aad_2020}, from which our three benchmark scenarios are selected.

In each case, the maximum significance is attained near a \ac{BDT} discriminator threshold of \(\textbf{xgb} \approx 0.95\). Higher LFV couplings (\(Z_{e\mu}\)) and suitable mixing angles lead to larger cross sections and thus higher significance.

%\rojo{add this?}FigureY illustrates the invariant mass distribution \(M_{e\mu}\) before the MVA selection, showing how the signal may appear as a peak above a more smoothly distributed SM background. Figure X shows a typical shape of the BDT output (\(\textbf{xgb}\)), where a tight cut on \(\textbf{xgb}\) can suppress a large fraction of the SM events while retaining a sizeable portion of the LFV signal.

%\begin{figure}[!htb]
%    \centering
    %\includegraphics[width=0.45\textwidth]{Memu_distribution.png}
 %   \caption{Illustration of the invariant mass \(M_{e\mu}\) for the signal $h \to e\mu$ and main backgrounds, prior to BDT cuts. A clear resonance shape near $m_{h}$ may be observed, distinguishing it from continuum processes. (Schematic figure.)}
%    \label{fig:invmass_emu_example}
%\end{figure}

\begin{figure}
    \centering
    \includegraphics[width=0.88\textwidth]{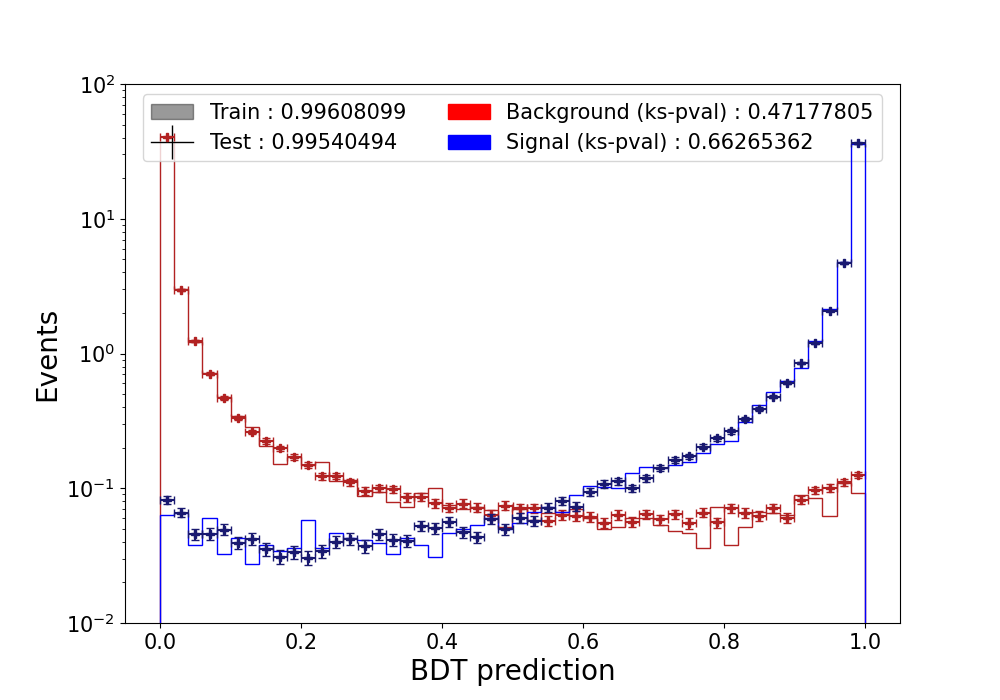}
    \caption{\ac{BDT} classifier distribution for \textbf{S6}, comparing signal (blue) and background (red) events in both training and test samples. The legend shows the KS p-values for each distribution.}
    \label{fig:classifier_distributions6}
\end{figure}

\begin{figure}
    \centering
    \includegraphics[width=0.9\textwidth]{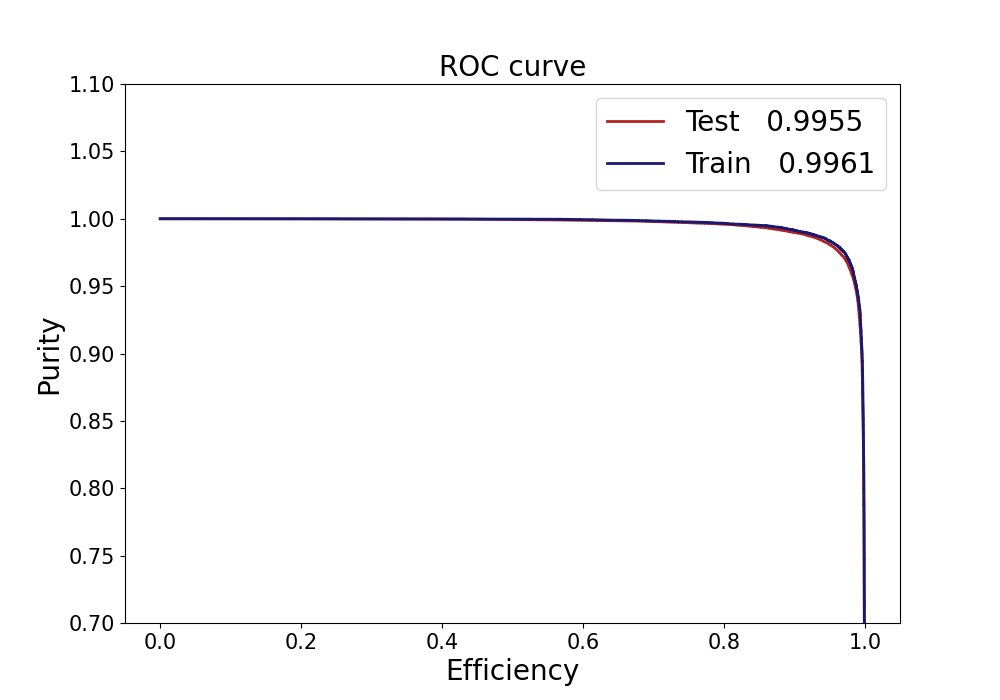}
    \caption{ROC curve for \textbf{S6}, illustrates purity vs.\ efficiency for both, the training (blue) and test (red) datasets. The AUC values are $\sim0.996$ (train) and $\sim0.996$ (test), reflecting strong discrimination.}
    \label{fig:rocCurves6}
\end{figure}

Figures~\ref{fig:classifier_distributions6} and~\ref{fig:rocCurves6} illustrate two key performance plots for \textbf{S6}. In Figure~\ref{fig:classifier_distributions6}, the \ac{BDT} prediction distributions for signal and background events in both training and test samples, along with their KS p-values. Furthermore, and in Figure~\ref{fig:rocCurves6}, the corresponding ROC curves for the training and test datasets, indicating the AUC and confirming that the classifier exhibits robust separation power without significant overfitting.

\begin{figure}
    \centering
    % First row: Scenario S4
    \includegraphics[width=0.8\textwidth]{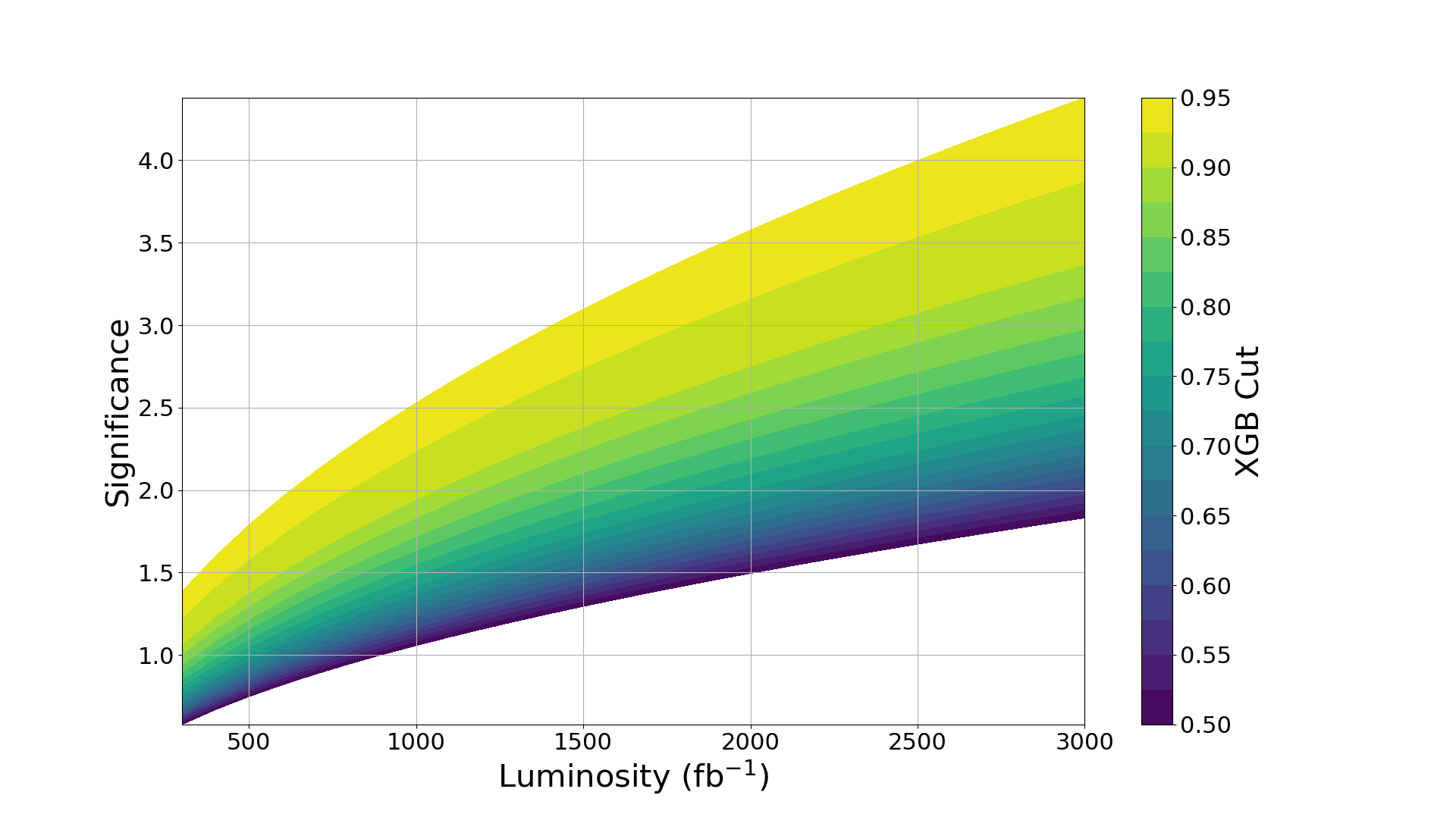}\\[1ex]
    \textbf{(a) S4}\\[2ex]
    % Second row: Scenario S5
    \includegraphics[width=0.8\textwidth]{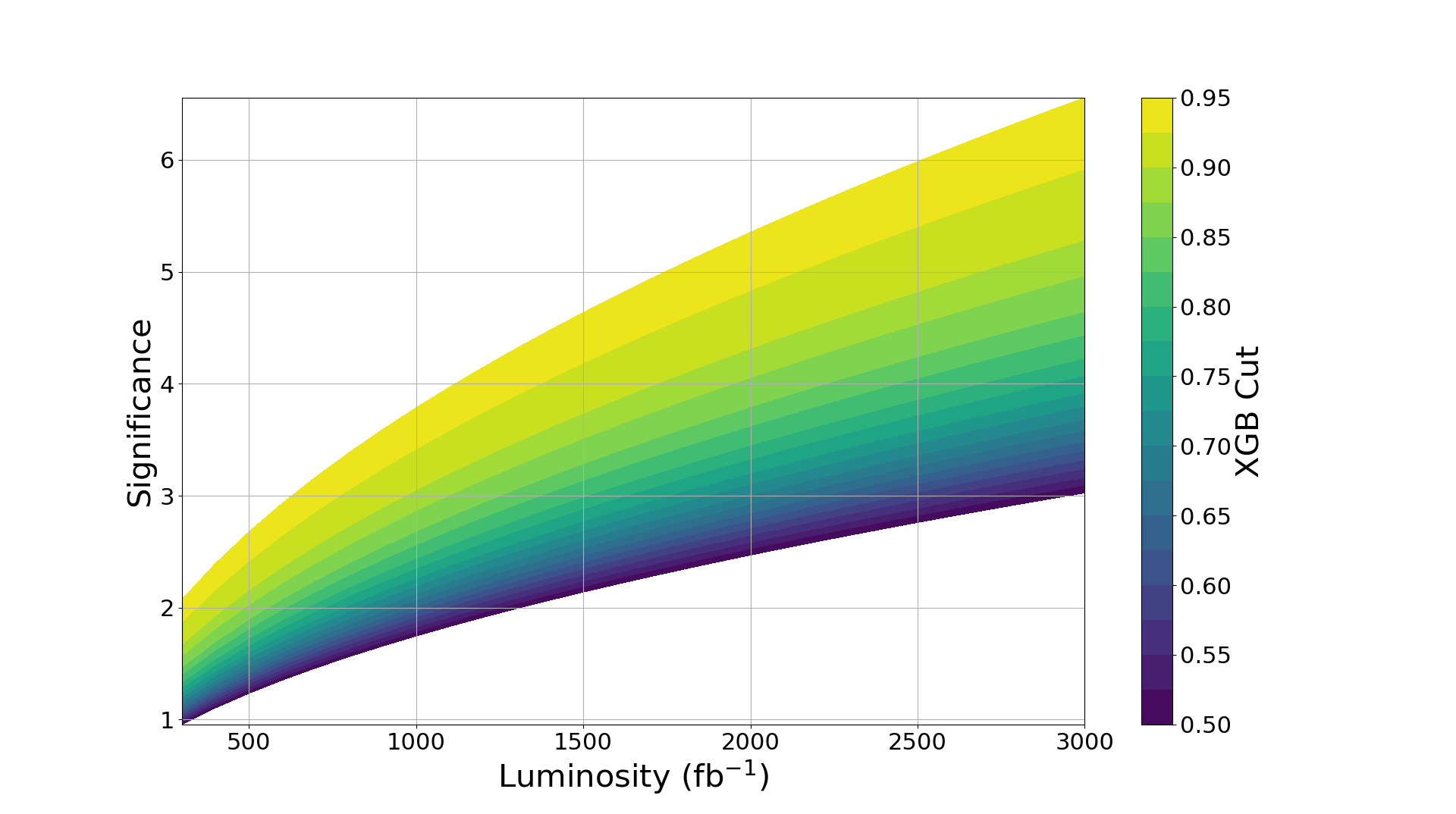}\\[1ex]
    \textbf{(b) S5}\\[2ex]
    % Third row: Scenario S6
    \includegraphics[width=0.8\textwidth]{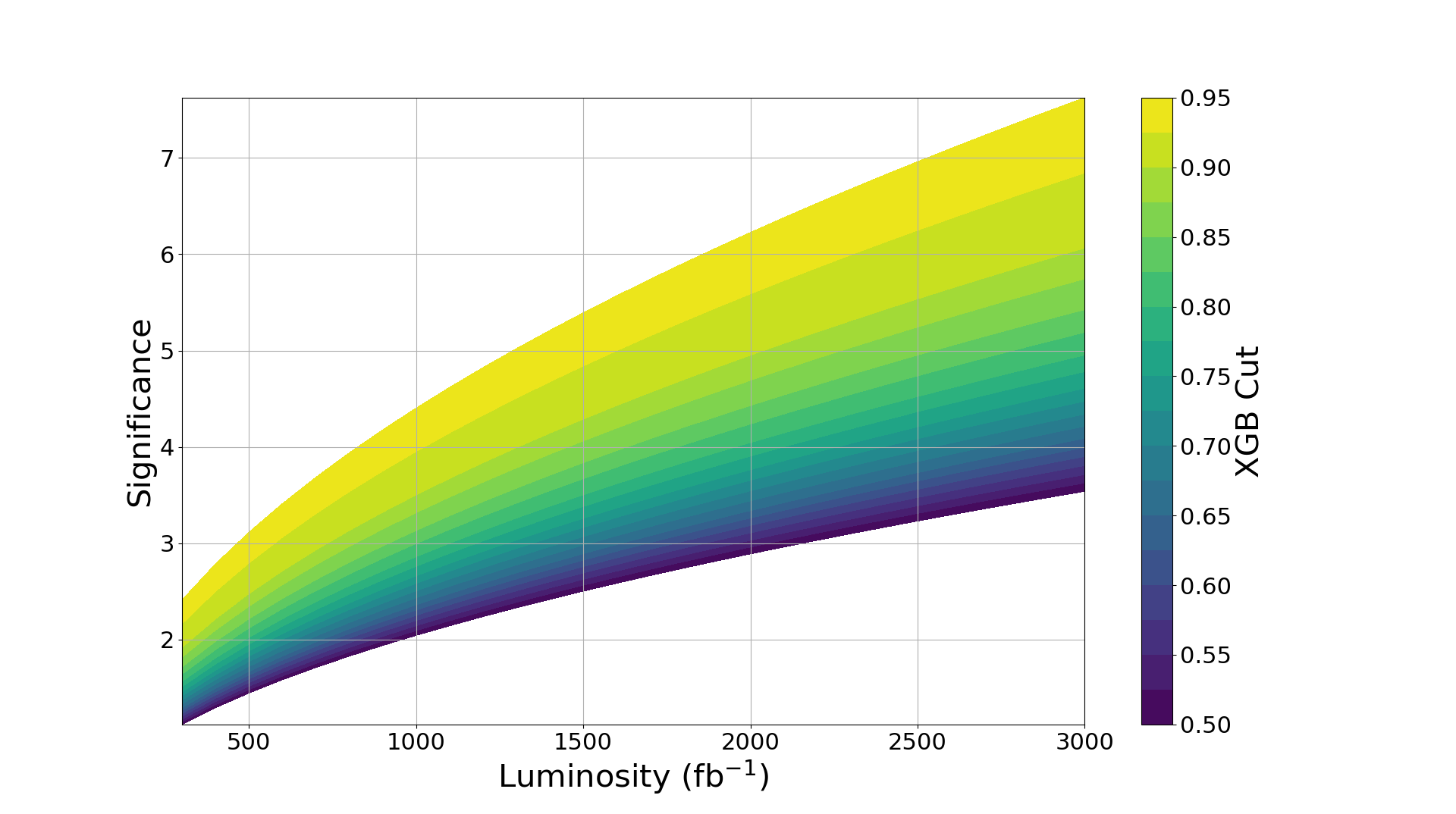}\\[1ex]
    \textbf{(c) S6}
    \caption{\textit{Signal Significance} as a function of the integrated luminosity and the XGB cut for the decay \(h \to e\mu\) in the scenarios: (a) $\textbf{S4}$, (b) $\textbf{S5}$, and (c) $\textbf{S6}$.}
    \label{SignHemu}
\end{figure}

Additionally, Fig.~\ref{SignHemu} shows how the \textit{Signal Significance} varies with the integrated luminosity and the \textbf{xgb} cut for \textbf{S4}, \textbf{S5} and \textbf{S6}. Finally, from the results obtained, only \textbf{S5} and \textbf{S6} would be feasible at the HL-LHC. Specifically, \textbf{S6} which can reach a $5\sigma$ discovery at around $1300\,\mathrm{fb}^{-1}$, while \textbf{S5} achieves a $3\sigma$ evidence level at $700\,\mathrm{fb}^{-1}$ and a $5\sigma$ at $1800\,\mathrm{fb}^{-1}$. Consequently, both scenarios could probe LFV in extended scalar sectors, complementing the flavour conserving searches presented in earlier sections.

Our collider analyses within both the 2HDM-III and FNSM frameworks demonstrate the potential to uncover new physics signatures BSM. By integrating analytical decay width calculations with advanced multivariate techniques, we have identified promising regions of parameter space where the predicted signals could be distinguished from substantial SM backgrounds.

\chapter{Conclusions and Future Work}
\label{chap:conclusions_future_work}
\noindent\rule{\linewidth}{0.4pt}
% Summary of the findings and their importance to particle physics.

In this thesis, we worked on BSM scenarios through comprehensive phenomenological studies, using proton-proton collisions at the LHC and its future upgrade, the HL-LHC, operating at a centre of mass energy of 14 TeV. Our research addressed three principal directions: the production and potential detection of a charged scalar Higgs pair \(H^-H^+\) decaying into the final state \(\mu\nu_{\mu}cb\) within the 2HDM-III framework; the Flavon scalar (\(H_F\)) decaying into a \(b\bar{b}h\), where the SM-like Higgs boson (\(h\)) subsequently decays into either a pair of photons (\(h \to \gamma\gamma\)) or a pair of \(b\)-quarks (\(h \to b\bar{b}\)); signals of LFV SM-like Higgs decays, specifically \(h\to e\mu\) also within the framework of the Flavon Model.

Initially, we analysed the production and subsequent detection potential of charged scalar boson pairs predicted by the 2HDM-III, specifically through the process $pp\to H^- H^+ \to \mu\nu_\mu cb$. After an exhaustive scan of the parameter space, we identified two highly promising scenarios for experimental verification at the HL-LHC, the first scenario (with $\tan\beta=\chi_{\mu\mu}=\chi_{cb}=5,\,\cos(\alpha-\beta)=0.01$) and the second scenario (with $\tan\beta=10,\,\chi_{\mu\mu}=\chi_{cb}=1,\,\cos(\alpha-\beta)=0.01$). It became evident from our analysis that the dominant contribution to the charged scalar pair production emerges from the decay of an on-shell heavy neutral scalar boson $H$. By employing Monte Carlo simulations and Machine Learning tools such as the \ac{BDT} algorithm, we efficiently suppressed the main background processes. This approach markedly improved the sensitivity compared to traditional kinematic analyses, reaching signal significances above $5\sigma$ for charged scalar masses in the range of $110 \lesssim M_{H^\pm}\lesssim 250$ GeV when the HL-LHC accumulates integrated luminosities of at least $300$ fb$^{-1}$ (assuming a neutral scalar mass of $M_H=500$ GeV). Conversely, for a heavier neutral scalar boson mass ($M_H=800$ GeV), higher luminosities ($\mathcal{L}_{\rm int}\geq 2400\,\mathrm{fb}^{-1}$) are required to reach comparable significance for charged Higgs masses between $180$ and $360$ GeV.

Furthermore, our research examined an extended scalar model based on the FN mechanism, predicting the existence of a Flavon neutral scalar particle ($H_F$), decaying dominantly into a SM-like Higgs boson plus a pair of $b$-quarks. Two decay channels of the resulting Higgs boson, $h\to b\bar{b}$ and $h\to \gamma\gamma$, were analysed under realistic scenarios obtained from theoretical and experimental constraints. At the HL-LHC (targeting an integrated luminosity of $3000\,\mathrm{fb}^{-1}$), we identified specific regions within the model parameter space, yielding signal significances up to $5\sigma$ for the $h\to b\bar{b}$ decay channel and about $2\sigma$ for the $h\to \gamma\gamma$ channel, thus reinforcing the potential for experimental verification of this theoretical scenario.

Finally, we directed our focus towards the LFV SM-like Higgs decay $h \to e\mu$, enabled by the presence of extended scalar sectors and associated LFV Yukawa couplings. The same analysis method allowed us to isolate scenarios which are accessible at the HL-LHC. In particular, one scenario (with parameters $\cos\alpha=-0.9548$, $v_s=2461.078$ GeV, $Z_{e\mu}=0.009781$ and a production cross section of $0.004404$ pb) stood out as the most promising, achieving a clear $5\sigma$ discovery potential at a luminosity of $1300\,\mathrm{fb}^{-1}$. Additionally, other scenario (with parameters $\cos\alpha=-0.8949$, $v_s=2198.396$ GeV, $Z_{e\mu}=0.005601$ and cross section of $0.003759$ pb) also proved viable, reaching a $3\sigma$ evidence level at $700\,\mathrm{fb}^{-1}$ and a $5\sigma$ significance threshold at $1800\,\mathrm{fb}^{-1}$. Hence, these scenarios significantly enrich the phenomenological landscape of flavour dynamics at the HL-LHC.

Throughout this research, we developed and utilised a robust Python based framework capable of processing LHCO files, extracting relevant physical observables, and preparing datasets suitable for advanced Machine Learning techniques. By leveraging the powerful capabilities of the XGB algorithm, our methodology effectively distinguished signal from background events, providing enhanced signal sensitivity beyond standard kinematic methods.

In conclusion, the outcomes presented here not only highlight the phenomenological prospects offered by both BSM scenarios at the LHC and HL-LHC, but also showcase the powerful combination of Monte Carlo simulation and Machine Learning techniques. Future developments of our analysis package will include integration with other widely adopted particle physics tools, as well as direct incorporation of Machine Learning training functionalities, aiming to further expand its utility.

\appendix

\chapter{Constraints} 

\label{V}

\begin{figure}
    \centering

    % Row 1
    \begin{subfigure}[b]{0.4\textwidth}
        \centering
        \includegraphics[width=\textwidth]{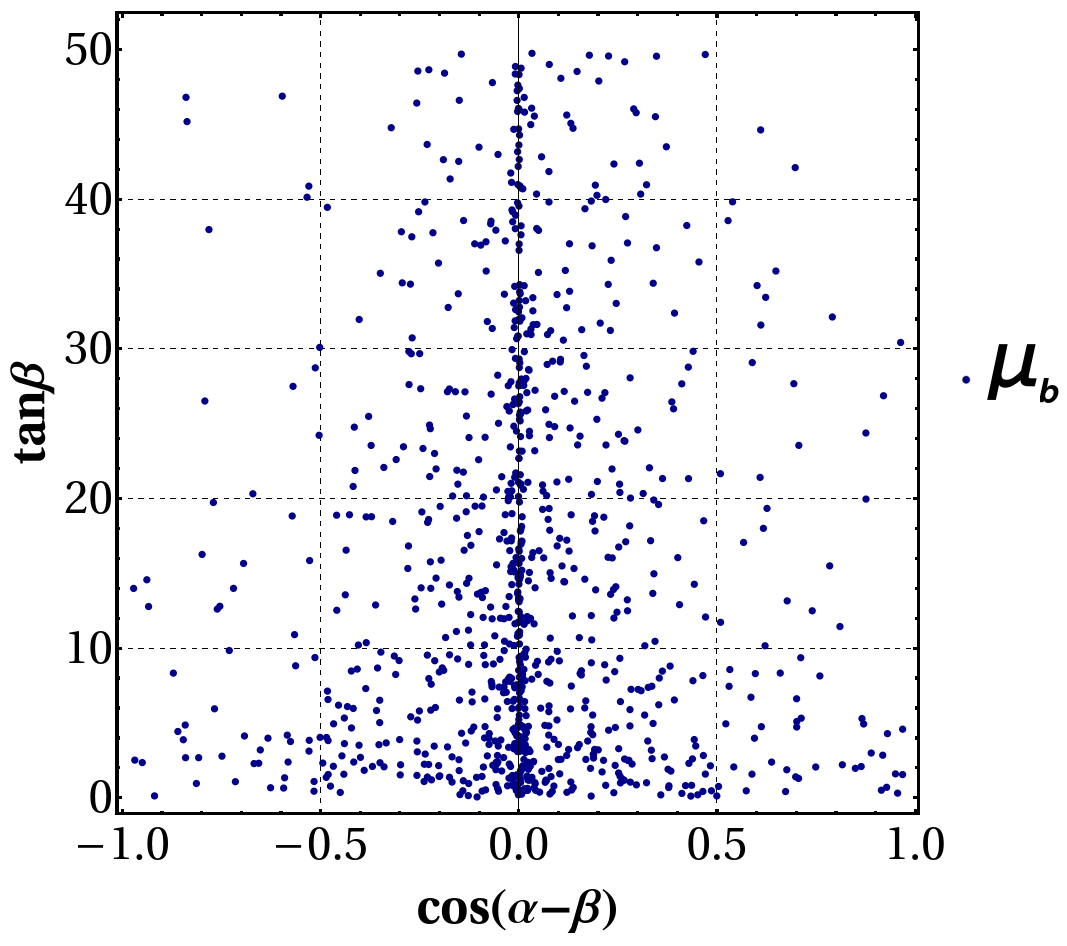}
        \caption{$\mu_b$}
    \end{subfigure}
    \hfill
    \begin{subfigure}[b]{0.4\textwidth}
        \centering
        \includegraphics[width=\textwidth]{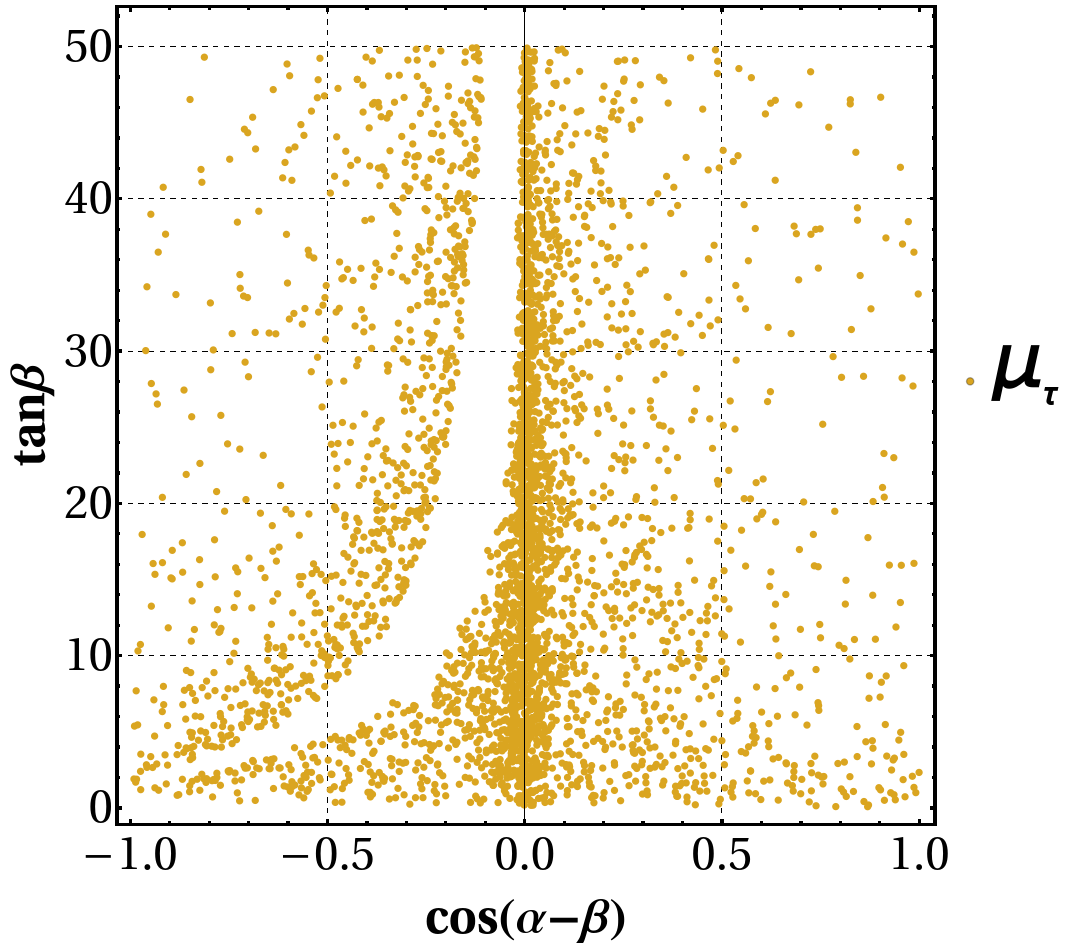}
        \caption{$\mu_{\tau}$}
    \end{subfigure}

    \vskip\baselineskip

    % Row 2
    \begin{subfigure}[b]{0.4\textwidth}
        \centering
        \includegraphics[width=\textwidth]{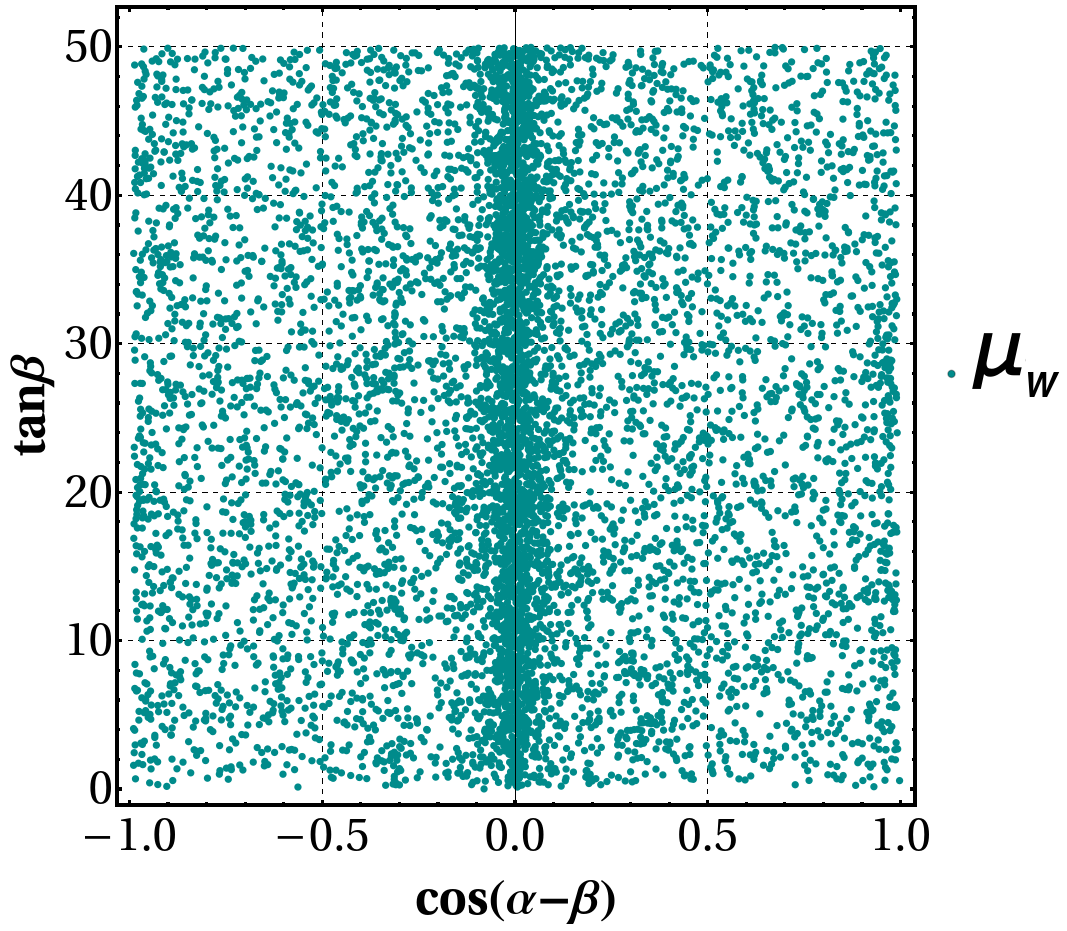}
        \caption{$\mu_W$}
    \end{subfigure}
    \hfill
    \begin{subfigure}[b]{0.4\textwidth}
        \centering
        \includegraphics[width=\textwidth]{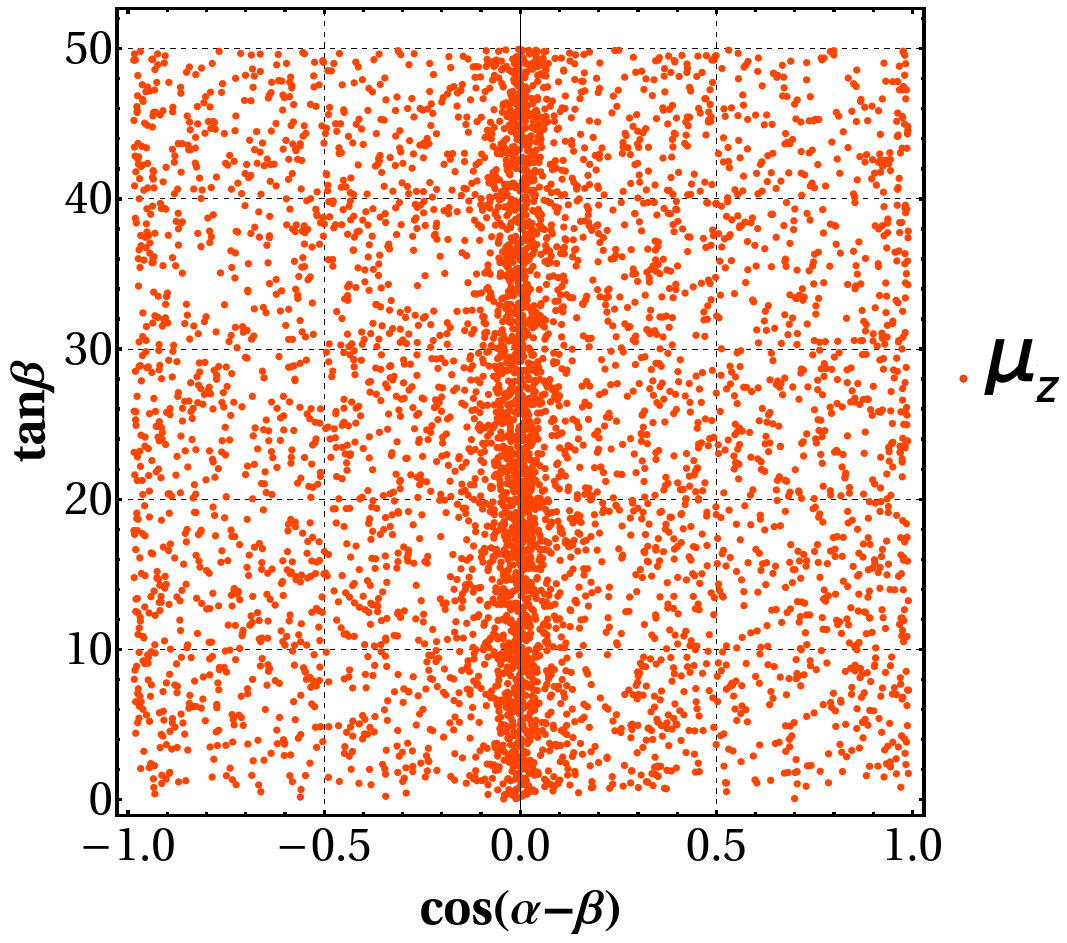}
        \caption{$\mu_Z$}
    \end{subfigure}

    \vskip\baselineskip

    % Row 3
    \begin{subfigure}[b]{0.4\textwidth}
        \centering
        \includegraphics[width=\textwidth]{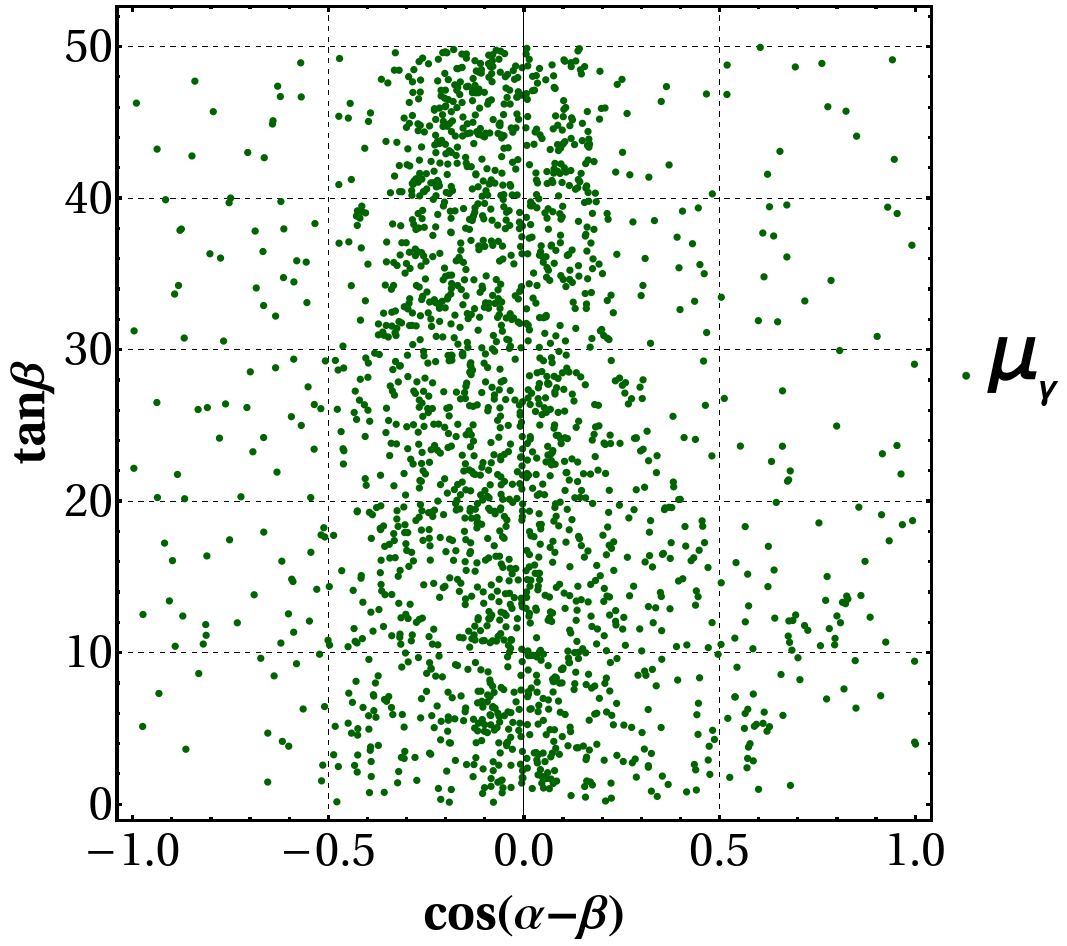}
        \caption{$\mu_{\gamma}$ ($114\leq M_{H^{\pm}}\leq 140$)}
    \end{subfigure}
    \hfill
    \begin{subfigure}[b]{0.4\textwidth}
        \centering
        \includegraphics[width=\textwidth]{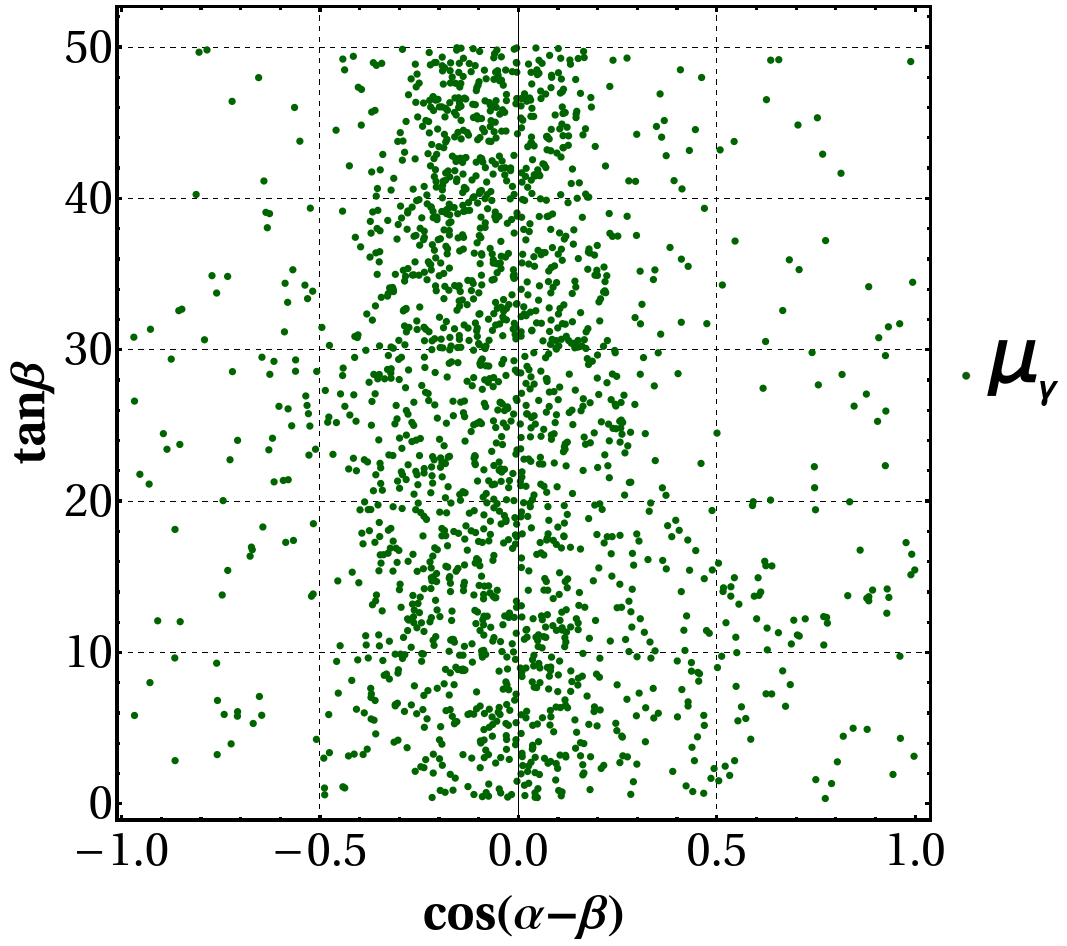}
        \caption{$\mu_{\gamma}$ ($m_t + m_b\leq M_{H^{\pm}}\leq 1000$)}
    \end{subfigure}

    \caption{Signal strengths: (a) $\mu_b$, (b) $\mu_{\tau}$, (c) $\mu_W$, (d) $\mu_Z$, (e) $\mu_{\gamma}$ ($114\leq M_{H^{\pm}}\leq 140$), and (f) $\mu_{\gamma}$ ($m_t + m_b\leq M_{H^{\pm}}\leq 1000$). In all cases, $50$K random points are generated.}
    \label{fig:all_mu_strengths}
\end{figure}

	By different constraints for each observable described in Sec. \ref{SecIII}, we show individual allowed regions, namely,
	\begin{itemize}
		\item Signal strengths
	\end{itemize}
	We present in Fig. \ref{fig:all_mu_strengths} individual  planes associated to each $\mu_X$ are $\cos(\alpha-\beta)-\tan\beta$. Those allowed by $\mu_X$ correspond to the coloured points. In the evaluation of the signal strengths $\mu_X$, the range of scanned parameters is shown in Tab. \ref{Scan_muX}.

\begin{figure}[!htb]
		\centering
		\includegraphics[width=6cm]{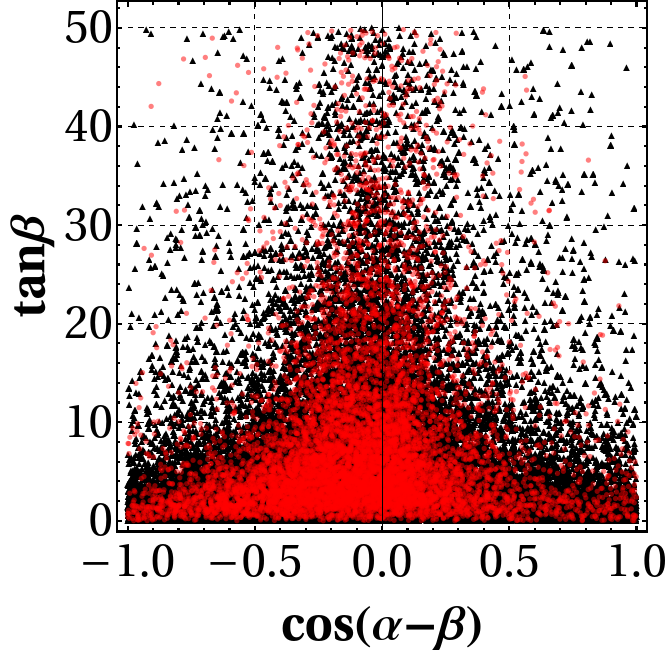}
		\caption{Allowed values by the measurement on $\mathcal{BR}(B_d^0\to\mu^+\mu^-)$ (black triangles) and by the upper limit on $\mathcal{BR}(B_s^0\to\mu^+\mu^-)$ (light red circles). For both cases we generate $50$K random points.}\label{INDBs}
	\end{figure}

	\begin{table}

		\begin{centering}
			\begin{tabular}{|c|c|}
				\hline 
				Parameter & Range\tabularnewline
				\hline 
				\hline 
				$\tan\beta$ & $[0,50]$\tabularnewline
				\hline 
				$\cos(\alpha-\beta)$ & $[-1,1]$\tabularnewline
				\hline 
				$\chi_{bb}$ & $[-1,1]$\tabularnewline
				\hline 
				$\chi_{tt}$ & $[-100,100]$\tabularnewline
				\hline 
				$\chi_{\tau\tau}$ & $[-1,1]$\tabularnewline
				\hline 
				$M_{H^{\pm}}$ (GeV) & $[114,140]$\tabularnewline
				\hline 
				$M_{H^{\pm}}$ (GeV) & $[m_{t}+m_{b},1000]$\tabularnewline
				\hline 
			\end{tabular}
			\par\end{centering}
            \caption{Range of scanned parameters for the signal strengths $\mu_{X}.$}\label{Scan_muX}
	\end{table}
	\begin{itemize}
		\item Decays $B_{s,d}^0\to \mu^+\mu^-$
	\end{itemize}
	The allowed points by experimental measurements on $\mathcal{BR}(B_{s,d}^0\to \mu^+\mu^-)$ are shown in Fig.~\ref{INDBs}. In Table~\ref{ScanB2mumu}, the displayed range of scanned parameters is
	
	\begin{table}[!htb]

		\begin{centering}
			\begin{tabular}{cc}
				\hline 
				Parameter & Range\tabularnewline
				\hline 
				\hline 
				$\tan\beta$ & $[0,50]$\tabularnewline
				\hline 
				$\cos(\alpha-\beta)$ & $[-1,1]$\tabularnewline
				\hline 
				$\chi_{\mu\mu}$ & $[-1,1]$\tabularnewline
				\hline 
				$\chi_{bd}$ & $[-1,1]$\tabularnewline
				\hline 
				$\chi_{db}$ & $[-1,1]$\tabularnewline
				\hline 
				$M_{A}$ (GeV) & $[100,1000]$\tabularnewline
				\hline 
				$M_{H}$ (GeV) & $[300,1000]$\tabularnewline
				\hline 
			\end{tabular}$\qquad$%
			\begin{tabular}{cc}
				\hline 
				Parameter & Range\tabularnewline
				\hline 
				\hline 
				$\tan\beta$ & $[0,50]$\tabularnewline
				\hline 
				$\cos(\alpha-\beta)$ & $[-1,1]$\tabularnewline
				\hline 
				$\chi_{\mu\mu}$ & $[-1,1]$\tabularnewline
				\hline 
				$\chi_{bs}$ & $[-1,1]$\tabularnewline
				\hline 
				$\chi_{sb}$ & $[-1,1]$\tabularnewline
				\hline 
				$M_{A}$ (GeV) & $[100,1000]$\tabularnewline
				\hline 
				$M_{H}$ (GeV) & $[300,1000]$\tabularnewline
				\hline 
			\end{tabular}
			\par\end{centering}
            \caption{Range of scanned parameters. On the left: $B_{d}^{0}\to\mu^{+}\mu^{-}$,
			on the right: $B_{s}^{0}\to\mu^{+}\mu^{-}$ }\label{ScanB2mumu}
	\end{table}
	
\begin{figure}
    \centering

    % First row: two subfigures
    \begin{subfigure}[b]{0.4\textwidth}
        \centering
        \includegraphics[width=\textwidth]{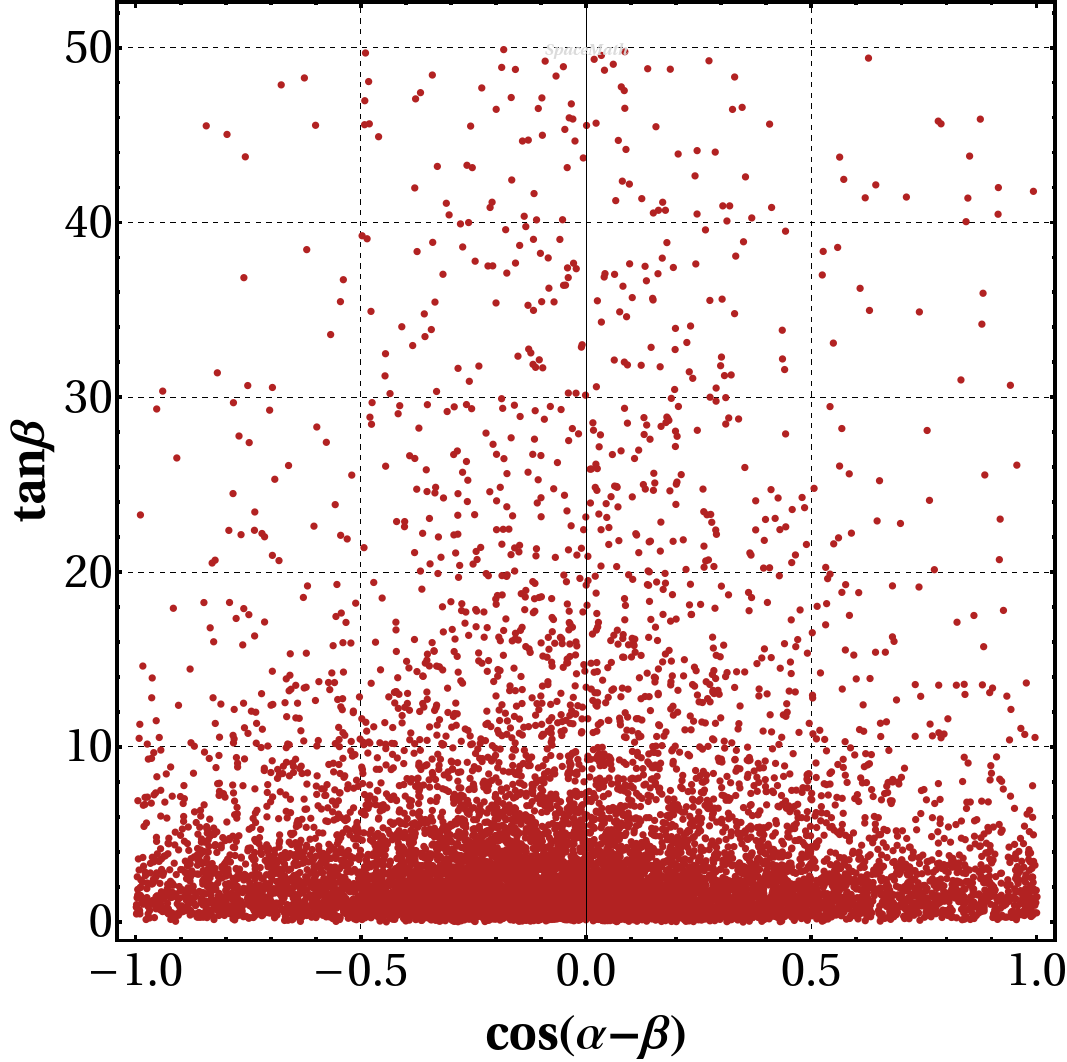}
        \caption{$\mu\to e\gamma$}
        \label{fig:mu-egamma}
    \end{subfigure}
    \hfill
    \begin{subfigure}[b]{0.4\textwidth}
        \centering
        \includegraphics[width=\textwidth]{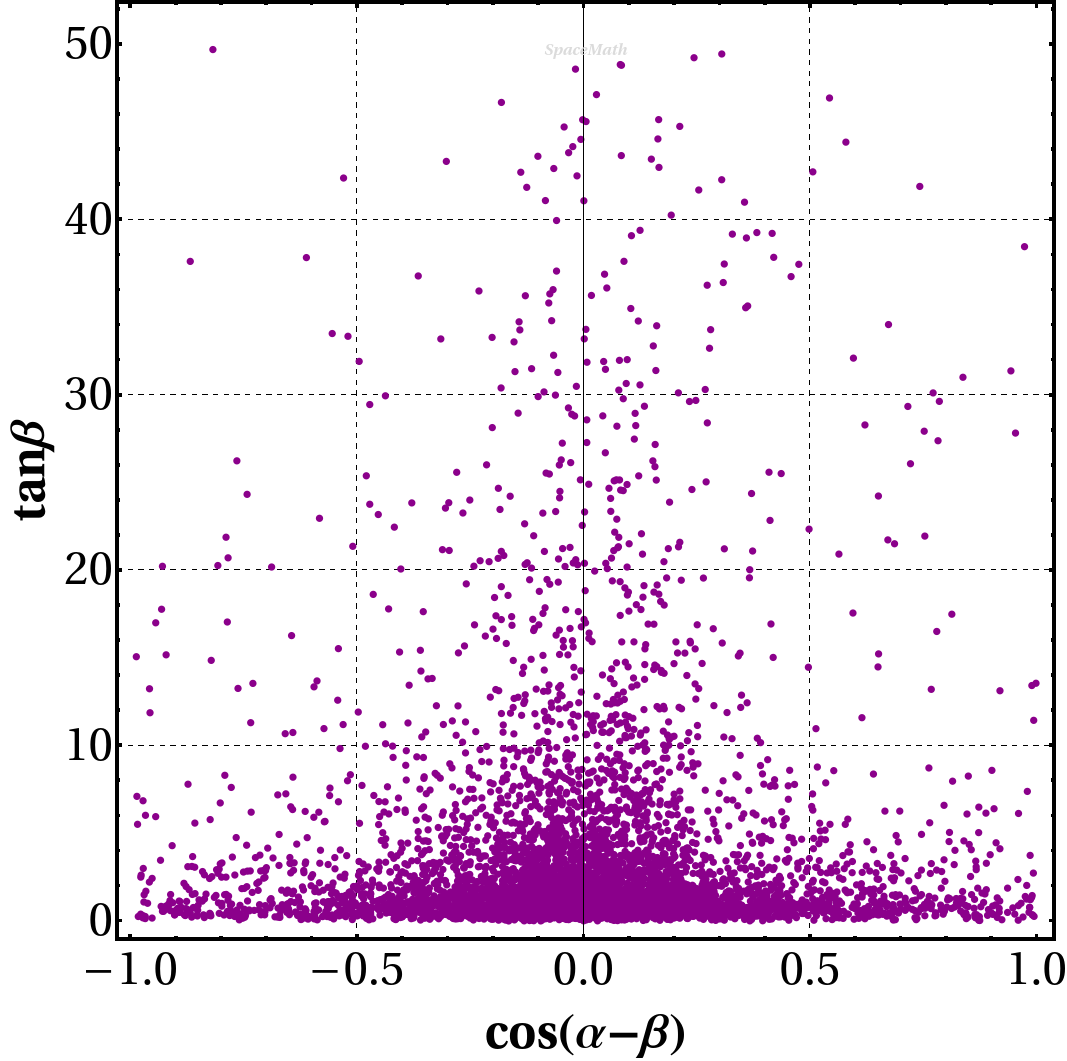}
        \caption{$\tau\to \mu\gamma$}
        \label{fig:tau-mugamma}
    \end{subfigure}

    \vskip\baselineskip % vertical space between rows

    % Second row: centered subfigure
    \begin{subfigure}[b]{0.4\textwidth}
        \centering
        \includegraphics[width=\textwidth]{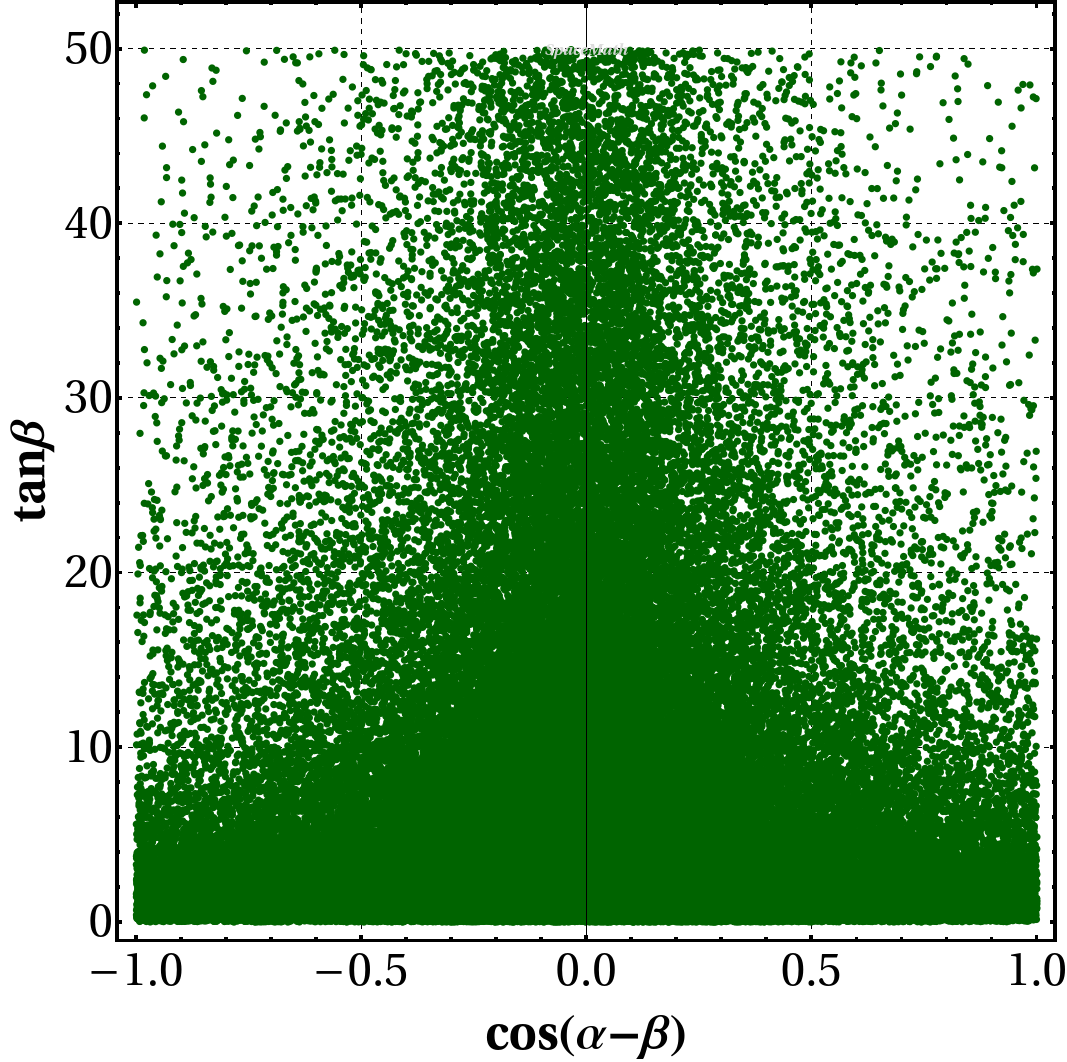}
        \caption{$\tau\to e\gamma$}
        \label{fig:tau-egamma}
    \end{subfigure}

    \caption{Radiative decays $\ell_i \to \ell_j \gamma$: (a) $\mu\to e\gamma$, (b) $\tau\to \mu\gamma$, (c) $\tau\to e\gamma$. For all cases, we generate 100K random points.}
    \label{fig:rad_decays}
\end{figure}

	\begin{itemize}
		\item $\ell_i\to \ell_j\gamma$
	\end{itemize}
	The corresponding allowed points that meet upper limits on $\mathcal{BR}(\ell_i\to \ell_j \gamma)$ are presented in Fig. \ref{fig:rad_decays}. Displayed in Table \ref{Scan_li-lj_Gamma} is the range of scanned parameters.
    
	\begin{table}[!htb]
	
	\begin{centering}
		\begin{tabular}{cc}
			\hline 
			Parameter & Range\tabularnewline
			\hline 
			\hline 
			$\tan\beta$ & $[0,50]$\tabularnewline
			\hline 
			$\cos(\alpha-\beta)$ & $[-1,1]$\tabularnewline
			\hline 
			$\chi_{\mu\mu}$ & $[-1,1]$\tabularnewline
			\hline 
			$\chi_{\mu e}$ & $[-1,1]$\tabularnewline
			\hline 
			$\chi_{tt}$ & $[-100,100]$\tabularnewline
			\hline 
			$M_{A}$ (GeV) & $[100,1000]$\tabularnewline
			\hline 
			$M_{H}$ (GeV) & $[300,1000]$\tabularnewline
			\hline 
			$M_{H^{\pm}}$ (GeV) & $[110,1000]$\tabularnewline
			\hline
		\end{tabular}$\qquad$%
		\begin{tabular}{cc}
			\hline 
			Parameter & Range\tabularnewline
			\hline 
			\hline 
			$\tan\beta$ & $[0,50]$\tabularnewline
			\hline 
			$\cos(\alpha-\beta)$ & $[-1,1]$\tabularnewline
			\hline 
			$\chi_{\tau\mu}$ & $[-100,100]$\tabularnewline
			\hline 
			$\chi_{\tau\tau}$ & $[-1,1]$\tabularnewline
			\hline 
			$\chi_{tt}$ & $[-100,100]$\tabularnewline
			\hline 
			$M_{A}$ (GeV) & $[100,1000]$\tabularnewline
			\hline 
			$M_{H}$ (GeV) & $[300,1000]$\tabularnewline
			\hline
			$M_{H^{\pm}}$ (GeV) & $[110,1000]$\tabularnewline
			\hline 
		\end{tabular}
	\par\end{centering}
    \vspace{1em}

	\begin{centering} 
		\begin{tabular}{cc}
			\hline 
			Parameter & Range\tabularnewline
			\hline 
			\hline 
			$\tan\beta$ & $[0,50]$\tabularnewline
			\hline 
			$\cos(\alpha-\beta)$ & $[-1,1]$\tabularnewline
			\hline 
			$\chi_{\tau e}$ & $[-1,1]$\tabularnewline
			\hline 
			$\chi_{\tau\tau}$ & $[-1,1]$\tabularnewline
			\hline 
			$\chi_{tt}$ & $[-100,100]$\tabularnewline
			\hline 
			$M_{A}$ (GeV) & $[100,1000]$\tabularnewline
			\hline 
			$M_{H}$ (GeV) & $[300,1000]$\tabularnewline
			\hline
			$M_{H^{\pm}}$ (GeV) & $[110,1000]$\tabularnewline
			\hline 
		\end{tabular}
	\par\end{centering}
    \caption{Range of scanned parameters. On the left: $\mu \to e \gamma$, on the right: $\tau \to \mu \gamma$, and below: $\tau \to e \gamma$}\label{Scan_li-lj_Gamma}
\end{table}

\begin{figure}[!h]
    \centering
    \begin{subfigure}{0.45\textwidth}
        \centering
        \includegraphics[scale=0.2]{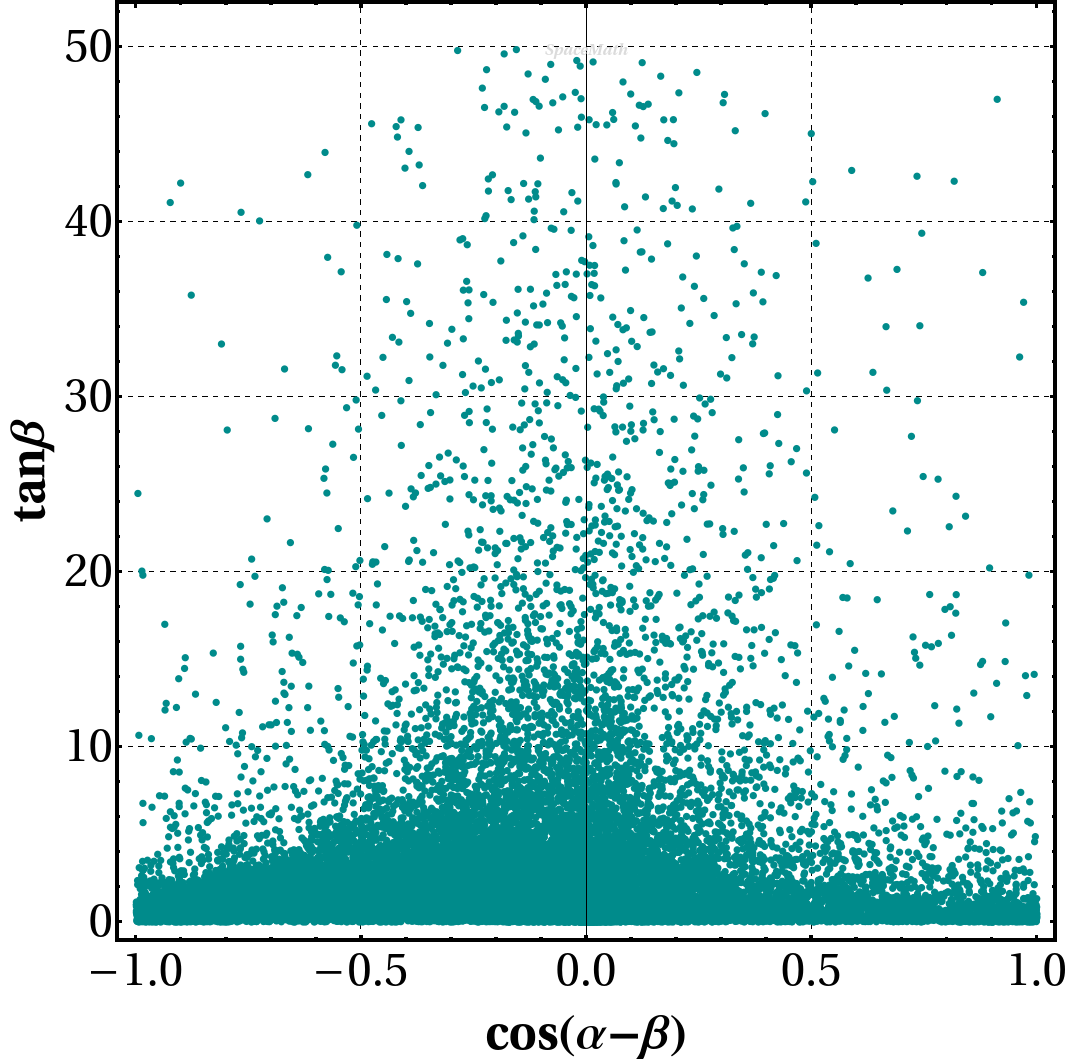}
        \caption{$\mu \to 3e$}
        \label{fig:mu3e}
    \end{subfigure}%
    \begin{subfigure}{0.45\textwidth}
        \centering
        \includegraphics[scale=0.2]{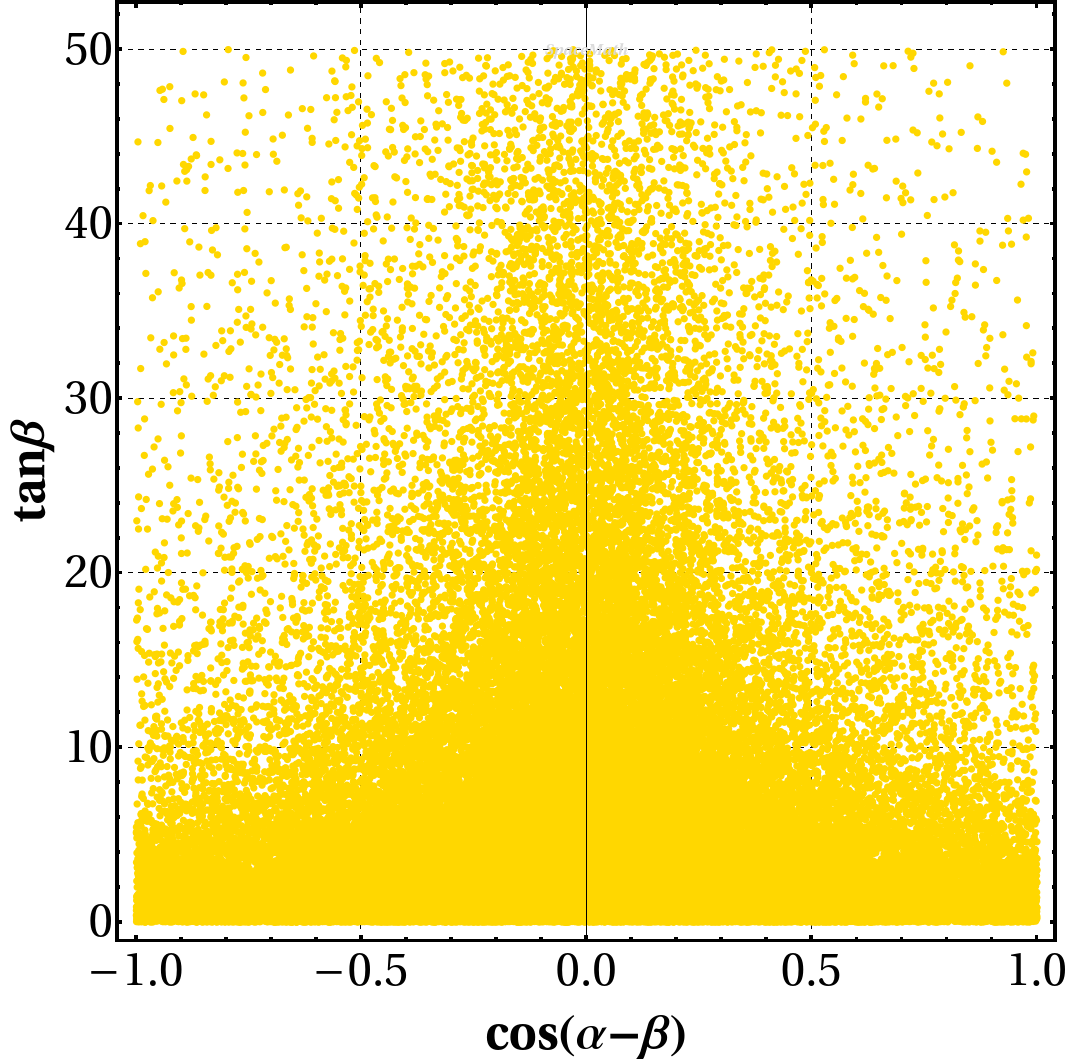}
        \caption{$\tau \to 3 \mu$}
        \label{fig:tau3mu}
    \end{subfigure}
    \caption{Decays $\ell_i\to \ell_j \ell_k \bar{\ell}_k$: (a) $\mu \to 3e$, (b) $\tau \to 3 \mu$. In all cases we generate $100$K random points.}
    \label{liljlklkDecays}
\end{figure}

	\begin{itemize}
		\item Decays $\ell_i\to \ell_j \ell_k \bar{\ell}_k$
	\end{itemize}
	We present in Fig. \ref{liljlklkDecays} the points that comply with the constraints on $\mathcal{BR}(\ell_i\to \ell_j \ell_k \bar{\ell}_k)$, as far as the decays $\ell_i\to \ell_j \ell_k \bar{\ell}_k$ are concerned. The intervals of the scan on the parameters are presented in Table \ref{Scanliljlklk} . 

\begin{figure}[H]\label{h-tau_mu}
		\centering
		\includegraphics[width=6cm]{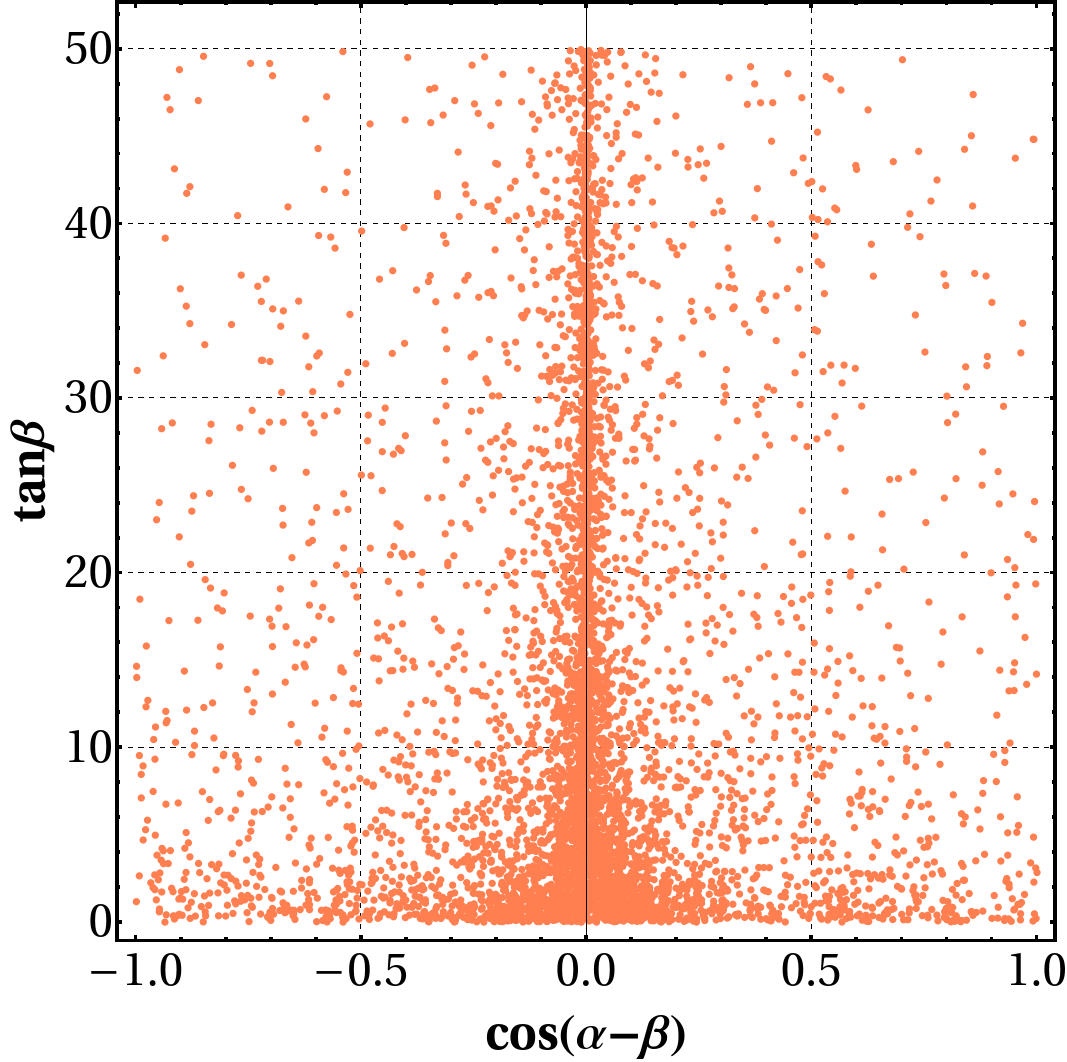}
		\caption{Allowed values by the upper limit on $\mathcal{BR}(h\to\tau\mu)$. We generate $50$K random points.}\label{INDhtaumu}
	\end{figure}

	\begin{table}[!htb]

		\begin{centering}
			\begin{tabular}{cc}
				\hline 
				Parameter & Range\tabularnewline
				\hline 
				\hline 
				$\tan\beta$ & $[0,50]$\tabularnewline
				\hline 
				$\cos(\alpha-\beta)$ & $[-1,1]$\tabularnewline
				\hline 
				$\chi_{ee}$ & $[-1,1]$\tabularnewline
				\hline 
				$\chi_{\mu e}$ & $[-1,1]$\tabularnewline
				\hline 
				$M_{A}$ (GeV) & $[100,1000]$\tabularnewline
				\hline 
				$M_{H}$ (GeV) & $[300,1000]$\tabularnewline
				\hline 
			\end{tabular}$\qquad$%
			\begin{tabular}{cc}
				\hline 
				Parameter & Range\tabularnewline
				\hline 
				\hline 
				$\tan\beta$ & $[0,50]$\tabularnewline
				\hline 
				$\cos(\alpha-\beta)$ & $[-1,1]$\tabularnewline
				\hline 
				$\chi_{\mu\mu}$ & $[-1,1]$\tabularnewline
				\hline
				$\chi_{\tau\mu}$ & $[-100,100]$\tabularnewline
				\hline 
				$M_{A}$ (GeV) & $[100,1000]$\tabularnewline
				\hline 
				$M_{H}$ (GeV) & $[300,1000]$\tabularnewline
				\hline
			\end{tabular}
			\par\end{centering}
            		\caption{Range of scanned parameters. On the left: $\mu \to 3e$, on the right: $\tau \to 3\mu$.}\label{Scanliljlklk}
	\end{table}
	\begin{itemize}
		\item Decays $h\to \ell_i \ell_j$
	\end{itemize}
	Only the allowed region coming from $h\to\tau\mu$ is presented because very weak bounds to $\tan\beta$ and $\cos(\alpha-\beta)$ are imposed by the process $h\to e\mu$. The scatter plot in the $\cos(\alpha-\beta)-\tan\beta$ plane is shown in Fig. \ref{INDhtaumu} shows the points allowed by the upper limit on $\mathcal{BR}(h\to\tau\mu)$.
	The range of scanned parameters is displayed in Table \ref{Scan_h-tau_mu}.
	\begin{table}[!htb]
	
		\begin{centering}
			\begin{tabular}{cc}
				\hline 
				Parameter & Range\tabularnewline
				\hline 
				\hline 
				$\tan\beta$ & $[0,50]$\tabularnewline
				\hline 
				$\cos(\alpha-\beta)$ & $[-1,1]$\tabularnewline
				\hline 
				$\chi_{\tau\mu}$ & $[-10,10]$\tabularnewline
				\hline 
			\end{tabular}
			\par\end{centering}
            	\caption{Range of scanned parameters associated to the decay $h\to\tau\mu$. }\label{Scan_h-tau_mu}
	\end{table}

\chapter{2HDM-III Variables and Parameters}
\label{VI}
In the Table~\label{VarBDT}, the variables used for training and testing the signal and background processes of the 2HDM-III framework are presented.	

\begin{table}[!htb]
    \centering
    \resizebox{\textwidth}{!}{%
    \begin{tabular}{|c|c|c|}
        \hline 
        \textbf{Rank} & \textbf{Variable} & \textbf{Description} \\
        \hline \hline
        1 & $P_{T}[b]$ & $b$-jet transverse momentum \\
        \hline
        2 & $\slashed{E}_T$ & Missing Energy Transverse \\
        \hline
        3 & $P_{T}[j_{1}]$ & Jet with the largest transverse momentum ($c$-jet) \\
        \hline
        4 & $M_{T}[\mu]$ & Transverse mass between the $\slashed{E}_T$ and the muon. \\
        \hline
        5 & $M_{inv}[b,c]$ & Invariant mass of the \textbf{$b$-jet} and \textbf{$c$-jet}. \\
        \hline
        6 & $\sum P_{T}$ & Sum of the moduli of the transverse momentum of the $b$-jet, leading jet and the muon \\
        \hline
        7 & $\eta_{b}$ & Pseudorapidity of the \textbf{$b$-jet} \\
        \hline
        8 & $P_{T}^{all}[jet]$ & Scalar sum of the transverse momentum of all jets. \\
        \hline
        9 & $\Delta\eta[j_{1},j_{2}]$ & Absolute value of the pseudorapidity separation between two jets \\
        \hline
        10 & $\eta[j_{1}]$ & Pseudorapidity of the leading jet \\
        \hline
        11 & $\slashed{E}_{tot}$ & Total transverse energy in the detector \\
        \hline
        12 & $\eta[j_{1}]\times\eta[j_{2}]$ & Product of the pseudorapidities of the two jets \\
        \hline
    \end{tabular}%
    }
    \caption{Variables used to train and test the signal and background events list. 
      The transverse mass is defined as  
      $M_{T}[\ell]=\sqrt{2P_{T}^{\ell}E_{T}(1-\cos\phi_{\ell \slashed{E}_T})}$.}
    \label{VarBDT}
\end{table}

\section*{Oblique parameters}\label{ObParam}
The $S,\,T,\,U$ oblique parameters in the 2HDMs framework are given by
\begin{align}
	\begin{array}{lll}
		S & = & \frac{1}{\pi m_{Z}^{2}}\left\{ \sin^{2}\left(\alpha-\beta\right)\left[\mathcal{B}_{22}\left(m_{Z}^{2};m_{Z}^{2},m_{h}^{2}\right)-M_{Z}^{2}\mathcal{B}_{0}\left(m_{Z}^{2};m_{Z}^{2},m_{h}^{2}\right)+\mathcal{B}_{22}\left(m_{Z}^{2};M_{H}^{2},M_{A}^{2}\right)\right]\right.\\
		& + & \cos^{2}\left(\alpha-\beta\right)\left[\mathcal{B}_{22}\left(m_{Z}^{2};m_{Z}^{2},M_{H}^{2}\right)-m_{Z}^{2}\mathcal{B}_{0}\left(m_{Z}^{2};m_{Z}^{2},M_{H}^{2}\right)+\mathcal{B}_{22}\left(m_{Z}^{2};m_{h}^{2},M_{A}^{2}\right)\right]\\
		& - & \left.\mathcal{B}_{22}\left(m_{Z}^{2};M_{H^{\pm}}^{2},M_{H^{\pm}}^{2}\right)-\mathcal{B}_{22}\left(m_{Z}^{2};m_{Z}^{2},m_{h}^{2}\right)+m_{Z}^{2}\mathcal{B}_{0}\left(m_{Z}^{2};m_{Z}^{2},m_{h}^{2}\right)\right\} ,
	\end{array}
\end{align}

\begin{align}
	\begin{array}{lll}
		T & = & \frac{1}{16\pi m_{W}^{2}s_{W}^{2}}\left\{ \sin^{2}\left(\alpha-\beta\right)\left[\mathcal{F}\left(M_{H^{\pm}}^{2},M_{H}^{2}\right)-\mathcal{F}\left(M_{H}^{2},M_{A}^{2}\right)+3\mathcal{F}\left(m_{Z}^{2},m_{h}^{2}\right)-3\mathcal{F}\left(m_{W}^{2},m_{h}^{2}\right)\right]\right.\\
		& + & \cos^{2}\left(\alpha-\beta\right)\left[\mathcal{F}\left(M_{H^{\pm}}^{2},m_{h}^{2}\right)-\mathcal{F}\left(m_{h}^{2},M_{A}^{2}\right)+3\mathcal{F}\left(m_{Z}^{2},M_{H}^{2}\right)-3\mathcal{F}\left(m_{W}^{2},M_{H}^{2}\right)\right]\\
		& + & \left.\mathcal{F}\left(M_{H^{\pm}}^{2},M_{A}^{2}\right)-3\mathcal{F}\left(m_{Z}^{2},m_{h}^{2}\right)+3\mathcal{F}\left(m_{W}^{2},m_{h}^{2}\right)\right\} ,
	\end{array}
\end{align}

\begin{align}
	\begin{array}{lll}
		U & = & \mathcal{H}\left(m_{W}^{2}\right)-\mathcal{H}\left(m_{Z}^{2}\right)\\
		& + & \frac{1}{\pi m_{W}^{2}}\left\{ \cos^{2}\left(\alpha-\beta\right)\mathcal{B}_{22}\left(m_{W}^{2};M_{H^{\pm}}^{2},m_{h}^{2}\right)+\sin^{2}\left(\alpha-\beta\right)\mathcal{B}_{22}\left(m_{W}^{2};M_{H^{\pm}}^{2},M_{H}^{2}\right)\right.\\
		& + & \left.\mathcal{B}_{22}\left(m_{W}^{2};M_{H^{\pm}}^{2},M_{A}^{2}\right)-2\mathcal{B}_{22}\left(m_{W}^{2};M_{H^{\pm}}^{2},M_{H^{\pm}}^{2}\right)\right\} \\
		& - & \frac{1}{\pi m_{Z}^{2}}\left\{ \cos^{2}\left(\alpha-\beta\right)\mathcal{B}_{22}\left(m_{Z}^{2};m_{h}^{2},M_{A}^{2}\right)+\sin^{2}\left(\alpha-\beta\right)\mathcal{B}_{22}\left(m_{Z}^{2};M_{H}^{2},M_{A}^{2}\right)\right.\\
		& - & \left.\mathcal{B}_{22}\left(m_{Z}^{2};M_{H^{\pm}}^{2},M_{H}^{2}\right)\right\} ,
	\end{array}
\end{align}	

where
\begin{align}
	\begin{array}{lll}
		\mathcal{H}\left(m_{V}^{2}\right) & \equiv & \frac{1}{\pi m_{V}^{2}}\left\{ \sin^{2}\left(\alpha-\beta\right)\left[\mathcal{B}_{22}\left(m_{V}^{2};m_{V}^{2},m_{h}^{2}\right)-m_{V}^{2}\mathcal{B}_{0}\left(m_{V}^{2};m_{V}^{2},m_{h}^{2}\right)\right]\right.\\
		& + & \cos^{2}\left(\alpha-\beta\right)\left[\mathcal{B}_{22}\left(m_{V}^{2};m_{V}^{2},M_{H}^{2}\right)-m_{V}^{2}\mathcal{B}_{0}\left(m_{V}^{2};m_{V}^{2},M_{H}^{2}\right)\right]\\
		& - & \left.\mathcal{B}_{22}\left(m_{V}^{2};m_{V}^{2},m_{h}^{2}\right)+m_{V}^{2}\mathcal{B}_{0}\left(m_{V}^{2};m_{V}^{2},m_{h}^{2}\right)\right\} ,
	\end{array}
\end{align}	

\begin{equation}\label{key}
	\mathcal{F}\left(m_{1}^{2},m_{2}^{2}\right)=\frac{1}{2}\left(m_{1}^{2}+m_{2}^{2}\right)-\frac{m_{1}^{2}m_{2}^{2}}{m_{1}^{2}-m_{2}^{2}}\log\left(\frac{m_{1}^{2}}{m_{2}^{2}}\right),
\end{equation}

\begin{equation}\label{key}
	\mathcal{B}_{22}\left(q^{2};m_{1}^{2},m_{2}^{2}\right)\equiv B_{22}\left(q^{2};m_{1}^{2},m_{2}^{2}\right)-B_{22}\left(0;m_{1}^{2},m_{2}^{2}\right),
\end{equation}

\begin{equation}\label{key}
	\mathcal{B}_{0}\left(q^{2};m_{1}^{2},m_{2}^{2}\right)\equiv B_{0}\left(q^{2};m_{1}^{2},m_{2}^{2}\right)-B_{0}\left(0;m_{1}^{2},m_{2}^{2}\right),
\end{equation}

\begin{align}
	\begin{array}{lll}
		B_{22}\left(q^{2};m_{1}^{2},m_{2}^{2}\right) & = & \frac{1}{6}\left[A_{0}\left(m_{2}^{2}\right)+2m_{1}^{2}B_{0}\left(q^{2};m_{1}^{2},m_{2}^{2}\right)+\left(m_{1}^{2}-m_{2}^{2}+q^{2}\right)B_{1}\left(q^{2};m_{1}^{2},m_{2}^{2}\right)\right.\\
		& - & \left.\frac{q^{2}}{3}+m_{1}^{2}+m_{2}^{2}\right],
	\end{array}
\end{align}
in which $A_0,\,B_0,\,B_1$ are scalar Passarino-Veltman functions \cite{PASSARINO1979151}.

\bibliographystyle{JHEP}
\bibliography{bibliography}

\end{document}